\documentclass[aps,rmp,reprint,amsmath,amssymb,superscriptaddress,longbibliography]{revtex4-2} 

\usepackage{bm,upgreek,appendix,soul,xcolor,graphicx, physics, chemformula, hyperref, tikz, pgfplots, changepage, cancel}
\usepackage[scaled]{helvet} 
\usepackage{pdfpages}

\hypersetup{
    colorlinks=true,
    linkcolor=black,
    urlcolor=blue,
    filecolor=blue,
    citecolor=blue!60!black,
    }

\let\hide\iffalse

\usepackage[T1]{fontenc}

\newcommand{\beq}{\begin{equation}}
\newcommand{\eeq}{\end{equation}}
\newcommand{\beqn}{\begin{eqnarray}}
\newcommand{\eeqn}{\end{eqnarray}}

\usepackage[capitalise]{cleveref}

\colorlet{MAGENTA}{magenta}
\colorlet{ORANGE}{orange}
\colorlet{BLUE}{blue}
\colorlet{BLACK}{black}

\newcommand{\rev}[1]{\textcolor{black}{#1}}   

\def\kb{{k_{\rm B}}}

\def\bk{{\bf k}}
\def\br{{\bf r}}
\def\bx{{\bf x}}
\def\bq{{\bf q}}
\def\bQ{{\bf Q}}
\def\bn{{\bf n}}

\def\bR{{\bf R}}

\def\bv{{\bf v}}

\def\a{\alpha}
\def\b{\beta}
\def\w{\omega}
\def\k{\kappa}
\def\e{\epsilon}
\def\ve{\varepsilon}
\def\btau{\boldsymbol{\tau}}

\def\D{\partial}
\def\d{\delta}
\def\<{\langle}
\def\>{\rangle}

\def\had{\hat{a}^\dagger}
\def\ha{\hat{a}^{\vphantom{\dagger}}} 

\def\hb{\hat{b}^{\vphantom{\dagger}}} 

\def\hcd{\hat{c}^\dagger}
\def\hc{\hat{c}^{\vphantom{\dagger}}} 

\def\hx{\hat{x}}
\def\hpp{\hat{p}}

\def\hS{\hat{S}^{\vphantom{\dagger}}} 

\def\hbp{\hat{\bf p}}
\def\bP{{\bf P}}

\def\hn{{\hat{n}}}
\def\hT{{\hat{T}}}
\def\hU{{\hat{U}}}

\def\hH{{\hat{H}}}
\def\hp{{\hat{\psi}}}
\def\hp{{\hat{\psi}^\dagger}}

\def\hp{{\hat{\psi}}}
\def\hpd{{\hat{\psi}^\dagger}}
\def\hne{{\hat{n}_{\rm e}}}
\def\hnn{{\hat{n}_{\rm n}}}

\def\dtau{\Delta{\tau}}
\def\dhtau{\Delta{\hat{\tau}}}

\def\dhbtau{\Delta{\hat{\boldsymbol{\tau}}}}

\newcommand{\fourr}{{ \br_{\rm e}, \br_{\rm h}; \br_{\rm e}', \br_{\rm h}'}}

\newcommand{\ifc}{C_{\kappa \alpha p, \kappa' \alpha' p'} }
\newcommand{\iifc}{C^{-1}_{\kappa \alpha p, \kappa' \alpha' p'} }

\def\cG{{\mathcal{G}}}

\newcommand{\fro}{Fr\"ohlich}

\usepackage[capitalise]{cleveref}

\def\utoden{Oden Institute for Computational Engineering and Sciences, 
         The University of Texas at Austin, Austin, Texas 78712, USA}
\def\utphysics{Department of Physics, The University of Texas at Austin, 
         Austin, Texas 78712, USA}
\def\ubc{Department of Physics, University of the Basque 
         Country UPV/EHU, 48940 Leioa, Basque Country, Spain}

\makeatletter
\AtBeginDocument{\let\LS@rot\@undefined}
\makeatother

\begin{document}

\title{Polarons from first principles}

\author{Zhenbang Dai}
  \thanks{These authors contributed equally to this work.}
  \affiliation{\utoden}
  \affiliation{\utphysics}

\author{Jon Lafuente-Bartolome}
  \thanks{These authors contributed equally to this work.}
  \affiliation{\ubc}

\author{Feliciano Giustino}
  \affiliation{\utoden}
  \affiliation{\utphysics}

\date{\today}

\begin{abstract}
This article reviews recent theoretical developments in the \textit{ab initio} study of polarons in materials. The polaron is an emergent quasiparticle that arises from the interaction between electrons and phonons in solids, and consists of an electron or a hole accompanied by a distortion of the crystal lattice. 
Recent advances in experiments, theory, and computation have made it possible to investigate these quasiparticles with unprecedented detail, reigniting the interest in this classic problem of condensed matter physics. Recent theoretical and computational advances include \textit{ab initio} calculations of polaron spectral functions, wavefunctions, lattice distortions, and transport and optical properties. These developments provide new insight into polaron physics, but they have evolved somewhat independently from the earlier effective Hamiltonian approaches that laid the foundation of the field. This article aims to bridge these complementary perspectives by placing them within a single unified conceptual framework.
To this end, we start by reviewing effective Hamiltonians of historical significance in polaron theory, \textit{ab initio} techniques based on density functional theory, and many-body first-principles approaches to polarons. After this survey, we outline a general field-theoretic framework that bridges between these diverse approaches to polaron physics. For completeness, we also review recent progress in the study of exciton polarons and self-trapped excitons and their relations to polarons. Beyond the methodology, we discuss recent applications to several classes of materials that attracted attention in the context of polaron physics.   
\end{abstract}


\maketitle

\tableofcontents{}

\section{Introduction}\label{Sec:Introduction}

\begin{figure}[t]
\label{Fig:polaron_illustration}   \includegraphics[width=0.99\linewidth]{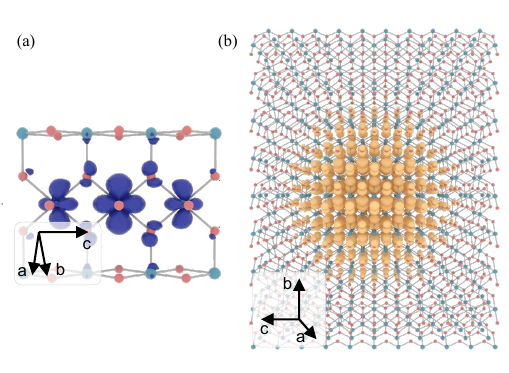}
        \caption{
        Illustration of small polarons (a) and large polarons (b) in rutile \ch{TiO2}, as obtained from first-principles calculations. The isosurfaces represent the square modulus of the polaron wavefunction. Ti and O are in blue and red, respectively. }
\end{figure}

A charge carrier propagating through a crystal may induce distortions in the surrounding lattice through the electron-phonon interaction. The emergent quasiparticle formed by this carrier and its induced lattice distortion is referred to as a polaron~\cite{Alexandrov_Mott_1996, Alexandrov_2008, Emin_2013, Devreese_2020}. In materials with weak electron-phonon couplings, polarons behave like conventional Bloch waves, only with heavier effective masses; conversely, in systems with strong interactions, polarons take the form of localized wavepackets and profoundly alter transport, electrical, and optical properties of the host material, see Fig.~\ref{Fig:polaron_illustration}. 

The polaron concept is almost as old as condensed matter physics itself, with the first mention appearing in a short, 500-words article by \textcite{Landau_1933}. Since then, the study of polarons has led to foundational developments in condensed matter theory, including the seminal works by \textcite{Pekar_1946}, \textcite{Frohlich_Zienau_1950}, \textcite{Alexandrov_Mott_1996}, \textcite{Feynman_1955}, \textcite{Holstein_1959a}, \textcite{Su_Heeger_1979}, \textcite{Lee_Pines_1953}, and many others. 
On the experimental front, polarons inspired the search for superconducting oxides and eventually led to the discovery of high-temperature superconductivity by \textcite{Bednorz_Muller_1986,Bednorz_Muller_1988}; they contributed to the understanding of colossal magnetoresistive manganites~\cite{Millis_Shraiman_1995,Roder_Bishop_1996,Teresa_Arnold_1997}; and they have been connected with the exceptional optoelectronic properties of halide perovskites~\cite{Miyata_Zhu_2017}.
Recently, high-resolution angle-resolved photoelectron spectroscopy (ARPES) experiments have identified the fingerprints of polarons in a broad array of transition metal oxides \cite{Moser_Grioni_2013, Chen_Asensio_2015, Wang_Baumberger_2016, Cancellieri_Strocov_2016, Verdi_Giustino_2017, Riley_King_2018} and two-dimensional materials \cite{Chen_Asensio_2018, Kang_Kim_2018}; ultrafast time-resolved X-ray diffraction experiments elucidated the formation and dynamics of polarons in halide perovskites \cite{Guzelturk_Lindenberg_2021}; new developments in ultrafast electron diffraction measurements identified the structural distortions that accompany polarons in thermoelectric materials \cite{deCotret_Siwick_2022}; and single-electron scanning tunneling miscroscopy and spectroscopy (STM/STS) experiments demonstrated the possibility of writing and erasing individual polarons in two-dimensional materials \cite{Liu_Wu_2023}.

On the application front, polarons play important roles in artificial photosynthesis, organic electronics, neuromorphic computing, lighting, and display technology. For example, the oxygen evolution reaction in solar-driven water splitting is mediated by surface polarons \cite{Reticcioli_Franchini_2019,Reticcioli_Franchini_2017}; the current modulation in electrochemical synapses for physical neural networks is achieved via a polaronic insulator-to-metal transition \cite{Onen_DelAlamo_2022,Yao_Yildiz_2020}; and the record quantum efficiency of light-emitting devices (LEDs) based on halide perovskites originates from self-trapped excitons, which are the excitonic counterpart of self-trapped polarons \cite{Luo_Tang_2018}. Polarons are also ubiquitous in quantum materials research; for example, spin polarons were recently observed in MoTe$_2$/WSe$_2$ moir\'e bilayers \cite{Tao_Mak_2024}; polaronic moir\'e crystals were reported in WSe$_2$/WS$_2$ superlattices \cite{Arsenault_Zhu_2024}; and magnetic polarons \cite{Huang_Huang_2017} and trimerons \cite{Baldini_Gedik_2020} were shown to be at the origin of the Vervey transition in magnetite.

On the front of theory and computation, several advances in \textit{ab initio} calculations have enabled detailed comparison with experiments. For example, the development of the cumulant expansion method led to first-principles calculations of polaron spectral functions including the incoherent phonon satellites observed in ARPES experiments \cite{Verdi_Giustino_2017,Riley_King_2018,Nery_Gonze_2018,Garcia-Goiricelaya_Eiguren_2019,Antonius_Louie_2020}, and to quantifying their impact on transport properties \cite{Zhou_Bernardi_2019}; a reciprocal-space formulation of the polaron problem enabled the study of large and small polarons on the same footing, including visualizations of wavefunctions and lattice distortions \cite{Sio_Giustino_2019a,Sio_Giustino_2019b}; a Wannier-function description of polarons made efficient calculations of small polarons possible \cite{Lee_Bernardi_2021}. Developments in the theory and methods for electron-phonon interactions have increased the accuracy and reliability of these calculations in three- and two-dimensional materials, for example by including quadrupole corrections \cite{Brunin_Hautier_2020, Brunin_Hautier_2020b, Jhalani_Bernardi_2020, Park_Bernardi_2020, Ponce_Stengel_2023}, Berry-phase effects \cite{Ponce_Stengel_2023}, quasiparticle GW calculations of electron-phonon couplings \cite{Li_Louie_2019,Li_Louie_2021}, and efficient data-compression schemes \cite{Luo_Bernardi_2024}. New formal developments include an \textit{ab initio} many-body theory of polarons at all couplings \cite{Lafuente_Giustino_2022a,Lafuente_Giustino_2022b}, the \textit{ab initio} theories of exciton-phonon couplings \cite{Antonius_Louie_2022} and exciton polarons \cite{Dai_Giustino_2024a,Dai_Giustino_2024b}, and the demonstration of \textit{ab initio} Diagrammatic Monte Carlo (DMC) calculations for polarons \cite{Luo_Bernardi_2025}.
At the same time, polaron calculations based on density-functional theory (DFT) significantly improved with the development of effective techniques for mitigating the sensitivity to the exchange and correlation functional, for example via generalized Koopman's correction schemes \cite{Kokott_Scheffler_2018, Falletta_Pasquarello_2022a,Falletta_Pasquarello_2022b}; and with the adoption of machine learning techniques such as neural network potentials \cite{Birschitzky_Franchini_2025}.
\rev{Some of these recent developments in \textit{ab initio} methods for polarons have been implemented in several computational codes, such as \textsc{EPW}~\cite{Lee_Giustino_2023}, \textsc{Perturbo}~\cite{Zhou_Bernardi_2021b}, \textsc{Abinit}~\cite{Verstraete_Zwanziger_2025}, \textsc{Exciting}~\cite{Gulans_Draxl_2014}, \textsc{QuantumEspresso}~\cite{Giannozzi_Wentzcovitch_2009}, and \textsc{VASP}~\cite{Kresse_Furthmuller_1996}}.

Over the years, approaches based on effective Hamiltonian \cite{Hahn_Franchini_2018,Prokofev_Svistunov_1998,Mishchenko_Svistunov_2000,Berciu_2006} and those based on density functional theory (DFT) \cite{Pham_Deskins_2020,Reticcioli_Franchini_2022} have each provided valuable insights into distinct aspects of polaron physics. However, these methods typically rely on different approximations and emphasize different physical pictures, making their interrelations not always transparent. More recently, there has been a gradual convergence between many-body techniques applied to model Hamiltonians and fully \textit{ab initio} many-body theories of polarons \cite{Lafuente_Giustino_2022a,Luo_Bernardi_2025}. One of the goals of this review is to clarify the relationships between these approaches and help bridge the gap by identifying a common conceptual framework.

Comprehensive introductions to polaron physics based on model Hamiltonians can be found in the textbook by \textcite{Emin_2013}, and in the classic references by \textcite{Alexandrov_Mott_1996,Alexandrov_Devreese_2010,Alexandrov_2008} and \textcite{Devreese_2020}. These resources remain foundational and have shaped our understanding of polarons.
A more recent and already very popular review article by \textcite{Franchini_Diebold_2021} provides a comprehensive overview of experimental and theoretical progress. The present article is meant to complement these references by providing a detailed and pedagogical account of \textit{ab initio} methods, their relations to effective Hamiltonian approaches, and the underlying algorithms with their advantages and limitations. 

\rev{To keep the presentation concise, the present review article focuses on the theory of polarons in pristine crystals. The study of polaron-defect interaction is an important topic but would require a thorough discussion~\cite{Birschitzky_Franchini_2024, Reticcioli_Franchini_2017, Setvin_Diebold_2014, Gerosa_Galli_2018, Janotti_Walle_2014, Seo_Galli_2018, Pastor_Walsh_2022}. }

The review is organized as follows. In Sec.~\ref{Sec:History} we build intuition on polaron physics by summarizing models of historical significance, from the Landau-Pekar model to the Holstein model. In Sec.~\ref{Sec:DFTPolaron}, we review DFT-based methods for calculating polarons, and we highlight the current challenges faced by these approaches. Section~\ref{Sec:ManyBody} focuses on many-body \textit{ab initio} approaches to polarons starting from the standard second-quantized electron-phonon Hamiltonian. In Sec.~\ref{Sec:FormalTheory} we outline a general many-body theory of polarons based on the Hedin-Baym equations; this framework clarifies what are the single-particle energies, phonon frequencies, and electron-phonon coupling matrix elements that enter the second-quantized Hamiltonian in \textit{ab initio} many-body  calculations of polarons. Section~\ref{Sec:PolaronEquations} reviews practical approximations to the Hedin-Baym framework, leading to the variational \textit{ab initio} polaron equations, \rev{and in Section~\ref{Sec:Connection} we establish connections among the \textit{ab initio} methods of calculating polarons.} These methods are generalized to the description of exciton polarons and self-trapped excitons in Sec.~\ref{Sec:ExcitonPolaron}. 
In Sec.~\ref{Sec:Applications}, we discuss representative applications of these methods to real materials, and make the connection with experimental observations, and in Sec.~\ref{Sec:transport} we focus on the transport properties of polarons.
Section~\ref{Sec:BrokenSymmetry} briefly addresses the issue of broken translational symmetry in polaron physics. 
In Sec.~\ref{Sec:Conclusions} we provide our conclusions and identify a few open questions. The Supplemental Material contains details of derivations and a more in-depth discussion of the open questions~\cite{SI}.

\section{Effective Hamiltonian approaches to polarons}\label{Sec:History}

Effective Hamiltonian approaches to polarons aim to isolate and analyze specific physical mechanisms by introducing idealized representations of coupled electron-phonon systems. These models can be derived from the general many-body electron-ion Hamiltonian (cf.\ Sec.~\ref{Sec:FormalTheory}) by making controlled approximations about the nature of the electronic and vibrational states involved and their mutual interactions. While these models are not intended for quantitative predictions of real materials, they played a central role in shaping our conceptual understanding of polarons for almost a century. Furthermore, these models continue to serve as testing ground for new electronic structure methods and even for emerging quantum hardware \cite{Macridin_Harnik_2018,Kumar_Barrios_2025}.

The \textcite{Frohlich_1954} and \textcite{Holstein_1959a} Hamiltonians represent the two foundational models of polaron physics, and capture the essential features of large and small polarons, respectively. The former describes a free electron interacting with longitudinal optical phonons in a polar crystal. The latter considers a tight-binding model of an electron on a discrete lattice, interacting locally with a transverse optical phonon; in this model, phonons modulate the onsite electron energy. Most theoretical approaches to polarons can be traced back to these models or their generalizations. For example, the Landau-Pekar model describes the \fro\ polaron at strong electron-phonon coupling \cite{Buimistrov_Pekar_1957a,Buimistrov_Pekar_1957b}. The Lee-Low-Pines transformation constitutes an exact reformulation of the \fro\ Hamiltonian that enables analytical and numerical solutions across different coupling strengths \cite{Lee_Pines_1953}. The path-integral method by \textcite{Feynman_1955} provides a powerful variational approach to investigate the \fro\ polaron at all couplings. 
The Lang-Firsov method \cite{Lang_Firsov_1963} is a canonical transformation of the Holstein Hamiltonian that reduces the interacting electron-phonon system to a system of non-interacting polarons and phonons. The Peierls/Su-Schrieffer-Heeger (PSSH) Hamiltonian shares the same discrete lattice structure of the Holstein model, but has phonons modulating the inter-site hopping energy via bond length fluctuations~\cite{Barisic_Friedel_1970,Su_Heeger_1979,Peierls_1991}. 

Together, these effective Hamiltonians and the methods developed to study them form a broad theoretical framework that continues to inform our understanding of polaron \rev{energetics and transport properties~\cite{Landau_Peka_1948, Lee_Pines_1953, Feynman_1955, Volovik_Edelshtein_1973}}, and to inspire new methods. In the following, we review the Landau-Pekar model (Sec.~\ref{Sec:LandauPekar}), the \fro\ model and the associated Lee-Low-Pines transformation (Sec.~\ref{Sec:FrohlichLLP}), the Feynman approach (Sec.~\ref{Sec:Feynman}), and the Holstein model and the related Lang-Firsov transformation (Sec.~\ref{Sec:HolsteinLF}).

\subsection{Landau-Pekar model}\label{Sec:LandauPekar}

The Landau-Pekar model \cite{Landau_1933, Pekar_1946} is historically significant as being the first model to describe the concept of polaron formation. The model describes an excess electron in a uniform and isotropic dielectric continuum, and can be understood as a semiclassical approximation of the \fro\ model discussed in Sec.~\ref{Sec:FrohlichLLP}. In the Landau-Pekar model, the electron wavefunction is assumed to span many crystal unit cells, so that the atomistic details of the crystal are inconsequential and can be neglected \cite{Devreese_2020}. The interaction of the excess electron with the valence manifold is described via the effective-mass approximation, and the interaction with the ionic lattice is described by means of classical electrostatics. The total energy reads:
\begin{equation} \label{Eq:LP_toten_initial}
    E 
    = 
    \frac{\hbar^2}{2m^*}\!\! \int\!\! d\br |\nabla \psi|^2
    + \frac{1}{2}\! \int\!\! d\br \, \mathbf{E}\cdot\mathbf{D}~,
\end{equation}
where $\hbar$ is the reduced Planck constant, $\psi(\br)$ is the electron wavefunction at the position $\br$, $\mathbf{E}(\br)$ is the self-consistent electric field, and $\mathbf{D}(\br)$ is the electric displacement field.

The displacement field is related to the self-consistent electric field by $\mathbf{D}=\epsilon_0\epsilon^0\mathbf{E}$, where $\epsilon_0$ is the vacuum permittivity, and $\epsilon^0$ is the static dielectric constant. The displacement field can be expressed in terms of the excess electron density via the Gauss law,
$\mathbf{D} = (e/4\pi) \nabla \! \int d\br' \, |\psi(\br')|^2/|\br\!-\!\br'|$, where $e$ is the electron charge, 
so that the electrostatic energy in Eq.~\eqref{Eq:LP_toten_initial} becomes:
\begin{equation} \label{Eq:elec_energy_to_psi}
    \frac{1}{2} \!\int \!d\br \, \mathbf{E}\cdot\mathbf{D}
    =
    \frac{1}{2} \frac{e^2}{4\pi\epsilon_0}\frac{1}{\epsilon^0} \int d\br d\br' \, \frac{|\psi(\br)|^2|\psi(\br')|^2}{|\br-\br'|} ~.
\end{equation}
Since the screening of the excess electron by the valence manifold is already accounted for via the effective mass $m^*$ in Eq.~\eqref{Eq:LP_toten_initial},
this contribution needs to be removed from Eq.~\eqref{Eq:elec_energy_to_psi}. This step amounts to replacing $1/\epsilon^0$ by $1/\epsilon^0 - 1/\epsilon^\infty$, where $\epsilon^\infty$ is the high-frequency, electronic permittivity. With this modification, the total energy in Eq.~\eqref{Eq:LP_toten_initial} becomes a functional of the electron wavefunction:
\begin{equation} \label{Eq:LP_totel_funcpsi}
    E = \!\frac{\hbar^2}{2m^*} \!\!\int \!\!d\br \, |\nabla \psi(\br)|^2 
     - \frac{1}{2}\frac{e^2}{4\pi\epsilon_0\k} \!\!\int\!\! d\br d\br' \frac{|\psi(\br)|^2|\psi(\br')|^2}{|\br-\br'|}~,
\end{equation}
having defined $1/\k = 1/\epsilon^\infty \!-\! 1/\epsilon^0$. 
Variational optimization of Eq.~\eqref{Eq:LP_totel_funcpsi} with respect to $\psi$, under the normalization constraint $\int d\br |\psi(\br)|^2 = 1$, leads to the following Schr\"odinger-like equation for the polaron wavefunction:
\begin{equation} \label{Eq:schro_LP}
    -\frac{\hbar^2}{2m^*} \nabla^2 \psi(\br)
    - \frac{e^2}{4\pi\epsilon_0} \frac{1}{\k} \int d\br' \frac{\,\,|\psi(\br')|^2}{|\br-\br'|}\psi(\br)
    = \varepsilon \, \psi(\br)~.
\end{equation}
In this equation, the polaron eigenvalue $\ve$ corresponds to the Lagrange multiplier associated with the wavefunction normalization. The second term on the left hand side is the ``self-trapping'' potential: a localized electron induces a lattice response that generates a potential well, which in turn further localizes the electron. 

\begin{figure}
  \centering    \includegraphics[width=0.99\linewidth]{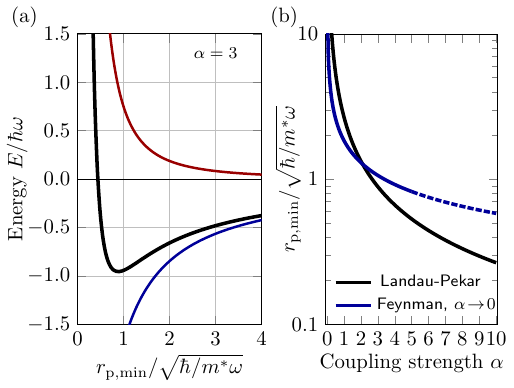}
  \caption{(a) Effective potential energy surface of the Landau-Pekar polaron, Eq.~\eqref{Eq:LP_E_vs_rp}. The black curve is $E(r_{\rm p})$, the blue curve is the attractive potential energy term, the red curve is the repulsive kinetic energy term. (b) Radius of the Landau-Pekar polaron (black line), from Eq.~\eqref{Eq:r_LP_alpha}. For comparison, we also show the radius of the Feynman model (Sec.~\ref{Sec:Feynman}) in the small-$\a$ limit (blue line), from Eq.~\eqref{Eq:radius_feynm1}. \rev{At large $\a$, the Feynman result (not shown) coincides with the Landau-Pekar result.}
  }
  \label{Fig:polaron_radius}
\end{figure}

The self-consistent solution of Eq.~\eqref{Eq:schro_LP} results from the competition between kinetic and potential energies in Eq.~\eqref{Eq:LP_totel_funcpsi}: minimization of the kinetic energy favors delocalization, while minimization of the lattice polarization energy favors localization. To visualize this competition, it is expedient to consider a normalized Gaussian trial wavefunction, $\psi(\br) = \pi^{-3/4} r_{\mathrm{p}}^{-3/2} \mathrm{exp}(-|\br|^2/2r^2_{\mathrm{p}})$, where $r_p$ is the variational parameter representing the polaron radius. This wavefunction is appropriate for three-dimensional bulk solids; two-dimensional (2D) materials are discussed in Sec.~\ref{Sec:2d_materials}. Using this expression in Eq.~\eqref{Eq:LP_totel_funcpsi}, one finds the total energy \cite{Devreese_2020}:
\begin{equation} \label{Eq:LP_E_vs_rp}
    E(r_{\mathrm{p}})= \frac{3\hbar^2}{4m^*}\frac{1}{r_{\mathrm{p}}^2} - \frac{1}{\sqrt{2\pi}}\frac{e^2}{4\pi\epsilon_0 \k }\frac{1}{r_{\mathrm{p}}}~.
\end{equation}
The competition between kinetic and potential energies is illustrated in Fig.~\ref{Fig:polaron_radius}(a). The minimum of this energy is:
\begin{equation} \label{Eq:LP_minE}
    \frac{E_{\mathrm{min}}}{E_{\mathrm{Ha}}} = -\frac{1}{6\pi} \frac{m^*/m_{\rm e}}{\k^2}~,
\end{equation}
and is reached at the radius:
\begin{equation}\label{Eq:LP_radius}
 \frac{r_{\rm p, min}}{a_0} = \frac{3\sqrt{2\pi}}{2} \frac{\k}{m^*/m_{\rm e}}~,
\end{equation}
where $m_{\rm e}$, $E_{\mathrm{Ha}}$, and $a_0$ are the free electron mass, the Hartree energy, and the Bohr radius, respectively. The eigenvalue of Eq.~\eqref{Eq:schro_LP} corresponding to this minimum is $\ve_{\rm min} = 3 E_{\rm min}$.
Equations~\eqref{Eq:LP_minE}-\eqref{Eq:LP_radius}
provide an intuitive understanding of how the polaron energy and size depend on basic materials properties such as $m^*$, $\epsilon^\infty$, and $\epsilon_0$:
heavier effective masses and/or stronger ionic dielectric response (i.e., smaller $\k$) lead to more stable and more localized polarons, cf.\ Fig.~\ref{Fig:polaron_radius}(b); this trend is consistent with atomic-scale \textit{ab initio} calculations of polarons.
While the prefactors in these equations depend on the specific choice of the trial wavefunction, the scaling trends carry general validity \cite{Miyake_1976}. 

The Landau-Pekar model provides an accurate solution for the more sophisticated Fr\"ohlich model (Sec.~\ref{Sec:FrohlichLLP}) in the limit of strong electron-phonon couplings. Furthermore, as we discuss in Sec.~\ref{Sec:DFTPolaron}, since this model describes ions within the adiabatic and classical approximations, it directly connects with DFT calculations of polarons.

In the polaron literature, it is common to express the energy and the radius of the Landau-Pekar polaron in terms of the dimensionless Fr\"ohlich coupling constant $\alpha$, which is defined as:
\begin{equation} \label{Eq:Frohlich_alpha}
    \alpha = \frac{e^2}{4\pi\epsilon_0\hbar}\sqrt{\frac{m^*}{2\hbar\omega}}\frac{1}{\k}~,
\end{equation}
where $\omega$ is the frequency of a longitudinal-optical phonon in the \fro\ model.
By combining Eqs.~\eqref{Eq:LP_minE}-\eqref{Eq:LP_radius} and \eqref{Eq:Frohlich_alpha}, the total energy and the radius of the Landau-Pekar polaron take the well-known forms \cite{Devreese_2020}:
\begin{eqnarray} \label{Eq:Etot_LP_alpha}
    E_{\mathrm{min}} &=& -\frac{\alpha^2}{3\pi}  \,\hbar \omega~, \\
  r_{\mathrm{p,min}} &=& \frac{3\sqrt{\pi}}{2}\sqrt{\frac{\hbar}{m^* \omega}}\,\frac{1}{\a}~. \label{Eq:r_LP_alpha}
\end{eqnarray}
We emphasize that the there is no notion of retardation in the Landau-Pekar model, and the 
frequency $\w$ in these expressions only appears as a result of the definition of $\a$ in Eq.~\eqref{Eq:Frohlich_alpha}. The above estimates were subsequently refined by \textcite{Buimistrov_Pekar_1957a,Buimistrov_Pekar_1957b} by solving the \fro\ polaron problem discussed in the next section, and using a variational minimization of both electron and phonon degrees of freedom.

\subsection{\fro\ model and Lee-Low-Pines transformation}\label{Sec:FrohlichLLP}

\subsubsection{\fro\ model and perturbative expansion}
The Landau-Pekar model of Sec.~\ref{Sec:LandauPekar} can be conceptualized as being obtained from the more general \fro\ model under the adiabatic and classical approximations for the ionic displacements. In its simplest form, the \fro\ Hamiltonian reads:
  \begin{eqnarray}\label{Eq:froh_eph_H}
    \hH &=& \sum_{\bk} \ve_\bk \hcd_\bk \hc_\bk 
      + \hbar\w \sum_\bq (\had_\bq\ha_\bq + 1/2) \nonumber \\
    &+& \frac{1}{\sqrt{N_p}} \sum_{\bk,\bq} g(|\bq|) \hcd_{\bk+\bq} \hc_\bk 
    (\ha_\bq+ \had_{-\bq})~,
  \end{eqnarray}
where $\hcd_\bk$/$\hc_\bk$ are the creation/annihilation operators for a single electron in a normalized plane wave with wavevector $\bk$, $\had_\bq$/$\ha_\bq$ are the creation/annihilation operators for a phonon with wavevector $\bq$ and frequency $\omega$; $\ve_\bk = \hbar^2 |\bk|^2/2m^*$ is the dispersion relation for a free electron with effective mass $m^*$; and $g$ is the \fro\ electron-phonon matrix element \cite{Sio_Giustino_2022}:
\begin{equation}
  \label{Eq:froh_eph}
  g(q) = i\hbar\w \frac{\sqrt{\a} }{q/q_0}, \qquad
  q_0^2 = \frac{4\pi}{\Omega} \sqrt{\frac{\hbar}{2m^*\w}},
\end{equation}
where $q=|\bq|$ and $\Omega$ is the volume of the primitive unit cell. In Eq.~\eqref{Eq:froh_eph_H}, the summations run over a uniform grid of wavevectors commensurate with a periodic Born-von K\'arm\'an (BvK) supercell containing $N_p$ primitive unit cells of the crystal, and the total energy refers to the entire BvK supercell. We here follow the same notation as in Appendix~A of \textcite{Giustino_2017}. The \fro\ matrix element in Eq.~\eqref{Eq:froh_eph} should really be periodic in reciprocal space Refs.~(\citeauthor{Verdi_Giustino_2015}, \citeyear{Verdi_Giustino_2015};  \citeauthor{Sjakste_Mauri_2015}, \citeyear{Sjakste_Mauri_2015}), but this is inconsequential for the following analysis. 

The Hamiltonian in Eq.~\eqref{Eq:froh_eph_H} describes a single electron interacting with a dispersionless longitudinal optical phonon in a isotropic material. It is obtained from the general electron-phonon coupling Hamiltonian of Eq.~\eqref{Eq:epi-hamilt} by considering only one parabolic electron band, one longitudinal optical phonon branch, and by replacing the complete matrix element with  Eq.~\eqref{Eq:froh_eph}.
The Landau-Pekar model is obtained from this Hamiltonian by considering (i) an adiabatic trial wavefunction $|\Psi\> = |\psi\>|\chi\>$, where $|\psi\>$ and $|\chi\>$ describe the electron and phonon sectors, respectively; and (ii) by restricting the vibrational solutions to coherent states, $|\chi\> = \exp \sum_\bq (\a_\bq \had_\bq - \a^*_\bq \ha_\bq)|0_{\rm ph}\>$, where $|0_{\rm ph}\>$ represents the phonon vacuum and the parameters $\a_\bq$ are to be determined. Coherent states are essentially displaced Gaussian wavefunctions, and serve as proxies for classical atomic displacements; they are eigenstates of the phonon annihilation operator: $\ha_\bq |\chi\>= \a_\bq |\chi\>$ \cite{Tannoudji_Laloe_1977}. By taking the expectation value of Eq.~\eqref{Eq:froh_eph_H} over $|\Psi\>$ and minimizing with respect to the parameters $\a_\bq$, one recovers Eq.~\eqref{Eq:LP_totel_funcpsi}, except for the zero-point energy which is absent in the Landau-Pekar model (cf. Supplemental Note 1). 

In the {\fro} approach, the ground state energy of the polaron is obtained by applying second-order
perturbation theory to Eq.~\eqref{Eq:froh_eph_H}; the first line of the equation serves as the unperturbed Hamiltonian, and the second line is the perturbation. The ground state of the unperturbed Hamiltonian corresponds to an electron at the band bottom and no phonons, $\hcd_0|0\>$, where $|0\>$ denotes the electron and phonon vacuum. 
The first order correction to the energy vanishes identically due to the action of $\ha_\bq$ and $\had_{-\bq}$ on the phonon vacuum. The second order correction is evaluated via Rayleigh–Schr\"odinger perturbation theory in Fock space \cite{Kittel_1987}, and is given by:
  \begin{equation}\label{Eq:Fro-RS}
  E_{\rm min} = -\frac{1}{N_p}\sum_\bq \frac{|g(q)|^2}{\hbar^2 q^2/2m^*+\hbar\w}~.
  \end{equation}
Upon taking the continuum limit for the $\bq$-mesh and extending the integration domain to $\mathbb{R}^3$, one finds \cite{Frohlich_1954}: 
  \begin{equation}\label{Eq:emin_fro}
  E_{\rm min} = -\a \hbar\w~,
  \end{equation}
which is the ground-state energy of the \fro\ polaron at small $\a$. Unlike in the Landau-Pekar model, to first order in $\a$ the perturbative polaron solution is not localized, and consists of a coherent superposition of coupled electron-phonon states \cite{Verdi_Giustino_2017,Mahan_2000}.

\subsubsection{Lee-Low-Pines transformation} \label{Sec:LLP_transform}

One alternative approach to determine the energy of the \fro\ Hamiltonian in Eq.~\eqref{Eq:froh_eph_H} is the Lee-Low-Pines canonical transformation method \cite{Lee_Pines_1953}. In this method, the electronic degrees of freedom are integrated out by centering the atomic displacement field around the electron position. To see how this works, it is convenient to rewrite the electronic operators appearing in Eq.~\eqref{Eq:froh_eph_H} in terms of the electron position $\hat{\br}$ and momentum $\hbp$.  This can be achieved by introducing the field operators $\hat{\psi}(\br) = \sum_\bk \psi_\bk \,\hc_\bk$ and $\hat{\psi}^\dagger(\br) = \sum_\bk \psi^*_\bk \,\hcd_\bk$, where $\psi_\bk(\br)=(N_p \Omega)^{-1/2} e^{i \bk \cdot \br}$; and noting that $\sum_\bk \ve_\bk \hcd_\bk \hc_\bk= 
-\int \,d\br\, \hat{\psi}^\dagger(\br)\,(\hbar^2\nabla^2/2m^*)\,\hat{\psi}(\br)
=\hbp^2 / 2m^*$ since we assumed $\ve_\bk = \hbar^2|\bk|^2/2m^*$; similarly,
$\sum_\bk \hcd_{\bk+\bq} \hc_\bk = \int d\br \,\exp(i\bq\cdot \br) \,\hat{\psi}^\dagger(\br) \hat{\psi}(\br) = \exp(i\bq\cdot \hat{\br})$. 
The result is:
  \begin{equation}\label{Eq:froh_eph_x}
    \hH 
    = \frac{\hbp^2}{2m^*} 
      + \hbar\w \!\sum_\bq (\had_\bq\ha_\bq + \frac{1}{2})
     + \frac{1}{\sqrt{N_p}}\!\sum_{\bq} 
       g(q) \ha_\bq e^{i\bq\cdot \hat{\br}} + \textrm{h.c.}~.
  \end{equation}
Starting from this expression, Lee-Low-Pines perform a translation of the electron coordinate via the operator $\exp(i \hbp \cdot \hat{\br})$. Since the Hamiltonian is translationally invariant and therefore the total momentum $\hbp + \sum_\bq \hbar\bq \,\had_\bq\ha_\bq$ is a good quantum number, say $\bP$, one can write $\hbp = \bP - \sum_\bq \hbar\bq \,\had_\bq\ha_\bq$; this suggests defining the canonical transformation as:
\begin{equation}\label{Eq:llp_can_tr}
    \hH' = \hU^{-1}\hH \hU, \quad
    \hU = \exp \left[i 
    \left(\bP/\hbar - \!{\sum}_\bq \bq \,\had_\bq\ha_\bq\right)\!\cdot \!\hat{\br}\right]~.
  \end{equation}
Upon applying this transformation to Eq.~\eqref{Eq:froh_eph_x}, the first two terms remain unchanged because everything commutes; this is immediately seen by noting that $\hn_\bq = \had_\bq\ha_\bq$ is the phonon number operator. The third term in Eq.~\eqref{Eq:froh_eph_x} is evaluated by writing the transformation explicitly and factoring out all commuting terms; what remains is the sum
$
N_p^{-\frac{1}{2}}\!\sum_{\bq} g(q)\, e^{i\bq\cdot \hat{\br}} 
e^{i\hn_\bq\bq\cdot \hat{\br}} \ha_\bq e^{-i \hn_\bq\bq\cdot  \hat{\br}} 
$
and its Hermitian conjugate. The operator product $e^{i\hn_\bq\bq\cdot \hat{\br}} \ha_\bq e^{-i \hn_\bq\bq\cdot  \hat{\br}}$ can be evaluated using the Campbell identity: 
$e^{-\hat{S}} \hat{O} e^{\hat{S}} = \hat{O} + [\hat{O}, \hat{S}]+1/2[[\hat{O}, \hat{S}], \hat{S}]+\cdots$, yielding $\ha_\bq e^{-i \bq\cdot  \hat{\br}}$. Putting together, the transformed Hamiltonian reads:
\begin{eqnarray}
    \label{Eq:llp_H_1}
    \hH'
    &=&\frac{1}{2m^*} \left(\bP - {\sum}_\bq \hn_\bq\,\hbar\bq \right)^2 \nonumber \\
    &+& \hbar \w\sum_\bq (\hn_\bq + 1/2)
    \!+\! N_p^{-\frac{1}{2}}\!
    \sum_\bq g(q)\, \ha_\bq + \textrm{h.c.}~. \hspace{10pt}
\end{eqnarray}
In this form, the electronic degrees of freedom have been absorbed into the translation. The second line now describes the Hamiltonian of shifted harmonic oscillators with modified ladder operators $\hb_\bq = \ha_\bq + N_p^{-\frac{1}{2}}g^*(q)/\hbar\w$ \cite{Tannoudji_Laloe_1977}; this observation suggests attempting a variational solution with a trial state consisting of a product of coherent phonons.
To this end, Lee-Low-Pines choose again a trial coherent state, $|\chi\> = \exp \sum_\bq (\a_\bq \had_\bq - \a^*_\bq \ha_\bq)|0\>$, so that $\ha_\bq |\chi\>= \a_\bq |\chi\>$ and the expectation value of Eq.~\eqref{Eq:llp_H_1} can be expressed as a function of the coefficients $\a_\bq$. For $\bP=0$, the minimization of the energy with respect to these coefficients yields the same result as in Eq.~\eqref{Eq:emin_fro} to linear order in the coupling strength $\a$; for finite total momentum $\bP$, the variational energy at the lowest order in $\a$ is: 
\begin{equation}
    E_{\rm min}({\bf P}) = -\alpha \hbar \w + \frac{|\bP|^2}{2(1+\a/6)m^*},
\end{equation}
implying an enhancement of the effective mass by a factor $1+\a/6$, as in \fro's perturbative solution. Detailed, step-by-step derivations of the Lee-Low-Pines model are provided by \textcite{Devreese_2020}.

As compared to \fro's perturbative solution, the conceptual strength of the Lee-Low-Pines transformation is that it lends itself more naturally to a variational solution, and therefore it provides well-defined upper bounds to the formation energy. Comparison with the more accurate DMC solutions indicate that the Lee-Low-Pines solution is very accurate up to coupling strengths $\a \lesssim 6$ \cite{Cataudella_Peroni_2007}. Beyond this value, the Landau-Pekar solution of Sec.~\ref{Sec:LandauPekar} and the Feynman solution discussed in the next section provide better variational estimates as compared to full-blown DMC calculations (a discussion of the DMC method is provided in Sec.~\ref{Sec:DMC}).

\rev{The generalization of the Fr\"ohlich electron-phonon matrix elements in Eq.~\eqref{Eq:froh_eph} for \textit{ab initio} calculations was developed in Refs.~(\citeauthor{Verdi_Giustino_2015}, \citeyear{Verdi_Giustino_2015};  \citeauthor{Sjakste_Mauri_2015}, \citeyear{Sjakste_Mauri_2015}), and constitutes the principal coupling mechanism for large polaron formation in materials (cf. Sec.~\ref{Sec:Applications}).
As an intermediate level between the original Fr\"ohlich model and full \textit{ab initio} calculations, \textcite{Miglio_Gonze_2020} introduced an extension of Eq.~\eqref{Eq:froh_eph_H} that accounts for anisotropic and degenerate electron bands as well as multiple phonon branches, while neglecting band and phonon dispersions.
Using this generalized Fr\"ohlich model, \textcite{deMelo_Verstraete_2023} analyzed over a thousand ionic materials, showing that many host polarons within the perturbative regime captured by the model, whereas a significant fraction exhibit stronger coupling or localization requiring a full \textit{ab initio} treatment.}

\begin{figure}
\centering
\includegraphics[width=0.90\linewidth]{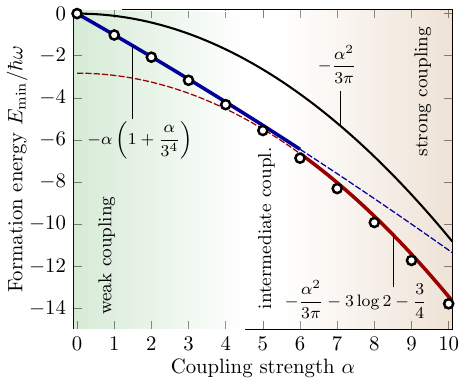}
  \caption{Formation energy of the \fro\ polaron as a function of the electron-phonon coupling strength $\a$. The energy zero corresponds to a free electron in the absence of lattice distortion. The black line is the Landau-Pekar solution, Eq.~\eqref{Eq:Etot_LP_alpha}, while the blue line and red line are the asymptotic expansions of Feynman's solution at weak coupling and strong coupling, respectively, Eqs.~\eqref{Eq:Emin_feynm_alpha1}-\eqref{Eq:Emin_feynm_alpha2}. The weak-coupling expansion coincides with \fro's and LLP's solution, the strong-coupling expansion coincides with the Landau-Pekar solution apart from a constant shift. Black circles are DMC data from \textcite{Hahn_Franchini_2018}. The regions of weak, intermediate, and strong coupling are marked qualitatively.}
  \label{Fig:frohlich}
\end{figure}

\subsection{Feynman path-integral approach} \label{Sec:Feynman}

The method developed by \textcite{Feynman_1955} to tackle the \fro\ polaron problem is historically significant as one of the early applications of imaginary-time path integrals, and for yielding estimates of the ground state-energy with accuracy comparable to numerical DMC methods.

The path-integral approach is based on the observation that the ground-state energy of a given Hamiltonian, $E_{\rm min}$, can be obtained as the zero-temperature limit of its Helmholtz free energy, $F$: $E_{\rm min}=\lim_{\b\rightarrow \infty} F(\b)$, where $\b = 1/\kb T$, $\kb$ is Boltzmann's constant and $T$ is the temperature. This relation comes from the definition of the free energy in terms of the canonical partition function $Z$, $F = -\b^{-1} \log Z$, and the fact that the term $e^{-\b E_{\rm min}}$ in $Z$ dominates over all others in the large-$\b$ limit \cite{Landau_Lifshitz_1980}. If one could compute $Z(\b)$ for the \fro\ Hamiltonian in Eq.~\eqref{Eq:froh_eph_H}, then its large-$\b$ limit would provide the ground state energy. \citeauthor{Feynman_1955} evaluates this partition function using imaginary-time path integrals:
  \begin{equation}\label{Eq:Z_feynman}
    Z(\b) = \int \!dx \!\int_{x(0) = x}^{x(\beta\hbar) = x} \!\mathcal{D}[x(\tau)] \,\, e^{-S_{\rm E}/\hbar}~,
  \end{equation}
where $x$ is a short-hand notation for all the degrees of freedom of the \fro\ model, that is the electron coordinates and the amplitudes of the harmonic oscillators. $\int \mathcal{D}[x(\tau)]$ indicates the integral over all trajectories of the system that start at $x$ at the fictitious time $\tau=0$ and return to the same point at $\tau=\hbar\b$; the outer integral $\int dx$ is over all such possible points; and the quantity $S_{\rm E}$ in the exponential is the Euclidean action of the system evaluated along the trajectory $x(\tau)$ \cite{Feynman_Roach_1965}.

The evaluation of the Euclidean action is performed starting from the \fro\ Hamiltonian in the form provided by Eq.~\eqref{Eq:froh_eph_x}, making the transition to the corresponding classical Lagrangian, and performing the Wick rotation \cite{Lancaster_Blundell_2014}. Since the second and third terms of Eq.~\eqref{Eq:froh_eph_x} correspond to a sum of forced harmonic oscillators, one can use the standard result for the imaginary-time path integral of the forced oscillator [cf.\ Eqs.~(8.136) and (8.138) of \citeauthor{Feynman_Roach_1965}], and then carry out the sum over $\bq$ by taking the continuous limit $N_p\rightarrow\infty$. The result is: 
  \begin{equation}\label{Eq:Feynm-SE1}
  S_{\rm E} = \!\!\int_0^{\b\hbar}\!\!\!\!\!\!\!d\tau\! \left[\frac{1}{2}m^* |\dot\br(\tau)|^2 
   - \frac{\a}{2\sqrt{2}}\frac{(\hbar \w)^{3/2}}{m^{*,1/2}}\!\!
    \int_0^{\b\hbar} \!\!\!\!\!\!\!d\tau'  \,\frac{e^{-\w |\tau-\tau'|}}{|\br(\tau)-\br(\tau')|}\!\right]\!\!.
  \end{equation}
This action does not contain the phonon degrees of freedom, which have been integrated out, and only depends on the electron trajectory $\br(\tau)$ \rev{and the electron velocity $\dot\br(\tau)$}. The integrand has the intuitive meaning of the Euclidean Lagrangian of an electron that feels a retarded and nonlocal Coulomb self-attraction mediated by the lattice polarization. 

By using Eq.~\eqref{Eq:Feynm-SE1} inside Eq.~\eqref{Eq:Z_feynman}, in principle one should be able to evaluate the partition function and thereafter the ground-state energy of the polaron. To make further progress without resorting to numerical methods, Feynman employs instead a variational approach that provides an upper bound to the polaron energy.
In this approach, a trial action $S'_{\rm E}$ is designed to mimic Eq.~\eqref{Eq:Feynm-SE1} while at the same time being amenable to exact analytical evaluation. Using the Jensen-Feynman
inequality \cite{Feynman_Roach_1965}, $F \le F'+(\hbar\b)^{-1}\<S_{\rm E}-S'_{\rm E}\>_{S'_{\rm E}}$,
and taking the large-$\beta$ limit, one obtains:
\begin{equation}\label{Eq:feynm-jens}
    E_{\rm min} \le E_{\rm min}'+\lim_{\b \rightarrow \infty} \frac{1}{\hbar\b}\<S_{\rm E}-S'_{\rm E}\>_{S'_{\rm E}}~.
\end{equation}
In this expression, $E_{\rm min}'$ is the ground-state energy corresponding to the trial action, and $\<\cdots\>_{S'_{\rm E}}$ denotes the average over trajectories, weighted by the Boltzmann factor $e^{-S'_{\rm E}/\hbar}$ (cf. Sec. 11.1 of \citeauthor{Feynman_Roach_1965}).

Since the potential term in the Lagrangian appearing in Eq.~\eqref{Eq:Feynm-SE1} is reminiscent of a potential well, \textcite{Feynman_1955} proposes to replace it via a parabolic potential, yielding the trial action:
\begin{equation}\label{Eq:Feynm-SE2}
    S'_{\rm E} = \!\!\int_0^{\b\hbar}\!\!\!\!\!\!\!d\tau\! \left[\frac{1}{2}m^* |\dot\br(\tau)|^2 
    + a\!\!\int_0^{\b\hbar} \!\!\!\!\!\!\!d\tau'  \,e^{-b |\tau-\tau'|}|\br(\tau)-\br(\tau')|^2\right]\!.
\end{equation}
The quantities $a$ and $b$ appearing in the second term are variational parameters that need to be optimized to make Eq.~\eqref{Eq:Feynm-SE2} as close as possible to Eq.~\eqref{Eq:Feynm-SE1}, and to provide the lowest upper bound to the ground-state energy. With this trial action, the right-hand side of Eq.~\eqref{Eq:feynm-jens} can be evaluated explicitly; after moving to dimensionless variational parameters $v$ and $w$ such that $a = m\w^2 w(v^2-w^2)/8$ and $b=\w w$, the result takes the form \cite{Feynman_1955}:
  \begin{equation}\label{Eq:feynm-final}
   \frac{E_{\rm min}}{\hbar \w} \le \frac{3(v-w)^2}{4v} - \frac{\a v}{\sqrt{\pi}} 
    \int_0^\infty \!\!\!\!\!\!\frac{e^{-u}du}{\sqrt{w^2 u+\displaystyle\frac{v^2-w^2}{v}(1-e^{-vu})}}.
  \end{equation}
The weak-coupling and strong-coupling limits of this expression are:
\begin{eqnarray}\label{Eq:Emin_feynm_alpha1}
  \mbox{small }\a:&&\,\,  E_{\rm min}/\hbar\w \le -\a(1-\a/3^4), \\
  \mbox{large }\a:&&\,\, E_{\rm min}/\hbar\w \le -\a^2/3\pi-3(\log 2+1/4)~, \label{Eq:Emin_feynm_alpha2}
\end{eqnarray}
and are shown in Fig.~\ref{Fig:frohlich}. These values reduce to the \fro\ limit of Sec.~\ref{Sec:FrohlichLLP} at small $\a$, and to the Landau-Pekar limit of Sec.~\ref{Sec:LandauPekar} at large $\a$, and interpolate in between.
These expressions yield energies that are in very close agreement with DMC calculations \cite{Mishchenko_Svistunov_2000,Hahn_Franchini_2018}, with errors as small as a few percent when one uses Eq.~\eqref{Eq:Emin_feynm_alpha1} for $\a \lesssim 5$ and Eq.~\eqref{Eq:Emin_feynm_alpha2} for $\a \gtrsim 5$. For these reasons, Feynman's approach is referred to as an ``all-coupling'' method.

This approach was also used to estimate the radius $r_p$ of the polaron as a function of coupling strength $\a$. Calculations by \textcite{Schultz_1959} and \textcite{Peeters_Devreese_1985} give the following limits at weak- and strong-coupling, respectively:
\begin{eqnarray}
  \mbox{small }\a:&&\,\, r_{\mathrm{p,min}} = \frac{3\sqrt{3}}{2\sqrt{2}}\sqrt{\frac{\hbar}{m^* \w}} \frac{1}{\sqrt{\a}}, \label{Eq:radius_feynm1}\\
  \mbox{large }\a:&&\,\,  r_{\mathrm{p,min}} = \frac{3\sqrt{\pi}}{2}\sqrt{\frac{\hbar}{m^* \w}} \frac{1}{\a}~. \label{Eq:radius_feynm2}
\end{eqnarray}
Both expressions show how the polaron size decreases as the electron-phonon coupling increases. We note that the second equation coincides with the Landau-Pekar result in Eq.~\eqref{Eq:r_LP_alpha}. These trends are shown in Fig.~\ref{Fig:polaron_radius}(b).

\subsection{Holstein model and Lang-Firsov transformation} \label{Sec:HolsteinLF}

The \textcite{Holstein_1959a,Holstein_1959b} polaron model provides an alternative viewpoint to the \fro\ model of Sec.~\ref{Sec:FrohlichLLP}, by emphasizing local electron-phonon couplings.

The Holstein Hamiltonian is formally similar to Eq.~\eqref{Eq:froh_eph_H}, except that (i) the free electron band is replaced by a nearest-neighbor tight-binding model on a $d$-dimensional cubic lattice, and (ii) the electron-phonon coupling matrix element is a momentum-independent constant, $g$:
  \begin{eqnarray}\label{Eq:Holstein1}
    \hH &=& \sum_{\bk} \ve_\bk \hcd_\bk \hc_\bk 
      + \hbar\w \sum_\bq (\had_\bq\ha_\bq + 1/2) \nonumber \\
    &+& \frac{g}{\sqrt{N_p}} \sum_{\bk,\bq} \hcd_{\bk+\bq} \hc_\bk 
    (\ha_\bq+ \had_{-\bq})~, \\
    \ve_\bk &=& -2t \sum_{\a = 1}^d \cos(k_\a a)~. \label{Eq:Holst-bands}
  \end{eqnarray}
Here, $t$ is the hopping energy parameter, $d$ is the dimension of the lattice, and $a$ is the lattice parameter. As in the \fro\ model, the phonon frequency is independent of momentum; however, this frequency represents here a transverse-optical phonon (TO) instead of the LO phonon in the \fro\ model. 
Historically, Holstein introduced this model to describe a linear chain of identical diatomic molecules, where electrons on each molecule feel the local vibration of the dimer and are connected electronically to their nearest neighbors. Therefore, in this case, the electron-phonon interaction does not involve long-range polarization fields as in the \fro\ and Landau-Pekar models, but rather local vibrations that change the electron site energy.

The ``local'' nature of the Holstein model is more easily appreciated by transforming Eq.~\eqref{Eq:Holstein1} into a real-space representation. We employ the following convention for the discrete Fourier transform of the operators:
$\hc_\bk = N_p^{-1/2}\sum_i e^{i\bk\cdot \bR_i} \hc_i$ and
$\ha_\bq = N_p^{-1/2}\sum_i e^{i\bq\cdot \bR_i} \ha_i$.
Upon using these definitions inside Eq.~\eqref{Eq:Holstein1}, the Hamiltonian becomes:
  \begin{equation}\label{Eq:Holstein2}
    \hH = -t\sum_{\<ij\>} \hcd_i\hc_j
      + \hbar\w \sum_i (\had_i\ha_i + 1/2) 
      +  g\sum_i \hn_i (\ha_i + \had_{i})~,
  \end{equation}
where $\<ij\>$ indicates that the sum is over all $i$ and $j$ indices that are nearest neighbors on the lattice, and we introduced the electron number operator, $\hn_i = \hcd_i \hc_i$, which counts how many electrons reside at site $i$. This Hamiltonian represents the energy of a cubic BvK supercell containing $N_p$ sites. In this form, it is seen that the electron-phonon interaction takes place between electron and phonons on the same site. Equation~\eqref{Eq:Holstein2} is especially useful for tackling the Holstein polaron problem via correlated methods such as density matrix renormalization group \cite{Jeckelmann_White_1998}\rev{, exact diagonalization (\citeauthor{Bonca_Batistic_1999}, \citeyear{Bonca_Batistic_1999}; \citeauthor{Barisic_2004}, \citeyear{Barisic_2004})} and neural quantum states \cite{Mahajan_Reichman_2025}, and represents the starting point for building the Holstein-Hubbard Hamiltonian \cite{Beni_Kanamori_1974}. This model continues to serve as the standard benchmark in the development of new analytical \cite{Berciu_2006} and numerical methods \cite{Mitric_Tanaskovic_2022} for polaron physics.

The energy scales of the Holstein model are the electron hopping energy $t$, the phonon energy $\hbar\w$, and the electron-phonon coupling $g$. It is expedient to combine these energies into a dimensionless electron-phonon coupling strength, $\lambda$ \cite{Goodvin_Sawatzky_2006}:
 \begin{equation}\label{Eq:holst-lambda}
     \lambda = \frac{g^2}{2d t \hbar\w}~.
 \end{equation}
With this definition, the weak and strong coupling regimes correspond to $\lambda \ll 1$ and $\lambda \gg 1$, respectively. 

\begin{figure}
\includegraphics[width=0.9\linewidth]{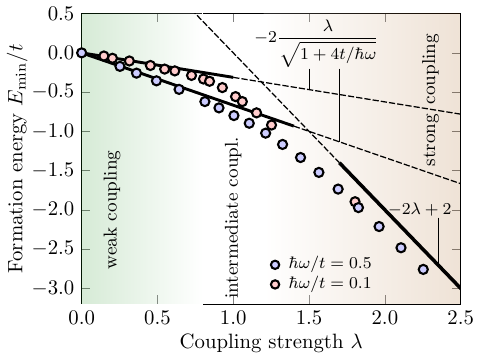}
  \caption{Formation energy of the Holstein polaron in one dimension, as a function of the Holstein electron-phonon coupling strength $\lambda$ and for two different values of the ``adiabaticity parameter'' $\hbar\w/t$. The energy zero corresponds to a band electron without lattice distortion. The black lines correspond to the weak coupling and strong coupling expressions in Eqs.~\eqref{Eq:Holstein_weak} and \eqref{Eq:Holstein_strong}, respectively. The disks are DMC data from \citeauthor{Macridin_Sawatzky_2003} (\citeyear{Macridin_Sawatzky_2003}, Fig.~3.6), for the adiabaticity parameters $\hbar\w/t = 0.1$ (red) and 0.5 (blue). The color shading marks qualitatively the weak-, intermediate-, and strong-coupling regimes.}
  \label{Fig:holstein}
\end{figure}

The simplest strategy to examine the weak coupling limit is to employ Rayleigh–Schrödinger perturbation theory in Fock space, as already seen in Sec.~\ref{Sec:FrohlichLLP} for the \fro\ model. We can directly use Eq.~\eqref{Eq:Fro-RS}, after replacing $g(q)$ by the constant $g$, and $\hbar^2 q^2/2m^*$ by $\ve_{\bk+\bq}-\ve_{\bk}$ from Eq.~\eqref{Eq:Holst-bands}, evaluated at $\bk=0$:
  \begin{equation}
  E_{\rm min} = \ve_{\bk=0}-\frac{g^2}{N_p}\sum_\bq \frac{1}{\ve_{\bq}+2dt+\hbar\w}~.
  \end{equation}
Evaluation of the integral in the one-dimensional case ($d=1$) leads to:
  \begin{equation} \label{Eq:Holstein_weak}
  \frac{E_{\rm min}}{t} = -2\frac{\lambda}{\sqrt{1+4t/\hbar\w}}~,
  \end{equation}
showing that the polaron formation energy increases linearly with the coupling strength $\lambda$. 
\rev{Here and in the following, we have referenced the energy minimum $E_{\rm min}$ to $\ve_{\bk=0}$, so that it has the meaning of formation energy.}
This  expression \rev{can be found in \cite{Marsiglio_1995}}, and it is compared to DMC calculations \cite{Macridin_Sawatzky_2003} in Fig.~\ref{Fig:holstein}. A similar expression for the two-dimensional case can be found in \cite{Li_Marsiglio_2010}.

Equation \eqref{Eq:Holstein_weak} describes the weak coupling limit because, as for Eq.~\eqref{Eq:Fro-RS}, it is obtained perturbatively starting from the delocalized state $\hcd_\bk|0\>$. The opposite limit $\lambda \gg 1$ can be analyzed by considering that, at strong coupling, the oscillator displacements are large, hence a semiclassical approximation may be adequate. In this spirit, \textcite{Holstein_1959a} rewrites Eq.~\eqref{Eq:Holstein2} in terms of the position and momentum of each oscillator, $\hx_i = \sqrt{\hbar/2m\w}\,(\had_i+\ha_i)$, $\hpp_i = i\sqrt{\hbar\w m/2}\,(\had_i-\ha_i)$, where $m$ is the oscillator mass. In the limit of large masses, the kinetic energy of the oscillators vanishes, and the position operators can be replaced by classical amplitudes. The resulting Hamiltonian acts on the electronic degrees of freedom and depends parametrically on the displacements:
  \begin{equation}\label{Eq:Holstein_classical}
    \hH = -t\sum_{\<ij\>} \hcd_i\hc_j +  
      \frac{1}{2}m\w^2\sum_i x_i^2    
      +  g\sqrt{2m\w/\hbar}\,\sum_i \hn_i x_i~.
  \end{equation}
The ground state of this Hamiltonian can be found variationally, following the same procedure as for the Landau-Pekar model. To this end, one first expresses the electronic state in the site basis, $|\psi\>=\sum_i c_i\, \hcd_i|0\>$ where the coefficients $c_i$ are to be determined; then one takes the expectation value $E=\<\psi|\hH|\psi\>$, and minimizes this energy with respect to the displacements $x_i$ and the coefficients $c_i$, subject to the normalization constraint $\sum_i |c_i|^2=1$. These steps lead to the following energy functional of the electronic coefficients:
 \begin{equation}
 E = -t\sum_{\<ij\>} c^*_i c_j -2d t \lambda \sum_i  |c_i|^4 ~.
 \end{equation}
This expression is reminiscent of the Landau-Pekar functional in Eq.~\eqref{Eq:LP_totel_funcpsi}, as well as the \textit{ab initio} polaron equations in Sec.~\ref{Sec:plrn_eq_reciprocal}.
The minimization of this functional using the method of Lagrange multipliers yields a nonlinear discrete Schr\"odinger-type equation for the coefficients $c_i$ (here for the one-dimensional case), which is the counterpart of Eq.~\eqref{Eq:schro_LP} in the Landau-Pekar model:
 \begin{equation}\label{Eq:Holstein_discrete}
 - (c_{i+1}+c_{i-1})  -  4\lambda   |c_i|^2 c_i= \frac{\ve}{t} \,c_i~.
 \end{equation}
Owing to normalization, we must have $|c_i|<1$, therefore for $\lambda \gg 1$ the potential term dominates and the equation simplifies to $- 4\lambda c_i  |c_i|^2 = (\ve/t) c_i$. It is immediate to verify that the lowest energy is obtained when a single site is occupied and all the others are empty, e.g., $c_0=1$ and $c_i=0,\, i\ne 0$; in this case, the total \rev{formation energy (referred to the system without electron-phonon coupling) reads
  \begin{equation} \label{Eq:Holstein_strong}
  \frac{E_{\rm min}}{t} = -2\lambda + 2~.
  \end{equation}
  }
This expression indicate that, in the Holstein model, the energy of the polaron scales linearly with the coupling strength $\lambda$ even at strong coupling. In Fig.~\ref{Fig:holstein}, the simplified limit provided by Eq.~\eqref{Eq:Holstein_strong} is compared with DMC data from \textcite{Macridin_Sawatzky_2003}. Extensive comparisons between various methods for calculating the energy of the Holstein polaron can be found in \cite{Romero_Lindeberg_1998}.

\textcite{Holstein_1959a} also proposes an alternative strategy to solve Eq.~\eqref{Eq:Holstein_discrete}, which provides a useful relation between the polaron size and the coupling strength. In that approach, one converts the hopping terms on the left hand side of Eq.~\eqref{Eq:Holstein_discrete} into a continuous kinetic energy; the resulting continuous nonlinear Schr\"odinger equation admits the exact solution $\psi(x) = (2l)^{-1/2}\sech(x/l)$, where the characteristic size is given by:
\begin{equation}
 l = \frac{a}{\lambda}~.
\end{equation}
This result is entirely analogous to the strong coupling limits of the Landau-Pekar and Feynman models, Eqs.~\eqref{Eq:r_LP_alpha}, \eqref{Eq:radius_feynm2}. Despite this useful result, the continuous version of the Holstein model carries an incorrect $\lambda^2$ scaling of the energy at strong coupling, therefore the discrete version given by Eq.~\eqref{Eq:Holstein_discrete} is to be preferred when comparing to numerical methods.

\subsubsection{Lang-Firsov transformation} \label{Sec:LF_transform}

A popular approach to investigate the Holstein model and its applications is the \textcite{Lang_Firsov_1963} canonical transformation. The approach consists of rewriting the Hamiltonian of Eq.~\eqref{Eq:Holstein2} in such a way as to absorb the electron-phonon coupling terms into phonon-dressed hopping terms, thereby transforming the problem into one of non-interacting polarons and phonons. While this approach does not necessarily improve over the simple weak- and strong-coupling estimates for the ground-state energy in Eqs.~\eqref{Eq:Holstein_weak} and \eqref{Eq:Holstein_strong}, respectively, it becomes advantageous in the study of polaron transport \cite{Ortmann_Hannewald_2009}, in variational treatments of more realistic Hamiltonians \cite{Hohenadler_vonderLinden_2007}, and in \textit{ab initio} calculations \cite{Lee_Bernardi_2021}.

The starting point of the Lang-Firsov transformation is to observe that the phonon and electron-phonon terms in Eq.~\eqref{Eq:Holstein2} can be rewritten by completing the squares:
\begin{equation}\label{Eq:Holstein3}
 \hbar\w \had_i\ha_i  
      +  g \hn_i (\ha_i + \had_{i}) = 
      \hbar\w \!\left(\ha_i\!+\!\frac{g}{\hbar\w}\hn_i\right)^{\!\!\dagger}\!\!\left(\ha_i\!+\!\frac{g}{\hbar\w}\hn_i\right)-\frac{g^2}{\hbar\w} \hn_i~.
\end{equation}
In this form, one recognizes shifted harmonic oscillators; the shifts $(g/\hbar\w) \hn_i$ can now be eliminated by moving to a new frame where each oscillator is in its ground state. This is accomplished by performing a translation of the phonon coordinates via the exponential of the phonon momentum operator, $\exp(-i\hat{p}_i u /\hbar)$, and setting the translation to $u = (\hbar/2m\w)^{1/2}(2g/\hbar\w)\hn_i$. The resulting unitary transformation is \cite{Lang_Firsov_1963}:
\begin{equation} \label{eq:LF-transform}
 \hU= e^{\hat{S}}, \qquad \hat{S} = \frac{g}{\hbar\w}{\sum}_i\hn_i(\had_i-\ha_i)~.  
\end{equation}
By applying this transformation to Eq.~\eqref{Eq:Holstein2}, one obtains the transformed Hamiltonian $\hH' = \hU^{-1}\hH \hU$:
  \begin{equation}\label{Eq:Holst_Hprime}
    \hH' = -t\sum_{\<ij\>}  \hat{D}^\dagger_i \hat{D}_j^{\phantom{\dagger}}\hcd_i \hc_j 
      + \hbar\w \sum_i (\had_i\ha_i + 1/2) 
      -2dt\lambda~,
  \end{equation}
where the $\hat{D}$ operators are the same displacement operators as in the theory of coherent states \cite{Tannoudji_Laloe_1977}, $\hat{D}_i = \exp\big[-(g/\hbar\w)(\had_i-\ha_i)\big]$.
In Eq.~\eqref{Eq:Holst_Hprime}, the electron-phonon coupling term has been eliminated, but the hopping term is now dressed by phonons. The constant $-2dt\lambda$ on the right hand side represents the energy lowering upon polaron formation, and coincides with the strong-coupling estimate in Eq.~\eqref{Eq:Holstein_strong}. 

The form of the transformed Hamiltonian in Eq.~\eqref{Eq:Holst_Hprime} suggests to interpret $\hat{D}^{\phantom{\dagger}}_i \hc_i$ and its hermitian conjugate as polaron annihilation and creation operators, respectively \cite{Ortmann_Hannewald_2009}. These operators play an important role in studying the dynamics of the Holstein polaron, and constitute the starting point for discussing temperature-activated hopping transport in the Holstein model, see \textcite{Lang_Firsov_1963} as well as \textcite{Mishchenko_Cataudella_2015} for modern state-of-the-art calculations.

By evaluating the expectation value of $\hH'$ on the phonon vacuum $|0_{\rm ph}\>$, one finds: 
  \begin{equation}\label{Eq:Holst-teff1}
    \<0_{\rm ph}|\hH'|0_{\rm ph}\> = -t_{\rm eff}\sum_{\<ij\>} \hcd_i\hc_j 
      + N_p \frac{\hbar\w}{2} -2dt\lambda~,
  \end{equation}
where the effective hopping parameter $t_{\rm eff}$ is given by
\begin{equation}\label{Eq:Holst-teff2}
  t_{\rm eff} = t\, e^{-(g/\hbar\w)^2}~.
\end{equation}
Equations~\eqref{Eq:Holst-teff1}-\eqref{Eq:Holst-teff2} explicitly show the phonon dressing of the electron hopping in the approach by \citeauthor{Lang_Firsov_1963}; such dressing leads to band narrowing and effective mass renormalization.

In the context of \textit{ab initio} calculations, the canonical transformation method has been adapted and generalized to the study of small polarons in real materials by \textcite{Lee_Bernardi_2021, Luo_Bernardi_2022}; these developments are discussed in Sec.~\ref{Sec:bernardi_polaron}.

At the end of this brief survey of model Hamiltonian approaches to polarons, we note that several additional techniques and models are currently being explored, whose discussion lies beyond the scope of this review. For example, the momentum average approximation provides an elegant and accurate method to compute the Green's function of the Holstein polaron and its generalization \cite{Berciu_2006,Goodvin_Sawatzky_2006,Berci_Goodvin_2007}, effectively capturing both weak and strong coupling regimes.

Beyond the Holstein model, several other important polaron models have been developed to capture additional physical mechanisms. Notable examples include the PSSH model and the Jahn-Teller model. The PSSH model describes coupling between electrons and bond-length fluctuations~\cite{Barisic_Friedel_1970,Su_Heeger_1979,Marchand_Stamp_2010}; this model recently received renewed attention due to the possibility of bipolaron formation \cite{Sous_Berciu_2018} and their superconducting pairing \cite{Zhang_Svistunov_2023}. The Jahn-Teller model focuses on the coupling of electronic degeneracies to local lattice distortions \cite{Bersuker_2006}. For detailed discussions, we refer the reader to the comprehensive reviews offered in \cite{Alexandrov_Devreese_2010,Cataudella_Peroni_2007,Ceulemans_2022}.

\section{Polarons from density functional theory calculations}\label{Sec:DFTPolaron}

The effective Hamiltonian approaches reviewed in Sec.~\ref{Sec:History} provide an ideal testbench for advanced many-body methods, and are especially useful to systematically explore the model parameters over wide ranges. One limitation of these methods is that they rely on simplifying assumptions on the electronic structure, lattice dynamics, and electron-phonon couplings, which make it somewhat challenging to perform quantitative comparison with experiments on polarons in materials.

Quantitative, atomic-scale study of polarons can now be performed using DFT and related methods \cite{Franchini_Diebold_2021}. Historically, DFT approaches to polarons have evolved largely independently of effective Hamiltonians approaches, and are closer in spirit to the study of defects in solids \cite{Freysoldt_VandeWalle_2014}. The key advantage of DFT is that it captures essential aspects of polaron physics, e.g., including Fr\"ohlich, Holstein, and PSSH-type couplings, without relying on empirical parameters and without the need to choose \textit{a priori} which interactions to retain in the model. Furthemore, DFT approaches naturally incorporate nonlinearities in both the lattice dynamics and the electron-phonon interactions, which are hard to capture using effective Hamiltonians.

Unsurprisingly, the advantages of DFT-based approaches are not without drawbacks. The main limitations of such approaches are (i) that the lattice is described within the \textit{adiabatic} and \textit{classical} approximation, therefore the quantum nature of vibrations and non-adiabatic effects are not included in standard treatments; (ii) that calculations in BvK supercells include interactions between periodic replica of charged polarons, which require special handling; and (iii) that in DFT calculation the polaron experiences a spurious interaction with itself via the self-consistent field, which needs to be removed. In this section, we review current DFT approaches to polarons, and techniques to overcome the above limitations. In particular, we discuss polaron energetics in Secs.~\ref{Sec:DFT-energy} and \ref{Sec:paths_barriers}, supercells and self-interaction error in Sec.~\ref{Sec:DFT_difficulty}, hybrid functionals and Hubbard-corrected DFT functionals in Sec.~\ref{Sec:DFTPolaron_hubbard_hybrid}, and recent proposals for self-interaction corrections in Sec.~\ref{Sec:DFTPolaron_SIC}. We conclude this section by reviewing, in Sec.~\ref{Sec:ml_approach}, machine learning methods for augmenting DFT simulations of polarons.

\subsection{Polaron energy and wavefunction}\label{Sec:DFT-energy}

\begin{figure}
    \centering    \includegraphics[width=0.9\linewidth]{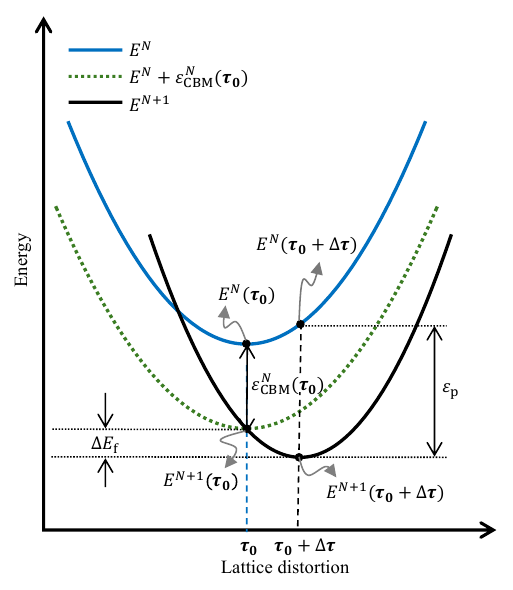}
    \caption{
    Schematic representation of the energies involved in the polaron formation process \rev{in the dilute limit, i.e. infinitely large supercells}. The solid blue line represents the $N$-electron system, and the solid black line represents the $(N + 1)$-electron system. The dashed green line represents a fictitious $(N + 1)$-electron system in which an electron has been added to the conduction band minimum ($\ve_{\rm CBM}$). The excitation energy from the $(N + 1)$-electron ground state (at $\btau_0 + \Delta \btau$) to the distorted $N$-electron state is the electron addition or removal energy $\ve_{\rm p}$. The polaron formation energy, that is the energy gained by the system when a delocalized electronic state becomes localized in a polaronic state, is represented by $\Delta E_\mathrm{f}$.}
    \label{Fig:polaron_energetics}
\end{figure}

Calculations of polarons using DFT are most often performed by adding or removing one electron in the BvK supercell, and by optimizing the resulting system. In the following we will distinguish between the insulating, charge-neutral system with $N$ electrons, and the system with an excess electron or hole, $N\pm 1$. To ensure that the relaxation of the $(N\pm 1)$-electron system yields a localized polaron, one needs to start from a slightly distorted structure so as to break translational symmetry, \rev{as well as to overcome the possible energy barrier for polaron formation~\cite{Yuan_Bevan_2019}}. This is achieved, for example, by manually modifying one or more bond lengths. Different initial distortions may lead to different local energy minima, therefore the search for the most stable polaron state requires testing multiple initial guesses. 
Once structural relaxation is complete, the Kohn-Sham orbital associated with the excess charge corresponds to the polaron wavefunction, and the total energy difference with respect to the system with excess charge and undistorted structure yields the polaron formation energy.
In the literature one finds different naming conventions for polaron energetics. Below, we make an attempt at clarifying and standardizing these conventions. 

We use $\btau$ to indicate the entire set of atomic coordinates $\{\tau_{\k p \a}\}$, where $\tau_{\k p \a}$ denotes the $\a$-th component of the coordinate of atom $\k$ in unit cell $p$ of the BvK supercell. Furthermore, we employ $\btau_0$ to denote the set of coordinates of the ground state of the $N$-electron system, and $\Delta \btau$ to indicate the lattice distortion accompanying the polaron in the $(N\pm 1)$-electron system.

The \textit{polaron formation energy}, $\Delta E_\mathrm{f}$, describes the energy lowering upon polaron formation. $\Delta E_\mathrm{f}$ is measured with respect to the ground state of the $(N\pm 1)$-electron system, in the undistorted lattice of the $N$-electron system.
Sometimes, \textit{polaron trapping energy} and \textit{polaron binding energy}~\cite{Neukirch_Tretiak_2018, Kokott_Scheffler_2018} are also used to indicate the formation energy~\cite{Elmaslmane_McKenna_2018}. 
Two definitions of $\Delta E_\mathrm{f}$ are found in the literature:
\begin{align}
\label{Eq:dft_eform_expr1}
    \Delta E_\mathrm{f} &= E^{N+1}(\btau_0+\Delta \btau)
    -
    E^{N}(\btau_0) - \varepsilon^N_\mathrm{CBM}(\btau_0),
\end{align}
and
\begin{align}
\label{Eq:dft_eform_expr2}
    \Delta E_\mathrm{f} &= E^{N+1}(\btau_0+\Delta \btau)
    -
    E^{N+1}(\btau_0),
\end{align}
where we have considered the electron polaron as an example. Here, $E^N$ and $E^{N+1}$ indicate the DFT total energies of the $N$-electron system and the $(N+1)$-electron system, respectively, and $\varepsilon^{N}_\mathrm{CBM}$ is the eigenvalue of the Kohn-Sham state at the conduction band minimum (CBM), evaluated for  the $N$-electron system. The coordinates $\btau_0$ and $\btau_0+\Delta\btau$ indicate whether the calculation refer to the undistorted crystal or to the polaron state, respectively.

In general, Eqs.~\eqref{Eq:dft_eform_expr1} and \eqref{Eq:dft_eform_expr2} are not equivalent, as the quantity $E^{N+1}(\btau_0)$ typically differs from $E^{N}(\btau_0) + \varepsilon^{N}_\mathrm{CBM}(\btau_0)$ \rev{for finite supercells} due to the presence of a double-counting term in approximate DFT functionals \cite{Giustino_2014}.
However, under certain conditions, the two expressions can be matched by tuning certain parameters in the DFT calculation. As discussed in Secs.~\ref{Sec:DFT_difficulty} and \ref{Sec:DFTPolaron_hubbard_hybrid}, this tunability can be leveraged in Hubbard-corrected DFT or hybrid functional calculations to improve polaron energetics and wavefunctions.

One concept that is often source of ambiguity in the polaron literature is the \textit{polaron eigenvalue}~\cite{Sio_Giustino_2019b, Dai_Giustino_2024b, Lafuente_Giustino_2022a}, $\varepsilon_{\rm p}$, which is also referred to as polaron level~\cite{Falletta_Pasquarello_2022a} or polaron energy~\cite{Lee_Bernardi_2021, Luo_Bernardi_2022}. This quantity describes the energy of the single-particle Kohn-Sham (KS) (or generalized Kohn-Sham) eigenstate corresponding to the polaron, and differs from the polaron formation energy. Physically, the polaron eigenvalue provides an approximate measure of the energy lowering that occurs when an excess electron is added to the $N$-electron system in the polaron structure $\btau_0+\Delta \btau$,
\begin{align}
    \label{Eq:plrn_eigval_def}
    E^{N+1}(\btau_0 + \Delta \btau) - E^{N}(\btau_0 + \Delta \btau).
\end{align}
This energy differs from the polaron formation energy, since the latter includes the energy cost of deforming the lattice from $\btau_0$ to $\btau_0+\Delta \btau$. Generally, the polaron formation energy is smaller in magnitude than the polaron eigenvalue; this effect is also seen in the Landau-Pekar model, where the eigenvalue is found to be $3\times$ the formation energy, cf.\ Sec.~\ref{Sec:LandauPekar}. We note that Eq.~\eqref{Eq:plrn_eigval_def} differs from the polaron eigenvalue due to the Hartree, exchange and correlation double-counting terms \cite{Giustino_2014}.

Another quantity that is a close relative of the polaron eigenvalue is the \textit{vertical} (or optical) charge transition level \cite{Deak_Frauenheim_2011} employed in the theory of defects in solids~\cite{Freysoldt_VandeWalle_2014, Alkauskas_Walle_2014, Bouquiaux_Gonze_2023}. This quantity is given by:
\begin{align}
\label{Eq:vertical_ctl}
    &\varepsilon^\mathrm{v} (0/\!\!-1)
    =
    E^{N+1}(\btau)
    -
    E^{N}(\btau)~.
\end{align}
The choice of $\btau$ depends on whether the charge state is transitioning from $0$ to $-1$ or the other way around. When going from $0$ to $-1$, $\btau$ should be set as $\btau_0$; for the $-1$ to $0$ transition, $\btau=\btau_0 + \Delta \btau$, which coincides with Eq.~\eqref{Eq:plrn_eigval_def}. 
It is common in the literature to report vertical charge transition levels instead of polaron eigenvalues, considering transitions from charged to neutral state~\cite{Elmaslmane_McKenna_2018, Dai_Giustino_2024c};
one exception is found in \cite{Deak_Frauenheim_2011}, where $\btau$ is chosen to $\btau_0$.
We also point out that, when the eigenvalue is evaluated via GW quasiparticle calculations \cite{Hybertsen_Louie_1986}, it has the formal meaning of additon/removal energy, and therefore it coincides with the charge transition level in Eq.~\eqref{Eq:vertical_ctl}, cf.\ also Eq.~\eqref{Eq:gw_etot}.

A related concept is the \textit{adiabatic} (or thermodynamic) charge transition level, which reflects the energy required to add or remove a charge carrier while allowing for full lattice relaxation. This quantity is defined as~\cite{Deak_Frauenheim_2011}:
\begin{align}
\label{Eq:adiabatic_ctl}
    \varepsilon^\mathrm{a} (0/\!\!-1)
    &= E^{N+1}(\btau_0+\Delta \btau)
    -
    E^{N}(\btau_0).
\end{align}
The evaluation of this quantity is complicated by the fact that the electrostatic potentials in the supercells for the $N$ and $(N+1)$-electron system are generally not aligned to the same reference. To overcome this challenge, electrostatic level alignment techniques have been developed in the context of defect calculations~\cite{Falletta_Pasquarello_2020,Freysoldt_VandeWalle_2014}. We elaborate more on this aspect in Sec.~\ref{Sec:periodicimages}.

Figure~\ref{Fig:polaron_energetics} illustrates this nomenclature using an idealized Franck-Condon diagram. Vertical transitions refer to vertical arrows in the figure, whereby the atomic position remain unchanged; while adiabatic transitions involve a change of atomic configuration.

\subsection{Polaron migration and hopping barriers} \label{Sec:paths_barriers}

\begin{figure}
    \centering
    \includegraphics[width=0.9\linewidth]{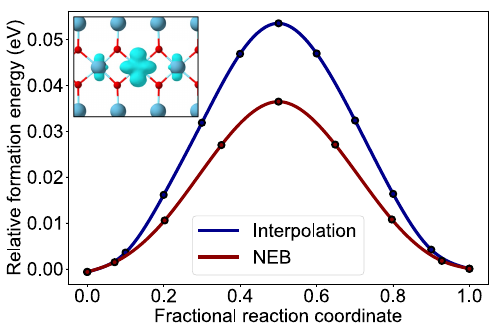}
    \caption{Polaron migration and hopping barriers computed for the electron polaron in rutile titanium dioxide, using linear interpolation (blue curve) and NEB (red curve). The inset shows the charge density isosurface of the small electron polaron in this material. The data points and the charge density isosurface are extracted from \textcite{Morita_Walsh_2023}. 
    }
    \label{fig:hopping_barrier}
\end{figure}

Once the ground state energy and wavefunction of the polaron are obtained, one may further ask: is the polaron confined to a given lattice site, or is it a mobile species that can hop to adjacent sites by overcoming trapping potential? To answer this question, one needs to determine the path for polaron hopping in the atomic configuration space, and the corresponding energy landscape along this path.

The simplest procedure to approximate the polaron hopping path is by linearly interpolating between the initial polaron structure $\Delta \btau_i$ and the final polaron structure $\Delta \btau_f$ \cite{Dai_Giustino_2024a, Lafuente_Giustino_2024, Deskins_Dupuis_2009}. This approach provides a simple and computationally-efficient estimate of the upper bound of the polaron \textit{hopping barrier}; the latter is defined as the maximum of the potential energy landscape along the path, referred to the energy of the initial configuration.

A more accurate approach to modeling polaron hopping consists in determining the migration barriers by exploring all possible paths, until a minimum-energy path is identified; this is accomplished via the nudged elastic band methods (NEB). The NEB was introduced by \textcite{Henkelman_Jonsson_2000} to study chemical reactions within transition state theory~\cite{Norskov_Bligaard_2014}. A key property of the minimum energy path is that, for any configuration along the path, the atomic forces have only components along the path~\cite{Henkelman_Jonsson_2000}. Accordingly, the central idea of NEB is to minimize the force perpendicular to a trial energy path, while keeping NEB images along the path separated via by artificial springs. The ingredients for performing NEB calculations are the forces acting on each atom, which can be obtained from DFT calculations via the Hellmann-Feynman theorem, and an externally controlled spring constant.  Following this idea, NEB was successfully employed to study polaron hopping in many systems, e.g., \ch{TiO2}~\cite{Morita_Walsh_2023} and metal halides~\cite{Tygesen_Garcia-Lastra_2023, Loftager_Garcia-Lastra_2016}. Today, NEB calculations represent the state-of-the-art tool for computing polaron migration barriers in DFT.

The migration barriers thus obtained are usually coupled to lattice models to investigate the transport properties of small polarons. These aspects are discussed in Sec.~\ref{sec:smallpolhop}. It is important to note that the discussion of polaron hopping within the context of DFT is based on the premise that the nuclear dynamics can be described within the adiabatic and classical approximations. However, in many systems of practical interest, it is often necessary to include quantum nuclear effects and nonadiabatic effects, for example by means of non-adiabatic molecular dynamics (NAMD) \cite{Prezhdo_2021}. 

\subsection{Difficulties with DFT calculations of polarons: supercells, periodic images, and self-interaction}
\label{Sec:DFT_difficulty}

\subsubsection{Supercell size}
Even though DFT calculations of polarons can be performed using widely available software, the size of the supercell required to host the polaron wavefunction often poses a challenge. In fact, DFT calculations exhibit $O(N^3)$ scaling with respect to the number $N$ of atoms~\cite{Martin_2020}, and the study of small polarons may require supercells with 100-1000 atoms. The study of intermediate and large polarons may require tens of thousands of atoms, which is impractical for explicit DFT calculations. 

In Sec.~\ref{Sec:ml_approach}, we review recent developments in accelerating these calculations using machine-learned force fields. Furthermore, in Sec.~\ref{Sec:PolaronEquations}, we discuss recent progress on \textit{ab initio} approaches to polarons that circumvent the need for large supercells by employing density-functional perturbation theory (DFPT). These techniques complement standard DFT approaches by enabling polaron calculations in large systems.

\subsubsection{Interaction between periodic images}\label{Sec:periodicimages}

When performing DFT calculations of polarons in BvK supercells, the polaron and its periodic images experience spurious Coulomb interactions, despite the compensating background charge that is introduced to maintain charge neutrality in the system.

In order to correctly capture the energetics of a polaron in isolation, it is common practice to perform  calculations for increasingly large BvK supercells, and to extrapolate the results to the limit of infinitely-extended supercell~\cite{Kokott_Scheffler_2018, Sio_Giustino_2019b, Freysoldt_VandeWalle_2014}.
It can be shown that, in three dimensions, the polaron formation energy for a periodically-arranged array of point charges scales with the linear size $L$ of the supercell as:
\begin{align}
    \label{Eq:makov-payne}
    \Delta E_{\rm f}(L) = \Delta E_{\rm f}(\infty) + a_1 \frac{1}{L} + a_3 \frac{1}{L^3},
\end{align}
with $\Delta E_{\rm f}(\infty)$ being the formation energy of the polaron in isolation, and $a_1$, $a_3$ being the coefficients that depend on the crystal structure~\cite{Makov_Payne_1995}. This equation shows how the intercept of the plot of $\Delta E_{\rm f}$  versus $L^{-1}$ yields the desired polaron formation energy in the dilute limit. While this expression is useful and elegant, we emphasize that it is valid in the limit of large $L$, therefore one often needs to reach relatively large supercells before observing the asymptotic trend expressed by Eq.~\eqref{Eq:makov-payne}~\cite{Sio_Giustino_2023},
which further adds to the computational complexity.

A more efficient alternative to direct calculations on increasingly large supercells consists of applying electrostatic corrections schemes, which are designed to eliminate the spurious interactions between periodic images of the polaron. These methods find their roots in the study of charged defects in supercells, where one faces similar issues. In these correction schemes, the polaron charge density is typically replaced by a model density so that the long-range electrostatic interaction energy can be expressed in closed form and removed from the DFT total energy. A popular expression for such a correction is~\cite{Freysoldt_Walle_2009}:
\begin{equation}
    \label{Eq:model_correction}
    E_\mathrm{corr} 
    =
    \!\!\int_\Omega d\br
    \left[
    \frac{1}{2}(n_m + n_0)
    (\Tilde{V}^\mathrm{lr}-V^\mathrm{lr})
    +
    n_0 V^\mathrm{lr} 
    \right] -
    \Delta,
\end{equation}
where $n_m(\br)$ is the model charge density, and is often taken as point charge or a Gaussian distribution, $n_0$ is the uniform compensating background charge, and $V^\mathrm{lr}(\br)$ is the electrostatic potential generated by the model charge density taken in isolation: $V^\mathrm{lr}(\br) = (4\pi \ve_0 \ve^0)^{-1}\int d {\bf r} \,n_m(\br')/\abs{\br-\br'}$. 
The quantity $\Tilde{V}^\mathrm{lr}(\br)$, on the other hand, represents the superposition of the potentials $V^\mathrm{lr}(\br)$ from all periodically-repeated supercells, reflecting the BvK boundary conditions that are employed in DFT calculations. $\Delta$ is a parameter that needs to be adjusted so that $V^\mathrm{lr}(\br)$ matches the actual electrostatic potential in DFT calculations far away from the polaron charge center; this adjustment is necessary because the energy reference in the model is generally not the same as in the DFT calculation~\cite{Freysoldt_Walle_2009, Freysoldt_VandeWalle_2014, Freysoldt_Jorg_2018}.

It should be noted that, in the above expression, the electrostatic interaction between the localized ionic charge generated from the polaronic lattice distortion is absorbed into the static dielectric constant $\ve^0$. This treatment does not fully capture the short-range screening in the vicinity of the polaron core; to incorporate these effects, more refined schemes have been successfully demonstrated~\cite{Kokott_Scheffler_2018}. Furthermore, in the study of vertical charge transition levels, a slight modification of Eq.~(\ref{Eq:model_correction}) has been proposed to account for the absence of lattice relaxation during the transition~\cite{Falletta_Pasquarello_2020}. 

\begin{figure}
  \begin{center}
      \includegraphics[width=\columnwidth]{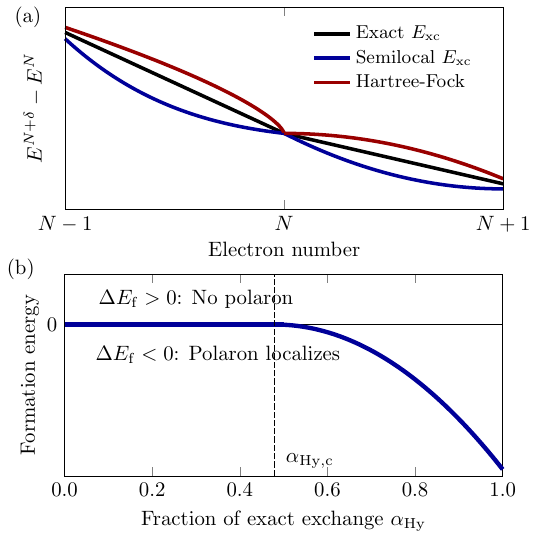}
      \vspace{-10pt}
  \end{center}
  \caption{(a) The total energy in exact DFT is a piece-wise linear function of the number of electrons (black). Semilocal DFT functionals exhibit a concave behavior vs.\ electron number, while nonlocal Hartree-Fock functionals are convex. These behaviors are related to the SIE and lead to incorrect polaron energetics and localization, see for example \cite{Elmaslmane_McKenna_2018}. (b) The effect of hybrid functionals on the polaron formation energy can intuitively be understood in terms of the Landau-Pekar model: the SIE is compensated by increasing the fraction of exact exchange [in this example, $\alpha_{\rm Hy,c} = 0.48$ corresponds to MgO, see \textcite{Kokott_Scheffler_2018}, whereas the model to generate this plot is from \textcite{Sio_Giustino_2023}].
 }
  \label{Fig:piece_wise_linearity}
\end{figure}

\subsubsection{Polaron self-interaction}

In addition to the spurious interaction between a polaron and its periodic images, DFT calculations also introduce spurious Hartree and exchange-correlation interactions between the electron or hole density of the polaron and itself, within the same BvK supercell. This issue is a particular manifestation of the general self-interaction error (SIE) of DFT \cite{Perdew_Zunger_1981}; it is especially pronounced in the case of small polarons where the electron or hole density is highly localized. We note that the SIE is distinct from the interaction between periodic images of Sec.~\ref{Sec:periodicimages}.

If one could hypothetically perform calculations using exact DFT, then the Hartree self-interaction would exactly cancel the exchange and correlation self-interaction, so that the total energy would be a piece-wise linear function of the fractional electron charge, as shown in Fig.~\ref{Fig:piece_wise_linearity}(a)~\cite{Mori_Yang_2008, Kronik_Baer_2012}.
However, in practical DFT calculations, which employ approximate exchange-correlation functionals, such a cancellation does not occur, and the SIE causes a deviation of the total energy curve from piece-wise linearity. The practical consequence of this deviation is that semilocal functionals tend to underestimate the formation energy of the polaron, and in many cases a localized polaron ground state cannot even be found.

The impact of SIE on polaron calculations via DFT can qualitatively be rationalized by invoking the Landau-Pekar model discussed in Sec.~\ref{Sec:LandauPekar}. If we temporarily rename the Landau-Pekar total energy of Eq.~\eqref{Eq:LP_toten_initial} as $E_{\rm LP}$, and we consider only Hartree self-interaction for simplicity, the energy in the presence of SIE reads \cite{Sio_Giustino_2023}:
  \begin{equation}\label{eq:SIE_LP}
  E_{\rm SIE} = E_{\rm LP} + 
  \frac{1}{2}\frac{e^2}{4\pi\varepsilon_0} \int \! d{\bf r}^\prime
  \frac{\,\,|\psi({\bf r}^\prime)|^2}{|{\bf r}-{\bf r}^\prime|},
  \end{equation}
where the last term is the self-interaction energy. This term is of the same form as the electron-lattice interaction energy in the Landau-Pekar model, Eq.~\eqref{Eq:elec_energy_to_psi}, therefore the ground state polaron energy can be evaluated as in Sec.~\ref{Sec:LandauPekar}, except that the potential energy is multiplied by the extra prefactor $1\!-\!\kappa$. Since $\epsilon^0\ge \epsilon^\infty \ge 1$, we have $\kappa>1$, hence the trapping potential of the Landau-Pekar model becomes a repulsive potential in the presence of SIE; as a result, the polaron becomes unstable. 

Hybrid functionals and Hubbard-corrected DFT, as will be discussed in the next section, mitigate the SIE but do not
remove it completely. To see this, we can rewrite Eq.~\eqref{eq:SIE_LP} by adding an exchange term as in hybrid-functional DFT: 
  \begin{equation}
   E_{\rm Hy} = E_{\rm SIE} - 
  \alpha_{\rm Hy} \frac{1}{2}\frac{e^2}{4\pi\varepsilon_0} \int \! d{\bf r}^\prime 
  \frac{\,\,|\psi({\bf r}^\prime)|^2}{|{\bf r}-{\bf r}^\prime|},
  \end{equation}
where $\alpha_{\rm Hy}$ is the fraction of exact exchange in the hybrid functional.
With this correction term, one finds that polaron localization only occurs for 
$\alpha_{\rm Hy}$ exceeding a critical value $\alpha_{\rm Hy,c} = 1 - 1/\epsilon^\infty+1/\epsilon^0$.
Beyond this point, the formation energy increases with $\alpha_{\rm Hy}$, signaling incomplete removal
of the SIE \cite{Sio_Giustino_2023}. These trends are shown schematically in Fig.~\ref{Fig:piece_wise_linearity}(b).

In practice, when using common DFT functionals such as the local-density approximation (LDA)~\cite{Ceperley_Alder_1980, Perdew_Zunger_1981} or the generalized-gradient approximation (GGA)~\cite{Perdew_Ernzerhof_1996}, the SIE typically hinders or prevents polaron formation. To circumvent this limitation, several strategies have been developed to mitigate or eliminate the SIE; these strategies are reviewed in Sec.~\ref{Sec:DFTPolaron_hubbard_hybrid} and Sec.~\ref{Sec:DFTPolaron_SIC}.

\subsection{Hybrid functionals and Hubbard-corrections}
\label{Sec:DFTPolaron_hubbard_hybrid}

\subsubsection{Hybrid functionals}

A popular method to mitigate the SIE in DFT calculations of polarons is to employ hybrid functionals \cite{Heyd_Ernzerhof_2003,Perdew_Bruke_1996}, as discussed for example by \textcite{Elmaslmane_McKenna_2018}.

Hybrid functionals have become extremely popular across the board in DFT calculations, as they typically deliver superior accuracy in total energies as compared to standard semilocal functionals. In calculations of electronic band structures, hybrid functionals are often employed to overcome the band gap problem of DFT~\cite{Kronik_Baer_2012}. At a qualitative level, since semilocal exchange-correlation functionals tend to underestimate band gaps, while Hartree-Fock calculations tend to overestimate them, it is natural to expect that by using a fraction of the exact exchange potential, 
\begin{align}
    \label{Eq:hartree_pot} 
    \alpha_{\rm Hy}\hat{V}_\mathrm{x}
    =
    -\alpha_{\rm Hy}\frac{e^2}{4\pi\e_0} 
    \sum_{n\bk,\sigma}
    f^\sigma_{n\bk}
    \frac{\ket{\psi^\sigma_{n\bk}} \bra{\psi^\sigma_{n\bk}}}
    {|\br-\br'|},
\end{align}
one might be able to obtain band gaps in better agreement with experiments with an appropriate choice of $\alpha_{\rm Hy}$. In the above expression, the $\ket{\psi^\sigma_{n\bk}}$ are the eigenstates of the KS
Hamiltonian with band index $n$, crystal momentum $\bk$, and spin index $\sigma$, and $f^\sigma_{n\bk}$ are the corresponding occupations. 
Following a similar line of reasoning, the Hartree self-interaction from semilocal DFT functionals is also partially canceled when including a fraction of exact exchange as in Eq.~\eqref{Eq:hartree_pot}. Therefore one would expect that the use of hybrid functionals could mitigate the SIE, in line with the schematic in Fig.~\ref{Fig:piece_wise_linearity}(b).

These heuristic remarks can be placed on more rigorous ground starting from 
generalized Kohn-Sham (GKS) theory, whereby the many-body system is mapped to an auxiliary system governed by nonlocal potentials~\cite{Kronik_Baer_2012}. In this framework it can be shown that the piece-wise linearity condition can be restored via a judicious choice of the nonlocal potential~\cite{Falletta_Pasquarello_2022a}, thereby effectively removing the SIE.

Several hybrid functionals, e.g., PBE0~\cite{Perdew_Bruke_1996} and HSE06~\cite{Heyd_Ernzerhof_2003}, have been adopted to calculate polarons in materials, from alkali halides, to transition metal oxides and halide perovskites; some of these applications are reviewed in Sec.~\ref{Sec:Applications}. In such calculations, it is often the case that the fraction $\alpha_{\rm Hy}$ of exact exchange is treated as an adjustable parameter. However, different choices may lead to qualitatively different results, for example a localized solution vs.\ a fully delocalized state, as illustrated schematically in Fig.~\ref{Fig:piece_wise_linearity}(b) \cite{Kokott_Scheffler_2018}. 
It is therefore critical to be able to determine the most appropriate fraction of exact exchange. 

A simple strategy would be to choose the $\alpha_{\rm Hy}$ that makes the calculated band gap match the experimental value; however, this choice typically does not yield accurate polaron formation energies~\cite{Elmaslmane_McKenna_2018}. A more refine\rev{d} approach would consist of enforcing piece-wise linearity \rev{or Koopmans' condition; in exactly solvable models, tuning the fraction of exact exchange to meet this condition has been shown to yield electron densities and quasiparticle energy gaps in close agreement with the reference solutions \cite{Elmaslmane_Godby_2018}~\cite{Falletta_Pasquarello_2022a}.} The nonlocal exchange potential poses a challenge for DFT calculations using planewave basis sets, since their computational cost scales as $\mathcal{O}(N^4)$~\cite{Lin_2016, Szabo_Ostlund_1996}. In practice, performing accurate polaron calculations with hybrid functionals, including extrapolations for finite-size corrections, remains  challenging.

\subsubsection{Hubbard-corrected DFT}

A computationally more affordable approach to investigate polaron formation is Hubbard-corrected DFT, or DFT+$U$ for short~\cite{Liechtenstein_Zaanen_1995}. The idea of DFT+$U$ is to incorporate Hubbard physics in a orbitally-selective manner, so as to take into account correlation effects beyond semilocal DFT functionals. Typically, the correction is applied to localized $d$- or $f$-orbitals in order to counter the delocalization error of DFT, which is a consequence of the SIE~\cite{Mori_Yang_2008}. The $U$ parameters in this method are meant to quantify the magnitude of the onsite Coulomb repulsion energy, and tend to favor electron localization.

Although many variants of the DFT+$U$ method have been developed~\cite{Himmetoglu_Cococcioni_2014}, one of the most popular versions is the simplified rotationally-invariant scheme of \textcite{Dudarev_Sutton_1998}. In this method, the energy functional is written as:
\begin{eqnarray}
    \label{Eq:DFT+U}
    &&\hspace{-10pt}E^\mathrm{{DFT+U}}[n_{\rm e}(\br),\{n^{I\sigma}_{\a\b}\}]
    =\nonumber \\
    &&
    E^\mathrm{{DFT}}[n_{\rm e}(\br)]
    +
    \sum_I \frac{U_I}{2}\left(
    \sum_{\a,\sigma} n^{I\sigma}_{\a\a} 
    + \sum_{\a \b, \sigma} n^{I\sigma}_{\a\b} n^{I\sigma}_{\b\a} 
    \right)\!\!,\hspace{15pt}
\end{eqnarray}
where $n_{\rm e}(\br)$ is the electron density, and the $n^{I\sigma}_{\a\b}$ are density matrices over the manifold of atomic orbitals $\phi^{I}_{\a}(\br)$ associated with the atom $I$ in the spin channel $\sigma$:
\begin{equation}
    n^{I\sigma}_{\a \b}
    =\sum_{n\bk} f^\sigma_{n\bk}
    \bra{\psi^\sigma_{n\bk}} \ket{\phi^{I}_{\a}}
    \bra{\phi^{I}_{\b}} \ket{\psi^\sigma_{n\bk}}.
\end{equation}
To see how the energy functional of Eq.~\eqref{Eq:DFT+U} mitigates the SIE, we take the functional derivative with respect to the KS state to find the corresponding potential:
\begin{equation}
    \label{Eq:DFT+U_potential}
    \hat{V}^{\mathrm{DFT}+U} = \hat{V}^\mathrm{DFT} 
    \!+\!\sum_{I, \a, \a'} \!U_I \!\left( \frac{1}{2} - n^{I\sigma}_{\a \a'} \right) 
    \ket{\phi^{I}_{\a'}} \bra{\phi^{I}_{\a}}.
\end{equation}
The second term on the right-hand side is a nonlocal operator that can be thought of as a local approximation to the Fock exchange potential in Eq.~(\ref{Eq:hartree_pot}). In this sense, DFT+$U$ can serve as an efficient, albeit approximate, strategy to incorporate nonlocal exchange in polaron calculations, and thereby to reduce the SIE. Since the computational cost of DFT+$U$ is comparable to that of a standard DFT calculation, this approach enjoys considerable popularity in polaron calculations. We discuss some of its applications in Sec.~\ref{Sec:Applications}.

\begin{figure}
    \centering
    \includegraphics[width=0.9\linewidth]{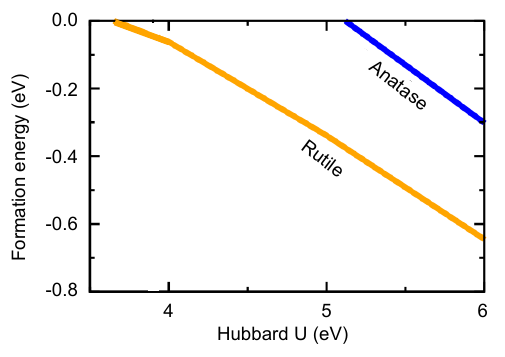}
    \caption{ Polaron formation energies of anatase (hole) and rutile (electron) as a function of $U$ in the DFT+$U$ scheme. Adapted from \textcite{Setvin_Diebold_2014}.}
    \label{Fig:eform_vs_u}
\end{figure}

As in the case of hybrid functionals, the reliance on adjustable parameters in DFT+$U$ calculations leads to some difficulties~\cite{Wang_Bevan_2024}, since different choices of $U$, or of the orbital manifold to which $U$ is applied, may lead to qualitatively different pictures of polaron formations and dynamics, as illustrated in Fig.~\ref{Fig:eform_vs_u}~\cite{Setvin_Diebold_2014}. Several methods have been developed to determine  Hubbard parameters from first principles, e.g., the constrained random phase approximation~\cite{Aryasetiawan_Lichtenstein_2004}\rev{,} the linear response method~\cite{Cococcioni_DeGironcoli_2005}\rev{, and the enforcement of the piecewise linearity of the total energy upon electron occupation \cite{Falletta_Pasquarello_2022c}}; however, general consensus on the most appropriate methodology is yet to be reached, especially for polaron studies~\cite{Himmetoglu_Cococcioni_2014}. These difficulties have motivated efforts to develop parameter-free DFT approaches to correct the SIE; the next section is devoted to such methods.

\subsection{Self-interaction corrections: one-body and many-body}
\label{Sec:DFTPolaron_SIC}

One of the earliest studies on self-interaction correction (SIC) is the seminal paper by \textcite{Perdew_Zunger_1981}, where the popular parametrization of the local density approximation was introduced. Here, the authors addressed the SIE by noting that, in an exact theory, the spurious self-interaction in the Hartree energy should be exactly canceled by the exchange-correlation functional for each state. Starting from this principle, they proposed a SIC approach that subtracts the Hartree and exchange-correlation energies from each individual orbital:
\begin{equation}\label{eq:PZ}
    \Delta E^\mathrm{PZ} = -\sum_{v\bk,\sigma} E_\mathrm{H}[n^\sigma_{v\bk}]
    -\sum_{v\bk,\sigma} E_\mathrm{xc}[n^\sigma_{v\bk}],
\end{equation}
where the $v$ runs over all the valence bands, and $E_\mathrm{H}$ and $E_\mathrm{xc}$ correspond to Hartree and exchange-correlation energies, respectively. However, this correction is orbital-dependent and therefore is not invariant under unitary rotations of the occupied manifold.

A related strategy was proposed by \textcite{Dabo_Cococcioni_2010}, in which the total energy is modified to enforce piecewise linearity with respect to electron addition or removal. The key idea is to eliminate the SIE by subtracting from the total energy the occupation-dependent contributions of the KS eigenvalues, and by adding back linear terms in the orbital occupations. This approach is designed to restore a more physical dependence of the energy on occupation, consistent with Koopmans' theorem. This methodology has since evolved into a broader class of Koopman's functionals \cite{Borghi_Marzari_2014}. Unlike Eq.~\eqref{eq:PZ}, Koopmans-compliant functionals mitigate the orbital dependence by variationally selecting a unique set of localized orbitals. While these methods have not been applied extensively to polaron studies, it will be interesting to see how they compare to alternative SIC schemes specifically designed for polarons, as we discuss below.

\textcite{Lany_Zunger_2009} introduced a ``polaron self-interaction correction'' (pSIC) that selectively eliminates the SIE of a hole polaron, while leaving the other valence electrons unaffected. The formalism is similar in spirit to the DFT+$U$ correction in Eq.~(\ref{Eq:DFT+U_potential}), except that the Hubbard potential vanishes in the absence of polaron:
\begin{align}
    \label{Eq:lany_zunger}
    \Delta \hat{V}^\mathrm{LZ} = 
    \sum_{I, \a} \lambda \left( 1 - \frac{n^{I\sigma}_{\a}} {n^{I\sigma,0}_{\a}}  \right) 
    \ket{\phi^{I}_{\a}} \bra{\phi^{I}_{\a}},
\end{align}
where $n^{I\sigma,0}_{\a}$ is the spin-resolved occupation of the local orbital $\a$ at the atomic site $I$ when the system is neutral, and $\lambda$ is a parameter used to enforce piecewise linearity. This correction compensates for the SIE in the presence of the polaron, and vanishes when ${n^{I\sigma}_{\a}} ={n^{I\sigma,0}_{\a}}$.
Even though $\Delta \hat{V}^\mathrm{LZ}$ is designed to target the polaron state, its use in self-consistent field calculations will necessarily alter also the other KS states, albeit the impact on these states is expected to be less significant than for the polaron.

In a similar spirit, \textcite{Falletta_Pasquarello_2022a, Falletta_Pasquarello_2024} proposed to address the SIE by adding a weak potential that favors localization. In their method, the potential is local and is chosen to be of the form:
\begin{equation}
    \label{Eq:falletta_pSIC}
    \Delta \hat{V}^\mathrm{FP} = q \gamma \frac{\hat{\partial V_\mathrm{xc}} }{\partial q},
\end{equation}
where $q=1$ or $-1$ for hole or electron polaron, respectively, and $\gamma$ is parameter that is used to enforce piecewise linearity. This parameter is determined self-consistently by monitoring the dependence of the polaron eigenvalue on the fractional charge $q$.

\textcite{Falletta_Pasquarello_2022a} also introduced a unified framework to distinguish ``one-body'' and ``many-body'' SIEs in DFT calculations. The former refers to the spurious Coulomb self-interaction of an electron, which is exactly canceled in Hartree-Fock theory by the exchange term. In contrast, the latter corresponds to the deviation of the total energy from piecewise linearity upon electron addition or removal; this is an exact condition for the true DFT exchange-correlation functional. The authors showed that, when piecewise linearity is restored, such as by tuning the fraction of exact exchange $\alpha_{\rm Hy}$ in hybrid functionals, the many-body self-interaction is suppressed. Furthermore, they demonstrated that, in the absence of orbital relaxation, the many-body and one-body SIEs coincide. This implies that the one-body self-interaction can be viewed as the limit of the many-body self-interaction in the absence of electronic relaxation.

In addition to the above polaron functionals, which rely on parameters to be determined \textit{a posteriori} [e.g., $\lambda$ in Eq.~(\ref{Eq:lany_zunger}) and $\gamma$ in Eq.~(\ref{Eq:falletta_pSIC})], several parameter-free functionals have been proposed in the literature. For example, \textcite{Sio_Giustino_2019b} proposed the following pSIC functional:
\begin{align}
    \label{Eq:sio_pSIC}
    &\hspace{-5pt}\Delta E^\mathrm{SVPG}[n_\uparrow+n_p, n_\downarrow] 
    \nonumber \\
    &=
    -E_\mathrm{H}[n_p]
    - \frac{1}{2}\big(
    E_\mathrm{xc}[n_\uparrow+n_p, n_\downarrow] - 2E_\mathrm{xc}[n_\uparrow, n_\downarrow]
    \nonumber \\
    &\phantom{=}+ E_\mathrm{xc}[n_\uparrow-n_p, n_\downarrow]
    \big),
\end{align}
where $n_\uparrow$, $n_\downarrow$, and $n_p$ represent the spin-up, spin-down, and polaron charge densities in the $(N+1)$-electron system, respectively. The Hartree term in this expression cancels exactly the Hartree self-interaction of the polaron; the exchange-correlation term is nothing but the finite-difference expression for the second derivative of the energy with respect to the polaron charge density,
\begin{align}
    \label{Eq:p_xc_SI}
    -\frac{1}{2} \int d\br d\br' 
    \frac{\delta ^2 E_\mathrm{xc}}{\delta n^\uparrow \delta n^\uparrow} n_p(\br) n_p(\br')~.
\end{align}
This terms cancels the exchange-correlation self-interaction up to second order in the DFT total energy functional. The advantage of this formulation is that, in the case of electron polarons, the polaron charge density is well defined and the correction does not suffer from orbital dependence as in the standard SIC method of Eq.~\eqref{eq:PZ}. The corresponding pSIC potential is obtained by differentiating Eq.~\eqref{Eq:sio_pSIC} with respect to the Kohn-Sham states and is provided in \cite{Sio_Giustino_2019b}.

Prior to \textcite{Sio_Giustino_2019b}, a similar pSIC scheme was introduced by \textcite{Avezac_Mauri_2005}, as follows:
\begin{align}
    \label{Eq:mauri_pSIC}
    &\hspace{-5pt}\Delta E^\text{DM}[n_\uparrow+n_p, n_\downarrow] 
    \nonumber \\
    &=
    -E_\mathrm{H}[n_p]
    - 
    E_\mathrm{xc}[n_\uparrow+n_p, n_\downarrow] + 2E_\mathrm{xc}[n_\uparrow, n_\downarrow].
\end{align}
By comparing this expression with Eq.~\eqref{Eq:sio_pSIC}, it is seen that the expansion of the exchange-correlation term to linear order, $-\int d\br (\partial E_\mathrm{xc}/\partial n^\uparrow ) n_p(\br)$, leads to a cancellation of the exchange-correlation potential in the KS equations. This feature can in turn modify the resulting band structure. For this reason, it is advisable to carefully examine the band structures when performing polaron calculations using Eq.~\eqref{Eq:mauri_pSIC}.

Another interesting pSIC scheme worth mentioning is that proposed by \textcite{Sadigh_Aberg_2015}. Unlike in Eqs.~(\ref{Eq:lany_zunger})-(\ref{Eq:mauri_pSIC}), this scheme does not require the modification of the KS Hamiltonian, therefore it can be used without any change to existing codes.
In this method, the total energy of the system with the polaron, $E^{N+1}(\btau^0+\Delta \btau)$, is approximated as the sum of the energy of the neutral system in the distorted structure, $E^{N}(\btau^0+\Delta \btau)$, and the eigenvalue of the frontier orbital, $\ve^N_\mathrm{CBM}(\btau^0+\Delta \btau)$. This approximation is analyzed in detail in Sec.~\ref{Sec:Connection}, and can be derived either from Eq.~\eqref{Eq:sio_pSIC} or from a many-body GW framework (cf.\ Sec.~\ref{Sec:GW-dft}).
In the approach of \textcite{Sadigh_Aberg_2015}, the polaron formation energy is written as:
\begin{align}
\label{Eq:dft_eform_expr3}
    \Delta E_{\rm f}&= E^{N}(\btau_0+\Delta \btau)
    - E^{N}(\btau_0)
    \nonumber \\
    &+\ve^N_\mathrm{CBM}(\btau_0+\Delta \btau) -\ve^N_\mathrm{CBM}(\btau_0).
\end{align}
This expression already contains SIC in the same form as in Eq.~\eqref{Eq:sio_pSIC}, except that the correction is implicit and calculations are performed using standard DFT functionals. In practical calculations, \textcite{Sadigh_Aberg_2015} employ Janak's theorem~\cite{Janak_1978} to express $\ve^N_\mathrm{CBM}$ as the derivative of the DFT total energy with respect to the electron occupation:
\begin{align}
    \label{Eq:janak}
    \ve^N_\mathrm{CBM}(\btau_0+\Delta \btau) = 
    \frac{\partial E^{\tilde{N}}(\btau_0+\Delta \btau)}{\partial \tilde{N}} \bigg|_{\tilde{N}=N+\delta},
\end{align}
and then evaluate this derivative via finite differences. This approach is advantageous because it only requires standard DFT calculations of total energies; furthermore, the ionic forces needed to determine the polaron distortion are obtained via the Hellman-Feynman theorem as in standard DFT.

The pSIC schemes given by Eq.~(\ref{Eq:lany_zunger})-(\ref{Eq:dft_eform_expr3}) are most useful in the study of small polarons, but they become too expensive computationally in the case of intermediate and large polarons, which typically require large BvK supercells.
In Sec.~\ref{Sec:PolaronEquations}, we review methods to extend the core ideas of the pSIC to perform \textit{ab initio} calculations of large polarons that are free of SIE.

At the end of this section, it is worth to emphasize that more systematic benchmarks are needed to evaluate the applicability, accuracy, and stability of all the aforementioned pSIC schemes. A general consensus on which method should be employed going forward has not yet been reached.

\subsection{Machine-learning approaches to polarons}
\label{Sec:ml_approach}

In addition to computing polaron formation energies and hopping barriers, DFT approaches can be used to simulate polaron dynamics. Since polaron hopping is inherently a rare event (cf.\ Sec.~\ref{Sec:transport}), molecular dynamics (MD) simulations must span nanosecond timescales to capture the relevant physics; this requirement poses significant computational challenges. With the advent of machine-learned interatomic potentials (MLIPs) \cite{Behler_Parrinello_2007,Bartok_Czani_2010,Batzner_Kozinksy_2023}, such simulations have recently become feasible~\cite{Eckhoff_Jorg_2020, Deng_Ceder_2023, Birschitzky_Franchini_2025}.

\begin{figure}
    \centering
    \includegraphics[width=0.98\linewidth]{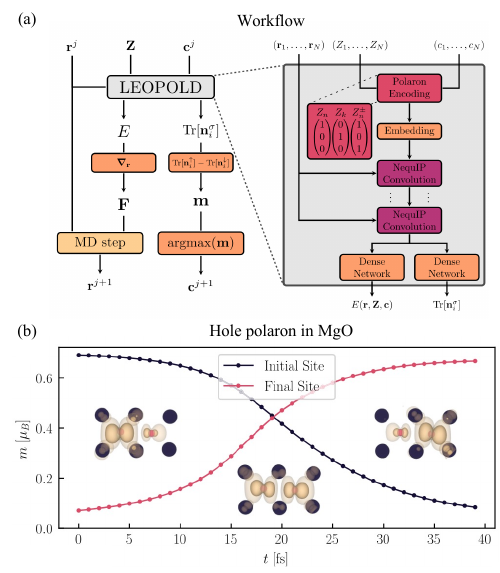}
    \caption{ML approach to polarons. (a) Workflow and model architecture from \textcite{Birschitzky_Franchini_2025}. On the left, a ML-MD loop uses the predicted energies $E$ and forces ${\bf F}$ from the LEOPOLD model to propagate the atomic positions $\br_j$. The site occupations ${\rm Tr}[\bn_i^\sigma]$ and magnetizations ${\bf m}$ are used to determine the charge state $c_i$ at the next step. On the right, the LEOPOLD architecture combines ionic positions $\br_j$, charges $Z_j$, and oxidation states $c_j$ as input. A polaron encoding appends one-hot vectors for nuclear charge $Z_j$ and charge state $c_j$ of each atom to represent the polaronic state of the system. These features are passed through NequIP \cite{Batzner_Kozinksy_2023} convolutions and dense layers to predict energeis and occupations. (b) Hopping event of small hole polarons in \ch{MgO}, as obtained from ML-MD with LEOPOLD, monitored via the evolution of the local magnetization ${\bf m}$. Polaron charge isosurfaces are shown for the initial, intermediate, and final states. From \cite{Birschitzky_Franchini_2025}.}
    \label{Fig:ml_polarons}
\end{figure}

\textcite{Birschitzky_Franchini_2025} were the first to employ a MLIP to study polaron dynamics \rev{that enforces charge conservation}, see Fig.~\ref{Fig:ml_polarons}(a). In their framework, the input features used to construct equivariant descriptors \cite{Zhang_E_2018,Batzner_Kozinksy_2023} include the atomic positions, atomic charges, and oxidation states.
Their model was trained to predict total energies, atomic forces, and, crucially, the occupations of localized orbitals associated with each atom. Since the spatial localization of polarons is governed by these orbital occupations, incorporating oxidation states as input features was essential to enable the model to learn how local charge distributions affect the atomic structure.

Their machine learning model is able to reproduce the polaronic structures and energies for electron polarons in rutile \ch{TiO2} and hole polarons in rocksalt \ch{MgO}. Furthermore, they could perform MD simulations of polaron hopping events in these systems, as shown in Fig.~\ref{Fig:ml_polarons}(b). From these simulations, they extracted the mean squared displacements as a function of time, and from there the polaron mobilities via the Einstein-Smoluchowski relation, cf.\ Sec.~\ref{Sec:transport}.

The ability to perform large-scale, long-time simulations using MLIPs for polarons offers several advantages. For example, in the case of large polarons, the simulation supercell must be sufficiently large to capture both the polaron wavefunction and the associated lattice distortion; here MLIPs could enable calculations on very large simulation cells at much lower cost than standard DFT. However, since the training data for the MLIP are typically generated using calculations in small supercells, it will be important to investigate whether these models can reliably generalize to large polarons with long-range structure. 

MLIPs can also be used to access complex systems that are beyond the scope of explicit DFT simulations, such as for example multiple polarons in the same supercell. Along this direction, \textcite{Birschitzky_Franchini_2024} developed a machine learning model to study multi-polaron configurations near defects at the (110) surface of rutile \ch{TiO2}. \rev{This model was later extended to be more efficient and automated~\cite{Yalcin_Reticcioli_2025}}. To describe such as complex system, the authors coarse-grained the simulation cell into small regions, each characterized by descriptors indicating whether it contains a polaron, a defect, or is pristine. These descriptors were used as input to a neural network trained on DFT+$U$ calculations. This type of coarse-grained featurization significantly reduces the dimensionality of the problem and makes calculations tractable, albeit at the cost of sacrificing some of the atomistic details. Yet, this approach reproduced experimentally observed polaron physics at the rutile \ch{TiO2} (110) surface, as imaged by STM. This success demonstrates the potential of such ML-based approaches to tackle polarons in complex chemical environments.

\section{Fock-space \textit{ab initio} approaches to polarons} \label{Sec:ManyBody}

The strength of the DFT-based methods for modeling polarons, discussed in Sec.~\ref{Sec:DFTPolaron}, lies in their ability to capture the atomistic details of coupled electrons and ions in real materials. However, these approaches also face fundamental limitations beyond the technical challenges outlined in Sec~\ref{Sec:DFT_difficulty}. In fact, since these methods are based on DFT, like every DFT calculation they also employ the Born-Oppenheimer approximation and treat ions as classical particles. As a result, DFT calculations of polarons fail to capture the quantum nature of phonons and the effects of non-adiabatic electron-phonon couplings. Addressing these shortcomings requires a fully quantum many-body treatment of the polaron problem.

The vast majority of many-body approaches to polarons begin with the electron-phonon Hamiltonian in the occupation number representation or Fock space \cite{Grimvall_1981,Mahan_2000}, which we here write following the notation of \textcite{Giustino_2017}:
\begin{eqnarray}\label{Eq:epi-hamilt}
  \hH &=& \sum_{n\bk} \ve_{n\bk} \hcd_{n\bk} \hc_{n\bk} +
      \sum_{\bq\nu} \hbar\w_{\bq\nu} (\had_{\bq\nu} \ha_{\bq\nu}+1/2) \nonumber \\
      &+& \!N_p^{-\frac{1}{2}}\!\! \sum_{\substack{\bk,\bq \\ m n \nu }} \!g_{mn\nu}(\bk,\bq) \,
      \hcd_{m\bk+\bq} \hc_{n\bk}(\ha_{\bq\nu}+\had_{-\bq\nu}) ~.
\end{eqnarray}
This expression is the \textit{ab initio} counterpart of Eq.~\eqref{Eq:froh_eph_H}, the difference being that we now include the electron band index $n$, the phonon branch index $\nu$, and the dependence of the electron-phonon matrix elements on these quantities. In particular, $g_{mn\nu}(\bk,\bq)$ represents the probability amplitude for the electronic transition from from the state $n\bk$ to the state $m\bk+\bq$, mediated by the phonon $\bq\nu$. The Hamiltonian refers to a periodic BvK supercell containing $N_p$ primitive unit cells, and a uniform grid of wavevectors commensurate with the periodicity of this supercell is assumed.

All the model Hamiltonians discussed in Sec.~\ref{Sec:History} can be obtained by making specific choices for the parameters $\ve_{n\bk}$, $\w_{\bq\nu}$, and $g_{mn\nu}(\bk,\bq)$ entering Eq.~\eqref{Eq:epi-hamilt}. We say that polarons calculations based on Eq.~\eqref{Eq:epi-hamilt} are \textit{ab initio} when all these parameters are determined from first principles. Typically, the electron energies are obtained by means of DFT or GW calculations \cite{Hybertsen_Louie_1986}, and phonon frequencies and electron-phonon matrix elements are from DFPT calculations \cite{Baroni_Gianozzi_2001}.

However, it should be pointed out that Eq.~\eqref{Eq:epi-hamilt} is still an effective Hamiltonian, since it is based on the following assumptions:
(i) electron-electron interactions are incorporated in the single-particle energies $\ve_{n\bk}$, so that electrons can be identified as well-defined quasiparticles in the absence of electron-phonon coupling;
(ii) phonons are described within the adiabatic and harmonic approximations; 
(iii) only linear electron-phonon coupling is retained; 
and (iv) the phonon frequencies and the electron-phonon matrix elements already incorporate electronic screening at the mean-field level.
In Sec.~\ref{Sec:FormalTheory} we go beyond these approximations, and we show how a self-consistent theory of polarons can be developed starting from the most general electron-ion Hamiltonian; this development is useful to clarify the assumptions underlying Eq.~\eqref{Eq:epi-hamilt}, and the origin of its parameters $\ve_{n\bk}$, $\w_{\bq\nu}$, and $g_{mn\nu}(\bk,\bq)$.

In the following sections, we review several many-body methods aimed at investigating polarons starting from the electron-phonon Hamiltonian in Eq.~\eqref{Eq:epi-hamilt}. 

\subsection{Canonical transformation methods}\label{sec:canonical}

As we have seen in Secs.~\ref{Sec:LLP_transform} and \ref{Sec:LF_transform}, a common strategy in the development of effective models is to introduce a unitary transformation $\hU$ such that the transformed Hamiltonian $\hH' = \hU^{-1}\hH \hU$ is more amenable to approximations in either the weak or the strong coupling limits.
Recently, these ideas have been extended to the study of realistic systems by applying unitary transformations to the more general electron-phonon Hamiltonian given in Eq.~\eqref{Eq:epi-hamilt}. 
We review these approaches in the following sections.

\subsubsection{\textit{Ab initio} Lang-Firsov transformation} \label{Sec:bernardi_polaron}

The generalization of the Lang-Firsov transformation method to an \textit{ab initio} setting has been developed by \textcite{Hannewald_Hafner_2004, Lee_Bernardi_2021}. 
The first step in this approach involves transforming the electronic degrees of freedom in Eq.~\eqref{Eq:epi-hamilt} from the Bloch basis to the Wannier basis.
Following a similar notation as in \textcite{Marzari_Vanderbilt_2012},
the Wannier states are related to the Bloch states by the following transformation:
\begin{equation} \label{Eq:Wannier2Bloch}
    \psi_{n\bk} = \sum_{m\bR} e^{i\bk \cdot \bR} U^{\dagger}_{mn\bk} w_{m\bR} ~,
\end{equation}
where $\bR$ denotes a lattice vector in the BvK supercell.
The unitary matrices $U_{mn\bk}$ are arbitrary in principle; in practice, they are typically chosen so as to obtain maximally-localized Wannier functions (MLWFs) \cite{Marzari_Vanderbilt_2012}.
Similarly, the electron annihilation operators can be transformed to real-space as:
\begin{align} \label{Eq:c_k2j_transform}
    \hc_{n\bk} &= \sum_{m\bR} e^{-i\bk\cdot\bR} U_{nm\bk} \hc_{m\bR} ~,
\end{align}
so that $\hc_{m\bR}$ represents the annihilation operator of an electron in the Wannier function $m$ at site $\bR$. Using a composite index $i$ for the pair $m,\bR$, the electron-phonon Hamiltonian in Eq.~\eqref{Eq:epi-hamilt} can be written as:
\begin{eqnarray}\label{Eq:epi-hamilt_wann}
  \hH &=& \sum_{ij} \ve_{ij} \hc_{i} \hcd_{j} +
      \sum_{\bq\nu} \hbar\w_{\bq\nu} (\had_{\bq\nu} \ha_{\bq\nu}+1/2) \nonumber \\
      &+& \!N_p^{-\frac{1}{2}}\!\! \sum_{ij, \bq \nu } \! g_{ij\nu}(\bq) \,
      \hcd_{i} \hc_{j}(\ha_{\bq\nu}+\had_{-\bq\nu}) ~.
\end{eqnarray}
In this expression, $\ve_{ij}=N_p^{2}\bra{w_{j}} \hH_{\rm e} \ket{w_{i}}$ represent the matrix elements of the electronic Hamiltonian in the Wannier representation. 
The electron-phonon matrix elements in mixed space are related to the conventional momentum-space matrix element by~\cite{Giustino_2017}:
\begin{align} \label{Eq:g_Bloch2Wann}
   & g_{m'\bR',n'\bR,\nu}(\bq) =
   \sum_{mn\bk} e^{i[(\bq+\bk)\cdot\bR'-\bk\cdot\bR]} U^{\dagger}_{m'm\bk+\bq}  U_{nn'\bk} \nonumber\\
    & \hspace{2.1cm}\times g_{mn\nu}(\bk,\bq)~.
\end{align}
The canonical transformation employed by \textcite{Lee_Bernardi_2021} is
$\hU = \exp$$(-\hS)$, with $\hS$ given by:
\begin{equation} \label{Eq:bernardi_trans}
    \hS = N_p^{-\frac{1}{2}} \sum_{ij,\bq \nu} B_{\bq\nu, ij} \hcd_{i} \hc_{j}(\had_{\bq\nu} - \ha_{-\bq\nu})~,
\end{equation}
where the coefficients $B_{\bq\nu,ij}$ are to be determined. This transformation is inspired by the Lang-Firsov transformation for the Holstein model given by Eq.~\eqref{eq:LF-transform} of Sec.~\ref{Sec:LF_transform}. A similar transformation was proposed earlier by \textcite{Hannewald_Hafner_2004}, \rev{except that the electronic energies were obtained from a DFT-parametrized tight-binding model.}

Although the canonical transformation of Eq.~\eqref{Eq:bernardi_trans} can in principle be carried out explicitly using the Baker-Campbell-Hausdorff identity \cite{Hannewald_Hafner_2004}, the transformed hopping and electron-phonon matrices contain phonon field operators, whose evaluation remains challenging. As in the Lang-Firsov transformation for the Holstein model (Sec.~\ref{Sec:LF_transform}), one can simplify by taking the expectation value of these operators on the phonon vacuum of the displaced oscillator, or perform a canonical average to include temperature effects. In practice, one applies the canonical transformation $\hU$ and then takes the average over the phonon states. The result is \cite{Luo_Bernardi_2022}:
\begin{eqnarray}\label{Eq:bernardi_plrn_hamil}
  \hH' &=& \sum_{mn} E_{ij} \hcd_{i} \hc_{j} +
      \sum_{\bq\nu} \hbar\w_{\bq\nu} (\had_{\bq\nu} \ha_{\bq\nu}+1/2) \nonumber \\
      &+& \!N_p^{-\frac{1}{2}}\!\! \sum_{\bq ij \nu } \!G_{ij\nu}(\bq) \,
      \hcd_{i} \hc_{j}(\ha_{\bq\nu}+\had_{-\bq\nu}) ~.
\end{eqnarray}
In this expression, the energies $E_{ij}$ are the counterparts of the renormalized hopping parameters in the Holstein model, and are given by:
\begin{align} \label{Eq:plrn_ene_bernardi}
    E_{ij} = &~ \left[ \ve_{ij} - \frac{2}{N_p} \sum_{\bq \nu} 
    \frac{g_{ii\nu}(-\bq) g_{ij\nu}(\bq)}{\hbar\omega_{\bq\nu}} \right] e^{-\lambda_{ij}} \nonumber \\
    &~ + \frac{1}{N_p} \sum_{\bq \nu}   \frac{|g_{ii\nu}(\bq)|^2}{\hbar \omega_{\bq\nu}} \delta_{ij},
\end{align}
where the coefficients $\lambda_{ij}$ are the band-narrowing factors:
\begin{equation} \label{Eq:narrow_factor}
    \lambda_{ij} = \frac{1}{N_p} ( n_{\bq \nu} + 1/2) |B_{\bq\nu, ii} - B_{\bq\nu, jj}|^2 ~,
\end{equation}
and $n_{\bq\nu}$ denotes the phonon Bose-Einstein occupation factors. The quantities $G_{ij\nu}(\bq)$ in Eq.~\eqref{Eq:bernardi_plrn_hamil} are renormalized electron-phonon matrix elements, and are given by:
\begin{equation}
    G_{ij\nu}(\bq) = \frac{e^{-\lambda_{ij}} g_{ij\nu}(\bq)}{\hbar\omega_{\bq\nu}} - B_{\bq\nu, ij} ~,
\end{equation}
By choosing the coherent displacement parameters $B_{\bq\nu, ij}$ so that $G_{ij\nu}(\bq)=0$ in this
expression, one is able to eliminate the electron-phonon coupling term in the second line of the transformed Hamiltonian in Eq.~\eqref{Eq:bernardi_trans} \cite{Lee_Bernardi_2021}. As a result, the polaron formation energy can be computed directly from the on-site energies $E_{ii}$ given by Eq.~\eqref{Eq:plrn_ene_bernardi}. In these derivations, it is assumed that onsite electron-phonon couplings dominate, so that $B_{\bq\nu, ij}$ can be taken to be diagonal in $i,j$,
and that the carrier density is sufficiently low for polaron-polaron interactions to be neglected; we refer to \textcite{Luo_Bernardi_2022} for a thorough discussion and extensive numerical benchmarks. 

The \textit{ab initio} Lang-Firsov transformation was first applied in the context of organic crystals \cite{Hannewald_Hafner_2004}, where the electron-phonon coupling is primarily local and the Holstein limit is approached. Using this approach, the authors investigated the temperature dependence of the electronic bandwidth in these materials. Subsequent work by \textcite{Ortmann_Hannewald_2009} extended this canonical transformation to the current operators\rev{, providing an improved analytic treatment of the current-current correlation function beyond the narrow-band approximation, and thereby enabling} calculations of the temperature-dependent mobility via the Kubo formula (cf.\ Sec.~\ref{Sec:transport}).
Recent work by \textcite{Fetherolf_Berkelbach_2020} combined this approach with a semiclassical treatment of non-local electron-phonon couplings from low-energy phonons to calculate the mobility, the optical conductivity, and the spectral function of molecular crystals, thereby capturing contributions from both polaron effects and dynamic disorder.

The canonical transformation approach was recently employed by \textcite{Lee_Bernardi_2021} to investigate polarons in ionic crystals, including alkali halides and alkali metal oxides. Using MLWFs as ansatz for the polaron wave function in Eq.~\eqref{Eq:plrn_ene_bernardi}, the authors demonstrated that, for the case of small polarons, the choices $B_{\bq \nu, mm} = g_{mm\nu}(\bq) \rev{/ \hbar \w_{\bq \nu}}$ and $\lambda_{mn} = \delta_{mn}$ yield reliable polaron energies in real materials. A subsequent study by \textcite{Luo_Bernardi_2022} extended this work, and used Eq.~\eqref{Eq:plrn_ene_bernardi} to compute polaron band structures and finite-temperature effects. This study is instructive as it also provides a formal comparison with the variational approach by \textcite{Sio_Giustino_2019b}. We elaborate on the connections between these two methods in Sec.~\ref{Sec:Connection_Denny_Bernardi}.

\subsubsection{All-coupling variational canonical transformation}\label{Sec:Reichman}

While the \textit{ab initio} Lang-Firsov-based method discussed in Sec.~\ref{Sec:bernardi_polaron} works well to study small polarons at strong coupling, it is not designed to describe band structure renormalization in the weak-coupling limit (cf.\ Sec.~\ref{Sec:LF_transform}). To overcome this limitation, \textcite{Robinson_Reichman_2025} proposed an \textit{ab initio} generalization of a related canonical transformation, first proposed by \textcite{Nagy_Markos_1989} in the context of the Fr\"ohlich model (cf.\ Sec.~\ref{Sec:FrohlichLLP}).
This transformation can be thought of as a combination of a Lang-Firsov transform (Sec.~\ref{Sec:LF_transform}) that shifts the oscillators in real space, and a Lee-Low-Pines transform (Sec.~\ref{Sec:LLP_transform}) that shifts the phonon momentum. The transformation is $\hU = e^{\hS}$, with $\hS$ given by:
\begin{equation} \label{Eq:NagyMarkos_transform}
  \hS = \sum_{n\nu, \bk \bq} B_{\bq\nu} \hcd_{n \bk+d_{\bq}\bq} \hc_{n \bk}(\had_{\bq\nu} - \ha_{-\bq\nu})~,
\end{equation}
with the shift parameters $d_{\bq}$ and $B_{\bq\nu}$ to be determined. The Lang-Firsov and Lee-Low-Pines transformations are recovered by setting $d_\bq=0$ or $d_\bq=1$, respectively.
One notable case in which this transformation can be carried out explicitly is the Fr\"ohlich Hamiltonian in Eq.~\eqref{Eq:froh_eph_H}. For this case, \textcite{Nagy_Markos_1989} showed that Eq.~\eqref{Eq:NagyMarkos_transform} yields very good agreement with Feynman's path integral results at all coupling strengths. 

In order to apply the transformation of Eq.~\eqref{Eq:NagyMarkos_transform} to the general electron-phonon Hamiltonian, Eq.~\eqref{Eq:epi-hamilt}, \textcite{Robinson_Reichman_2025} employed the trial polaron wavefunction $|\Psi\rangle = |\psi\rangle_{\mathrm{el}}|0\rangle_{\mathrm{ph}}$, where $\ket{0}_{\mathrm{ph}}$ represents the phonon vacuum. Using this variational ansatz, and performing an expansion of the electron part in terms of KS states, $\ket{\psi}_{\mathrm{el}} = N_p^{-1/2} \sum_{n\bk} A_{n\bk} \ket{\psi_{n\bk}}$, the ground-state energy of the polaron becomes:
\begin{eqnarray} \label{Eq:Reichman_E_2}
    E \!&= &\!\frac{1}{N_p} \!\sum_{n\bk} \!\bigg[ \varepsilon_{n\bk} \!-\! \sum_{\bq\nu} \frac{|B_{\bq\nu}|^2}{2} \xi_{n\bk,\bq} \bigg] |A_{n\bk}|^2 \!+\!\! \sum_{\bq\nu} \hbar\omega_{\bq\nu} |B_{\bq\nu}|^2\!\! \nonumber \\
    &-& \frac{1}{N_p}\! \sum_{\substack{mn\nu\\ \bk,\bq}}  g_{mn\nu}(\bk, \bq) B_{-\bq\nu} A_{n\bk+\bq}^{*} A_{m\bk+d_{\bq}\bq} + \mathrm{c.c.} ~,\!\!\!
\end{eqnarray}
where the energy shift $\xi_{n\bk,\bq}$ is given by: 
\begin{equation}\label{Eq:Reichman_xi}
    \xi_{n\bk,\bq} = 2 \varepsilon_{n\bk} - \varepsilon_{n\bk-d_{\bq}\bq} - \varepsilon_{n\bk+d_{\bq}\bq} ~,
\end{equation}
and the normalization condition $\langle \psi \ket{\psi}_{\mathrm{el}}=1$ has been assumed. In deriving Eq.~\eqref{Eq:Reichman_E_2}, the authors retained terms up to second order in $B_{\bq\nu}$, in line with the harmonic approximation underlying Eq.~\eqref{Eq:epi-hamilt}.

By numerically minimizing Eq.~\eqref{Eq:Reichman_E_2}, \textcite{Robinson_Reichman_2025} computed the ground state energy of electron and hole polarons in LiF from first principles, and obtained good agreement with other all-coupling \textit{ab initio} many-body methods, cf.\ Fig.~\ref{Fig:manybody_comparison}.

It is also interesting to note that, by setting $d_\bq=0$ in Eqs.~\eqref{Eq:Reichman_E_2} and \eqref{Eq:Reichman_xi}, Eq.~\eqref{Eq:Reichman_E_2} reduces exactly to the \textit{ab initio} polaron equations of \textcite{Sio_Giustino_2019a}, which we review in Sec.~\ref{Sec:plrn_eq_reciprocal}. Therefore, the method of \textcite{Robinson_Reichman_2025} can be thought of as a generalization of the method of \textcite{Sio_Giustino_2019a} that improves the ground state energy at weak and intermediate coupling.
    
\subsection{Green's function methods}

The approaches discussed in Sec.~\ref{sec:canonical} focus on the minimization of the polaron energy using a variational many-body wavefunction. Another family of many-body methods concentrates on the excitation spectrum of polarons, and focuses on computing the polaron Green's function. The single-electron Green’s function is defined as \cite{Fetter_Walecka_2003}: 
\begin{equation} \label{Eq:green_el_N}
  G_{n\bk,n'\bk'}(t, t') = -\frac{i}{\hbar} \< \, \hT\, \hc_{n\bk}(t) \hcd_{n'\bk'}(t') \, \> ~.
\end{equation}
In this expression, $\hT$ is Wick's time-ordering operator, and $\< \cdots\>$ denotes the expectation value over the many-body ground state of the charge-neutral $N$-electron system. $\hcd_{n\bk}(t)$ is the fermionic operator that creates an electron in the KS state $n\bk$ at the time $t$. For $t>t'$, the time-ordered Green's function of Eq.~\eqref{Eq:green_el_N} carries the meaning of probability amplitude for an electron added to the system in the state $n'\bk'$ at time $t'$ to be found in the state $n\bk$ at the later time $t$; for $t'<t$ the same function describes the propagation of a hole \cite{Onida_Rubio_2002}.

Green's functions describe a variety of properties of many-body systems, e.g., the ground state density and  energy, electron addition and removal energies, and experimental observables such as the photoemission and optical spectra. These functions play key roles in theories of electron-electron interactions in materials \cite{Onida_Rubio_2002}, and in the study of electron-phonon interactions \cite{Giustino_2017}. In this section, we review Green's function methods for polarons that take the effective Hamiltonian in Eq.\eqref{Eq:epi-hamilt} as the starting point.
In Sec.~\ref{Sec:FormalTheory}, we place these methods in a broader context by outlining a self-consistent Green's function framework starting from the bare electron-ion Hamiltonian, Eq.~\eqref{Eq:gen_ham}.

\subsubsection{Spectral functions and the cumulant expansion method} \label{Sec:Cumulant}

Some of the earliest many-body field-theoretic studies of polarons focused on calculating the single-particle spectral function for the coupled electron–phonon system:
\begin{equation} \label{Eq:spectral_func}
    A_{n\mathbf{k},n\bk}(\omega) = -\frac{1}{\pi} \mathrm{Im} \, G^{\mathrm{ret}}_{n\mathbf{k},n\bk}(\omega) ~,
\end{equation}
where $G^{\mathrm{ret}}$ is the ``retarded'' version of the Green's function introduced in Eq.~\eqref{Eq:green_el_N}; unlike the time-ordered Green's function, the poles of $G^{\mathrm{ret}}$ reside in the lower half of the complex plane. The spectral function carries the meaning of momentum-resolved many-body density of electronic states, and is directly proportional to measured angle-resolved photoemission spectra (ARPES) (cf.\ Sec.~\ref{Sec:Connection}). 
Electron-phonon interactions lead to kinks and satellites in the momentum-resolved spectral function, as observed in the ARPES spectra of numerous transition metal oxides and two-dimensional materials.

Starting from the work by \textcite{Engelsberg_Schrieffer_1963},
\textit{ab initio} calculations of phonon-induced mass renormalization or ``kinks'' in ARPES spectra via the Green's function have been performed for a number of systems, e.g., metallic surfaces (\citeauthor{Eiguren_Tosatti_2003}, \citeyear{Eiguren_Tosatti_2003}; \citeauthor{Eiguren_Claudia_2008}, \citeyear{Eiguren_Claudia_2008}; \citeauthor{Eiguren_Echenique_2009}, \citeyear{Eiguren_Echenique_2009}), superconductors \cite{Giustino_Louie_2008, Li_Louie_2019, You_Li_2025}, and two-dimensional materials \cite{Mazzola_Wells_2017, Garcia-Goiricelaya_Eiguren_2019}. 
These studies employ the Migdal approximation \cite{Migdal_1958}, which retains only the Fan-Migdal electron self-energy and neglects vertex corrections (cf.\ Sec.~\ref{Sec:FormalTheory} for a complete derivation).
In the language of Feynman diagrams, this corresponds to including only non-crossing electron–phonon self-energy diagrams. 

\begin{figure}
\centering
\includegraphics[width=0.9\linewidth]{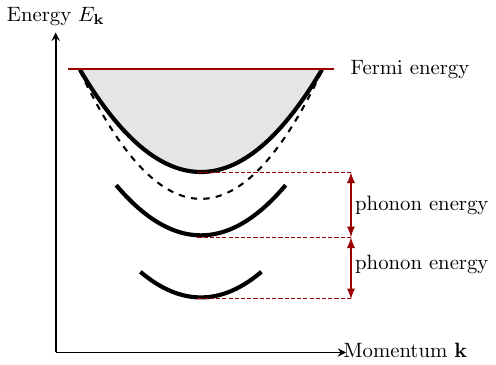}
\caption{Schematic illustration of the satellite structure induced by the electron-phonon
interaction at the conduction band bottom of a polar semiconductor or insulator. The scheme
refers, e.g., to an $n$-doped oxide with a Fermi energy comparable to the characteristic
phonon energy. 
The dashed line indicates the band structure without electron-phonon effects; 
the solid lines show the renormalized band minimum,  as well as two phonon sidebands. In ARPES
experiments, these satellite correspond to the emission of one electron plus one phonon,
or one electron plus two phonons. The energy separation between the main band and the sidebands 
is an integer multiple of the phonon energy. From \textcite{Kandolf_Giustino_2022}.}
\label{fig:kandolf}
\end{figure}

During the past decade, the ability to perform ARPES measurements on oxides with highly tunable carrier densities ~\cite{Moser_Grioni_2013, Cancellieri_Strocov_2016, Chen_Asensio_2015, Wang_Baumberger_2016, Yukawa_Matsuda_2016, Riley_King_2018, Mao_Wang_2020, Xiang_Shao_2023, Lee_Shen_2014, Krsnik_Mishchenko_2020} and on two-dimensional materials~\cite{Chen_Asensio_2018, Kang_Kim_2018} has led to the observation of additional replica bands that have been interpreted as signatures of polarons; these features are shown schematically in Fig.~\ref{fig:kandolf}. In this regime, the phonon energy become comparable to, or even exceeds, the Fermi energy, and the Migdal approximation no longer holds.

To investigate these features and their relation to polarons, several groups employed the so-called cumulant expansion approach, which was originally developed to describe electron-plasmon interactions \cite{Langreth_1970, Aryasetiawan_Karlsson_1996}. In this method, the time-evolution of the Green's function is expressed as an exponential function of an auxiliary quantity, $C_{n\mathbf{k},n\bk}(t-t')$, called the cumulant function: %
\begin{equation} \label{Eq:cumulant}
    G_{n\mathbf{k},n\bk}(t-t') = G_{n\mathbf{k},n\bk}^{0}(t-t') e^{C_{n\mathbf{k},n\bk}(t-t')} ~,
\end{equation}
where the subscript 0 denotes the non-interacting Green's function.
An approximate cumulant function can be derived by comparing the lowest-order expansion of the Green's function in Eq.~\eqref{Eq:cumulant} to the expansion of the Dyson equation for the Green's function (discussed in Sec.~\ref{Sec:FormalTheory}). With such a function, the exponential in Eq.~\eqref{Eq:cumulant} automatically generates \rev{a subset of} higher-order electron-phonon coupling diagrams \cite{Gunnarsson_Schonhammer_1994, Gumhalter_Giustino_2016}; this improvement leads to the correct description of the multiple satellites shown schematically in Fig.~\ref{fig:kandolf}. One important advantage of this method is that it requires simple post-processing of the standard Fan-Migdal self-energy \cite{Giustino_2017}, with negligible computational overhead.

\begin{figure}
    \centering
    \includegraphics[width=0.99\linewidth]{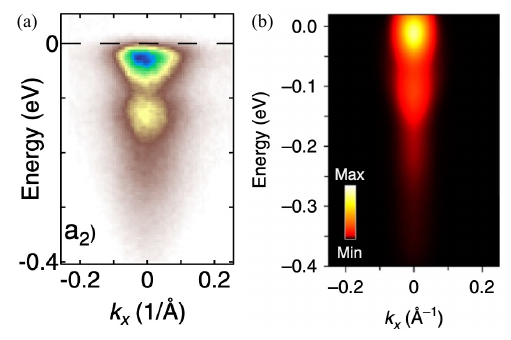}
    \caption{
    Examples of \textit{ab initio} calculations of polaronic satellite structures using cumulant expansion.(a) ARPES spectra of doped anatase \ch{TiO2}. From~\textcite{Moser_Grioni_2013}. (b) \textit{Ab initio} calculations of the electron spectral function using the cumulant expansion method. From \textcite{Verdi_Giustino_2017}. }
    \label{Fig:cumulant_tio2}
\end{figure}

The first application of the cumulant expansion method in the context of \textit{ab initio} electron-phonon calculations was reported by \textcite{Story_Rehr_2014}, where the authors investigated the spectral functions of  elemental metals using Fermi-surface averaged Eliashberg functions. Momentum-resolved spectral functions within the cumulant expansion approach were computed by \textcite{Verdi_Giustino_2017} for TiO$_2$ (Fig.~\ref{Fig:cumulant_tio2}), and achieved good agreement with measured spectra. Subsequent work by \textcite{Riley_King_2018} rationalized the multiple satellites observed in doped EuO as originating from both electron-phonon and electron-plasmon interactions. Detailed assessments of the reliability of the cumulant expansion method have been performed by \textcite{Nery_Gonze_2018}, with applications to the Fr\"ohlich model and representative polar materials; and by \textcite{Antonius_Louie_2020}, who applied the cumulant expansion method to compute the spectral functions of doped SrTiO$_3$ and ZnO. More recently, a related approach has been applied by \textcite{deAbreu_Verstraete_2022} to investigate many-body spectral functions of non-polar materials, leading to the identification of phonon-induced features in the bands dubbed ``nonpolaron'' signatures~\cite{Emin_Bussac_1994}.
The cumulant expansion method has also been combined with DFPT+$U$ electron–phonon matrix elements (cf. Sec.~\ref{Sec:DFTPolaron_hubbard_hybrid}) to compute the spectral function of \rev{the prototypical Mott insulator CoO~\cite{Zhou_Bernardi_2021} and} the parent cuprate La$_2$CuO$_4$, revealing strong Fr\"ohlich-type hole-phonon couplings and pronounced phonon sidebands \cite{Chang_Bernardi_2025}.

A finite-temperature cumulant method was developed by \textcite{Zhou_Bernardi_2019} and by \textcite{Chang_Bernardi_2022} to investigate carrier transport beyond the quasiparticle regime; here, the authors combined the cumulant spectral functions with the Kubo formula to predict temperature-dependent electron mobilities in SrTiO$_3$ and in crystalline naphtalene, respectively.
Self-consistent Green's function calculations similar in spirit to the cumulant method have also been reported recently, in the study of carrier transport  in \rev{two-dimensional and bulk materials with weak to strong electron-phonon coupling \cite{Lihm_Ponce_2025, Lihm_Ponce_2025b}.}
The use of the cumulant expansion method to model large polaron transport is discussed in more detail in Sec.~\ref{Sec:transport}.

At the end of this section, it is perhaps worth noting that the replica bands observed in doped oxides and related materials do not necessarily indicate the formation of localized polarons and cannot be used to estimate the ``size'' of a polaron. In fact, the width of the satellites bands in momentum space (Fig.~\ref{fig:kandolf}) is merely related to the size of the Fermi surface, and therefore reflects the density of carriers rather than the extent of the polaron wavepacket in real space.

\subsubsection{Diagrammatic Monte Carlo approach} \label{Sec:DMC}

\begin{figure}
    \centering
    \includegraphics[width=1.0\linewidth]{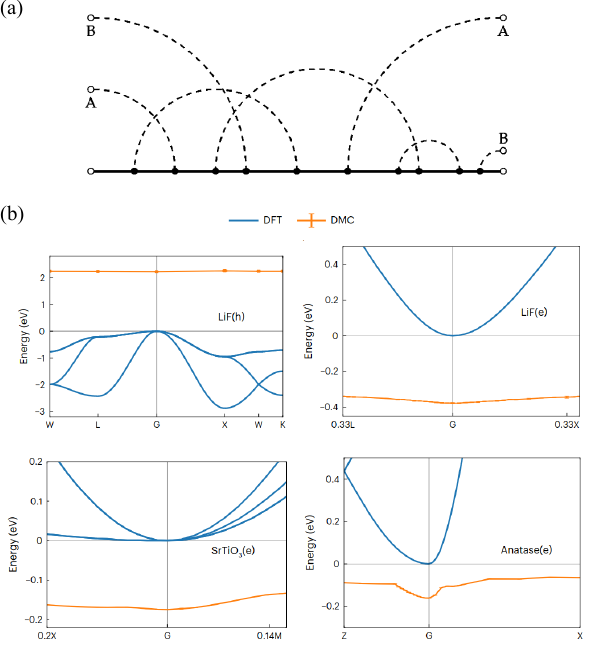}
    \caption{Diagrammatic Monte Carlo method 
    (a) Example electron-phonon diagram used in the DMC method. Solid lines represent the free electron Green's functions, dashed lines are the free phonon Green's functions, and dots  denote electron-phonon vertices. In these cases there are two oscillators in addition to the electron, indicated by A and B; therefore there are three external lines. From \textcite{Mishchenko_Svistunov_2000} (b) Band structure of hole polarons in LiF (top left), electron polarons in LiF (top right), electron polarons in SrTiO$_3$ (bottom left), and electron polarons in anatase TiO$_2$ (bottom right). Non-interacting DFT band structures are represented by the blue lines, while polaron energies obtained from \textit{ab initio} DMC are shown in orange. From \textcite{Luo_Bernardi_2025}.}
    \label{Fig:diagmc_fig}
\end{figure}

The DMC method is a powerful many-body technique to study electron-phonon interactions and polarons ({\citeauthor{Prokofev_Svistunov_1998}, \citeyear{Prokofev_Svistunov_1998}}; {\citeauthor{Mishchenko_Svistunov_2000}, \citeyear{Mishchenko_Svistunov_2000}}; {\citeauthor{Hahn_Franchini_2018}, \citeyear{Hahn_Franchini_2018}}). This method aims to compute the zero-temperature one-electron/$M$-oscillators equilibrium Green's function starting from the imaginary time representation:
\begin{align}\label{eq:G_dmc}
    G(\mathbf{p}, \{\mathbf{q}_\nu\},\tau)
    =&
    \langle  \hat{a}_{\mathbf{q}_M}(\tau) \dots \hat{a}_{\mathbf{q}_1}(\tau) \hat{c}_{\mathbf{k}}(\tau) \nonumber \\
    &\times \,
    \hat{c}_{\mathbf{k}}^{\dagger}(\tau) \hat{a}_{\mathbf{q}_1}^{\dagger}(0) \dots \hat{a}_{\mathbf{q}_M}^{\dagger}(0)
    \rangle ~.
\end{align}
As in Eq.~\eqref{Eq:green_el_N}, the expectation value is taken over the electron and phonon vacuum, which corresponds to the charge-neutral many-body ground state with $N$ electrons in the valence. The fermionic and bosonic operators are the same as in Sec.~\ref{Sec:History}, and their time dependence is to be intended in the Heinsenberg picture, e.g., $\hat{a}_{\bq\nu}(\tau) = e^{\hat{H}\tau} \hat{a}_{\bq\nu} \, e^{-\hat{H}\tau}$, where \rev{$\tau$} is the imaginary time. In Eq.~\eqref{eq:G_dmc}, the momenta $\bq_\nu$ for $\nu=1,2,\dots, M$ refer to the oscillators, and $\mathbf{p}$ denotes the total momentum of the excitation, $\mathbf{p} = \mathbf{k} + \sum_\nu \mathbf{q}_\nu$; in this sum, $\bk$ is the usual electron momentum. We note that this expression relies on the conservation of the total crystal momentum, which is a consequence of the translational invariance of Eq.~\eqref{Eq:epi-hamilt} (cf.\ Secs.~\ref{Sec:LLP_transform} and \ref{Sec:BrokenSymmetry}).

The physical meaning of $G(\mathbf{p}, \{\mathbf{q}_\nu\},\tau)$ is similar to that of Eq.~\eqref{Eq:green_el_N}, except that in this case one considers the probability amplitude for the propagation of a simultaneous electron and phonon excitation of the system. Using the resolution of identity for the electronic eigenstates of the $(N+1)$-electron system from Eq.~\eqref{Eq:epi-hamilt}, one can express the Green's function in the Lehmann spectral representation: 
\begin{equation} \label{Eq:one_e_N_ph_Greens}
    G(\mathbf{p}, \{\mathbf{q}_\nu\},\tau)
    \!=\!
    \sum_{\lambda} \! 
    Z_{\lambda}(\mathbf{p}, \{\mathbf{q}_\nu\})
    \, e^{-\left[E_{\lambda}^{N+1}(\mathbf{p})-E_{0}^{N}\right]\tau/\hbar}\!\!,
\end{equation}
where
\begin{equation}
    Z_{\lambda}(\mathbf{p}, \{\mathbf{q}_\nu\})
    =
    |\langle \Psi_{\lambda}^{N+1}(\mathbf{p}) | \hat{a}_{\mathbf{q}_M}^{\dagger} \dots \hat{a}_{\mathbf{q}_1}^{\dagger} \hat{c}_{\mathbf{k}}^{\dagger} | \Psi_{0}^{N} \rangle|^2
\end{equation}
serves as a generalized quasiparticle weight, and quantifies the overlap between the polaron eigenstate $\Psi_{\lambda}^{N+1}(\mathbf{p})$ and a state with one free electron with momentum $\mathbf{k}$ and one phonon in each oscillator channel.
This representation shows how the ground state energy of the polaron and its spectral decomposition can be obtained from the long-time limit ($\tau\rightarrow \infty$) of Eq.~\eqref{Eq:one_e_N_ph_Greens}, as for the case of Feynman's polaron in Sec.~\ref{Sec:Feynman}.

The Green's function in Eq.~\eqref{Eq:one_e_N_ph_Greens} admits an expansion in terms of an infinite set of Feynman diagrams \cite{Mattuck_1976}:
\begin{equation} \label{Eq:G_expand_diagrams}
   G(\mathbf{p}, \{\mathbf{q}_\nu\},\tau)
   =
   \sum_{n=0}^{\infty} \sum_{\xi_n} \!\int \!\mathcal{D}_{n,\xi_n} (\mathbf{p}, \{\mathbf{q}_\nu\}, \tau; \mathbf{u}) \, d\mathbf{u} ~,
\end{equation}
where $n$ denotes the expansion order, $\xi_n$ indexes topologically distinct diagrams $\mathcal{D}_{n,\xi_n}$ of the same order, and $\mathbf{u}$ represents the collection of all internal integration variables, i.e., the crystal momentum and the time of every line. For the electron-phonon Hamiltonian in Eq.~\eqref{Eq:epi-hamilt}, the diagrams $\mathcal{D}_{n,\xi_n}$ correspond  to time-ordered products of non-interacting electron Green's functions, non-interacting phonon Green's function, and electron-phonon matrix elements. The order $n$ of a diagram equals the number of electron-phonon matrix elements appearing in it.

In DMC, the Green's function $G$ in Eq.~\eqref{Eq:G_expand_diagrams} is interpreted as a probability distribution function over the external variables $(\mathbf{p}, \{\mathbf{q}_\nu\},\tau)$; the distribution is evaluated stochastically using Markov chain Monte Carlo techniques.
In practice, the sampling process is carried out using a Metropolis-Hastings algorithm,
whereby a sequence of updates is proposed to sample the entire space of Feynman diagrams, e.g., increase or lower the expansion order, swap electron-phonon vertices to change the diagram topology, and move the electron-phonon vertices along the imaginary time axis. Whether a proposed update is accepted or rejected depends on the relative statistical weight of the new diagram, which is given by the corresponding numerical value of $\mathcal{D}_{n,\xi_n}$. By collecting statistics over many Monte Carlo updates, one can evaluate the Green’s function; from there, polaron properties can be extracted with controlled statistical uncertainty.
For example, \textcite{Mishchenko_Svistunov_2000} introduced a set of estimators to collect the polaron energy, effective mass, quasiparticle weights, and phonon distribution functions.

DMC was first applied to polarons by \textcite{Prokofev_Svistunov_1998} and \textcite{Mishchenko_Svistunov_2000} in the context of the Fr\"ohlich model (see Sec.~\ref{Sec:FrohlichLLP}). This was the first highly accurate numerical method to compute the polaron formation energy and related properties at all coupling strengths; these calculations have also served as an accurate reference to benchmark Feynman's method, cf.\ Sec.~\ref{Sec:Feynman}.

This methodology has been extended to study the optical conductivity in the Fr\"ohlich model \cite{Mishchenko_Svistunov_2003}, the temperature-dependent mobility in the Holstein model \cite{Mishchenko_Cataudella_2015} (cf.\ Secs.~\ref{Sec:HolsteinLF} and \ref{Sec:transport}),
Fr\"ohlich polarons in two dimensions \cite{Hahn_Franchini_2018} (cf.\ Sec.~\ref{Sec:2d_materials}),
polaron models with \rev{nonlinear} electron-phonon interactions \cite{Ragni_Mishchenko_2023, Ragni_Mishchenko_2025} (cf.\ Sec.~\ref{Sec:higher_order_effect})\rev{, exciton polarons and their optical signatures \cite{Burovski_Mishchenko_2008, Mishchenko_Fehske_2018} (cf.\ Sec.~\ref{Sec:ExcitonPolaron})}, and a two-dimensional Holstein model at finite polaron density \cite{Mishchenko_Prokofev_2014} (cf. Supplementary Note 2), among others.

Beyond model Hamiltonians, recently \textcite{Luo_Bernardi_2025} demonstrated the first \rev{implementation} of the DMC method to the complete electron-phonon Hamiltonian, Eq.~\eqref{Eq:epi-hamilt}, with electron bands, phonon dispersions, and electron-phonon matrix elements obtained from \textit{ab initio} calculations.
This advance was made possible by two \rev{theoretical and} computational developments: a data-compression technique introduced by \textcite{Luo_Bernardi_2024}, which enables the efficient evaluation of the electron-phonon matrix elements,
and a matrix-product formalism that mitigates the sign problem arising in DMC when the Green's function describes multiple electron bands. This method was employed to calculate the polaron formation energy, effective mass, and phonon distribution in materials across the coupling spectrum, from weak coupling (electrons in LiF, anatase TiO$_2$, and SrTiO$_3$) to strong coupling (holes in LiF). The method was further extended to compute the finite-temperature mobility of polarons in anatase and rutile TiO$_2$, generalizing the work by \textcite{Mishchenko_Nagaosa_2019} to \textit{ab initio} electron-phonon couplings (cf.\ Sec.~\ref{Sec:transport}).

\rev{Before moving to the next section, it is worth mentioning that other Monte Carlo methods, such as the path-integral Monte Carlo approach~\cite{Alexandrou_Rosenfelder_1990,DeRaedt_Lagendijk_1982, Kornilovitch_1998, Alexandrov_Kornilovitch_1999, Kornilovitch_1999}, have been employed to study polarons at the level of effective Hamiltonians (Secs.~\ref{Sec:FrohlichLLP} and \ref{Sec:HolsteinLF}), but their application to \textit{ab initio} calculations of polarons has not been explored. This method could offer another interesting direction for future first-principles calculations.}

\subsubsection{Self-consistent Green's function method} \label{Sec:selfcon_Green_2022}

Instead of summing over infinitely many Feynman's diagrams, an alternative strategy is to obtain the interacting Green's function of the polaron using a self-consistent set of equations, in the same spirit as Hedin's approach to interacting electron systems (\citeauthor{Hedin_1965}, \citeyear{Hedin_1965}; \citeauthor{Kato_1960}, \citeyear{Kato_1960}; \citeauthor{Hedin_Lundqvist_1969}, \citeyear{Hedin_Lundqvist_1969}) which forms the basis for the GW method \cite{Hybertsen_Louie_1986,Onida_Rubio_2002}. 

The self-consistent Green's function method for polarons was developed by \textcite{Lafuente_Giustino_2022a, Lafuente_Giustino_2022b}, starting from the effective \textit{ab initio} Hamiltonian in Eq.~\eqref{Eq:epi-hamilt}. In this section we review this methodology, while in Sec.~\ref{Sec:FormalTheory} we make a further step by showing how this method can be derived from the most general electron-ion Hamiltonian. This extra step is useful to clarify the nature of the electron-phonon matrix elements that enter Eq.~\eqref{Eq:epi-hamilt}.

\textcite{Lafuente_Giustino_2022b} derived the equation of motion for the electron Green's function starting from Eq.~\eqref{Eq:epi-hamilt}. Within the Heisenberg picture, the equation of motion for the fermionic operators is:
\begin{align} \label{Eq:eqmo_cnk}
     i\hbar \frac{d}{d t} \hat{c}_{n\mathbf{k}}(t) = &
    \varepsilon_{n\mathbf{k}} \hat{c}_{n\mathbf{k}}(t) 
    + (2\,M_{0}\omega_{\mathbf{q},\nu}/\hbar N_p)^{1/2} \!\nonumber \\
    \times &\sum_{n' \mathbf{q} \nu } \! g_{nn'\nu}(\mathbf{k-q},\mathbf{q}) \,
    \hat{c}_{n'\mathbf{k-q}}(t) \, \hat{u}_{\mathbf{q}\nu}(t)~,
\end{align}
having introduced for brevity the normal-mode coordinate operator:
\begin{equation} \label{Eq:uqnu}
    \hat{u}_{\mathbf{q}\nu} = (\hbar/2M_{0}\omega_{\mathbf{q},\nu})^{1/2} \, ( \hat{a}_{\mathbf{q}\nu}+ \hat{a}_{-\mathbf{q}\nu}^\dagger )~,
\end{equation}
where $M_0$ is a reference mass, e.g., the mass of the proton. This relation can be combined with Eq.~\eqref{Eq:green_el_N} to obtain an equation of motion for the Green's function:
\begin{align} \label{Eq:eqmo_green}
        & \left ( i\hbar \partial/\partial t - \varepsilon_{n\mathbf{k}} \right ) G^{N+1}_{n\mathbf{k},n'\mathbf{k'}} (t,t') = 
        \delta(t-t') \, \delta_{n\mathbf{k},n'\mathbf{k'}} \nonumber \\
        &\hspace{25pt} - \frac{i}{\hbar} (2\,M_{0}\,\omega_{\mathbf{q}\nu}/N_p\hbar)^{1/2} \sum_{n'' \mathbf{q} \nu } g_{nn''\nu}(\mathbf{k-q},\mathbf{q}) \nonumber \\
        &\hspace{25pt} \times 
        \langle \, \hat{T} \, \hat{u}_{\mathbf{q}\nu}(t) \, \hat{c}_{n''\mathbf{k-q}}(t) \, \hat{c}^\dagger_{n'\mathbf{k'}}(t') \, \rangle^{N+1} ~.
\end{align}
In this expression, the superscript $N+1$ indicates that the expectation values are taken over the ground-state of the $(N+1)$-electron system. This choice is important to ensure that the self-energies derived below account for the presence of the polaron even at the lowest order \cite{Lafuente_Giustino_2022b}.

In the remainder of this section, we drop the superscript $N+1$ to keep a light notation, but it is intended that every expectation value is evaluated with respect to the $(N+1)$-electron system. 

The last term on the right-hand side of Eq.~\eqref{Eq:eqmo_green} can be handled using Schwinger's variational technique (\citeauthor{Schwinger_1951}, \citeyear{Schwinger_1951}; \citeauthor{Kato_Namiki_1960}, \citeyear{Kato_Namiki_1960}). In this approach, the Hamiltonian is augmented with an external force that acts on the oscillators:
\begin{equation} \label{eq:source_term}
    \hat{H}_{\mathrm{ext}}(t) = {\sum}_{\mathbf{q\nu}} F_{\mathbf{q}\nu}(t) \, \hat{u}_{\mathbf{q}\nu}(t) ~.
\end{equation}
This term is key to obtaining a set of self-consistent equations for the electron Green's function through functional derivatives with respect to the fictitious forces $F_{\mathbf{q}\nu}(t)$. These forces are sent to $0^+$ at the end of the derivation; they are infinitesimal but nonzero in order to break translational symmetry and enable electron localization, as discussed in Sec.~\eqref{Sec:BrokenSymmetry}. With these premises, \textcite{Lafuente_Giustino_2022b} derive the following Dyson's equation for the Green's function:
\begin{equation} \label{Eq:Dyson_inverted}
    G^{-1}_{n\mathbf{k},n'\mathbf{k'}}(t,t')
    =
    (G^{0})^{-1}_{n\mathbf{k},n'\mathbf{k'}}(t,t') - \Sigma_{n\mathbf{k},n'\mathbf{k'}}(t,t') ~,
\end{equation}
where the self-energy $\Sigma_{n\mathbf{k},n'\mathbf{k'}}(t,t')$ consists of two terms:
\begin{equation} \label{Eq:selfen_combined}
    \Sigma_{n\mathbf{k},n'\mathbf{k'}}(t,t') = \Sigma^{\mathrm{FM}}_{n\mathbf{k},n'\mathbf{k'}}(t,t') + \Sigma^{\rev{\mathrm{LP}}}_{n\mathbf{k},n'\mathbf{k'}}(t,t')~.
\end{equation} 
The first term is the standard Fan-Migdal self-energy \cite{Engelsberg_Schrieffer_1963}, which leads to band-structure renormalization effects \cite{Allen_Heine_1976, Miglio_Gonze_2020}, as well as the phonon-induced kinks and satellites discussed in Secs.~\ref{Sec:Cumulant}. This self-energy reads:
\begin{align} \label{Eq:selfen_FM_kq}
    & \Sigma^{\mathrm{FM}}_{n\mathbf{k},n'\mathbf{k'}}(t,t')
    =~
    i N_p^{-1/2} \!
    \int \! dt'' dt''' \nonumber \\
    & \times \!\!\!\sum_{\substack{n'' \mathbf{k''}  n'''\mathbf{k}'''\\ \nu \nu' \mathbf{q}'}}\!\!\!
    g_{n n'' \nu}(\mathbf{k}'',\mathbf{k-k''}) \, G_{n''\mathbf{k}'',n'''\mathbf{k'''}}(t,t''') \nonumber \\
    & \times \Gamma_{n'''\mathbf{k'''},n'\mathbf{k'},\mathbf{q}'\nu'}(t''',t',t'')
    \,D_{\mathbf{q}'\nu',\mathbf{k-k''}\nu}(t'',t) ~,
\end{align}
where $\Gamma$ is the so-called vertex function \cite{Giustino_2017}, and $D$ is the phonon Green's function; see Eq.~\eqref{Eq:green_disp} in Sec.~\ref{Sec:FormalTheory} for a definition of the phonon Green's function.

The second term in Eq.~\eqref{Eq:selfen_combined} is a static contribution originating from finite oscillator displacements in the polaronic ground state:
\begin{align} \label{Eq:selfen_P_effective_ham}
    \Sigma^{\rev{\mathrm{LP}}}_{n\mathbf{k},n'\mathbf{k'}}(t,t')
    =&~ N_p^{-1/2}
    \sum_{\nu } \, g_{nn'\nu}(\mathbf{k'},\mathbf{k-k'}) \nonumber \\
    & ~ \times \langle   \hat{u}_{\mathbf{k-k'}\nu}(t)  \rangle \delta(t-t') ~\rev{.}
\end{align}
\rev{
In \textcite{Lafuente_Giustino_2022b}, this term was referred to as the
``polaronic'' self-energy. 
Here we adopt the more specific term ``Landau-Pekar'' self-energy,
and denote it by $\Sigma^{\mathrm{LP}}$, 
to avoid ambiguity with the broader historical use of the term ``polaron'' to denote a Bloch electron dressed by lattice vibrations.
This terminology emphasizes that $\Sigma^{\mathrm{LP}}$ describes the static
contribution generated by the finite lattice distortions associated with
carrier localization in a self-trapped polaron configuration. 
As shown in Sec.~\ref{Sec:plrn_eq_reciprocal}, this self-energy leads to the \textit{ab initio} DFT polaron equations. 
In the continuum approximation, and for a carrier interacting with a single
longitudinal-optical phonon mode, \textcite{Sio_Giustino_2019b} showed that these equations reduce to the conventional
Landau-Pekar model (see Sec.~\ref{Sec:LandauPekar}). 
Therefore, $\Sigma^{\mathrm{LP}}$ can be regarded as the \textit{ab initio}
generalization of the static lattice-polarization self-energy underlying
the Landau-Pekar theory of self-trapped polarons.
}

In order to obtain a self-consistent expression in Eq.~\eqref{Eq:Dyson_inverted}, one needs to relate the expectation value of the oscillator displacements $\langle   \hat{u}_{\mathbf{k-k'}\nu}(t) \rangle$ to the electron Green's function. 
This is achieved by writing the equation of motion for the normal-mode coordinate operator starting from its Heisenberg time evolution:
\begin{align} \label{Eq:eqmo_uqv}
    \frac{d^{2}}{dt^2}
    \frac{1}{\omega_{\mathbf{q}\nu}^{2}}
    \hat{u}_{\mathbf{q}\nu}(t) 
    &=
    \frac{1}{\omega_{\mathbf{q}\nu}} [[\hat{u}_{\mathbf{q}\nu}(t),\hat{H}],\hat{H}] \nonumber \\
    &=
    - \hat{u}_{\mathbf{q}\nu}(t)
    - \frac{2}{\hbar\omega_{\mathbf{q}\nu}}
    \sqrt{\frac{\hbar}{2M_0 N_p \omega_{\mathbf{q}\nu}}} \nonumber \\
    & \times \sum_{n\mathbf{k}}
    g_{n'n\nu}^*(\mathbf{k}, \mathbf{q})
    \hat{c}_{n\mathbf{k}}^{\dagger}(t) \hat{c}_{n'\mathbf{k}+\mathbf{q}}(t)
\end{align}
This equation admits the static solution:
\begin{align} \label{Eq:u_from_g}
    \langle \hat{u}_{\bq\nu}\rev{\cancel{(t)}} \rangle 
    = &~
    \frac{2 }{\hbar\omega_{\mathbf{q}\nu}}
    \sqrt{\frac{\hbar}{2 N_p M_0\omega_{\mathbf{q}\nu}}} \nonumber \\
    & \times \sum_{n\mathbf{k}}
    g_{n'n\nu}^*(\mathbf{k}, \mathbf{q})
    \langle  \hat{c}_{n\mathbf{k}}^{\dagger}\rev{\cancel{(t)}} \hat{c}_{n'\mathbf{k}+\mathbf{q}}\rev{\cancel{(t)}} \rangle ~.
\end{align}
Combining Eqs.~\eqref{Eq:green_el}, \eqref{Eq:selfen_P_effective_ham}, and \eqref{Eq:eqmo_uqv}, one finds the following self-consistent expression for the \rev{Landau-Pekar} self-energy:
\begin{eqnarray} \label{Eq:P_from_G}
    &&\Sigma^{\rev{\mathrm{LP}}}_{n\mathbf{k},n'\mathbf{k'}}(t,t')
    =
    - \delta(t-t') 
    \, \frac{2}{N_p} \! \sum_{n'' n''' \mathbf{k''} \nu}
    \!\!\!\!\!\frac{g_{n n' \nu}(\mathbf{k}',\mathbf{k-k'}) }{\hbar\omega_{\mathbf{k-k'},\nu}} 
    \nonumber \\
    &&\times \!\! \left[ -i\hbar \, G_{n'''\mathbf{k''+k-k'},n''\mathbf{k}''}(t,t^{+}) \right]
    \! g^{*}_{n''' n'' \nu}(\mathbf{k}'',\mathbf{k-k'}).\,\,\,
\end{eqnarray}
In order to use Eqs.~\eqref{Eq:Dyson_inverted}, \eqref{Eq:selfen_combined}, \eqref{Eq:selfen_FM_kq}, and \eqref{Eq:P_from_G} in \textit{ab initio} calculations, \textcite{Lafuente_Giustino_2022b} employ the Lehmann representation for the electron Green's functions, and obtain a self-consistent generalized eigenvalue problem; these aspects are discussed in Sec.~\ref{Sec:MBPE}. 
The total energy of the polaronic ground state is obtained in terms of the Dyson orbitals and the eigenvalues through the Galitskii-Migdal formula~\cite{Galitskii_Migdal_1958} applied to Eq.~\eqref{Eq:epi-hamilt} \cite{Lafuente_Giustino_2022b}. 
In their work, this scheme was applied study the impact of polaron formation of the phonon-induced band gap renormalization of LiF. These calculations showed that the polaronic contribution can be significant in cases of strong electron-phonon coupling, such as in the valence band of LiF.
The authors also applied their method to the Fr\"ohlich model (cf.\ Sec.~\ref{Sec:FrohlichLLP}), and found that this method yields accurate polaron formation energies at all coupling strengths.

\subsection{Comparison between many-body Fock-space approaches}

\begin{figure*}
    \centering
    \includegraphics[width=0.8\linewidth]{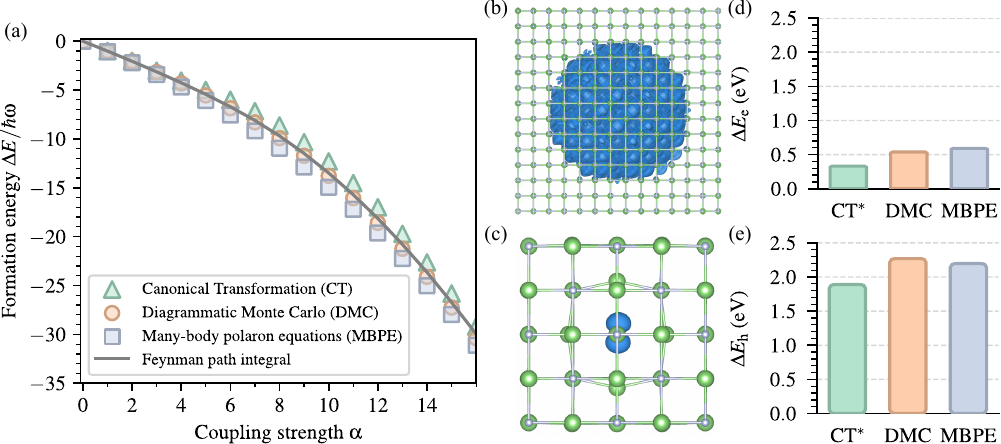}
    \caption{
Comparison between \textit{ab initio} Fock-space approaches to polarons. 
    (a) Formation energy of the Fr\"ohlich polaron, as a function of the electron-phonon coupling strength $\alpha$ (cf.\ Sec.~\ref{Sec:FrohlichLLP}). 
    Green triangles are from the canonical transformation (CT) approach of  \textcite{Nagy_Markos_1989}; 
    orange circles denote data from the DMC method ({\citeauthor{Prokofev_Svistunov_1998}, \citeyear{Prokofev_Svistunov_1998}; \citeauthor{Mishchenko_Svistunov_2000}, \citeyear{Mishchenko_Svistunov_2000}}), as calculated by \textcite{Hahn_Franchini_2018} (cf.\ Sec.~\ref{Sec:DMC});
    blue squares are from the self-consistent Green's function method (``many-body polaron equations'', MBPE) of \textcite{Lafuente_Giustino_2022b}; the gray line is from Feynman's path integral approach \cite{Feynman_1955} (cf.\ Sec.~\ref{Sec:Feynman}), as reported in \cite{Rosenfelder_Schreiber_2001}.
    Panels (b)-(e) show the applications of these methods to \textit{ab initio} calculations of polarons in LiF:
    (b) wavefunction of the large electron polaron, and (c) wavefunction of the small hole polaron.
    (d) Formation energy of the electron in LiF, as obtained from CT, DMC, and MBPE (green, orange, and blue bars, respectively).
    (e) Same as (d) but for hole polarons in LiF.
    CT data are from \cite{Robinson_Reichman_2025}, DMC data are from \cite{Luo_Bernardi_2025}, and MBPE data are from \cite{Lafuente_Giustino_2022a}. The asterisk indicates that the CT results from \cite{Robinson_Reichman_2025} do not include the Debye-Waller contribution to the self-energy; we anticipate that including this term will bring all three approaches in very close agreement.
    (b) and (c) are reproduced from \cite{Lafuente_Giustino_2022a}.
    }
    \label{Fig:manybody_comparison}
\end{figure*}

At the end of this section on Fock-space \textit{ab initio} methods for polarons, it may be useful to compare the approaches discussed in Secs.~\ref{Sec:Reichman}, \ref{Sec:DMC}, and \ref{Sec:selfcon_Green_2022}. Figure~\ref{Fig:manybody_comparison}(a) illustrates how these methods perform for the Fr\"ohlich model (cf.\ Sec.~\ref{Sec:FrohlichLLP}); we also compare with Feynman's path integral solution which is nearly a closed-form solution and is known to be very accurate. 

All three methods recover the expected behavior in both the weak- and strong-coupling limits, and they all exhibit excellent agreement with each other and with Feynman’s results throughout the entire coupling spectrum. The DMC method of \textcite{Prokofev_Svistunov_1998} agrees most closely with Feynman's results, and is expected to be the most accurate approach; the canonical transformation method of \textcite{Robinson_Reichman_2025} tends to slightly underestimate the polaron formation energy at intermediate coupling; while the self-consistent Green's function method of \textcite{Lafuente_Giustino_2022b} tends to slightly overestimate this energy in the same regime. 

We speculate that the slight overestimation of the self-consistent Green's function method may be the result of the approximate solution provided by \textcite{Lafuente_Giustino_2022b} rather than an inherent feature of the theory. In fact, in this work the authors solved the problem for the \rev{Landau-Pekar} self-energy first, and then added the Fan-Migdal self-energy in perturbation theory. We speculate that a complete self-consistent solution where the polaron and Fan-Migdal terms are treated on the same footing should deliver formation energies that are indistinguishable from DMC and Feynman's results.

The three methods discussed in this section have been used to calculate the formation energies of electron and hole polarons in the prototypical alkali halide LiF, as shown in Fig.~\ref{Fig:manybody_comparison}(b)-(e).
All three approaches predict the formation of large electron polarons with relatively small formation energies, and small hole polarons with large formation energies.
Despite significant methodological differences, the results are consistent across these approaches; in our view, this convergence represents an important milestone in the study of polarons from first principles.

We also note that both \textcite{Lafuente_Giustino_2022a} and \textcite{Luo_Bernardi_2025} included the Debye-Waller self-energy in the calculations for LiF summarized in Figs.~\ref{Fig:manybody_comparison}(d) and (e). This self-energy is discussed in Sec.~\ref{Sec:FormalTheory}, and corresponds to the second-order variation of the KS potential taken to first order in perturbation theory. The Debye-Waller self-energy is absent from Eq.~\eqref{Eq:epi-hamilt}, but is known to be important in calculations of temperature-dependent and quantum zero-point renormalization of band structures \cite{Giustino_Cohen_2010,Ponce_Gonze_2015,Lihm_Park_2020}. 

The Debye-Waller self-energy arises naturally in the \textit{ab initio} formulation of the self-consistent Green’s function method starting from the bare electron-ion Hamiltonian, cf.\ Sec.~\ref{Sec:FormalTheory}.

\section{Field-theoretic approach to polarons based on the Hedin-Baym equations}\label{Sec:FormalTheory}

In this section we develop a general theory of polarons derived from the first-principles field-theoretic Hamiltonian for interacting electrons and ions. Our aim is to formulate a unified conceptual framework that encompasses the effective Hamiltonians introduced in Sec.~\ref{Sec:History}, the density functional theory approaches discussed in Sec.~\ref{Sec:DFTPolaron}, and the Fock-space many-body methods reviewed in Sec.~\ref{Sec:ManyBody}. As we show in Sec.~\ref{Sec:field_to_Fock}, this formulation provides a rigorous justification for the Fock-space electron-phonon Hamiltonian introduced in Eq.~\eqref{Eq:epi-hamilt}, and enables the unambiguous determination of the parameters entering this effective Hamiltonian. In particular, this analysis leads to a rigorous definition of the electron-phonon matrix elements.

The derivation follows a similar strategy as that used to obtain the Hedin-Baym equations for coupled electron-phonon systems~\cite{Giustino_2017}, but extends it to account for the possibility of polaronic self-trapping. The key ideas follow closely the self-consistent Green's function approach outlined in Sec.~\ref{Sec:selfcon_Green_2022}, except that now the starting point is the most general electron-ion Hamiltonian with bare Coulomb interactions.

The derivation proceeds in four steps: (a) Definition of the \textit{ab initio} electron-ion Hamiltonian within the harmonic approximation; (b) formulation of the self-consistent Hedin-Baym equations; (c) identification of the \rev{Landau-Pekar} self-energy; and (d), connection between the polaron displacements and the electron Green’s function. These steps are outlined in Secs.~\ref{Sec:elionham}-\ref{Sec:manybody_disp} below, respectively.

\subsection{Harmonic electron-ion Hamiltonian}\label{Sec:elionham}

Here we introduce the field-theoretic electron-ion Hamiltonian, and we perform a second-order Taylor expansion of the ionic displacement operators near the equilibrium positions of the $N$-electron system; this steps corresponds to performing the harmonic approximation at the operator level, \rev{which implicitly assumes small atomic displacements in the formation of polarons;
the consequence of ignoring lattice anharmonicity is discussed in Sec.~\ref{Sec:higher_order_effect}}. In this section and in Secs.~\ref{Sec:HedinBaym}-\ref{Sec:manybody_disp}, we align with the notation of \textcite{Giustino_2017}.

The nonrelativistic Hamiltonian describing a system of interacting electrons and ions is:
\beq \label{Eq:gen_ham}
  \hH = \hT_{\textrm{e}} + \hT_{\textrm{n}} + \hU_{\textrm{ee}} + \hU_{\textrm{nn}} + \hU_{\textrm{en}},
\eeq
where $\hT_{\textrm{e}}$ is the electronic kinetic energy:
\beq \label{Eq:kin_e}
  \hT_{\textrm{e}} = -\frac{\hbar^2}{2m_{\textrm{e}}}\int\! d\bx \, \hp^\dagger(\bx) \, \nabla^2 \, \hp(\bx),
\eeq
the $\hT_{\textrm{n}}$ is the ionic kinetic energy:
\beq \label{Eq:kin_n}
  \hT_{\textrm{n}} = -\sum_{\k\a p}\frac{\hbar^2}{2M_\k}\frac{\D^2}{\D \tau_{\k\a p}^2},
\eeq
$\hU_{\textrm{ee}}$ is the electron-electron interaction:
\beq \label{Eq:u_ee}
  \hU_{\rm ee} =\! \frac{1}{2} \!\int\! d\br \int\! d\br'\, \hne(\br)\left[\hne(\br') \!-\!
                 \d(\br-\br')\right]v(\br,\br'),
\eeq
$\hU_{\textrm{nn}}$ is the ion-ion interaction:
\beq \label{Eq:u_nn}
  \hU_{\rm nn} = \frac{1}{2}\!\!\sum_{\,\,\k' p' \ne \k p}
                 \!\!\!\!Z_\k Z_{\k'} \, v( \btau^0_{\k p}\!+\!\Delta\hat{\btau}_{\k p},
                 \btau^0_{\k' p'}\!+\!\Delta\hat{\btau}_{\k' p'}),
\eeq
and $\hU_{\textrm{en}}$ is the electron-ion interaction:
\beq \label{Eq:u_en}
  \hU_{\rm en} = \int\! d\br \int \! d\br' \, \hne(\br) \hnn(\br') v(\br,\br').
\eeq
In these expressions, $\hpd(\bx)$ is the electron field creation operator, $\bx = (\br,\sigma)$ denotes both the electron position $\br$ and its spin $\sigma$, and $v(\br t, \br' t')=\d(t-t')~e^2/(4\pi\varepsilon_{0}|\br-\br'|)$ is the bare Coulomb interaction. The ionic coordinates are defined as in Sec.~\ref{Sec:DFT-energy}, and $M_\k$ is the atomic mass of ion $\kappa$. Throughout this section, we assume that the limit of infinite BvK supercell has already been taken, therefore all spatial integrals extend to the whole space. In this formulation, the equilibrium positions $\btau^0_{\k p}$ are external parameters that can be determined, for example, from X-ray diffraction measurements or from the DFT energy minimization of the $N$-electron system, without excess charge. $\dhbtau_{\k p}$ represent the ionic displacement operators. In Eqs.~\eqref{Eq:u_ee}-\eqref{Eq:u_en},  the operators $\hne(\br)$ and $\hnn(\br)$ denote the electron density and the ionic density, respectively; these operators are defined as:
\begin{align}
  \hne(\br) &= {\sum}_{\sigma}\hpd(\bx)\hp(\bx),  \label{Eq:e_density} \\
  \hnn(\br) &= - {\sum}_{\k p} Z_\k\, \d(\br-\btau^0_{\k p}-\dhbtau_{\k p}), \label{Eq:n_density}
\end{align}
where $Z_\k$ is the ionic charge.
To second order in the atomic displacements, Eq.~(\ref{Eq:n_density}) takes the form:
\begin{align} \label{Eq:ndensity_harm}
  \hnn(\br) &= \, n_{\rm n}^0(\br)  +
	         {\sum}_{\k p} Z_\k\, \dhbtau_{\k p}\cdot \nabla \d(\br-\btau_{\k p}^0) \nonumber \\
                 &-\frac{1}{2}{\sum}_{\k p} Z_\k\, \dhbtau_{\k p} \cdot \nabla \nabla
	         \d(\br-\btau_{\k p}^0) \cdot \dhbtau_{\k p},
\end{align}
where $n_{\rm n}^0(\br)$ is the \textit{classical} density of ionic charge at the equilibrium
sites $\btau_{\k p}^0$, i.e., a sum of Dirac delta functions.
Substituting Eq.~\eqref{Eq:ndensity_harm} in Eqs.~(\ref{Eq:u_nn}) and (\ref{Eq:u_en}), we obtain the ion-ion and the electron-ion
interaction energies in the harmonic approximation: 
\begin{eqnarray} \label{Eq:unn_harm}
	&&\hspace{-10pt}\hU_{\rm nn} =  \, \frac{1}{2} \sum_{\k p} Z_{\k} \sideset{}{'}\sum_{\k' p'}
    Z_{\k'} \int\! d\br \, d\br' \, v(\br,\br') \Big[ \d(\br-\boldsymbol{\tau}_{\k' p'}^0)\nonumber \\
    &&\hspace{-10pt}- \Delta\hat{\boldsymbol{\tau}}_{\k' p'}\!\cdot \!\nabla \d(\br-\boldsymbol{\tau}_{\k' p'}^0)+ \frac{1}{2}\Delta\hat{\boldsymbol{\tau}}_{\k' p'} \!\cdot\! \nabla \nabla \d(\br\!-\!\boldsymbol{\tau}_{\k' p'}^0)\! \cdot\! \Delta\hat{\boldsymbol{\tau}}_{\k' p'} \Big] \nonumber \\
    &&\hspace{-10pt}\times \Big[ \d(\br-\btau_{\k p}^0) - \dhbtau_{\k p}\cdot \nabla \d(\br-\btau_{\k p}^0) \nonumber \\
    &&\hspace{-10pt}+ \frac{1}{2}\dhbtau_{\k p} \!\cdot\! \nabla \nabla \d(\br-\btau_{\k p}^0)\! \cdot\! \dhbtau_{\k p} \Big] ~,\\
	&&\hspace{-10pt}\hU_{\rm en} =  - \int\! d\br \, \hne(\br) \,  {\sum}_{\k p}  Z_\k \, v(\br,\btau^0_{\k p}) \nonumber \\
	               &&\hspace{-10pt} \,\,+ \int\! d\br \,\, \hne(\br) \, \int\! d\br' \,
		       {\sum}_{\k p} Z_\k \,\, \dhbtau_{\k p} \!\cdot \!\nabla' \d(\br'-\btau^0_{\k p})  v(\br,\br') \nonumber \\
		       &&\hspace{-10pt} - \frac{1}{2} \int\! d\br \, \hne(\br) \int\! d\br' \,
		       {\sum}_{\k p} \!Z_\k 
		       \sum_{\a\b} \dhtau_{\k p \a} \dhtau_{\k p \b}\nonumber \\
		       &&\hspace{-10pt}\,\,\times \nabla_{\a}' \nabla_{\b}' \, \d(\br'-\btau^0_{\k p})
                       v(\br,\br') ~. \label{Eq:uen_harm}
\end{eqnarray}
where the primed summation indicates that the term $\k'p' = \k p$ is omitted. 
By combining Eqs.~\eqref{Eq:gen_ham}-\eqref{Eq:u_ee} and \eqref{Eq:unn_harm}-\eqref{Eq:uen_harm}, one obtains the  
non-relativistic field-theoretic electron-ion Hamiltonian \rev{in} the harmonic approximation.

\subsection{Hedin-Baym equations for polarons} \label{Sec:HedinBaym}

Here we use the electron-ion Hamiltonian of Sec.~\ref{Sec:elionham} to derive the self-consistent Hedin-Baym equations for the electron and phonon Green's functions. The procedure is similar to that outlined in~\cite{Giustino_2017}; the key difference is that, here, it is advantageous to work with an electron Green's function defined over the ground state of the $(N+1)$-electron system:
\begin{equation} \label{Eq:green_el}
  G^{N+1}(\bx t, \bx' t') = -\frac{i}{\hbar} \< \, \hT\, \hp(\bx t) \hpd(\bx' t') \, \>^{N+1} ~.
\end{equation}
In this expression, we use the superscript $N+1$ to keep in mind that the expectation value is taken over the ground state of the system with excess charge; in the following, we employ the same notation for all operators, e.g., $\<\,\hat{O}\>^{N+1}=\bra{N+1} \hat{O}\ket{N+1}$.

The choice of working with $G^{N+1}$, instead of the conventional $G^N$ as in Eq.~\eqref{Eq:green_el_N}, is motivated by the observation that the expectation value of the ionic displacement operators evaluated on the $(N+1)$-electron state can be nonzero when a polaron forms. The physical picture is that of an insulator or semiconductor with filled valence bands, and with an extra electron in the conduction bands. In this section we focus on the case of an excess electron for conciseness; in the case of an excess hole, 
\rev{that is, a $(N-1)$-electron system,}
the derivation proceeds along the same lines. Admitting nonvanishing displacements in the ground state is useful to clearly identify the polaron contribution to the electron self-energy, and to connect to the DFT-based approaches discussed in Sec.~\ref{Sec:DFTPolaron}; the meaning of this choice is discussed further in Sec.~\ref{Sec:BrokenSymmetry}.

The counterpart of Eq.~\eqref{Eq:green_el} for the ionic degrees of freedom is the displacement-displacement correlation function, also loosely referred to as the phonon Green's function:
\begin{equation} \label{Eq:green_disp}
  D^{N+1}_{\k p \a,\k' p' \a'}(t,t') 
	= -\frac{i}{\hbar} \< \, \hT\, \Delta \hat{\tau}_{\k p \a}(t) \,\Delta\hat{\tau}_{\k' p' \a'}(t') \, \>^{N+1}.
\end{equation}
This function describes the probability amplitude to find the ion \rev{$\k$} in the unit cell \rev{$p$} displaced by the amount \rev{$\Delta{\tau}_{\k p \a}$} at the time $t$, after the ion \rev{$\k'$} in cell \rev{$p'$} has been displaced by the amount \rev{$\Delta{\tau}_{\k' p' \a'}$} at the earlier time $t'$. 

Starting from the definition of electron and phonon Green's functions in Eqs.~\eqref{Eq:green_el} and \eqref{Eq:green_disp}, and using the harmonic electron-ion Hamiltonian derived in Sec.~\ref{Sec:elionham}, one can obtain a set of equations for the electron Green's function, following the same steps as in \cite{Giustino_2017}. For brevity, here we only mention that the derivation is based on writing equations of motion for $G^{N+1}(\bx t, \bx' t')$, and introducing a self-consistency condition via the Schwinger functional derivative technique \cite{Schwinger_1951}. The result is:
\begin{eqnarray}
   && \left[ \, i\hbar \, \D/\D t_1 + \frac{\hbar^2}{2m_{\mathrm{e}}}\nabla^2(1) -V_{\mathrm{tot}}^{N+1}(1) \, \right] G^{N+1}(12) \nonumber \\
      && \hspace{1cm}- \Sigma^{N+1}(13) G^{N+1}(32)  = \d(12) , \label{Eq:dyson} 
\end{eqnarray}
\vspace{-15pt}
\begin{multline}
\Sigma^{N+1}(12) = i\hbar G^{N+1}(13) \Gamma^{N+1}(324) \\
\times \left[ W_{\mathrm{e}}^{N+1}(41^+) + W_{\mathrm{ph}}^{N+1}(41^+) \right] , \label{Eq:selfen}
\end{multline}
\begin{multline}
\Gamma^{N+1}(123) \!=\! \d(12)\d(13) \!+ \left[\frac{\d\Sigma^{N+1}(12)}{\d G^{N+1}(45)}\right] \\
\times G^{N+1}(46) G^{N+1}(75) \Gamma\rev{^{N+1}}(673) , \label{Eq:vertex} 
\end{multline}
\begin{equation}
W^{N+1}_{\mathrm{e}}(12) = v(12) + v(13) \, P^{N+1}_{\mathrm{e}}(34) W^{N+1}_{\mathrm{e}}(42) , \vspace{3pt}\label{Eq:Wel}
\end{equation}
\vspace{-15pt}
\begin{equation}
P^{N+1}_{\mathrm{e}}(12) \!=\! -i\hbar \sum_{\sigma_1} G^{N+1}([1]3) G^{N+1}(4[1^+]) \Gamma^{N+1}(342) , \label{Eq:Pelec}
\end{equation}
\vspace{-15pt}
\begin{equation}
\e_{\mathrm{e}}^{N+1}(12) = \d(12) - v(13) P^{N+1}_{\mathrm{e}}(32) , \label{Eq:epsel}
\end{equation}
\vspace{-15pt}  
\begin{eqnarray}
    W_{\mathrm{ph}}^{N+1}(12) &=& D^{N+1}_{\k \a p,\k' \a' p'}(t_3 t_4)
    \nonumber \\
        & \times &\left(\e_{\mathrm{e}}^{-1}\right)^{N+1}\!\!(13) \nabla_{3,\a} V_{\k}(\br_3\,-\,\boldsymbol{\tau}_{\k p}^0) \nonumber \\
        & \times &\left(\e_{\mathrm{e}}^{-1}\right)^{N+1}\!\!(24) \nabla_{4,\a'} V_{\k'}(\br_4\!-\!\boldsymbol{\tau}_{\k' p'}^0).\,\,\, \label{Eq:Wph}
\end{eqnarray}
In these equations, a compact notation for the variables has been used, for example $(\bx t)$ or $(\br t)\rightarrow 1$, $(\bx' t')$ or $(\br' t')\rightarrow 2$, $(\br  t\!+\!\eta)\rightarrow 1^+$, and so on; below we adopt this shorthand notation whenever is convenient. Moreover, summations over repeated subscripts and integrals over repeated numbered indices are implied in Eqs.~\eqref{Eq:dyson}-\eqref{Eq:Wph}, except for numbers within square brackets.
Beyond the electron and phonon Green's functions, the new quantities appearing in Eqs.~\eqref{Eq:dyson}-\eqref{Eq:Wph} are: the Coulomb potential generated by the total charge density:
\begin{equation} \label{Eq:Vtot}
	V^{N+1}_{\mathrm{tot}}(1) = \int\! d2 \, v(12) \, \< \hn(2) \>^{N+1} ~;
\end{equation}
the electron self energy, $\Sigma^{N+1}$, which is discussed in Sec.~\ref{Sec:polselfen}; the vertex function $\Gamma^{N+1}$; the screened Coulomb interaction resulting from the electronic polarization, $W^{N+1}_{\mathrm{e}}$; the electronic polarizability, $P^{N+1}_{\mathrm{e}}$; the electronic dielectric matrix $\epsilon_\mathrm{e}^{N+1}$; and the screened Coulomb interaction resulting from the ionic polarization, $W_{\mathrm{ph}}^{N+1}$. Auxiliary quantities include the total charge density, $\hn = \hn_{\rm e} + \hn_{\rm n}$, and the bare ionic potential, $V_{\k}(\br) = -Z_\k e^2(4\pi\epsilon_{0})/|\br|$.

The total charge density required in Eq.~\eqref{Eq:Vtot} can be obtained from the electron Green's function, the phonon Green's function, and the expectation values of the displacement operators:
\begin{equation} \label{Eq:echarge_dens}
  \< \hne(1) \>^{N+1} = -i\hbar \sum_{\sigma_1} G^{N+1}(11^+) ,
\end{equation}
\vspace{-15pt}
\begin{eqnarray} \label{Eq:ncharge_dens}
  \hspace{-0.5cm}&& \<\hnn(\br t)\>^{N+1} = 
   n_{\rm n}^0(\br) \nonumber \\
  && \hspace{1cm}+ {\sum}_{\k p} Z_\k\, \<\Delta\hat{\boldsymbol{\tau}}_{\k p}(t)\>^{N+1} \cdot \nabla \d(\br-\boldsymbol{\tau}_{\k p}^0) \nonumber \\
  && \hspace{1cm}-\,\frac{i\hbar}{2}\!\!\sum_{\k p, \a \a'}\!\! Z_\k 
  \frac{\D^2 \d(\br-\boldsymbol{\tau}_{\k p}^0)}{\D r_\a \D r_{\a'}}
  D^{N+1}_{\k \a p,\k \a' p}(t^+ t).\hspace{1cm}
\end{eqnarray}
Equations~\eqref{Eq:dyson}-\eqref{Eq:Wph} are a subset of the Hedin-Baym equations for coupled electrons and phonons at equilibrium (\citeauthor{Giustino_2017}, \citeyear{Giustino_2017}; \citeauthor{Hedin_Lundqvist_1969}, \citeyear{Hedin_Lundqvist_1969}; \citeauthor{Baym_1961}, \citeyear{Baym_1961}). A generalization of these equations to the non-equilibrium case was recently proposed by \textcite{Steffanucci_Perfetto_2023}; the present formalism is recovered from that of \textcite{Steffanucci_Perfetto_2023} when evaluating their equations in the equilibrium limit.

We emphasize that the second term on the right-hand side of Eq.~\eqref{Eq:ncharge_dens} contains the expectation value of the displacement operator; this term vanishes in a pristine, charge-neutral crystal with translational periodicity. However, in the presence of strong electron-phonon couplings, a broken-symmetry state might emerge and $\<\Delta\hat{\boldsymbol{\tau}}_{\k p}(t)\>^{N+1}$ maybe be nonzero, leading to a pinned polaron; the physical origin of broken symmetry is discussed in Sec.~\ref{Sec:BrokenSymmetry}.

In Sec.~\ref{Sec:polselfen} we decompose the electron self-energy $\Sigma^{N+1}$ in such a way as to clearly identify the contribution that leads to polaron formation; then, in Sec.~\ref{Sec:manybody_disp}, we close the loop by deriving self-consistent equations for the phonon Green's function and for the expectation values of the ionic displacements.

\subsection{Fan-Migdal, \rev{Landau-Pekar}, and Debye-Waller self-energies}\label{Sec:polselfen}

Here we isolate the electron-phonon interactions in the Hedin-Baym equations; this step allows us to identify the contributions of the polaron distortions to the electron self-energy.

We introduce an auxiliary system wherein nuclei are fixed at their equilibrium positions.
This can be realized by evaluating expectation values over a $(N+1)$-electron state with  \textit{ions clamped at the equilibrium configuration of the $N$-electron system}.
The corresponding equation of motion for the electron Green’s function in this case takes the form:
\begin{eqnarray} \label{Eq:green_eq_cn}
	&&\hspace{-16pt} \left[ i\hbar \frac{\D}{\D t_1} \!+\! \frac{\hbar^2}{2m_{\rm e}}\nabla^2(1)
	\!-\hspace{-5pt} \int d2 \, v(12)\,  n^{\mathrm{cn},N+1} (2) \right] G^{\mathrm{cn}, N+1}(12)\nonumber \\ 
    &&  
    - \int\,\,d3\,\Sigma_{\mathrm{e}}^{\mathrm{cn}, N+1}(13) \, G^{\mathrm{cn}, N+1}(32) 
    = \d(12).
\end{eqnarray}
where ``cn'' stands for {clamped nuclei}. 
In this expression, $n^{\mathrm{cn},N+1}$ is the ground-state expectation value of the total density of the $(N+1)$-electron system with the nuclei fixed at the equilibrium positions of the $N$-particle system.
In Eq.~\eqref{Eq:green_eq_cn},
$\Sigma_{\mathrm{e}}^{\mathrm{cn}, N+1} = i\hbar  G^{\mathrm{cn}, N+1} \Gamma^{\mathrm{cn},N+1}W_{\mathrm{e}}^{\mathrm{cn}, N+1}$ is the component of the self-energy from Eq.~\eqref{Eq:selfen} that arises from electron-electron in interactions.

The complete equation of motion, Eq.~\eqref{Eq:dyson}, can be written in terms of Eq.~\eqref{Eq:green_eq_cn} by means of the following Dyson equation:
\begin{align} \label{Eq:dyson_ep}
	&G^{N+1}(12) = G^{\mathrm{cn}, N+1}(12) \nonumber \\ 
    & \hspace{10pt}+ \int\! d(34)\, G^{\mathrm{cn}, N+1}(13)\, \Sigma_{\mathrm{ep}}^{N+1}(34) \, G^{N+1}(42),  
\end{align}
where $G^{\mathrm{cn}, N+1}$ plays the role of the non-interacting Green's function. In this form, all electron-phonon interactions are contained in the self-energy $\Sigma_{\mathrm{ep}}^{N+1}$; after some algebra, this self-energy can be expressed as the sum of three contributions \cite{Giustino_2017}:\footnote{The complete expression also contains a correction term arising from the fact that the full Green's function and the screened Coulomb interaction may slightly differ from their counterparts evaluated at clamped nuclei; this term is called $\Sigma^{\rm dGW}$ in \cite{Giustino_2017}.}
\begin{equation} \label{Eq:selfen_ep}
	\Sigma_{\mathrm{ep}}^{N+1} = \Sigma_{\mathrm{FM}}^{N+1}+\Sigma_{\rev{\mathrm{LP}}}^{N+1}+\Sigma_{\mathrm{DW}}^{N+1} ~.
\end{equation}
The term $\Sigma_{\mathrm{FM}}^{N+1}$ originates from Eq.~\eqref{Eq:selfen}, and corresponds to the Fan-Migdal self-energy, as in Eq.~\eqref{Eq:selfen_FM_kq}:
\begin{equation} \label{Eq:selfen_FM}
    \Sigma_{\mathrm{FM}}^{N+1}(12) = i\hbar \!\! \int \! \! d(34)\, G^{N+1}(13) \Gamma^{N+1}(324) W_{\mathrm{ph}}^{N+1}(41^+)  ~.
\end{equation}
The terms $\Sigma_{\rev{\mathrm{LP}}}^{N+1}$ and $\Sigma_{\mathrm{DW}}^{N+1}$ in Eq.~\eqref{Eq:selfen_ep} originate from the difference between the expectation value of the total density appearing in Eq.~\eqref{Eq:dyson}, $\< \hn \>^{N+1}$, and the expectation value of the density at clamped nuclei in Eq.~\eqref{Eq:green_eq_cn}, $n^{\mathrm{cn},N+1}$. In particular, $\Sigma_{\rev{\mathrm{LP}}}^{N+1}$ arises from the Taylor expansion of $\< \hn \>^{N+1}$ to first order in the ionic displacements:
\begin{equation} \label{Eq:selfen_P}
  \Sigma_{\rev{\mathrm{LP}}}^{N+1}(12) =
	\, \d(12) \sum_{\k \a p} \frac{\D V_{\mathrm{tot}}^{N+1}(1)}{\D\tau_{\k\a p}^0} \, \<\Delta \hat{\tau}_{\k \a p}(t_1) \>^{N+1} ~.
\end{equation}
This self-energy modifies the electronic excitation energy via the variation of the total potential associated with a displacement of the ions from the equilibrium positions of the $N$-electron system, as it occurs in a \rev{self-trapped }polaron state. 
For this reason, $\Sigma_{\rev{\mathrm{LP}}}^{N+1}$ \rev{is} referred to as the \rev{Landau-Pekar} self-energy \rev{(see also Eq.~\eqref{Eq:selfen_P_effective_ham})}.

\rev{Its diagrammatic structure,
shown in Fig.~\ref{Fig:diagrams}(c), 
is reminiscent of the Hartree contribution to the electron self-energy \cite{Martin_Ceperley_2016}, 
since both are tadpole-like static mean-field terms. 
The Hartree term is generated by the valence electronic density, 
whereas $\Sigma_{\mathrm{LP}}^{N+1}$ is generated by the electron density of the excess electron or hole. 
In contrast to the Hartree term, 
the self-trapping term is absent from the clamped-nuclei reference structure and appears only after electron addition or removal induces a static lattice distortion relative to the neutral equilibrium geometry. 
This distortion is determined by the excess charge density associated with the polaron, as shown in Eq.~\eqref{Eq:steady_state_disp}.}

It is interesting that a closely related self-energy has been identified in the study of electron-phonon interactions out of equilibrium; in that context, the self-energy is referred to as Ehrenfest self-energy \cite{Marini_Gonze_2015, Steffanucci_Perfetto_2023}. Here, we maintain the nomenclature $\Sigma_{\rev{\mathrm{LP}}}^{N+1}$ to emphasize that the \rev{Landau-Pekar} and Ehrenfest self-energies are not the same: the former contains the screened electron-phonon matrix elements via the variation of $V_{\mathrm{tot}}^{N+1}$ (see Sec.~\ref{Sec:field_to_Fock}), while the latter contains bare matrix elements~\cite{Steffanucci_Perfetto_2023}. 

The last term in Eq.~\eqref{Eq:selfen_ep}, $\Sigma_{\mathrm{DW}}^{N+1}$, is the Debye-Waller self-energy and arises from the second-order term in the Taylor expansion of $\< \hn \>^{N+1}$ with respect to the ionic displacements:
\begin{equation} \label{Eq:selfen_DW}
    \Sigma_{\mathrm{DW}}^{N+1}(12) = 
    i\hbar\, \d(12)\!\!\! \sum_{\substack{\k \a p\\\k' \a' p'}}\!\!\! \frac{1}{2}\frac{\D^2  V_{\mathrm{tot}}^{N+1}(1)}
    {\D\tau_{\k\a p}^0\D\tau_{\k'\a' p'}^0} D_{\k \a p,\k' \a' p'}^{N+1}(t_1^+,t_1\!).
\end{equation}
This term can be interpreted as a static correction to the crystal potential, arising from the ionic charge density fluctuation around the equilibrium positions.

\subsection{Phonon Green's function and polaron displacements} \label{Sec:manybody_disp}

In this section, we derive an explicit expression for the phonon Green's function in Eq.~\eqref{Eq:green_disp}, and complete the Hedin-Baym equations by relating the ionic displacements to the electron and phonon Green's functions.

To apply Schwinger's functional derivative technique \cite{Schwinger_1951}, we introduce an external source field coupled to the ionic displacements, similar to Eq.~\eqref{eq:source_term} but this time in real space: $\hH_{\mathrm{F}}(t) = \sum_{\k p} {\bf F}^{\mathrm{ext}}_{\k p}(t) \cdot \dhbtau_{\k p}(t)$.
In the derivation of the Hedin-Baym equations by \textcite{Giustino_2017}, ${\bf F}^{\mathrm{ext}}_{\k p}(t)$ is a formal device that is set to zero at the end. For polaronic states, however, this infinitesimal force must be retained to break translational symmetry and allow for localization (cf.\ Sec.~\ref{Sec:BrokenSymmetry}).

Following the same procedure as for the normal mode operators in Eq.~\eqref{Eq:eqmo_uqv}, we obtain the equation of motion for the displacement operators in Newton-like form: $M_\k d^2 \Delta\hat{\boldsymbol{\tau}}_{\k p}(t)/d t^2 = -(M_\k/\hbar^2)[[\Delta\hat{\boldsymbol{\tau}}_{\kappa p}(t),\hH],\hH]$. Using Eqs.~\eqref{Eq:gen_ham}-\eqref{Eq:uen_harm} and the expression for ${\bf F}^{\mathrm{ext}}_{\k p}(t)$ inside this equation, taking the expectation value over the $|N+1\>$ state, and approximating $\< \hn \dhbtau\> = \<\hn\> \<\dhbtau\>$ \cite{Gillis_1970}, we obtain:
\begin{align} \label{Eq:tau_eqmotion_main}
  & M_\k \frac{d^2}{d t^2} \< \Delta\hat{\boldsymbol{\tau}}_{\k p}(t) \>^{\!N\!+\!1}
  = {\bf F}^{\mathrm{ext}}_{\k p}(t) \nonumber \\
  &\!\!-\!\!  Z_\k\!\!\! \int\!\!\! d\br  d\br'\! \< \hn^{(\k p)}(\br t) \>^{\!N\!+\!1} v(\br, \br')  \nabla^\prime \d(\br'\!\!-\!\btau_{\k p}^0\!\!-\!\! \< \Delta\hat{\boldsymbol{\tau}}_{\k p}(t) \>^{\!N\!+\!1} ),
\end{align}
In this expression, $\hn^{(\k p)}$ denotes the total electronic and nuclear charge density excluding the contribution from nucleus $\k p$, and $\nabla^\prime$ indicates that the derivatives are taken with respect to the variable $\br'$.

\subsubsection{Phonon self-energy}

Taking the functional derivative of Eq.~\eqref{Eq:tau_eqmotion_main} with respect to the external force yields the equation of motion for the phonon Green's function:
\begin{align} \label{Eq:eqmo_D}
   & \hspace{-7pt}M_\k\frac{\D^2}{\D t^2} 
   D_{\k \a p, \k' \a' p'}^{N+1} (t t')= -\delta_{\k \a p,\k' \a' p'}\delta(tt')  \nonumber \\
   &\hspace{-7pt}-\!\!\!\!\sum_{\k'' \a'' p''}\int \!\!dt'' 
   \Pi_{\k \a p, \k'' \a'' p''}^{N+1}(t t'') D_{\k'' \a'' p'',\k' \a' p'}^{N+1}(t''t'), 
\end{align}
where $\Pi^{N+1}$ is the phonon self-energy:
\begin{eqnarray} \label{eq:ph_selfen_t}
	&&\hspace{-2pt}\Pi_{\k \a p, \k' \a' p'}^{N+1}(t t') =  \int\!\!d\br d\br' \left\{ 
	   Z_\k \nabla_{\!\a}\, [ \d(\br\!-\!\btau_{\kappa p}^0) \right. \nonumber \\
	   && \hspace{-10pt} - \< \Delta\hat{\boldsymbol{\tau}}_{\k p}(t) \>^{N+1} \cdot \nabla \d(\br - \boldsymbol{ \tau}_{\k p}^0)] W_{\mathrm{e}}^{N+1}(\br t, \br' t')\, \nonumber \\
	   && \hspace{-10pt} \times Z_{\k'}\nabla^{\prime}_{\a'} [ \d(\br'\!-\!\boldsymbol{\tau}_{\k' p'}^0) 
	    - \< \Delta\hat{\boldsymbol{\tau}}_{\k' p'}(t') \>^{N+1} \cdot \nabla' \d(\br' - \boldsymbol{ \tau}_{\k' p'}^0) ] \nonumber \\
	&& \hspace{-10pt} \left. +\,\delta_{\k p, \k' p'}\delta(t t')\, \nabla_\a \< \hn (\br)\>^{N+1}  v(\br,\br') 
   Z_{\k'} \nabla^\prime_{\!\a'} \d(\br'\!-\!\btau_{\kappa' p'}^0) \right\}. \nonumber \\ && 
\end{eqnarray}
The derivation of these equations follows the procedure of \textcite{Giustino_2017}, but retains terms proportional to $\< \Delta\hat{\boldsymbol{\tau}}_{\k p} \>$. The function $\Pi_{\k \a p, \k' \a' p'}^{N+1}(t t')$ carries the physical meaning of many-body matrix of interatomic force constants.
The main difference between Eq.~\eqref{eq:ph_selfen_t} and the result in \cite{Giustino_2017} is that, here, the derivatives of the screened Coulomb interaction between ions are evaluated for the distorted configuration.

\begin{figure*}[t]
        \includegraphics[width=1.0\linewidth]{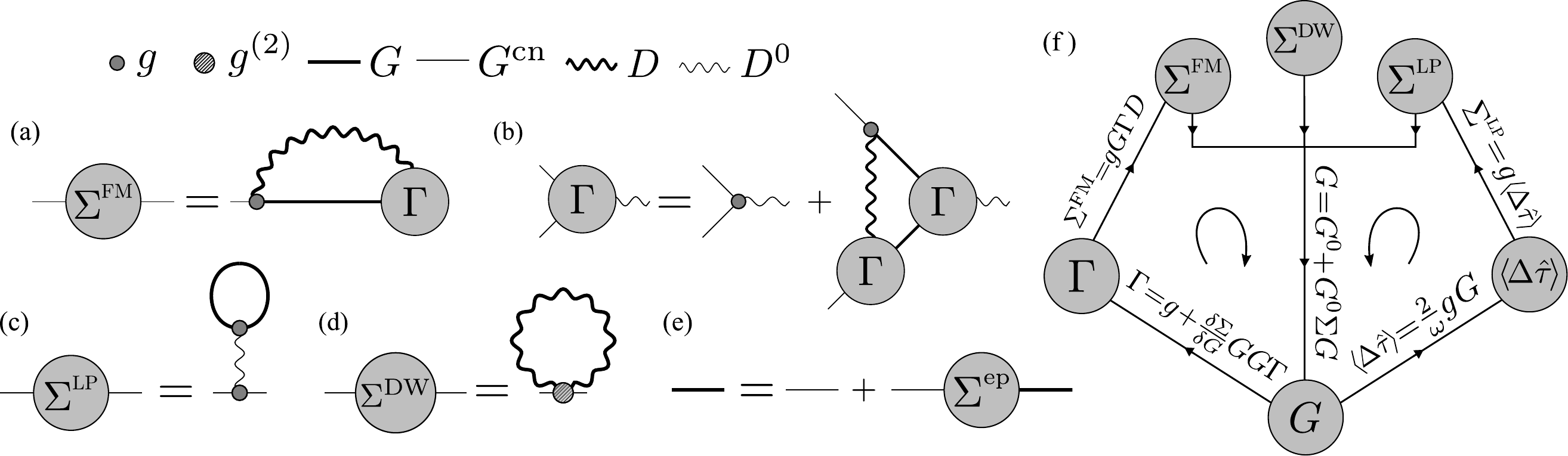}
        \caption{
        Diagrammatic representation of the self-consistent Green's function theory of polarons.
        Legends for the different parts of the diagrams are given in the upper left corner.
        (a) Fan-Migdal self-energy.
        (b) Self-consistent definition of the vertex function.
        (c) \rev{Landau-Pekar} self-energy.
        (d) Debye-Waller self-energy.
        (e) Dyson equation for the electron Green's function.
        (f) Schematic representation of the self-consistent solution.  
        The diagrammatic representation of the self-energy and Dyson equation for the phonon Green's function can be found in Fig.~1 of \textcite{Giustino_2017}.
        Adapted from \textcite{Lafuente_Giustino_2022b}.
        \label{Fig:diagrams}}
\end{figure*}

\subsubsection{Ionic displacements}

The equation of motion for the ionic displacements, Eq.~\eqref{Eq:tau_eqmotion_main}, contains the forces generated by all the electrons in the system. To achieve a practically useful many-body theory of polarons, it is advantageous to disentangle the force generated by the excess electron from the forces resulting from all the other electrons. 

To this end, we partition the charge density of the $(N+1)$-electron system into three contributions: (i) the total electron and ion density \rev{(excluding the contribution from nucleus $\k p$)} of the $N$-electron system in the distorted ionic configuration $\btau^{N+1}$ of the $(N+1)$-electron system, which we indicate by $\<\hn^{(\k p)}\>^N_{\btau^{N+1}}$; (ii) the charge density of the excess electron in the $(N+1)$-electron system, which we indicate by $\Delta n_{\rm e}$; and (iii) the difference between the density of all other electrons in the $(N+1)$-electron system and the density of the $N$-electron system, which we denote by $\Delta n_{\rm pol}$ (for ``polarization''):
 \begin{equation}\label{Eq:breakdown}
  \< \hn^{(\k p)}\>^{N+1} = \<\hn^{(\k p)}\>^N_{\btau^{N+1}} +   \Delta n_{\rm e} +
    \Delta n_{\rm  pol}~.   
\end{equation}
The charge density $\Delta n_{\rm e}$ of the excess electron is obtained from Eq.~\eqref{Eq:green_el}, by considering the poles lying above the chemical potential $\mu_N$ of the charge-neutral system (i.e., the conduction states in a gapped system)\footnote{\rev{For a hole polaron, the analogous quantity is defined as the
electron-density change associated with the missing electron. 
With the same time-ordered Green's-function convention,
\[
\Delta n_h(\mathbf r)
=
\frac{\hbar}{\pi}
\sum_\sigma
\int_{\mu_{N-1}/\hbar}^{\mu_N/\hbar}
d\omega\,
\operatorname{Im}
G^{N-1}_\sigma(\mathbf r,\mathbf r;\omega),
\]
where the integration window selects the empty hole-polaron state of the
\(N-1\)-electron system, 
i.e., the state associated with the missing electron 
in the valence manifold of a gapped system. 
}
}
: 
\begin{equation} \label{Eq:nexc_G}
    \Delta n_{\rm e}(\br) = \frac{\hbar}{\pi} \int_{\mu^{N}/\hbar}^{\mu^{N+1}/\hbar} \!\!\!d\omega \, \mathrm{Im}\left[ G^{N+1}(\br, \br ; \omega) \right],
\end{equation}
where we used the convention $f(\w) = \int dt f(t) e^{i\w t}$ for the Fourier transform between time and frequency domains, and summation over spin indices is implied.
The variation $\Delta n_{\rm pol}$ of the electron density of the $N$-electron system in response to the excess charge $\Delta n_{\rm e}$ can be written in terms of the polarizability $P_{\rm e}^N$, cf.\ Eq.~\eqref{Eq:Pelec} (to be used with $N$ instead of $N+1$). In fact, using Eq. (92) of \cite{Giustino_2017} one finds:
\begin{equation}
\Delta n_{\rm pol}(1) =  \!\!\int \!\!d(23) P^N_{\rm e}(12) v(23) [\Delta n_{\rm e}(3)+\Delta n_{\rm pol}(3)], 
\end{equation}
which combined with Eq.~\eqref{Eq:epsel} (written for $N$ instead of $N+1$), and using $W_{\mathrm{e}}(12)=W_{\mathrm{e}}(21)$ \cite{Hedin_1965} yields:
\begin{equation} \label{Eq:deltan_from_eps}
    \Delta n_{\rm e}(1) +
    \Delta n_{\rm pol}(1)
    =\int d2
    \left[ \e_{\mathrm{e}}^{N} \right]^{-1}\!(2,1)
    \, \Delta n_{\rm e}(2) ~.
\end{equation}
Using Eqs.~\eqref{Eq:breakdown} and \eqref{Eq:deltan_from_eps} inside Eq.~\eqref{Eq:tau_eqmotion_main}, we obtain: 
\begin{eqnarray} \label{Eq:eqmo_epsilon}
  && M_\k \frac{\D^2}{\D t^2} \< \dhtau_{\kappa \a p}(t) \>^{N+1} =
  {\bf F}^{\mathrm{ext}}_{\k p}(t)  - \int d\br'' dt'' \Delta n_{\rm e}(\br'' t'') \nonumber\\
  &&  \times \int\!\!d\br \, \left[ \e_{\mathrm{e}}^{N} \right]^{-1}\!(\br'' t'', \br t)
  \left.\frac{\D V_{\k}(\br)}{\D \btau_{\k p}}\right|_{\btau_{\kappa p}^0 + \<\dhbtau_{\kappa p}(t)\>^{N+1}} \nonumber \\
  && - \sum_{\k' \a' p'} \int dt' \, \Pi^{N, \btau^{N+1}}_{\k \a p, \k' \a' p'}(t, t') \<\dhtau_{\kappa' \a' p'}(t')\>^{N+1},  
\end{eqnarray}
where $V_\kappa$ is the bare Coulomb potential of ion $\kappa$, and the last line is obtained from the identity:
\begin{multline} \label{Eq:elastic_force_constants}
    Z_\k \int\!\!d\br\,d\br' \, \<\hn^{(\k p)}(\br t)\>^{N, \btau^{N+1}} v( \br, \br') \\
    \times \nabla^\prime \d(\br'\!-\!\btau_{\kappa p}^0 - \<\dhbtau_{\kappa p}(t)\>^{N+1}) \\
    =
    \sum_{\k' p' \a'} \int dt' \, \Pi^{N, \btau^{N+1}}_{\k p \a, \k' p' \a'}(t, t') \<\dhbtau_{\kappa' p' \a'}(t')\>^{N+1}~.
\end{multline}
The physical interpretation of Eq.~\eqref{Eq:eqmo_epsilon} is as follows: the ions obey a Newton-like equation, with the inertial term on the left hand side, and the forces on the right hand side. The forces comprise of (i) the external Schwinger force in the first line of Eq.~\eqref{Eq:eqmo_epsilon}; (ii) the force generated by the excess electron $\Delta n_{\rm e}$, in the second and third lines of Eq.~\eqref{Eq:eqmo_epsilon}; and (iii) the elastic restoring force of the lattice, in the third line of Eq.~\eqref{Eq:eqmo_epsilon}.

When interested in ground-state properties at equilibrium, we look for the steady-state solution of Eq.~\eqref{Eq:eqmo_epsilon} in the absence of external forces. In this case, the ionic displacements in the polaron ground state are: 
\begin{eqnarray} \label{Eq:steady_state_disp}
   &&\hspace{-10pt}\<\dhtau_{\k \a p}\>^{N+1} 
    \!\! = \!
   - \! \!\! \sum_{\k' \a' p'} \!\!\! \left[\Pi^N_{\btau^{N+1}}(\w\!=\!0)\right]^{-1}_{\k \a p, \k' \a' p'} 
   \!\! \int\! d\br \, \Delta n_{\rm e}(\br) \nonumber \\
   &&  \times \!\!\int\!d\br' \, \left[ \e_{\mathrm{e}}^{N} \right]^{-1}\!(\br,\br'; \w\!=\!0)
  \left.\frac{\partial V_{\k'}(\br')}{\partial \btau_{\k' p'}}\right|_{\btau_{\kappa' p'}^0 + \<\dhtau_{\kappa' \a' p'}\>^{N+1}}\!\!\!.\,\,\,
\end{eqnarray}
This result constitutes the many-body generalization of the atomic displacements in the \textit{ab initio} polaron equations of \textcite{Sio_Giustino_2019a}; it reduces to Eq.~(25) of that work when the phonon-self-energy is replaced by the DFT interatomic force constants and the screening is treated at the DFT level, cf.\ Sec.~\ref{Sec:plrn_eq_reciprocal}. 
In the following, when we do not indicate the frequency explicitly, we refer to $\w=0$ quantities.

We emphasize three important aspects of Eq.~\eqref{Eq:steady_state_disp}: (i) the charge density appearing in  this equation corresponds to the excess electron only;\footnote{We note that Eq.~(29) of \cite{Lafuente_Giustino_2022b} erroneously contains the total electron density; the total density of that equation should be replaced by the $\Delta n_{\rm e}$ of the present derivation.} (ii) the effect of all the other electrons is contained in the dielectric matrix and the interatomic force constants; (iii) the effective interaction between the excess electron and the ions is screened by all the other electrons.

Together with Eqs.~\eqref{Eq:dyson}-\eqref{Eq:Wph}, Eqs.~\eqref{Eq:selfen_FM}-\eqref{Eq:selfen_DW}, \eqref{eq:ph_selfen_t}, \eqref{Eq:nexc_G}, and \eqref{Eq:eqmo_epsilon} complete the generalization of the Hedin-Baym equations to account for electron localization and polarons within a self-consistent field-theoretic framework. A schematic diagrammatic representation of these relations is shown in Fig.~\ref{Fig:diagrams}.

\section{From Hedin-Baym equations to practical polaron calculations}\label{Sec:PolaronEquations}

In this section, we outline the steps needed to translate the Hedin-Baym theory of polarons of Sec.~\ref{Sec:FormalTheory} into a practical calculation method. The strategy is to make the transition from the general field-theoretical approach of Sec.~\ref{Sec:FormalTheory} to a reciprocal space formulation in a basis of DFT electron states and normal modes, and then from there obtain the DFT polaron equations.
In this section, all quantities are evaluated for the $(N+1)$-electron state unless otherwise stated; to unclutter the notation, in the following we omit the superscript $N+1$.

\subsection{Green's functions and self-energies in reciprocal space} \label{Sec:field_to_Fock}

To move from Green's functions to wavefunctions, we express the Green’s function in the frequency domain using the Lehmann representation \cite{Hedin_Lundqvist_1969}:
\beq \label{Eq:G_Lehmann}
    G(\mathbf{r},\mathbf{r}';\omega)
    =
    \sum_{s} \frac{f_{s}(\mathbf{r}) f^{*}_{s}(\mathbf{r'})}{\hbar \omega - \left[ \varepsilon_{s} + i\eta\,\mathrm{sgn}(\mu - \varepsilon_{s}) \right]} ~,
\eeq
where $\eta = 0^{+}$, the functions $f_s$ are Dyson orbitals, and the $\varepsilon_s$ denote electron addition/removal energies, as follows:
\beqn
      \varepsilon_{s} &=& E_{N+2,s}-E_{N+1}  \hspace{10pt} \mathrm{for}\quad \varepsilon_{s}\geq \mu \label{Eq:lehmann_e_1} ~, \\[4pt]
      \varepsilon_{s} &=& E_{N+1}-E_{N,s} \hspace{20pt}  \mathrm{for}\quad \varepsilon_{s}< \mu \label{Eq:lehmann_e_2} ~.
\eeqn
Here, $E_{N+1\pm1,s}$ is the energy corresponding to the ${|N+1\pm1,s\rangle}$ many-body eigenstate of the Hamiltonian in Eq.~(\ref{Eq:gen_ham}), and $E_{N+1}$ is the correspodning ground state energy.
The Dyson orbitals $f_s$ are obtained from the field operators as:
\beqn \label{Eq:DysonOrbitals}
      f_{s}(\mathbf{r}) &=& \langle \, N+1 \,| \, \hat{\psi}(\mathbf{r}) \, | \,N+2,s \, \rangle   \hspace{5pt}\mathrm{for} \quad \varepsilon_{s}\geq \mu \label{Eq:DysonOrbitals_1} ~, \\[4pt]
      f_{s}(\mathbf{r}) &=& \langle \, N,s \,| \, \hat{\psi}(\mathbf{r}) \, | \,N+1 \, \rangle \hspace{23pt}  \mathrm{for} \quad \varepsilon_{s}< \mu \label{Eq:DysonOrbitals_2} ~.
\eeqn
Using Eq.~\eqref{Eq:G_Lehmann}, Eq.~\eqref{Eq:dyson_ep} can be reformulated as a generalized nonlinear eigenvalue problem for the Dyson orbitals:
\begin{equation} \label{Eq:qp_equation_general}
     \varepsilon_{s}^{\rm cn} f_{s}(\br) + \int d\br' \,\Sigma_{\mathrm{ep}}(\br, \br'; \varepsilon_{s}) f_{s}(\br') 
     =
     \varepsilon_{s}  f_{s}(\br) ~,
\end{equation}
As in Eq.~\eqref{Eq:green_eq_cn}, here $\varepsilon_{s}^{\rm cn}$ refers to the $(N+1)$-electron system with the nuclei fixed at the equilibrium positions of the $N$-particle system. 
To make connection with DFT, we express these Dyson orbitals as linear combination of KS states:
\begin{equation} \label{Eq:DysonOrbitalExpansion}
    f_s(\mathbf{r}) = N_{p}^{-1/2} \sum_{n\mathbf{k}} A^s_{n\mathbf{k}} \, \psi_{n\mathbf{k}}(\mathbf{r}).
\end{equation}
Furthermore, the excess electron density $\Delta n_{\rm e}$ can be expressed using the topmost occupied Dyson orbital via Eqs.~\eqref{Eq:nexc_G} and \eqref{Eq:G_Lehmann}\footnote{\rev{
In the case of a hole-polaron state,
it appears as an empty pole of $G^{N-1}$,
corresponding to the process of adding an electron back to the $(N-1)$-electron system. 
Thus the relevant Dyson orbital is
$$
f_{s_0}^{h}(\mathbf r)
=
\langle N-1|\hat\psi(\mathbf r)|N,s_0\rangle ,
$$
and the electron-density change associated with the missing electron is
$$
\Delta n_h(\mathbf r)
=
-
|f_{s_0}^{h}(\mathbf r)|^2 .
$$
}}. 
We indicate this orbital via the index $s_0$:
\begin{equation} \label{Eq:Density_from_DysonOrb}
    \Delta n_{\rm e}(\br)
    =
    |f_{s_0}(\br)|^2.
\end{equation}
In this representation, the Fan-Migdal, \rev{Landau-Pekar}, and Debye-Waller self-energies derived in Sec.~\ref{Sec:FormalTheory} take the form:
\begin{eqnarray} 
    \label{Eq:Sigma_FM_Fourier}
    &&\hspace{-20pt}\Sigma^{\mathrm{FM}}_{n\mathbf{k},n'\mathbf{k'}}(\omega)
    =
    \frac{1}{N_{p}^{2}} \sum_{mm' \mathbf{q} \nu}
    g^{*}_{mn\nu}(\mathbf{k},\mathbf{q})
    \, g_{m'n'\nu}(\mathbf{k}',\mathbf{q}) \nonumber \\
    &&\hspace{-10pt}\times \sum_{s} A_{m\mathbf{k}+\mathbf{q}}^{s}  A_{m'\mathbf{k'}+\mathbf{q}}^{s,*} \nonumber \\
    &&\hspace{-10pt} \times \!\bigg[ \frac{\theta(\varepsilon_{s}-\mu)}{\hbar \omega - \varepsilon_{s} \!-\! \hbar \omega_{\mathbf{q}\nu} + i\eta}  + \frac{\theta(\mu\!-\!\varepsilon_{s})}{\hbar \omega \!-\! \varepsilon_{s} + \hbar \omega_{\mathbf{q}\nu} \!-\! i\eta} \bigg]\!, \hspace{10pt}
\end{eqnarray}
\begin{align} \label{Eq:SigmaPlrn}
    & \Sigma^{\rev{\mathrm{LP}}}_{n\mathbf{k},n'\mathbf{k'}} 
    = - \frac{2}{N_{p}^2} \sum_{\nu}
    \, g_{nn'\nu}(\mathbf{k'},\mathbf{k-k'}) \nonumber \\
    &\hspace{10pt}\times \!\!\!\!\sum_{{\mathbf k}''mm'}
    \!\!\! 
    A_{m'\mathbf{k}''+\mathbf{k-k'}}^{s_0}
    \, \frac{g_{m'm\nu}^{*}(\mathbf{k}'',\mathbf{k-k'})}{\hbar\omega_{\mathbf{k-k'}\nu}}
    A_{m\mathbf{k}''}^{s_0,*}
    ~,
\end{align}
\begin{equation} \label{Eq:DW_kq}
    \Sigma^{\mathrm{DW}}_{n\bk,n'\bk'} 
    =
    \delta_{\bk, \bk'}
    N_p^{-1} \sum_{\bq\nu} 
    g^{(2)}_{nn'\nu\nu}(\bk,\bq,-\bq) ~,
\end{equation}
and the displacements in the polaron state are given by:
\begin{eqnarray} \label{Eq:dtau_from_Ank}
    &&\hspace{-20pt}\< \dhtau_{\k\a p} \>
    = 
    - N_p^{-1} \sum_{nn'\bk, \bq\nu} A_{n\bk}^{s_0,*} A^{s_0}_{n'\bk+\bq} \nonumber \\
    && \hspace{-10pt}\times \frac{2}{\hbar\w_{\bq\nu}}
    (\hbar/2 M_\k \w_{\bq\nu})^{1/2} e_{\k\a\nu}(\bq) e^{i\bq\cdot\bR_p} 
    g^*_{n'n\nu}(\bk, \bq).\hspace{10pt}
\end{eqnarray}
In Eq.~\eqref{Eq:Sigma_FM_Fourier}, $\theta$ is the Heaviside function. 
To obtain Eqs.~\eqref{Eq:Sigma_FM_Fourier}-\eqref{Eq:dtau_from_Ank}, vertex corrections have been dropped, and the phonon self-energy of Eq.~\eqref{eq:ph_selfen_t} has been written within the adiabatic approximation, e.g., as obtained from DFPT \cite{Baroni_Gianozzi_2001}. Accordingly, we introduced the frequencies $\omega_{\bq\nu}$ and polarization vectors $e_{\k\a\nu}(\bq)$ to denote vibrational eigenmodes with wavevector $\bq$ and branch $\nu$. We also introduced the DFPT electron-phonon matrix elements \cite{Giustino_2017}: 
\begin{eqnarray} \label{Eq:dressed_g}
    && \hspace{-22pt}g_{mn\nu}(\bk, \bq) = N_p^{-2} \int \!d\br\, d\br'\,  \psi_{m\bk+\bq}(\br) 
    \, \epsilon_{\rm e}^{-1}(\br, \br') \nonumber \\
    &&\hspace{-20pt} \times \sum_{\kappa\alpha p} (\hbar/2M_\kappa\omega_{\mathbf{q}\nu})^\frac{1}{2}  e_{\kappa\alpha,\nu}(\mathbf{q})  e^{i\mathbf{q}\cdot\mathbf{R}_p} \frac{\partial V_{\kappa}(\br')}{\partial \tau_{\kappa\alpha p}} 
    \psi_{n\bk}(\br),\hspace{5pt}
\end{eqnarray}
and we assumed that the dielectric matrix does not change upon electron addition. In order to obtain the electron-phonon matrix element on the first line of Eq.~\eqref{Eq:SigmaPlrn}, we expressed the variation of the total potential in terms of the inverse dielectric matrix, using a reasoning similar to the one that leads to Eq.~\eqref{Eq:deltan_from_eps}.
The second-order electron-phonon matrix elements appearing in Eq.~\eqref{Eq:DW_kq} are given by:
\begin{eqnarray}
    \label{Eq:second_eph}
    &&g^{(2)}_{mn\nu\nu'}(\bk,\bq,\bq')
    = \!
    \frac{1}{2}\!\!
    \sum_{\substack{ \k \a p \\ \k' \a' p' }}\!\!
     (\hbar/2M_\kappa \omega_{\bq\nu} )^{1/2}
     (\hbar/2M_{\kappa'} \omega_{\bq'\nu'})^{1/2}\nonumber \\
     &&\hspace{10pt}\times
    \,e_{\kappa\alpha,\nu}(\bq)
    e^{i\bq\cdot\bR_p}
    \,e_{\kappa'\alpha',\nu'}(\bq')
    e^{i\bq'\cdot\bR_p'}
    \nonumber \\
    &&\hspace{10pt}\times
    \bra{m\bk+\bq + \bq'}
    \frac{\D^2 V_\mathrm{tot}}{\D \tau_{\k \a p}
    \D \tau_{\k' \a ' p'}} 
    \ket{n\bk},
\end{eqnarray}
where, similarly to Eq.~\eqref{Eq:dressed_g}, the second variation of the total potential can be expressed using the dielectric matrix: 
\begin{equation}
\hspace{-5pt}\frac{\D^2 V_\mathrm{tot}(\br)}{\D \tau_{\k \a p} \D\tau_{\k'\a' p'}} \!=\!
\frac{\D}{\D\tau_{\k' \a' p'}} \!\left[\int \!\!d\br' \epsilon_{\rm e}^{-1}(\br,\br')\frac{\D V_\k(\br')}{\D \tau_{\k \a p}} \right]\!.
\end{equation}
It should be noted that the electron-phonon coupling matrix elements appearing in Eqs.~\eqref{Eq:Sigma_FM_Fourier}-\eqref{Eq:dtau_from_Ank} are all screened. 

Equations~\eqref{Eq:SigmaPlrn} and \eqref{Eq:Sigma_FM_Fourier}, which have been derived starting from the Hedin-Baym equations, are equivalent to Eqs.~\eqref{Eq:selfen_FM_kq} and \eqref{Eq:P_from_G} once expressed via the Lehmann representation. This equivalence establishes a direct link between approaches based either on the bare electron-ion Hamiltonian in Eq.~\eqref{Eq:gen_ham} or on the Fock-space Hamiltonian in Eq.~\eqref{Eq:epi-hamilt}. The most important conceptual difference between these two approaches is that the Hedin-Baym equations allow one to express the electron-phonon coupling matrix elements in terms of the dielectric matrix, while in the Fock-space approach these matrix elements are taken as external parameters.

\subsection{Many-body polaron equations in reciprocal space} \label{Sec:MBPE}

We now outline a practical scheme for obtaining polaron eigenvalues and displacements from the self-energies outlined in Sec.~\ref{Sec:field_to_Fock}.
Substituting Eq.~\eqref{Eq:DysonOrbitalExpansion} in Eq.~\eqref{Eq:qp_equation_general}, and solving for the topmost occupied Dyson orbital, we find:
\beq \label{Eq:ManyBodyPolaronEquation}
    \sum_{n'\mathbf{k}'} H^{\mathrm{pol}}_{n\mathbf{k},n'\mathbf{k'}} A_{n'\mathbf{k}'}^{s_0}
    = \varepsilon_{s_0} A_{n\mathbf{k}}^{s_0} ~,
\eeq
where the effective Hamiltonian is given by:
\begin{equation} \label{Eq:ManyBodyPolaronHeff}
    H^{\mathrm{pol}}_{n\mathbf{k},n'\mathbf{k'}} \!=\! 
    \varepsilon^{\mathrm{cn}}_{n\mathbf{k}} \delta_{n\mathbf{k},n'\mathbf{k'}}
    + \Sigma^{\mathrm{FM}}_{n\mathbf{k},n'\mathbf{k'}}(\ve_{s_0})
    + \Sigma^{\rev{\mathrm{LP}}}_{n\mathbf{k},n'\mathbf{k'}}
    + \Sigma^{\mathrm{DW}}_{n\bk,n'\bk'} ~.
\end{equation}
Explicit expressions for the self-energies are given in Eqs.~\eqref{Eq:Sigma_FM_Fourier}-\eqref{Eq:DW_kq},
and the displacements can be obtained by using the solution of Eq.~\eqref{Eq:ManyBodyPolaronEquation} inside Eq.~\eqref{Eq:dtau_from_Ank}.

\begin{figure}[t!]
        \includegraphics[width=0.6\linewidth]{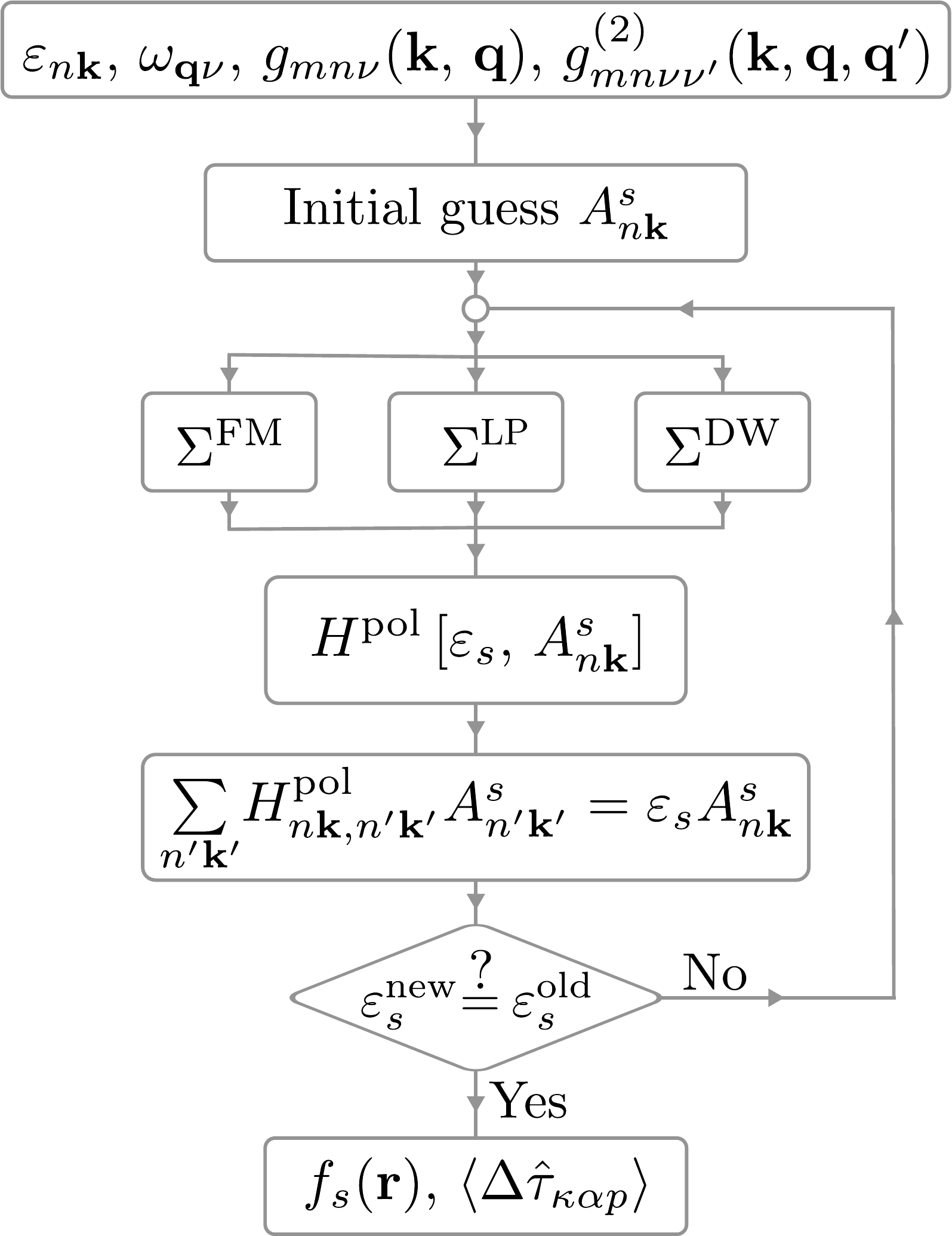}
        \caption{Schematic representation of the self-consistent procedure required to solve the many-body polaron equations, Eqs.~\eqref{Eq:ManyBodyPolaronEquation}-\eqref{Eq:ManyBodyPolaronHeff}.
        Adapted from \cite{Lafuente_Giustino_2022b}.
        \label{Fig:qp_eq_flow}}
\end{figure}

Equations~\eqref{Eq:ManyBodyPolaronEquation} and \eqref{Eq:ManyBodyPolaronHeff} define the \textit{ab initio} many-body polaron equations \cite{Lafuente_Giustino_2022b}.
Since the effective Hamiltonian matrix on the left-hand side of Eq.~\eqref{Eq:ManyBodyPolaronHeff} depends implicitly on the quasiparticle excitation energy $\varepsilon_{s}$ and amplitudes $A_{n\mathbf{k}}^{s}$ via the Fan-Migdal and \rev{Landau-Pekar} self-energies, Eq.~\eqref{Eq:ManyBodyPolaronEquation} constitutes a nonlinear self-consistent eigenvalue problem. The numerical solution of this problem can be performed iteratively, as illustrated schematically in Fig.~\ref{Fig:qp_eq_flow}.

\textcite{Lafuente_Giustino_2022b} proposed an approximate method to solve Eqs.~\eqref{Eq:ManyBodyPolaronEquation} and \eqref{Eq:ManyBodyPolaronHeff} via perturbation theory. These authors noted that, when the interacting Green's function $G$ is replaced by its non-interacting counterpart $G^{\mathrm{cn}}$, the Dyson orbital amplitudes in the Fan-Migdal self-energy, Eq.~\eqref{Eq:Sigma_FM_Fourier}, reduce to Kronecker deltas, $A_{n\bk}^{s}=\delta_{s,n\bk}$. As a result, $\Sigma^{\rm FM}$ becomes diagonal in the wavevector indices.
In this approximation, neither $\Sigma^{\mathrm{FM}}$ nor $\Sigma^{\mathrm{DW}}$ mix different wavevectors, therefore the only term that can drive electron localization by realizing a superposition of wavevectors is the \rev{Landau-Pekar} self-energy $\Sigma^{\rev{\mathrm{LP}}}$. This observation suggests looking for a solution to Eq.~\eqref{Eq:ManyBodyPolaronEquation} where only the polaron self-energy is retained, and then add the effects of the Fan-Migdal and Debye-Waller self-energies to this solution via perturbation theory. 
In practice, one performs two steps:
\begin{itemize}
    \item[(i)] Solve Eqs.~\eqref{Eq:ManyBodyPolaronEquation} and \eqref{Eq:ManyBodyPolaronHeff} by considering only the \rev{Landau-Pekar} self-energy:
    \begin{equation} \label{Eq:qp_eq_onlyP}
        \hspace{20pt}
        \sum_{n'\mathbf{k}'} ( \varepsilon_{n\mathbf{k}}^{\mathrm{cn}} \delta_{n\mathbf{k}, n'\mathbf{k'}}
        +
        \Sigma^{\rev{\mathrm{LP}}}_{n\mathbf{k}, n'\mathbf{k'}} )
        A_{n'\mathbf{k'}}^{\rev{\mathrm{LP}}}
        =
        \varepsilon^{\rev{\mathrm{LP}}} A_{n\mathbf{k}}^{\rev{\mathrm{LP}}} ~.
    \end{equation}
    \item[(ii)] Include the Fan-Migdal and Debye-Waller contributions to the quasiparticle excitation energy via first-order perturbation theory, starting from the solution of Eq.~\eqref{Eq:qp_eq_onlyP}:
    \begin{align} \label{Eq:epsilon_from_Ank}
        \hspace{21pt}&\varepsilon_{s_0} = \!\varepsilon^{\rev{\mathrm{LP}}}
         \!+\! \frac{1}{N_p}\!\!\sum_{nn'\mathbf{k}} A^{\rev{\mathrm{LP}},*}_{n\mathbf{k}} 
        \left[ \Sigma^{\mathrm{FM}}_{nn'\mathbf{k}}(\varepsilon_{s_0}) 
        + \Sigma^{\mathrm{DW}}_{nn'\mathbf{k}} \right]\!
        A^{\rev{\mathrm{LP}}}_{n'\mathbf{k}}.
    \end{align}
    \vspace{-15pt}
\end{itemize}
This last equation can be further simplified by taking the self-energies to be diagonal in the band indices, and by replacing the polaron eigenvalue inside $\Sigma^{\mathrm{FM}}_{nn'\mathbf{k}}(\varepsilon_{s_0})$ by the KS state at the band extremum; or else by performing a Taylor expansion as is commonly done for the GW self-energy when introducing the quasiparticle renormalization factor \cite{Hybertsen_Louie_1986}.

Equation~\eqref{Eq:epsilon_from_Ank} admits a clear physical interpretation: 
the first term corresponds to the energy gain resulting from electron localization; in the strong-coupling limit, this term reduces to a standard DFT calculation of polarons (cf.\ Sec.~\ref{Sec:plrn_eq_reciprocal}).
The second term in Eq.~\eqref{Eq:epsilon_from_Ank} is a weighted average of the Fan-Migdal and Debye-Waller self-energies, with the quasiparticle amplitudes serving as weights; in the limit of very large polaron, this term reduces to the standard phonon-induced energy shift in the Allen-Heine theory of band structure renormalization \cite{Allen_Heine_1976}.
Together, these two terms interpolate between the strong-coupling limit, which is captured by DFT polaron calculations, and the weak-coupling limit, which is well described by the Allen-Heine framework \cite{Lafuente_Giustino_2022a}.

\rev{This interpretation can be tested explicitly in the Fr\"ohlich model (see Sec.~\ref{Sec:FrohlichLLP}),
where Feynman's all-coupling variational solution provides a reference
benchmark (see Sec.~\ref{Sec:Feynman}). 
The comparison reported in Ref.~\cite{Lafuente_Giustino_2022a} shows very good overall agreement, 
despite the absence of vertex corrections. 
The decomposition in Eqs.~\eqref{Eq:epsilon_from_Ank} also clarifies the regimes in which different approximations are expected to be reliable. 
At weak coupling,
the excitation energy is dominated by the Fan-Migdal
contribution, 
explaining why nonadiabatic Allen-Heine approaches can
accurately predict band renormalizations in this regime
\cite{Miglio_Gonze_2020}. 
At strong coupling, in contrast, the static self-trapping contribution
dominates and can be related directly to the lattice distortion obtained
in adiabatic DFT calculations (see Sec.~\ref{Sec:Connect_recip}), 
thereby rationalizing their success for modelling very small polarons (see Sec.~\ref{Sec:DFTPolaron}).
The largest deviations from Feynman's solution occur at
intermediate coupling, indicating that vertex corrections are expected to
be most relevant in this crossover regime, 
although they do not alter the
qualitative evolution from a Fan-Migdal-dominated large-polaron regime
to a self-trapping-dominated small-polaron regime.}

\subsection{\textit{Ab initio} polaron equations in DFT} \label{Sec:plrn_eq_reciprocal}

The many-body polaron equations outlined in Sec.~\eqref{Sec:MBPE} map exactly into the \textit{ab initio} polaron equations introduced in the context of DFT by \textcite{Sio_Giustino_2019a, Sio_Giustino_2019b}.
In fact, if we neglect the contributions of the Fan-Migdal and the Debye-Waller self-energies in Eq.~\eqref{Eq:epsilon_from_Ank}, Eqs.~\eqref{Eq:SigmaPlrn} and \eqref{Eq:qp_eq_onlyP} can be recast into the coupled nonlinear system of equations:
\begin{align} 
    & \frac{2}{N_p} \sum_{\mathbf{q}m\nu} B_{\mathbf{q}\nu} \, g_{mn\nu}^*(\mathbf{k}, \mathbf{q}) A_{m\mathbf{k+q}} = (\varepsilon_{n\mathbf{k}}-\varepsilon) A_{n\mathbf{k}} ~, \label{Eq:polaron_eq_Denny_1} \\
    & B_{\mathbf{q}\nu} = \frac{1}{N_p} \sum_{mn\mathbf{k}} A_{m\mathbf{k+q}}^* \frac{g_{mn\nu}(\mathbf{k}, \mathbf{q})}{\hbar \omega_{\mathbf{q}\nu}}A_{n\mathbf{k}} ~.
    \label{Eq:polaron_eq_Denny_2}
\end{align}
These equations constitute the DFT \textit{ab initio} polaron equations of \textcite{Sio_Giustino_2019a, Sio_Giustino_2019b}. The auxiliary amplitudes $B_{\mathbf{q}\nu}$ 
appearing in Eq.~\eqref{Eq:polaron_eq_Denny_2} are related to the displacements of Eq.~\eqref{Eq:dtau_from_Ank} by the generalized Fourier transform:
\begin{equation} \label{Eq:Bqv2dtau}
    B_{\mathbf{q} \nu}^{*} =
    \sum_{\kappa \alpha p} \left(\!  \frac{M_{\kappa} \omega_{\mathbf{q}\nu}}{2 \hbar} \!\right)^{\!\!1/2}
    \!\!\!e^{*}_{\kappa \alpha, \nu} (\mathbf{q})
    \, e^{-i\mathbf{q}\cdot \mathbf{R}_{p}}
    \, \< \dhtau_{\kappa \alpha p} \> ~.
\end{equation}
To make the physical meaning of Eqs.~\eqref{Eq:polaron_eq_Denny_1}-\eqref{Eq:polaron_eq_Denny_2} apparent, one can rewrite them in \rev{real space} by recalling the relation between the polaron wavefunction and the quasiparticle amplitudes, Eq.~\eqref{Eq:DysonOrbitalExpansion}, and using the standard expressions for the electron-phonon matrix elements and the interatomic force constants \cite{Giustino_2017}. The result is: 
\begin{eqnarray}
    && \hat{H}^0_{\mathrm{KS}} \, \psi(\br)
    + \sum_{\k\a p}
    \frac{\partial V^0_{\mathrm{KS}}}{\partial \tau_{\k\a p}} \dtau_{\k\a p} \psi(\br) = \ve\, \psi(\br) ~,
    \label{Eq:polaron_eq_realspace_1}
    \\
    && \dtau_{\k\a p} = \!-\!\!\!\sum_{\k'\a'p'} \!\!(C^0)^{-1}_{\k\a p, \k'\a'p'} \!\!\int\!\! d\br \frac{\partial V_{\mathrm{KS}}}{\partial \tau_{\k'\a' p'}} |\psi(\br)|^2\!\!.\hspace{28pt}
    \label{Eq:polaron_eq_realspace_2}
\end{eqnarray}
In these expressions, $\hat{H}^0_{\mathrm{KS}}$, $V^0_{\mathrm{KS}}$, and $C^0$ are the DFT KS Hamiltonian, self-consistent potential, and interatomic force constants, all evaluated at the equilibrium ionic positions of the system without polaron; the integral spans the BvK supercell. Furthermore, the expectation value of the displacements, $\< \dhtau_{\kappa \alpha p} \>$, is being approximated by the classical DFT displacements, $\dtau_{\kappa \alpha p}$, and the Dyson orbital of the polaron state $f_{s_0}(\br)$ is replaced by the KS wavefunction $\psi(\br)$ with eigenvalue $\varepsilon$. 

In this review article, the \textit{ab initio} polaron equations have been derived as an approximation to the more general many-body polaron equations in Sec.~\ref{Sec:MBPE}.
In their original formulation, these equations were derived from an alternative starting point, namely by minimizing the self-interaction-free DFT functional $E^{\rm SVPG}$ in Eq.~\eqref{Eq:sio_pSIC}, within the harmonic approximation \cite{Sio_Giustino_2019b}. In this alternative formulation, the total energy reads:
\begin{eqnarray} \label{Eq:toten_aipe}
    &&\hspace{-10pt}E^{N+1}[\psi, \{\dtau_{\k\a p}\}] 
    =   
    E^{N}[\{\tau^0_{\k\a p}\}] \nonumber \\
    &&  + \frac{1}{2} \sum_{\substack{\k\a p \\ \k' \a' p'}}  (C^0)^{-1}_{\k\a p, \k'\a'p'} \dtau_{\k\a p} \dtau_{\k'\a' p'} \nonumber \\
    && \!+ \int \!\!d\br \, \psi^*(\br) \!\left[ \hat{H}^0_{\mathrm{KS}} + \sum_{\k\a p} \frac{\partial V^0_{\mathrm{KS}}}{\partial \tau_{\k\a p}} \dtau_{\k\a p} \right] \!\psi(\br).
\end{eqnarray}
Performing a variational minimization of Eq.~\eqref{Eq:toten_aipe} with respect to $\psi^*$ and $\dtau_{\k\a p}$, and taking into account the normalization of the wavefunction via the Lagrange multiplier $\varepsilon$, one arrives at Eqs.~\eqref{Eq:polaron_eq_realspace_1} and \eqref{Eq:polaron_eq_realspace_2}. Upon further expanding the polaron wave function in a KS basis, as in Eq.~\eqref{Eq:DysonOrbitalExpansion},
and inverting Eq.~\eqref{Eq:Bqv2dtau} to express the displacements as a linear combination of normal modes,
one arrives at Eqs.~\eqref{Eq:polaron_eq_Denny_1} and \eqref{Eq:polaron_eq_Denny_2}.
We note that a similar set of equations was recently reported by \textcite{Vasilchenko_Gonze_2022, Vasilchenko_Gonze_2025}, who also proceeded via the  variational minimization of \rev{the reciprocal-space counterpart of} Eq.~\eqref{Eq:toten_aipe}; in addition, these authors \rev{employed} a preconditioned conjugate gradient minimization method \rev{to ensure a robust and automated identification of distinct polaronic solutions \cite{Vasilchenko_Gonze_2024}, along with a filtering technique to efficiently handle large polarons. They also isolated the long-range component of the polaron energy and introduced a correction that improves the convergence of finite-grid calculations.}

Combining Eqs.~\eqref{Eq:DysonOrbitalExpansion}, \eqref{Eq:polaron_eq_Denny_1}, \eqref{Eq:Bqv2dtau} and \eqref{Eq:toten_aipe},
the polaron formation energy can be written as
\footnote{\rev{The equivalent expression for the formation energy of a hole polaron upon electron removal is
$$
\Delta E_{\mathrm{f}}
    =
    -\left(\ve - \ve_{\mathrm{VBM}}\right) + \frac{1}{N_p} \sum_{\bq\nu} |B_{\bq\nu}|^2 \hbar\w_{\bq\nu}
$$
}}:
\begin{equation} \label{Eq:Eform_aipe}   
    \Delta E_{\mathrm{f}}
    =
    \ve - \ve_{\mathrm{CBM}} + \frac{1}{N_p} \sum_{\bq\nu} |B_{\bq\nu}|^2 \hbar\w_{\bq\nu}~,
\end{equation}
where CBM refers to the conduction band minimum for electron polarons. This relation can also be understood as an approximation to the formation energy in the all-coupling variational canonical transformation approach, see Eq.~\eqref{Eq:Reichman_E_2} \cite{Robinson_Reichman_2025}.

These considerations show that the DFT-based \textit{ab initio} polaron equations can be recovered from the field-theoretic framework of Sec.~\ref{Sec:FormalTheory} by retaining only the \rev{Landau-Pekar} self-energy in Eq.~\eqref{Eq:qp_equation_general}, and by approximating the many-body phonon self-energy in Eq.~\eqref{Eq:steady_state_disp} via the Born-Oppenheimer interatomic force constants. Further connections between these two approaches are discussed in Sec.\ref{Sec:Connection}.

\section{Connection between many-body and DFT approaches to polarons}\label{Sec:Connection}

In Sec.~\ref{Sec:DFTPolaron} we reviewed current DFT-based approaches to polarons, and in Secs.~\ref{Sec:ManyBody} and \ref{Sec:FormalTheory} we reviewed \textit{ab initio} many-body approaches. In this section, we discuss how these approaches are related, clarify their respective domains of applicability, and identify remaining knowledge gaps that will require future theoretical and computational developments.

\begin{figure*}
    \centering
    \includegraphics[width=0.98\linewidth]{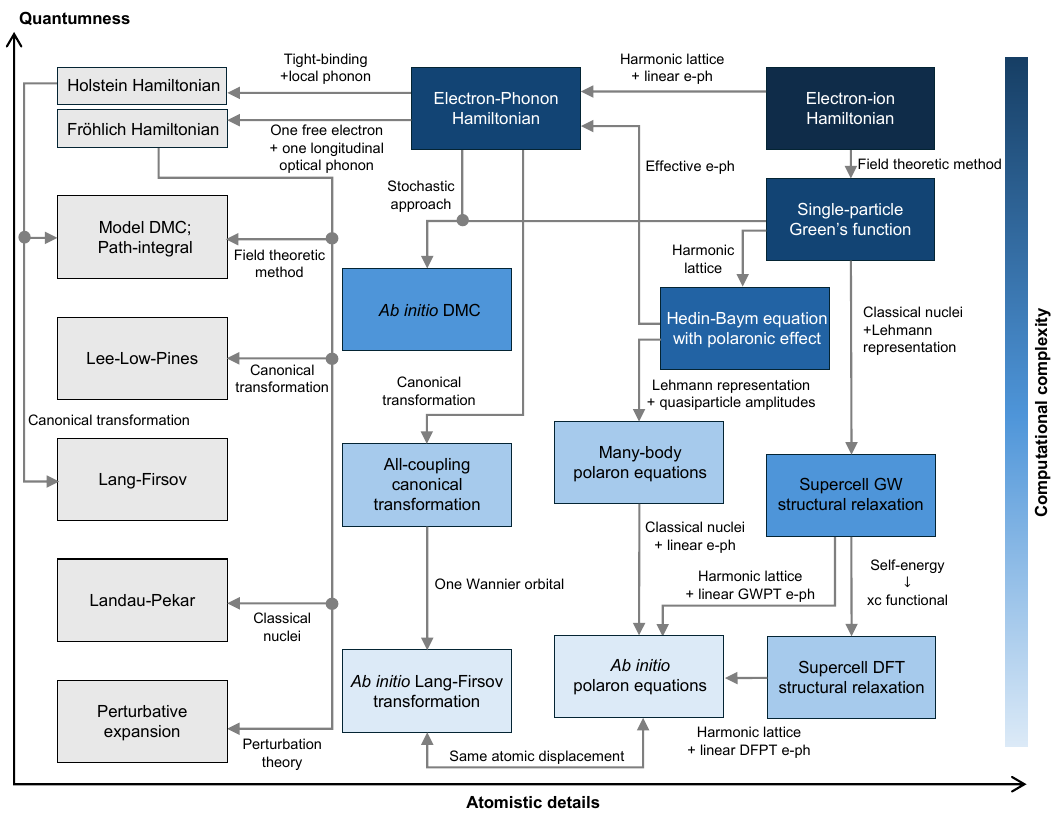}
    \caption{
    Mind map of models, theories, and computational methods for polarons reviewed in Secs.~\ref{Sec:History}-\ref{Sec:FormalTheory}. The shade of blue in each box represents a qualitative measure of the computational cost associated with the given \textit{ab initio} method. The effective Hamiltonian approaches reviewed in Sec.~\ref{Sec:History} are indicated in gray boxes to distinguish them from first-principles calculations. The horizontal axis quantifies the level of atomistic detail, ranging from effective Hamiltonians to Fock-space \textit{ab initio} methods to explicit DFT or self-consistent Green's functions methods. The vertical axis quantifies the level of description of the lattice degrees of freedom, ranging from the adiabatic and classical approximation to a complete quantum treatment of ionic motion. 
    }
    \label{Fig:jacob_ladder}
\end{figure*}

Figure~\ref{Fig:jacob_ladder} presents a mind map of current theories and methods to study polarons, from effective Hamiltonians to \textit{ab initio} methods. In the following section we discuss some of these connections, with a focus on the link between Green's function methods and DFT supercell calculations, the connection between GW calculations and the reciprocal-space polaron equations, and the connection between the polaron equations and canonical transformation approaches, respectively. At the end of this section, we also mention the challenges arising from nonlinear electron-phonon couplings and anharmonic effects.

\subsection{Connection between Green's function approaches and supercell calculations}\label{Sec:GW-dft}

In DFT calculation of polarons (Sec.~\ref{Sec:DFTPolaron}) and in  many-body Fock-space approaches (Sec.~\ref{Sec:ManyBody}), the polaronic state is the \textit{ground state} of the system with an excess electron or hole. On the other hand, Green's function methods such as DMC (Sec.~\ref{Sec:DMC}) view the polaron as a single-particle \textit{excitation} of the charge-neutral ground state. It is clear that these two viewpoints must be equivalent, since the Green's function contains the notion of creating an excess electron or hole in the $N$-particle ground state, leading to an $(N\pm 1)$-particle state. In this section, we aim to make this equivalence more apparent. 

To this end, we consider the equation of motion for the electron Green's function $G^{\rm cn}$ in the clamped-ion approximation, Eq.~(\ref{Eq:green_eq_cn}). However, unlike in Eq.~(\ref{Eq:green_eq_cn}), we now consider expectation values on the many-body charge-neutral state $\ket{N; \btau_0 + \Delta \btau}$; the $\Delta \btau$ are used to allow for possible distortions of the lattice. This choice introduces the adiabatic Born-Oppenheimer approximation and the classical approximation for the ionic displacements, precisely as in DFT.

After introducing the Lehmann representation, cf.\ Eqs.~(\ref{Eq:lehmann_e_1}) and (\ref{Eq:lehmann_e_2}), the total energy in the polaronic state for the configuration $\btau_0 + \Delta \btau$ reads:
\begin{equation}
    \label{Eq:gw_etot}
    E^{\rm N+1}(\btau_0 + \Delta \btau) = E^{\rm N}(\btau_0 + \Delta \btau) + \ve^{\rm N}_\mathrm{CBM}(\btau_0 + \Delta \btau).
\end{equation}
Here, $E^N(\btau_0 + \Delta \btau)$ is the ground-state total energy of the neutral $N$-electron system; it can formally be obtained from the Green's function via the Galitskii-Migdal formula~\cite{Galitskii_Migdal_1958, Holm_Aryasetiawan_2000}, or it can be approximated by the DFT total energy.
$\ve^{N}_\mathrm{CBM}$ is the lowest excitation energy; this energy is associated with a Dyson orbital $f_\mathrm{CBM}$, which can be identified with the polaron wave function.

In practice, when solving the Dyson equation Eq.~(\ref{Eq:green_eq_cn}) for $\ve^{N}_\mathrm{CBM}$ and $f^N_\mathrm{CBM}$, one typically employs the GW approximation for the electron self-energy~\cite{Hybertsen_Louie_1986,Onida_Rubio_2002}. Within this approximation, after transforming Eq.~(\ref{Eq:green_eq_cn}) to the frequency domain one obtains the generalized eigenvalue problem: 
\begin{align}
    \label{Eq:dyson_eq}
   &\left[
    \hH_{\rm KS}(\br) - V_{\rm xc}(\br)\right]f^N_\mathrm{CBM}(\br)
    \nonumber \\
    &+ \int d\br'\Sigma_{\rm GW}(\br, \br', \ve^{N,\rm GW}_\mathrm{CBM})
    f^N_\mathrm{CBM}(\br')
    =
    \ve^{N,\rm GW}_\mathrm{CBM} f^N_\mathrm{CBM}(\br),
\end{align}
where $\hH_{\rm KS}$ is the standard KS Hamiltonian, $V_{\rm xc}$ is the exchange and correlation potential, and $\Sigma_{\rm GW}$ is the GW self-energy, all evaluated for the charge-neutral, $N$-electron system. All quantities carry an implicit dependence on the configuration $\btau_0 + \Delta \btau$. Once $\ve^{N,\rm GW}_\mathrm{CBM}$ is determined by solving this equation in perturbation theory \cite{Hybertsen_Louie_1986}, Eq.~\eqref{Eq:gw_etot} can be used to find the polaron formation energy:
\begin{align}
    \label{Eq:plrn_etot_gw}
    \Delta E^{\rm GW}_{\rm f}
    &=
    E^N(\btau_0+\Delta \btau) - E^N(\btau_0)
    \nonumber \\
    &+\ve^{N,\rm GW}_\mathrm{CBM}(\btau_0 + \Delta \btau) - \ve^{N,\rm GW}_\mathrm{CBM}(\btau_0).
\end{align}
The close similarity between Eq.~(\ref{Eq:plrn_etot_gw}) and Eq.~(\ref{Eq:dft_eform_expr3}) provides a formal justification for the proposal by \textcite{Sadigh_Aberg_2015} to calculate polaron formation energies via the DFT frontier eigenvalues in the charge-neutral state. From this perspective, the approach of \textcite{Sadigh_Aberg_2015} can be viewed as an approximate version of Eq.~\eqref{Eq:plrn_etot_gw}, where the GW quasiparticle eigenvalues are replaced by DFT eigenvalues. Since the GW method is self-interaction free, this strategy effectively amounts to eliminating the SIE.

One can also show that the SVPG functional by \textcite{Sio_Giustino_2019b}, as given by Eq.~\eqref{Eq:sio_pSIC}, directly leads to an expression similar to Eq.~\eqref{Eq:plrn_etot_gw}. 
To see this, we follow \textcite{Dai_Giustino_2026a} and consider a gapped system with $N+1$ electrons, and we write the electron density as $n^{N+1} = n_{\rm val}^{N+1} + |\psi_{\rm CBM}^{N+1}|^2$. In the notation of Sec.~\ref{Sec:DFTPolaron_SIC}, $n_{\rm val}^{N+1}$ corresponds to $n^\uparrow+n^\downarrow$, and $|\psi_{\rm CBM}^{N+1}|^2$ corresponds to $n_p$. This partitioning is meaningful because the valence and conduction manifolds are separated by a gap. Using the self-interaction correction of Eq.~\eqref{Eq:plrn_etot_gw}, after a few algebraic steps the total energy of this system can be written as:
  \begin{equation}\label{Eq:svpg-gw}
    E^{\rm SVPG}[n^{N+1}] = E[n_{\rm val}^{N+1}] + \ve_{\rm CBM}[n_{\rm val}^{N+1}]~,
  \end{equation}
where each quantity is evaluated for the electron density indicated within square brackets. In this expression, terms of order three and higher in $|\psi_{\rm CBM}^{N+1}|^2$ are neglected. From this expression one obtains the polaron formation energy:
  \begin{eqnarray}\label{Eq:svpg-gw-ef}
   && \Delta E_{\rm f}^{\rm SVPG}= E[n_{\rm val}^{N+1},\btau_0+\Delta\btau] - E[n_{\rm val}^{N+1},\btau_0] \nonumber \\
     && \quad + \ve_{\rm CBM}[n_{\rm val}^{N+1},\btau_0+\Delta\btau] - \ve_{\rm CBM}[n_{\rm val}^{N+1},\btau_0]~.\hspace{10pt}
  \end{eqnarray}
The approach of \textcite{Sadigh_Aberg_2015} in Eq.~\eqref{Eq:dft_eform_expr3} is immediately obtained from this result by making the further approximation that $n_{\rm val}^{N+1}$ can be replaced by $n_{\rm val}^{N}$. This approximation is reasonable because the valence charge densities in the $N$-electron and $(N+1)$-electron systems should be similar, and the error resulting from their difference is partially canceled in Eq.~\eqref{Eq:svpg-gw-ef} by taking the difference between total energy and eigenvalue at the distorted configuration ($\btau_0+\Delta \btau$) and the undistorted crystal ($\btau_0$).
\rev{A similar analysis that leads to the same equation as Eq.~\eqref{Eq:svpg-gw-ef} can be found in \textcite{Falletta_Pasquarello_2025}.}

This further analysis confirms that the equations that govern polaron energetics have the same form in the GW method, Eq.~\eqref{Eq:dyson_eq}, and in self-interaction-corrected DFT, Eq.~\eqref{Eq:svpg-gw-ef}. Since one  expects the GW method to correct the DFT KS eigenvalues by approximately the same amount in both the undistorted crystal and the polaronic distortion, Eq.~\eqref{Eq:dft_eform_expr3} should provide fairly reliable estimates of the polaron energy. The residual difference between DFT and GW must be related to the changes in the GW corrections upon atomic displacements; this effect is precisely what is captured by the GW perturbation theory (GWPT) approach of \textcite{Li_Louie_2019}.

\subsection{Connection between GW calculations and reciprocal-space polaron equations} \label{Sec:Connect_recip}

To connect GW calculations of polarons as discussed in Sec.~\ref{Sec:GW-dft} to the polaron equations of Sec.~\ref{Sec:plrn_eq_reciprocal}, we expand the effective Hamiltonian in Eq.~\eqref{Eq:dyson_eq} to linear order in the ionic displacements:
\begin{eqnarray}
    &&\hH_\mathrm{KS}(\br) \d(\br-\br') +\sum_{\k\a p} \left[\frac{\D \Sigma_{\rm GW}(\br,\br',\ve^{N,\rm GW}_\mathrm{CBM})}{\D \tau_{\k\a p}}\right.
    \nonumber\\
    &&\hspace{30pt} \left. -\frac{\D V_{\rm xc}(\br)}{\D \tau_{\k\a p}}\d(\br-\br')\right]\Delta \tau_{\k\a p},
\end{eqnarray}
where all quantities are evaluated for the undistorted crystal at $\btau_0$. From this expression, using the decomposition in terms of KS states given in Eq.~(\ref{Eq:DysonOrbitalExpansion}), we arrive at the GW counterpart of the polaron equations given in Sec.~\ref{Sec:plrn_eq_reciprocal}:
\begin{equation}  
    \label{Eq:plrneq_gwpt}
    \ve_{n\bk} A_{n\bk}
    -\frac{2}{N}\sum_{\bq m \nu} B_{\bq\nu} 
    g^{{\rm GW},*}_{mn\nu}(\bk,\bq) 
    A_{m\bk+\bq} = \ve A_{n\bk}.
\end{equation}
The only difference between this GW version and the DFT version of Eq.~\eqref{Eq:polaron_eq_Denny_1} is that the electron-phonon matrix elements are now evaluated via the GWPT method~\cite{Li_Louie_2019}:
\begin{eqnarray}
    \label{Eq:gwpt_eph}
    g^{\rm GW}_{mn\nu}(\bk,\bq) 
    &=&
    g^\mathrm{DFT}_{mn\nu}(\bk,\bq) 
    -
    \bra{\psi_{m\bk+\bq}}  \Delta_{\bq\nu} V_{\rm xc} \ket{\psi_{n\bk}}
    \nonumber \\
    &+&
    \bra{\psi_{m\bk+\bq}}  \Delta_{\bq\nu} \Sigma_{\rm GW} \ket{\psi_{n\bk}}~.
\end{eqnarray}
As shown by Eq.~\eqref{Eq:Bqv2dtau}, the coefficients $B_{\bq\nu}$ in Eq.~\eqref{Eq:plrneq_gwpt} describe the ionic displacements in the polaronic state. To determine these coefficients self-consistently, one writes the ground-state total energy in the distorted configuration within the harmonic approximation:
\begin{equation}
    \label{Eq:harmonic_lattice}
    E^N(\btau_0+\Delta \btau) \!=\! E^N(\btau_0) 
    + \frac{1}{2} \!\!\!\sum_{\substack{\k \a p \\ \k' \a' p'}}\!\!\!
    \ifc \Delta \tau_{\k\a p} \Delta \tau_{\k'\a' p'},
\end{equation}
and minimizes Eq.~\eqref{Eq:plrn_etot_gw} with respect to the displacements. After transforming to the reciprocal space representation, one finds:
\begin{align}\label{Eq:plrneq_gwpt2}
    B_{\bq\nu} = \frac{1}{N_p}\sum_{mn\bk} A^*_{m\bk+\bq} 
    \frac{g^{\rm GW}_{mn\nu}(\bk,\bq, \ve)}{\hbar\omega_{\bq\nu}} A_{n\bk}.
\end{align}
Together, Eqs.~\eqref{Eq:plrneq_gwpt} and \eqref{Eq:plrneq_gwpt2} constitute the GW counterparts of the DFT polaron equations of \textcite{Sio_Giustino_2019a} reviewed in Sec.~\ref{Sec:plrn_eq_reciprocal}. 
When these equations are combined with Eq.~\eqref{Eq:plrn_etot_gw}, the polaron formation energy in Eq.~\eqref{Eq:Eform_aipe} is recovered.
Therefore we can say that the polaron equations of Sec.~\ref{Sec:plrn_eq_reciprocal} can be obtained in one of two ways: (i) either as an approximation to the field-theoretic approach of Sec.~\ref{Sec:FormalTheory}, or (ii) as approximation to the GW polaron equations presented in this section. The key difference between these approaches is that the former describes ionic dynamics by including quantum and nonadiabatic effects, while the latter employs the adiabatic and classical approximations.  

Overall, these connections provide ample opportunities for improving existing methods via increasingly sophisticated computational tools. To the best of our knowledge, these improvements have not yet been attempted in the polaron literature.

\subsection{Connection between the \textit{ab initio} polaron equations and the \textit{ab initio} canonical transformation approaches} \label{Sec:Connection_Denny_Bernardi}

The \textit{ab initio} polaron equations of Sec.~\ref{Sec:plrn_eq_reciprocal} can directly be related to the \textit{ab initio} Lang-Firsov transformation approach introduced in Sec.~\ref{Sec:bernardi_polaron}; this connection establishes a bridge between Green's function and canonical-transformation approaches to polaron theory. This link has been analyzed in detail by \textcite{Luo_Bernardi_2022}; here we summarize the main points of their analysis, and we identify additional points of contact.

Both approaches yield formally similar expressions for the polaron formation energy. Starting from Eq.~(\ref{Eq:plrn_ene_bernardi}) and taking the limit $\lambda_{mn}\!\to\!\delta_{mn}$, which is valid in the case of strongly polar materials, one finds: 
\begin{align}\label{Eq:connection_cano_gf_ene_1}
    \Delta E^{\rm LF}_{\rm f}
    &= \ve_{00} - \ve_\mathrm{CBM} - \frac{1}{N_p}\sum_{\bq\nu}\hbar\omega_{\bq\nu}\,\abs{\tilde{B}_{\bq\nu}}^2,
\end{align}
where $\ve_{00}=\bra{w_0}\hH_{\rm KS}\ket{w_0}$ and $\tilde{B}_{\bq\nu} = B_{\bq\nu,00}$, the latter being defined in Eq.~(\ref{Eq:bernardi_trans}).
On the other hand, rewriting the formation energy from the polaron equations, Eq.~\eqref{Eq:Eform_aipe}, in the Wannier basis yields:
\begin{eqnarray}
    \label{Eq:connection_cano_gf_ene_2}
    \Delta E^{\rm SVPG}_{\rm f}
    &=&
    {\sum}_{mm',pp'} A_m(\bR_p) A^{*}_{m'}(\bR_{p'})\, \ve_{mp, m'p'} 
    \nonumber \\
    &-&\ve_\mathrm{CBM}
    -\frac{1}{N_p} \sum_{\bq \nu} \abs{B_{\bq\nu}}^2 \hbar \w_{\bq\nu}.
\end{eqnarray}
When the electron is so localized that it can be described by a single Wannier function,
$A_m(\bR_p)=\delta_{m0}\delta_{p0}$, Eqs.~\eqref{Eq:connection_cano_gf_ene_1} and \eqref{Eq:connection_cano_gf_ene_2} coincide. Thus, under the assumptions of (i) occupation of a single Wannier function and (ii) negligible hopping ($\lambda_{mn}\to\delta_{mn}$), the two formalisms yield identical formation energies. This conclusion is confirmed by direct numerical comparison, which we review in Sec.~\ref{Sec:app_alkali}.

In addition to the above points of contact, we note that the lattice displacements determined in each method are related. In fact, \textcite{Luo_Bernardi_2022} show that the displacement operator in the canonical transformation method is shifted by:
\begin{align} \label{Eq:disp_shifts}
    \Delta\hat{\tau}^0_{\k\a p} = & ~ -\frac{2}{N_p} \sum_{\bq\nu} \left(\frac{\hbar}{2M_{\k}\omega_{\bq\nu}}\right)^{1/2} e_{\k\a,\nu}(\bq) e^{i\bq\cdot\bR_p} \nonumber \\
    & \times \sum_{ij} B_{-\bq\nu, ji} \, \hcd_{j}\hc_{i} ~.
\end{align}
After setting $B_{\bq\nu, ij} = g_{ij\nu}(\bq)/\hbar\omega_{\bq\nu}$ \cite{Hannewald_Hafner_2004},
and assuming uncorrelated electrons and phonons,
this shift leads to the following expectation value of the normal-mode coordinate operator:
\begin{equation}
    \langle \hat u_{\bq\nu}\rangle = -\frac{2}{N_p^{3/2}}\!\left(\!\frac{\hbar}{2M_0\omega_{\bq\nu}}\!\right)^{1/2}
    \sum_{ij}\frac{g^*_{ij\nu}(\bq)}{\hbar\omega_{\bq\nu}}\,A^{*}_{j}A_i .
\end{equation}
When this expression is transformed into reciprocal space, one obtains precisely the displacements found via the polaron equations, cf.\ Eq.~\eqref{Eq:u_from_g}. Therefore, the \textit{ab initio} polaron equations can be thought of as obtained from the canonical transformation:
\begin{equation}
    \hU  = \exp\!\left[N_p^{-1/2}\sum_{\bq\nu}\tilde B^{*}_{\bq\nu}(\had_{\bq\nu}-\ha_{\bq\nu})\right]~,
\end{equation}
where 
\begin{equation}
    \tilde B_{\bq\nu} = \frac{1}{N_p}\sum_{ij}\frac{g_{ij\nu}(\bq)}{\hbar\omega_{\bq\nu}} A_i^{*}A_j .
\end{equation}
This procedure is entirely analogous to the way the Lang-Firsov transformation of the Fr\"ohlich Hamiltonian leads to the Landau-Pekar model (cf.\ Supplemental Note~1).

In summary, for polarons that can be described by a wave function with uncorrelated electrons and phonons, the energies and displacements obtained from the \textit{ab initio} Lang-Firsov approach and from the \textit{ab initio} polaron equations are equivalent.
This condition is satisfied for a localized polaron that can be represented by a single Wannier function.
Extensions of the canonical transformation method to handle multiple Wannier functions should be possible, and \textcite{Luo_Bernardi_2022} suggested that such extensions would be useful to capture electron-phonon correlation effects that are not included in the \textit{ab initio} polaron equations. An alternative route to improving the accuracy of both methods would be to directly employ the many-body versions of these theories, as discussed in Sec.~\ref{Sec:FormalTheory}.

\subsection{\rev{Capabilities and computational complexity of different approaches to polaron calculations}
}

The polaron approaches discussed above are closely related, but they differ substantially in their range of applicability, observables, and computational cost. It is therefore useful to summarize which methods are best suited to specific types of \textit{ab initio} polaron calculations.

If the main objective is to determine polaron energetics, such as formation energies, eigenvalues, or the accompanying atomic displacements, all the approaches discussed in this review can in principle be used. Supercell DFT and GW calculations, as well as the \textit{ab initio} polaron equations, provide a direct real-space description of the lattice distortion and charge localization. Canonical-transformation approaches provide similar information in a Wannier basis. Therefore, when the spatial structure of the polaron wave function or charge density is the central quantity of interest, all these approaches are generally more suitable.

A different situation arises when the target quantity is the electron spectral function, especially satellite features associated with phonon shake-up excitations. In this case, Green's-function approaches, including the cumulant expansion discussed in Sec.~\ref{Sec:Cumulant}, offer a unique advantage. These methods do not require the same adiabatic approximation as static total-energy approaches, and they provide direct access to the spectral function through the retarded Green's function [Eq.~\eqref{Eq:spectral_func}].

Computational cost provides a second important criterion. As illustrated schematically in Fig.~\ref{Fig:jacob_ladder}, methods that retain a more complete quantum description of the nuclei and a higher level of atomistic detail generally become more demanding. Among first-principles approaches, direct solutions of the Hedin-Baym equations (Sec.~\ref{Sec:HedinBaym}) and diagrammatic Monte Carlo (Sec.~\ref{Sec:DMC}) are formally among the most complete, since they seamlessly incorporate quantum nuclear effects and non-adiabatic effects. However, they are also the most computationally demanding: the former requires the self-consistent solution of coupled electron and phonon Green's functions, while the latter involves sampling a large number of Feynman diagrams. A systematic analysis of the computational complexity of these approaches has not yet been reported.

Supercell DFT and GW calculations have more favorable, but still significant, computational scaling. Standard implementations scale approximately as $O(N^3)$ for DFT and $O(N^4)$ for GW, where $N$ denotes the number of atoms in the BvK supercell. These methods are well suited to small polarons, for which relatively small supercells may be able to accurately describe the self-trapped particle. These approaches also naturally include lattice anharmonicity and nonlinear electron-phonon couplings, which can be important for small polarons as discussed in Sec.~\ref{Sec:higher_order_effect}. On the other hand, their cubic or quartic scaling makes them difficult to apply to large polarons, for which very large supercells are typically needed. In addition, the neglect of quantum nuclear and non-adiabatic effects in standard DFT and GW treatments may lead to significant errors in the case of large polarons, as discussed in Sec.~\ref{Sec:selfcon_Green_2022}.

In these cases, the \textit{ab initio} polaron equations discussed in Sec.~\ref{Sec:PolaronEquations} and the canonical transformation method discussed in Sec.~\ref{sec:canonical} provide more flexible alternatives. Although a formal scaling analysis of these methods has not been performed, typically the solution of the corresponding equations is computationally inexpensive as compared to the calculation of electron-phonon matrix elements via DFPT. Therefore, these approaches are generally much more efficient than large-supercell DFT or GW calculations.

In summary, the choice of method depends primarily on the desired observable and on the computational cost that can be afforded. Supercell methods are direct and physically transparent for small polarons; Green's-function approaches are most appropriate for spectral functions; canonical transformations are highly efficient for strongly localized polarons; and \textit{ab initio} polaron equations offer a practical compromise for small and large polarons alike. Whenever possible, applying more than one approach to the same system is desirable, since such comparisons can help gauge the sensitivity of the results to the underlying approximations. At any rate, it is clear that more systematic studies are needed to test the validity of different approaches and
to analyze the associated computational complexity.

\subsection{Anharmonicity and nonlinear electron-phonon couplings}
\label{Sec:higher_order_effect}

In Secs.~\ref{Sec:History}, \ref{Sec:ManyBody}, \ref{Sec:FormalTheory}, \ref{Sec:PolaronEquations}, and \ref{Sec:GW-dft}-\ref{Sec:Connection_Denny_Bernardi} we considered the harmonic approximation for the lattice dynamics and linear electron-phonon couplings.  
One may ask under which conditions these approximations are justified, and whether higher-order effects may play a role in certain materials classes. This question can quantitatively be addressed by comparing polaron energies obtained via supercell approaches, as in Eqs.~\eqref{Eq:dft_eform_expr3} and \eqref{Eq:plrn_etot_gw}, and the \textit{ab initio} polaron equations, Eq.~\eqref{Eq:polaron_eq_Denny_1}, or the canonical transformation method, Eq.~\eqref{Eq:plrn_ene_bernardi}. For example, one could compare the potential energy landscape in either approach by minimizing the electronic Hamiltonian at fixed geometry.

\begin{figure}
    \centering
    \includegraphics[width=0.99\linewidth]{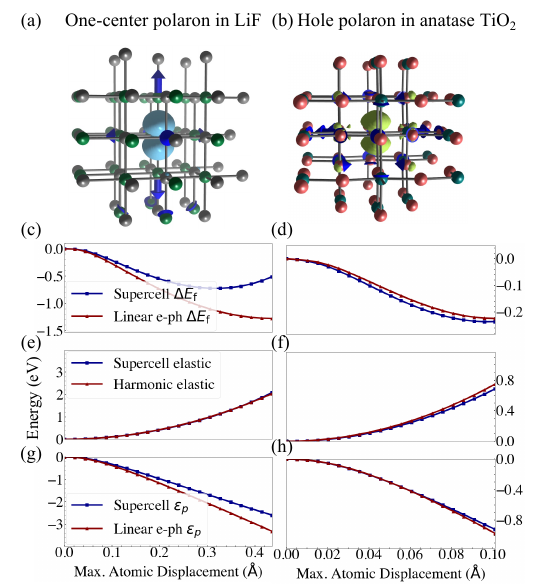}
    \caption{
    Polarons beyond the harmonic approximation and linear electron-phonon couplings.
    (a) Wavefunction and displacements of the small one-center hole polaron in LiF (Li, dark green; F, grey).
    (b) Wavefunction and displacements of the small hole polaron in anatase \ch{TiO2} (Ti, dark green; O, orange).
    (c) Polaron formation energy in LiF as a function of lattice distortion, using the supercell method of Eq.~\eqref{Eq:dft_eform_expr3} (blue) or the linear-response method of Eq.~\eqref{Eq:polaron_eq_Denny_1} (red). 
    (e),(g): Same as in (c), but for the elastic energy and the electronic energy contributions to the polaron formation energy, respectively. (d),(f),(h): Same as the adjacent panels on the left but for the hole polaron in anatase \ch{TiO2}. Data extracted from~\textcite{Dai_Giustino_2026a}
    }
    \label{Fig:higher_order}
\end{figure}

Figure~\ref{Fig:higher_order} shows such a comparison by \textcite{Dai_Giustino_2026a} for two prototypical cases, the small hole polaron in LiF and the small hole polaron in anatase \ch{TiO2}. 

In the case of LiF, Fig.~\ref{Fig:higher_order}(c) shows how linear electron-phonon coupling and harmonic approximation tend to overestimate the total energy in the distorted configuration; accordingly, the ionic displacements are also overestimated. A detailed analysis of each component of the total energy shows that the origin of such an overestimation is not the harmonic approximation, but rather the use of linear electron-phonon couplings, cf.\  Fig.~\ref{Fig:higher_order}(e) and (g). 

Unlike LiF, in the case of \ch{TiO2} Fig.~\ref{Fig:higher_order}(d) shows how linear electron-phonon coupling and harmonic approximation yield very good agreement with explicit supercell calculations. When decomposing the energy in terms of lattice and electronic components, Fig.~\ref{Fig:higher_order}(f) and (h) indicate that each component individually is well described by these approximations.

These two cases exemplify how the question on the validity of the harmonic approximation and linear electron-phonon coupling will probably need to be analyzed on a case-by-case basis. If linear electron-phonon couplings turned out to be insufficient to accurately describe polarons in a wide range of materials, this would have significant consequences for the large body of literature on polarons that relies on the linear approximation, ranging from effective Hamiltonians (Sec.~\ref{Sec:History}) all the way to the \textit{ab initio} DMC method (Sec.~\ref{Sec:DMC}) and the many-body polaron equations (Sec.~\ref{Sec:FormalTheory}).
\rev{Similarly, if the harmonic approximation fails to describe the lattice dynamics, then phonon-phonon interactions need to be considered when formulating Hamiltonians to study polarons. Alternatively, one would need to resort to \textit{effective} second-order interatomic force constants that account for anharmonicity}, as has been demonstrated, for example, by \textcite{Errea_Mauri_2013, Hellman_Simak_2013, Zacharias_Even_2023, Quan_Scheffler_2024}.

In the literature on effective Hamiltonians, recent studies introduced second-order electron-phonon couplings within effective models~\cite{Ragni_Mishchenko_2023, Klimin_Mishchenk0_2024}, and found a significant impact on the properties of polarons. Extending these studies to \textit{ab initio} calculations will require new methods to systematically compute electron-phonon matrix elements to second- and higher order. The challenge with these calculations is that the second-order matrix element in Eq.~\eqref{Eq:second_eph} necessitates evaluating the second-order variation of the KS potential, $\D^2 V_\mathrm{KS}/\D \tau_{\k \a p}\D \tau_{\k' \a ' p'}$. 
\rev{
The analytical expression of the long-range contribution to second-order variation has recently been established~\cite{Houtput_Tempere_2025}, and it is possible to access a subset of second-order electron-phonon coupling matrix elements approximately from the linear-order variation of the KS potential with the help of the rigid-ion approximation~\cite{Ponce_Gonze_2014, Ponce_Park_2025, Giustino_Cohen_2010}.
Nonetheless, calculating the full matrix of higher-order electron coupling} will require extending current density-functional perturbation theory methods \cite{Baroni_Gianozzi_2001} to higher-order derivatives.

We emphasize that these considerations are most relevant in the regime of very small polarons, where large lattice distortions amplify the role of higher-order terms. 
\rev{In fact, the maximum atomic distortion associated with the hole polaron in LiF is almost four times of that in anatase \ch{TiO2} [Fig.~\ref{Fig:higher_order}], which could rationalize why the linear electron-phonon approximation works less well in LiF, and there has been evidence showing that including the second-order electron-phonon interaction could significantly improve the agreement with the direct supercell calculations in this case. Preliminary work in this direction confirms that nonlinear
electron-phonon couplings start to dominate at large atomic displacements in LiF~\cite{Dai_Giustino_2026b}.}
For intermediate and large polarons, where ionic displacements are comparatively smaller, the harmonic approximation and linear electron-phonon couplings are expected to provide a robust and reliable description. Clearly, future work should focus on a systematic assessment of the validity of these approximations for a broad materials space (Sec.~\ref{Sec:Conclusions}). 

\section{Exciton polarons and self-trapped excitons}\label{Sec:ExcitonPolaron}

In this section, we review the progress in investigating exciton polarons and self-trapped excitons from first principles. Exciton polarons constitute a natural generalization of the concept of polarons to the case of neutral excitations of a solid. Much of the formalism reviewed in Secs.~\ref{Sec:History}-\ref{Sec:Connection} to study polarons can be generalized to investigate exciton polarons and self-trapped excitons.  

For completeness, we briefly review the standard Bethe-Salpeter approach for calculating excitons from first principles, without including exciton-phonon interactions. Then, we discuss how these interactions can be taken into account and how they lead to the formation of exciton polarons. In particular, we discuss supercell-based approaches as well as methods based on perturbation theory and linear exciton-phonon couplings. The presentation mirrors the discussion of polarons in Secs.~\ref{Sec:DFTPolaron} and \ref{Sec:PolaronEquations}.

\subsection{Excitons and the Bethe-Salpeter equation}
\label{Sec:bse_wannier}

Excitons are electron-hole pairs that are held together by their mutual Coulomb attraction. They are referred to as neutral excitations since the number of electrons does not change in the excited state. At a formal level, excitons corresponds to isolated poles of the two-particle correlation function~\cite{Onida_Rubio_2002, Sham_Rice_1966}. In \textit{ab initio} calculations for solids, excitons and their properties are computed using the Bethe-Salpeter equation (BSE)~\cite{Albrecht_Onida_1998,Rohlfing_Louie_1998}:
\begin{align}
    \label{Eq:bse_general}
    &\sum_{v'c'\bk'} 
    \big[
    (\ve_{c\bk+\bQ} - \ve_{v\bk}) \delta_{vv'}\delta_{cc'}\delta_{\bk \bk'}
    \nonumber \\
    &+
    \bra{v\bk, c\bk+\bQ} \hat{K}^\mathrm{eh} \ket{v'\bk', c'\bk'+\bQ}
    \big]
    a^{s\bQ}_{v'c'\bk'}
    =
    E_{s\bQ} a^{s\bQ}_{vc\bk}~.
\end{align}
The quantity within square brackets is the BSE Hamiltonian, $\hH_{\rm BSE}$; it contains the single-particle energies $\ve_{n\bk}$, which are typically obtained from GW calculations, and the interaction kernel $\hat K^\mathrm{eh}$. The latter consists of a direct screened Coulomb interaction, and a bare exchange interaction~\cite{Onida_Rubio_2002}. The matrix elements $\bra{v\bk, c\bk+\bQ} \hat{K}^\mathrm{eh} \ket{v'\bk', c'\bk'+\bQ}$ are evaluated over the BvK supercell, between the products of KS states $\psi_{v'\bk'}\psi_{c'\bk'+\bQ}$ and  $\psi_{v\bk}\psi_{c\bk+\bQ}$. The indices $v,v'$ and $c,c'$ run over valence and conduction states, respectively; this partitioning results from the Tamm-Dancoff approximation which is commonly employed in exciton calculations~\cite{Onida_Rubio_2002}.
The solution of Eq.~\eqref{Eq:bse_general} yields the exciton eigenvalues $E_{s\bQ}$, as well as the eigenvectors $a_{vc\bk}^{s\bQ}$. The latter represent the expansion coefficients of the exciton wavefunction in terms of KS electron-hole pairs:
\begin{align}
    \label{Eq:tda}
    \Psi_{s\bQ}(\br_{\rm e},\br_{\rm h}) 
    =
    \sum_{vc\bk} a_{vc\bk}^{s\bQ} 
    \psi_{v\bk}^*(\br_{\rm h}) \psi_{c\bk+\bQ}(\br_{\rm e})~.
\end{align}
In these expressions, $s$ denotes the exciton band index, and $\hbar\bQ$ is the exciton center-of-mass momentum. In Green's function's language, $\Psi_{s\bQ}$ is the Lehmann amplitude for the two-particle correlator.

Optical spectra computed within the \textit{ab initio} BSE framework often yield excellent agreement with experiments, both in terms of energies and intensities, and both for bound states and for the continuum portion of the spectrum~\cite{Onida_Rubio_2002}. The predictive power of this formalism makes it an ideal starting point to study interaction between excitons and phonons.

\subsubsection{Wannier exciton model}\label{Sec:Wannex}

As in the case of polarons, alongside \textit{ab initio} BSE calculations, exciton physics is also studied using several effective Hamiltonians. Here we briefly summarize the Wannier exciton model \cite{Wannier_1937}, since it provides a useful starting point to discuss exciton polarons. 

The Wannier model considers one parabolic valence band and one parabolic conduction band with effective masses $m_{\rm h}^*$ and $m_{\rm e}^*$, respectively; the electron and hole in this model experience an attractive Coulomb interaction, screened by the high-frequency dielectric constant of the host material \cite{Combescot_Shiau_2015,Mahan_2000}.
Within these approximations, the BSE Hamiltonian $\hH_{\rm BSE}$ of Eq.~\eqref{Eq:bse_general} simplifies into~\cite{Yu_Cardona_2010, Fuchs_Bechstedt_2008}:
\begin{eqnarray}
    \label{Eq:bse_wf}
    && \hspace{-15pt} H_{\mathrm{BSE}, v\bk c\bk+\bQ,v\bk' c\bk'+\bQ} 
    = \nonumber \\
    && \hspace{-15pt}
    \left(
    \frac{\hbar^2 |{\bk+\bQ}|^2}{2m_{\rm e}^*} + \frac{\hbar^2 |{\bk}|^2}{2m_{\rm h}^*} + E_{\rm g}
    \right)\delta_{\bk,\bk'}
    -  \frac{e^2}{\ve_0\ve^{\infty}}\frac{1}{N_p \Omega} \frac{1}{\abs{\bk'-\bk}^2}. \nonumber\\
\end{eqnarray}
In this expression, we employed the same notation as in Sec.~\ref{Sec:History}; $E_{\rm g}$ denotes the fundamental quasiparticle gap of the system. The eigenvalue problem for this effective Hamiltonian admits exact solutions~\cite{Yu_Cardona_2010}. To see this, we transform the exciton eigenvector $a_{\bk}^{s\bQ}$ into the Wannier representation $\Phi_s(\bR_{\rm e},\bR_{\rm h})$ via the double Fourier transform: 
\begin{align}
    \label{Eq:a_to_w_wannier}
    \Phi_s(\bR_{\rm e},\bR_{\rm h})
    &=
    \frac{1}{N_p^2}
    \sum_{\bk\bQ} e^{i(\bk+\bQ) \cdot \bR_{\rm e}}  e^{-i\bk \cdot \bR_{\rm h}} a_{\bk}^{s\bQ},
\end{align}
having omitted the subscripts $v,c$ because we only have one valence band and one conduction band. In this representation, the exciton wavefunction of Eq.~\eqref{Eq:tda} can be expressed as a linear combination of products of electron and hole Wannier functions, $w_{c\bR_{\rm e}}(\br_{\rm e})$ and $w^*_{v\bR_{\rm h}}(\br_{\rm h})$:
\begin{equation}
    \Psi_{s\bQ}(\br_{\rm e}, \br_{\rm h})
      = \!\!\sum_{\bR_{\rm e},\bR_{\rm h}}\!\! \Phi_s(\bR_{\rm e},\bR_{\rm h})
    w_{c\bR_{\rm e}}(\br_{\rm e}) w^*_{v\bR_{\rm h}}(\br_{\rm h}).
\end{equation}
Given the translational invariance of the Hamiltonian in Eq.~\eqref{Eq:bse_wf}, it is advantageous to recast these expressions in terms of center-of-mass coordinate, $\bR = (m_{\rm e}^* \bR_{\rm e} + m_{\rm h}^* \bR_{\rm h})/M$, and relative coordinate, $\br = \bR_{\rm e} - \bR_{\rm h}$, where $M = m_{\rm e}^* + m_{\rm h}^*$ is the total mass. In this reference frame, Eq.~\eqref{Eq:bse_wf} maps exactly into the Schr\"odinger equation for the hydrogen atom; its solutions correspond to the standard Rydberg series~\cite{Yu_Cardona_2010}. For completeness, we here report the energy and wavefunction of the lowest energy hydrogenic state:
\begin{equation}
    E_0(\bQ) = E_{\rm g} + \frac{\hbar^2 \abs{\bQ}^2}{2 M} 
    - \frac{\mu}{2} 
    \left(\frac{e^2}{4\pi\ve_0\ve^\infty  \hbar}\right)^2,
\end{equation}
\begin{equation}
    \Phi_0(\br, \bR; \bQ) = \sqrt{\frac{\Omega}{\pi a_0^3}}
    e^{i\bQ \cdot \bR}
    e^{-{\abs{\br}}/{a_0}},
\end{equation}
where $a_0 = {4\pi \ve^{\infty} \hbar^2} / {\mu e^2}$ is the exciton Bohr radius, and $\mu = (1/m_{\rm e}^* + 1/m_{\rm h}^*)^{-1}$ is the reduced effective mass. Going back to the Bloch representation, one obtains the BSE eigenvector corresponding to this state:
\begin{equation}
    \label{Eq:1s_state}
    a_{\bk}^{0,\bQ} = 8\sqrt{\frac{\pi a_0^3}{\Omega}}
    \left( 1+a_0^2\abs{\bk + m_{\rm h}^*\bQ/M}^2 
     \right)^{-2}.
\end{equation}
This expression is used in Sec.~\ref{Sec:ex-ph} to discuss analytical models of the exciton-phonon matrix elements that are needed to study exciton polarons. 

\subsection{Wannier/Landau-Pekar model of the exciton polaron}
\label{Sec:explrn_model}

Even though the exciton is a neutral excitation, the atomic-scale structure of its wavefunction gives rise to local charge fluctuations that couple with phonons. When this coupling is strong, the lattice distortion can produce a potential well, leading to exciton localization. The resulting excitation is called an exciton polaron. In the case of strong exciton-phonon coupling, the exciton polaron may be unable to escape this well, and one speaks of a self-trapped exciton (STE). STEs are often responsible for the Stokes shifts between the light absorption and emission energies in materials. In the following, we use the term ``exciton polaron'' to refer to both scenarios, with the understanding that the STE constitute the limiting case of an exciton polaron.

To gain insight into the mechanism of formation of exciton polarons, we discuss an effective model that blends together the Wannier exciton model of Sec.~\ref{Sec:Wannex} and the Landau-Pekar model of Sec.~\ref{Sec:LandauPekar} \cite{Dai_Giustino_2024b}. In this model, the electron-hole interaction is described via Eq.~\eqref{Eq:bse_wf}, and the interactions of the electron and the hole with the lattice are described via the second term in Eq.~\eqref{Eq:LP_totel_funcpsi}. Taken together, these contributions lead to the formation energy of the exciton polaron, referred to the quasiparticle band gap $E_{\rm g}$: 
\begin{align}
    \label{Eq:etot_density}
    \Delta E_{\mathrm{f}}^{\rm xp} = 
    &-\frac{\hbar^2}{2m_{\rm e}^*}  \sum_{\bR_{\rm e},\bR_{\rm h}} 
    |\nabla_{\bR_{\rm e}} \Phi(\bR_{\rm e},\bR_{\rm h})|^2
    \nonumber \\
    &-\frac{\hbar^2}{2m_{\rm h}^*} \sum_{\bR_{\rm e},\bR_{\rm h}}
    |\nabla_{\bR_{\rm h}}\Phi(\bR_{\rm e},\bR_{\rm h})|^2
    \nonumber \\
    &
    -\sum_{\bR_{\rm e},\bR_{\rm h}}
    \frac{e^2}{4\pi \ve_0 \ve^{\infty}} 
    \frac{|\Phi(\bR_{\rm e},\bR_{\rm h})|^2}{|\bR_{\rm e}-\bR_{\rm h}|}
    \nonumber \\
    &- 
    \frac{e^2}{8\pi\ve_0} \frac{1}{\kappa} 
    \sum_{\bR_{\rm e},\bR_{\rm e}'}
    \frac{n_{\rm e}(\bR_{\rm e})n_{\rm e}(\bR_{\rm e}')}{|\bR_{\rm e} - \bR_{\rm e}'|}
    \nonumber \\
    &- 
    \frac{e^2}{8\pi\ve_0} \frac{1}{\kappa} 
    \sum_{\bR_{\rm h},\bR_{\rm h}'}
    \frac{n_{\rm h}(\bR_{\rm h})n_{\rm h}(\bR_{\rm h}')}{|\bR_{\rm h} - \bR_{\rm h}'|}
    \nonumber \\
    &+
    2\frac{e^2}{8\pi\ve_0} \frac{1}{\kappa} 
    \sum_{\bR_{\rm e},\bR_{\rm h}}
    \frac{n_{\rm h}(\bR_{\rm h})n_{\rm e}(\bR_{\rm e})}{|\bR_{\rm h} - \bR_{\rm e}|}
    .
\end{align}  
The first three lines in this expression are the same as in the Wannier exciton model of Sec.~\ref{Sec:Wannex}, and so is the meaning of the function $\Phi(\bR_{\rm e},\bR_{\rm h})$. The last three lines contain the electron and hole densities, i.e., the two-particle density matrix marginalized over the electron or the hole variable:
\begin{equation}
    \label{Eq:charge_density}
    n_{\rm e}(\bR_{\rm e}) = \sum_{\bR_{\rm h}} |{\Phi(\bR_{\rm e},\bR_{\rm h})}|^2,\,\,\,
        n_{\rm h}(\bR_{\rm h}) = \sum_{\bR_{\rm e}} |{\Phi(\bR_{\rm e},\bR_{\rm h})}|^2.
\end{equation}
The lines in Eq.~\eqref{Eq:etot_density} containing these densities describe the interaction energy of the electron and hole with their induced lattice polarizations, respectively, and the interaction of the electron with the lattice polarization of the hole and vice-versa. 

Equation~\eqref{Eq:etot_density} can alternatively be interpreted as follows: if we eliminate the third and sixth lines, we have the formation energies of two independent electron and hole polarons; adding the third line introduces attractive interaction between these polarons; adding the sixth line accounts for the weakening of the lattice distortion from the partial cancellation of the electron and hole charge densities. From this point of view, the formation of an exciton polaron can be thought of as a two-step process: (i) the electron and the hole polarize the lattice to form polarons, and (ii) the polarons thus formed then bind together through their mutual Coulomb attraction to form the exciton polaron.

An expression similar to Eq.~\eqref{Eq:etot_density} was originally proposed by \textcite{Iadonisi_Basaani_1983} within the context of the Fr\"ohlich model and the canonical transformation method (Sec.~\ref{Sec:LLP_transform}); however, in their expression the center-of-mass momentum of the exciton is not included, and as a result the formation energy is overestimated. \rev{Their treatment mimics the study of an electron bound to a localized hole due to the presence of impurities or core holes, where the  excitons are naturally pinned, and one can therefore neglect dispersion of the exciton binding energy with respect to the center-of-mass momentum~\cite{Strinati_1987}. }

If we define the net charge density of the exciton as $\Delta n(\bR) = n_{\rm e}(\bR) - n_{\rm h}(\bR)$,  Eq.~\eqref{Eq:etot_density} can be rewritten more compactly by combining together the last three lines:
\begin{align}
    \label{Eq:etot_density2}
    \Delta E^{\rm xp}_{\mathrm{f}} = 
    &-\frac{\hbar^2}{2m_{\rm e}}  \sum_{\bR_{\rm e},\bR_{\rm h}} 
    |\nabla_{\bR_{\rm e}} \Phi(\bR_{\rm e},\bR_{\rm h})|^2
    \nonumber \\
    &-\frac{\hbar^2}{2m_{\rm h}}  \sum_{\bR_{\rm e},\bR_{\rm h}} 
    |\nabla_{\bR_{\rm h}} \Phi(\bR_{\rm e},\bR_{\rm h})|^2
    \nonumber \\
    &-\sum_{\bR_{\rm e},\bR_{\rm h}}
    \frac{e^2}{4\pi \ve_0 \ve^{\infty}} 
    \frac{|{\Phi(\bR_{\rm e},\bR_{\rm h})}|^2}{|{\bR_{\rm e}-\bR_{\rm h}}|}
    \nonumber \\
    &
    - \frac{e^2}{8\pi\ve_0} \frac{1}{\kappa} 
    \sum_{\bR,\bR'}
    \frac{\Delta n(\bR) \Delta n(\bR')}{|{\bR - \bR'}|}~.
\end{align}  
In this form, it is apparent that, unlike for the polaron, in the case of the exciton polaron it is the local charge density fluctuations that couple to lattice distortions. This suggests that, within this simplified Wannier/Landau-Pekar model, the formation energy of an exciton polaron should be smaller than the formation energies of the independent electron and hole polarons in the same material.

Using Eqs.~\eqref{Eq:etot_density} and \eqref{Eq:etot_density2}, it is possible to obtain simple variational estimates for the formation energy of exciton polarons. The reader is referred to \cite{Dai_Giustino_2024b} for an extended discussion of this aspect. We also mention that, while Eq.~\eqref{Eq:etot_density} has been introduced heuristically for pedagogical purposes, it can be derived starting from the \textit{ab initio} exciton polaron equations~\cite{Dai_Giustino_2024b}, which are reviewed in Sec.~\ref{Sec:explrn_eqns}.

\subsection{Supercell calculations of exciton polarons} \label{Sec:ExcitonPolaron_Supercell}

The effective Hamiltonians discussed in Secs.~\ref{Sec:bse_wannier} and \ref{Sec:explrn_model} are useful to gain insight into the mechanisms that drive the formation of exciton polarons, but are inadequate for quantitatively predictive calculations. In order to achieve predictive accuracy, several approaches have been developed over the past few decades. In this section we focus on approaches that rely on explicit supercell calculations; in Secs.~\ref{Sec:explrn_eqns} and \ref{Sec:ex-ph} we discuss methods that leverage a reciprocal space formulation of the problem, in analogy with the polaron equations of Sec.~\ref{Sec:PolaronEquations}.

Supercell-based methods can be grouped into three categories: wavefunction methods, DFT methods, and Green's function methods. A brief review of these methods is given below.

Wavefunction methods are the workhorse of quantum chemistry~\cite{Szabo_Ostlund_1996}, and aim to obtain highly accurate correlated wavefunctions in the excited state. From the wavefunction, one can determine atomic forces via the Hellman-Feynman theorem, and perform structural optimization in the excited state. So far, a variety of methods in this class have been employed to investigate structure and energetics of exciton polarons, including unrestricted Hartree-Fock (UHF), complete active space self-consistent field (CASSCF), and coupled-cluster calculations~\cite{Fisher_Stoneham_1990, VanGinhoven_Corrales_2003}.
These methods offer high accuracy but are primarily designed for finite systems and have unfavorable scaling with system size. In practice, they have been applied to the study of small molecular clusters carved out of extended solids, and structural optimization has been possible only within UHF~\cite{Fisher_Stoneham_1990, VanGinhoven_Corrales_2003}. Furthermore, constructing a model cluster to represent an extended crystal poses some challenges: for example, \textcite{VanGinhoven_Corrales_2003} found exciton polarons in $\alpha$-\ch{SiO2}, but these solutions are less stable than the extended exciton in the solid, in contrast to experiments.  
These practical limitations hinder the adoption of wavefunction methods in the study of exciton polarons in extended systems.

Unlike wavefunction methods, DFT-based approaches to exciton polarons focus on the evaluation of the electron density in the excited state. Work in this area can be further grouped into methods that employ standard DFT and those based on time-dependent DFT (TDDFT)~\cite{Runge_Gross_1984}.
In static DFT and non-magnetic systems, one can realize a state that approximates the exciton by considering the ground state of the spin triplet configuration; in this case, a DFT energy minimization with constrained spins yields a ground state with a hole in the valence bands and an electron in the conduction bands \cite{Song_Corrales_2000, Song_Jonsson_2000}. This method is typically referred to as $\Delta$SCF approach (for self-consistent field), and has been used to study exciton polarons and self-trapped excitons in a host of materials, from diamond and $\alpha$-\ch{SiO2} \cite{Song_Corrales_2000, Song_Jonsson_2000, Mackrodt_Dovesi_2022} to complex halide perovskites \cite{Luo_Tang_2018}. Recently, this approach was also used to investigate the dissociation dynamics of self-trapped excitons in LiF \cite{Ruiz_Lastra_2025}. An open question in this area is whether semilocal DFT exchange and correlation potentials can provide good approximations to the BSE kernel in Eq.~\eqref{Eq:bse_general}, given that this kernel contain long-range Coulomb terms~\cite{Onida_Rubio_2002}. 

\begin{figure}
    \centering
    \includegraphics[width=0.98\linewidth]{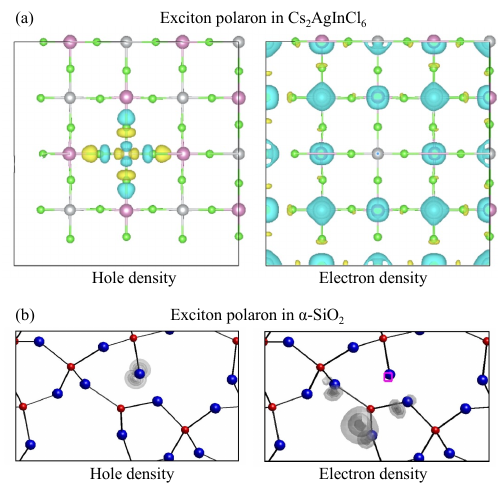}
    \caption{Calculations of exciton polarons via supercell-based approaches, using excited-state force obtained from TDDFT or BSE. (a) Charge density isosurfaces (blue and yellow surfaces) of the hole and the electron of the exciton polaron in \ch{Cs2AgInCl6}. From \textcite{Jin_Galli_2024}. (b) Charge density isosurfaces (grey shades) of the hole and the electron of the exciton polaron in $\alpha$-\ch{SiO2}. From \textcite{Ismail-Beigi_Louie_2005}.}
    \label{Fig:supercell_explrn}
\end{figure}

TDDFT provides a more rigorous way of computing excited states in DFT, and its linear-response version is popular in the study of optical absorption and exciton physics \cite{Onida_Rubio_2002, Jin_Galli_2023, Sitt_Chelikowsky_2007}. Since atomic forces are readily available within TDDFT~\cite{Jin_Galli_2023, Hutter_2003, Sitt_Chelikowsky_2007}, this method is also applicable to the study of exciton polarons. For example, TDDFT has been used to investigate the formation of exciton polarons in halide perovskites, cf.\ Fig.~\ref{Fig:supercell_explrn}(a), and quantitative agreement with experimental Stokes shift has been achieved \cite{Jin_Galli_2024}. 

In the case of Green's function methods, calculations of exciton polarons and self-trapped excitons are based on a generalization of the BSE approach (cf.\ Sec.~\ref{Sec:bse_wannier}) to obtain atomic forces in the excited state. In the Born-Oppenheimer approximation, the force on the ions can be obtained via the Hellman-Feynman theorem as the derivative of the total energy with respect to the ionic displacements. In the excited state, such a force consists of two contributions: the derivative of the ground-state energy, which is the standard DFT force, and the derivative of the excitation energy, $\partial E_{s\bQ}/\partial \tau_{\k \a p}$, which requires evaluating the variation of the BSE kernel with respect to ionic displacements. \textcite{Ismail-Beigi_Louie_2003} showed that, for an exciton $s\bQ$, the force due to the latter term can be expressed as:
\begin{align}
    \label{Eq:excited_force}
    F_{\k \a p}^{s\bQ}
    &=
    -\sum_{\substack{vc\bk \\ v'c'\bk'}} 
    a_{vc\bk}^{s\bQ,*} a_{v'c'\bk'}^{s\bQ}
    \nonumber \\
    &\times \frac{\partial}{\partial \tau_{\kappa \alpha p}} 
    \bra{v\bk c\bk+\bQ} \hat{H}^\mathrm{BSE} \ket{v'\bk' c'\bk'+\bQ},
\end{align}
where the derivative of the BSE Hamiltonian is evaluated perturbatively via DFPT. In subsequent work, \textcite{Ismail-Beigi_Louie_2005} applied this method to the study of self-trapped excitons in $\alpha$-\ch{SiO2}, and successfully reproduced the measured Stokes shift,  cf.\ Fig.~\ref{Fig:supercell_explrn}(b).

Since this method requires BSE calculations on supercells, its computational cost is significant and this has hindered its widespread adoption. Recently, following recent algorithmic advances in many-body perturbation theory methods, and especially the drastic acceleration provided by graphic processing units (GPUs) \cite{delBen_Deslippe_2020}, there has been a revival of interest in these techniques~\cite{Grande_Strubbe_2025}.

The main advantage of supercell-based methods is that they can capture anharmonic effects and nonlinear exciton-phonon couplings. The disadvantages are the computational cost of performing BSE calculations in large supercells, and the need to perform convergence tests with respect to supercell size. In fact, even though exciton polarons are neutral excitations and do not carry electric monopoles, they may carry \rev{dipoles,} quadrupoles, and higher-order terms; these long-range electrostatic interactions demand careful convergence tests, as for the DFT approaches to polarons reviewed in Sec.~\ref{Sec:DFT_difficulty}. In practice, these calculations are currently limited to highly localized self-trapped excitons.

\subsection{\textit{Ab initio} exciton polaron equations}
\label{Sec:explrn_eqns}

The use of large supercells in the study of exciton polarons can be avoided by adopting a reciprocal-space representation analogous to the \textit{ab initio} polaron equations of Sec.~\ref{Sec:PolaronEquations}. This approach was independently introduced by \textcite{Dai_Giustino_2024a, Dai_Giustino_2024b}, \textcite{Bai_Meng_2024, Yang_Draxl_2022}. Here, we follow the derivation of \textcite{Dai_Giustino_2024b}.

As in \textcite{Ismail-Beigi_Louie_2003}, one expresses the total energy of the charge-neutral, $N$-electron system in the presence of an exciton as the sum of the ground-state energy and the expectation value of the BSE Hamiltonian over the two-particle wavefunction $\Psi(\br_{\rm e}, \br_{\rm h})$:
\begin{align}
    \label{Eq:etot_rs}
    &E\left[ \Psi(\br_{\rm e},\br_{\rm h}), \btau_0 + \Delta \btau
    \right] = E^N(\btau_0)
    \nonumber \\
    &+\int_{\rm sc}
    \Psi^*(\br_{\rm e},\br_{\rm h}) \hH_{\mathrm{BSE}}(\fourr;\btau_0 + \Delta \btau) \Psi(\br_{\rm e}',\br_{\rm h}')
    d\br \nonumber \\
    &+\frac{1}{2}\!\! \sum_{\substack{\k p\a\\ \k' p' \a'}} \ifc \Delta \tau_{\k p \a} \Delta \tau_{\k' p' \a'},
\end{align}
where the lattice energy is written in the harmonic approximation, the matrix of interatomic force constants refers to the ground state (without exciton), and the integral extends over the BvK supercell. Equation~\eqref{Eq:etot_rs} is a variational energy in the wavefunction $\Psi$ and the atomic displacements $\Delta \tau_{\k p \a}$, in complete analogy with Eq.~\eqref{Eq:toten_aipe} for charged polarons.
Minimization of the energy can be performed using the method of Lagrange multipliers: one expands the BSE Hamiltonian to first order in the displacements, and sets to zero the derivatives $\delta E/\delta \Psi^*$ as well as  $\partial E/\partial \Delta \tau_{\k p \a}$, subject to the normalization constraint $\int_{\rm sc}|{\Psi(\br_{\rm e},\br_{\rm h})}|^2 d\br_{\rm e} d\br_{\rm h}=1$. 
After some algebra, one arrives at the following coupled nonlinear eigenvalue problem~\cite{Dai_Giustino_2024b}:
\begin{align}
    \label{Eq:minimize_psi}
    &\int_{\rm sc}\!\!
    H^0_{\mathrm{BSE}}(\fourr) \,\Psi(\br_{\rm e}',\br_{\rm h}')\,d\br_{\rm e}' d\br_{\rm h}' 
    \nonumber \\
    &+
    \sum_{\k p \a}
    \int_{\rm sc}
    \frac{\D \hat{H}^0_{\mathrm{BSE}}(\fourr)}{\D \tau_{\k p \a}}\,
    \Psi(\br_{\rm e}',\br_{\rm h}')\,
    d\br_{\rm e}' d\br_{\rm h}'\, \Delta \tau_{\k p \a}
    \nonumber \\
    &\hspace{5pt}= \ve \,\Psi(\br_{\rm e},\br_{\rm h}), \vspace{-15pt}
\end{align}
\begin{align}
    \label{Eq:minimize_tau}
    &\Delta \tau_{\k p \a} 
    = 
    -\!\!\sum_{\k'\a' p'}\iifc
    \nonumber \\
    &\hspace{2pt}\times\int_{sc} \!\!
    \Psi^*(\br_{\rm e},\br_{\rm h})  
    \frac{\D \hat{H}^0_{\mathrm{BSE}}(\fourr)}{\D \tau_{\k' p' \a'}}
    \Psi(\br_{\rm e}',\br_{\rm h}')\,d\br,
\end{align}
where $\hH^0_{\mathrm{BSE}}$ is the BSE Hamiltonian for the undistorted equilibrium system, and the eigenvalue $\ve$ is the Lagrange multiplier. These equations constitute the excitonic counterpart of Eqs.~\eqref{Eq:polaron_eq_realspace_1} and \eqref{Eq:polaron_eq_realspace_2} for charged polarons.

Equations~(\ref{Eq:minimize_psi}) and (\ref{Eq:minimize_tau}) are in real space, therefore their solution would still require the use of large supercells. To circumvent this challenge, one can employ a similar strategy as for polarons in Sec.~\ref{Sec:plrn_eq_reciprocal}, and express $\Psi(\br_{\rm e},\br_{\rm h})$ as a linear combination of exciton states $\Psi_{s\bQ}(\br_{\rm e},\br_{\rm h})$ of the undistorted, ground-state structure:
\begin{eqnarray}
    \label{Eq:exciton_basis}
    &&\Psi(\br_{\rm e},\br_{\rm h})
    =
    \frac{1}{\sqrt{N_p}}
    \sum_{s\bQ} A_{s\bQ}\, \Psi_{s\bQ}(\br_{\rm e},\br_{\rm h}).
\end{eqnarray}
Similarly, the displacements $\Delta \tau_{\k p \a}$ are expressed as a linear combination of normal vibrational modes of the undistorted structure, following Eq.~(\ref{Eq:Bqv2dtau}). The coefficients $A_{s\bQ}$ and $B_{\bq \nu}$ in Eqs.~\eqref{Eq:exciton_basis} and \eqref{Eq:Bqv2dtau} can be interpreted as the contributions of each exciton and each phonon of the undistorted structure to the exciton polaron, respectively.

Upon substituting Eqs.~\eqref{Eq:tda}, \eqref{Eq:exciton_basis}, and \eqref{Eq:Bqv2dtau} inside Eqs.~\eqref{Eq:minimize_psi}-\eqref{Eq:minimize_tau}, one arrives at the coupled nonlinear system of equations for the coefficients $A_{s\bQ}$ and $B_{\bq \nu}$:
\begin{eqnarray}
    \label{Eq:explrneqn}
    &&\sum_{s'\bQ'} 
    \bigg[ 
    E^0_{s\bQ}
    \delta_{ss'}
    \delta_{\bQ \bQ'} 
    -\frac{2}{N_p}\sum_{\nu}
    B_{\bQ-\bQ' \nu} \mathcal{G}_{ss'\nu}(\bQ',\bQ-\bQ')
    \bigg] 
    \nonumber \\ && \hspace{20pt}\times A_{s'\bQ'}
    = \ve A_{s\bQ},  \\
    &&B_{\bQ \nu}  
    = 
    \frac{1}{N \hbar \w_{\bQ \nu}} 
    \sum_{\substack{ss' \bQ'}}
    A_{s'\bQ'}^* A_{s\bQ'+\bQ}
 \mathcal{G}^*_{ss'\nu}( \bQ', \bQ), \label{Eq:bmat}
\end{eqnarray}
having introduced the exciton-phonon coupling matrix element $\cG_{ss'\nu}(\bQ,\bq)$:
\begin{align}
    \label{Eq:exphg}
    \cG_{ss'\nu}(\bQ,\bq)
    =&
    \sum_{vc\bk} a_{vc\bk}^{s\bQ+\bq*}  
    \Bigg[
    \sum_{c'}
    g_{cc'\nu} (\bk+\bQ,\bq)
    a_{vc'\bk}^{s'\bQ}
    \nonumber \\
    &-\sum_{v'}
    g_{v'v\nu} (\bk,\bq) 
    a_{v'c\bk + \bq}^{s'\bQ}
    \Bigg].
\end{align}
Here, the $a_{vc\bk}^{s\bQ}$'s are the BSE eigenvectors from Eq.~\eqref{Eq:bse_general}, and the $g_{cc'\nu} (\bk+\bQ,\bq)$'s are the standard electron-phonon coupling matrix elements. This expression for the exciton-phonon coupling matrix elements coincides with those obtained by \textcite{Chen_Bernardi_2020} and \textcite{Antonius_Louie_2022} in the study of exciton lifetimes and temperature dependence, respectively. Equations~\eqref{Eq:explrneqn} and \eqref{Eq:bmat} are referred to as the \textit{exciton polaron equations} \cite{Dai_Giustino_2024b}. 

Once Eq.~\eqref{Eq:explrneqn} and \eqref{Eq:bmat} are solved, the formation energy of the exciton polaron can \rev{be} computed according to the following expression:
\begin{align}
\label{Eq:exciton_etot}    
   \Delta E^{\rm xp}_{\mathrm{f}} =
   \ve - \ve_0
   + \frac{1}{N_p} \sum_{\bq \nu} |B_{\bq\nu}|^2 \hbar \w_{\bq\nu},
\end{align}
where $\ve$ is the eigenvalue in Eq.~\eqref{Eq:explrneqn}, and $\ve_0$ is the BSE eigenvalue of the lowest-lying bright exciton for the undistorted structure. It should be noted that Eq.~\eqref{Eq:exciton_etot} is in close analogy to the formation energy of charged polarons in Eq.~\eqref{Eq:Eform_aipe}.
The ingredients required in Eqs.~\eqref{Eq:explrneqn}-\eqref{Eq:exphg} can be computed from first principles within the crystal unit cell, therefore no supercell calculation is required with this approach \cite{Dai_Giustino_2024b, Yang_Draxl_2022,Bai_Meng_2024}. The capability of this approach is best illustrated by the identification of a large exciton polaron in anatase \ch{TiO2}, which is reviewed in Sec.~\ref{Sec:TMOs}.

\subsection{Exciton-phonon coupling matrix elements}
\label{Sec:ex-ph}

As shown in Sec.~\ref{Sec:explrn_eqns}, exciton-phonon coupling matrix elements naturally emerge in the derivation of the exciton-polaron equations. In addition to this derivation, which proceeds by directly differentiating the BSE Hamiltonian with respect to the atomic displacements, \textcite{Antonius_Louie_2022} showed that exciton-phonon coupling matrix elements can also be obtained by recasting the standard electron-phonon Hamiltonian in Eq.~\eqref{Eq:epi-hamilt} in terms of exciton creation/annihilation operators. This is achieved by combining the electron and hole creation/annihilation operators with the BSE eigenvectors; in this alternative representation, the exciton-phonon coupling matrix elements of Eq.~\eqref{Eq:exphg} appear naturally. 

To provide some intuition into the structure of exciton-phonon matrix elements, we consider the 
Fr\"ohlich model in Eq.~\eqref{Eq:froh_eph}, and the BSE eigenvectors for the Wannier model in Eq.~\eqref{Eq:1s_state}. After replacing inside Eq.~\eqref{Eq:exphg}, we obtain exciton-phonon couplings in the Wannier-Fr\"ohlich model following \textcite{Dai_Giustino_2024b}:
\begin{eqnarray}
    \label{Eq:exphg_froe}
    &&\hspace{-5pt}\cG^{\mathrm{WF}}(\bQ, \bq) 
    =i\hbar\w \frac{\sqrt{\a}\,q_0 }{|\bq|}\nonumber \\
    &&
    \hspace{10pt}\times \left[\frac{1}{ \left(1+ a_0^2 b^2|\bq|^2/4 \right)^2}
    -
    \frac{1}{ \left(1+a_0^2 a^2 |\bq|^2/4 \right)^2}
    \right]\!,\,\,\hspace{10pt}
\end{eqnarray}
where $q_0$ is given in Eq.~\eqref{Eq:froh_eph}, $a = m_{\rm e}^*/(m^*_{\rm e} + m_{\rm h}^*)$, and $b = m_{\rm h}^*/(m_{\rm e}^* + m_{\rm h}^*)$. 
From this expression, one sees that the electron-phonon coupling (first term in the square brackets) and the hole-phonon coupling (second term in the brackets) tend to compensate each other due to their opposite charge. Furthermore, Eq.~\eqref{Eq:exphg_froe} also shows that $\cG^{\mathrm{WF}}(\bQ, \bq)$ is proportional to $|\bq|$ in the limit $\bq \rightarrow 0$. This behavior is consistent with the fact that excitons are electrically neutral overall, therefore their interaction with the lattice does not carry the distinctive electric dipole term of the Fr\"ohlich interaction~\cite{Verdi_Giustino_2015}. This observation is consistent with the discussion of the Wannier/Landau-Pekar model in Sec.~\ref{Sec:explrn_model}.

\begin{figure}
\includegraphics[width=0.99\linewidth]{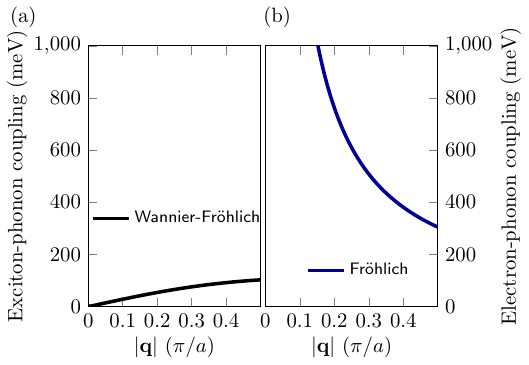}
  \caption{
  (a) Exciton-phonon coupling matrix elements in the Wannier-Fr\"ohlich model as a function of phonon wavevector $|\bq|$. (b) Electron-phonon coupling matrix elements in the Fr\"ohlich model as a function of phonon wavevector. In both panels, we use parameters corresponding to LiF, from Table~I of \textcite{Dai_Giustino_2024b}.
  }
  \label{Fig:wannier_ex_ph}
\end{figure}

Figure~\ref{Fig:wannier_ex_ph}(a) illustrates the exciton-phonon matrix elements in the Wannier-Fr\"ohlich model, for a choice of parameters corresponding to \ch{LiF}. For comparison, in panel (b) of the same figure we show the corresponding electron-phonon coupling matrix elements; this latter coupling diverges for $\bq \to 0$, as expected for the potential generated by a point dipole.

Using Eq.~\eqref{Eq:froh_eph} inside the exciton polaron equations, Eqs.~\eqref{Eq:explrneqn}-\eqref{Eq:exphg}, as well as the exciton polaron formation energy Eq.~\eqref{Eq:exciton_etot}, it can be shown that one recovers the Wannier/Landau-Pekar model discussed in Sec.~\ref{Sec:explrn_model}.

In addition to driving the formation of exciton polarons, exciton-phonon coupling matrix elements have also been used to study exciton lifetime \cite{Chen_Bernardi_2020}, thermal shift and broadening of exciton levels \cite{Antonius_Louie_2022}, and real-time exciton dynamics \cite{Cohen_Refaely-Abramson_2024}. Moreover, exciton-phonon coupling matrix elements constitute the starting point to go beyond the adiabatic and classical approximations introduced in Secs.~\ref{Sec:ExcitonPolaron_Supercell} and Secs.~\ref{Sec:explrn_eqns}, as discussed in Sec.~\ref{Sec:explrn_beyond}.

\subsection{Exciton polarons beyond the adiabatic approximation}
\label{Sec:explrn_beyond}

The methods for modeling exciton polarons reviewed in Secs.~\ref{Sec:explrn_model}-\ref{Sec:explrn_eqns} implicitly rely on the adiabatic Born-Oppenheimer approximation and on the description of ions as classical particles. These approximations are directly inherited from the underlying electronic structure framework, e.g., DFT or BSE. As a result of these approximations, nonadiabatic and quantum nuclear effects are neglected in these approaches. This situation is analogous to the DFT-based calculations of polarons reviewed in Sec.~\ref{Sec:DFTPolaron} and the \textit{ab initio} polaron equations of Sec.~\ref{Sec:plrn_eq_reciprocal}. 

As a result of these approximations, effects such as quantum zero-point renormalization, temperature dependence, and line broadening are not described by these theories. These effects are expected to become important at weak exciton-phonon coupling, where it is known that nonadiabatic Fan-Migdal contributions dominate the related physics of polarons~\cite{Miglio_Gonze_2020, Lafuente_Giustino_2022a, Lafuente_Giustino_2022b}.

Several efforts to include nonadiabatic and quantum effects have been made. For example, \textcite{Marini_2008} considered zero-point renormalization and thermal shifts of excitons by including the Fan-Migdal self-energy in the quasiparticle band structures entering the BSE. Furthermore, \textcite{Filip_Neaton_2021} and \textcite{Alvertis_Neaton_2024} incorporated quantum and nonadiabatic effects in the exciton-phonon coupling by considering how the phonon contribution to the screening modifies the BSE kernel in Eq.~\eqref{Eq:bse_general}. 

Despite significant progress in this area, a unified framework that can describe, at once, quantum and nonadiabatic effects, temperature dependence, line broadening, and the formation of exciton polarons is currently missing.

To achieve such a unified framework, the most immediate path forward would be to derive a set of many-body exciton polaron equations starting from the exciton-phonon Hamiltonian in second-quantized form. For example one could rewrite Eq.~\eqref{Eq:etot_rs} in terms of phonon operators and exciton creation and annihilation operators~\cite{Antonius_Louie_2022}, and then write the equations of motion for exciton and phonon Green's functions. From this point, the derivation steps would follow the strategy outlined in Sec.~\ref{Sec:ManyBody} or equivalently in \cite{Lafuente_Giustino_2022b}. Barring unforeseen difficulties, such an approach should deliver an all-coupling \textit{ab initio} theory of exciton polarons and self-trapped excitons, in analogy to the  case of charged polarons.

\section{Applications and connection to experiment}\label{Sec:Applications}

In this section, we highlight a few recent applications of the techniques discussed in Secs.~\ref{Sec:DFTPolaron}-\ref{Sec:ExcitonPolaron}. In Secs.~\ref{Sec:app_alkali} and \ref{Sec:TMOs}, we focus on systems that are well known for hosting polarons, namely alkali halides and metal oxides. In Sec.~\ref{Sec:app_halide_perov}, we discuss polarons in halide perovskites and their potential nontrival topology. In Sec.~\ref{Sec:2d_materials} we briefly review recent work on polarons in two-dimensional (2D) materials\rev{, and in Sec.~\ref{Sec:Organics} we discuss disordered and organic semiconductors.}
We leave out of this section several applications that have already been discussed throughout this article, in particular ML/AI calculations of polarons (Fig.~\ref{Fig:ml_polarons}), calculations of phonon sidebands in ARPES (Fig.~\ref{Fig:cumulant_tio2}), polaron band structures via DMC (Fig.~\ref{Fig:diagmc_fig}), and calculations of self-trapped excitons (Fig.~\ref{Fig:supercell_explrn}). 

The following set of examples is not exhaustive and is only meant to illustrate some of the methods reviewed in Secs.~\ref{Sec:DFTPolaron}-\ref{Sec:ExcitonPolaron}. We devote a separate section to the calculation of transport properties of polarons, cf.\ Sec.~\ref{Sec:transport}.

\subsection{Alkali halides}
\label{Sec:app_alkali}

\begin{figure}
    \centering
    \includegraphics[width=0.98\linewidth]{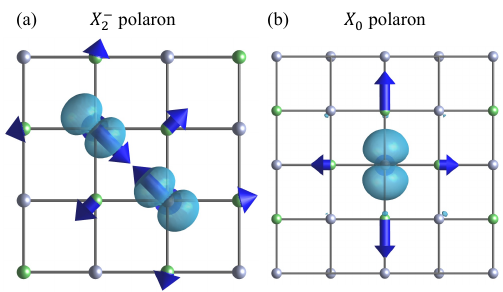}
    \caption{(a) Typical two-center $X_2^-$ polaron in alkali halides ($V_k$ center), for the case of LiF. Green and grey indicate Li and F, respectively. The isosurface is the polaron wavefunction, and the arrows are the atomic displacements. (b) One-center $X^0$ polaron in LiF.
    }
    \label{Fig:alkali_halides}
\end{figure}

Alkali halides are binary crystals \ch{AX}, with A = Li, Na, K, Rb, Cs, and X = F, Cl, Br, I. The crystal structure at ambient conditions is rocksalt for most combinations, Fig.~\ref{Fig:alkali_halides}, except for RbI and Cs halides which take the CsCl simple cubic structure. Valence states are heavy holes derived from the halogen $p$ states, while conduction states are light electrons derived from alkali $s$ states. The heavy holes are expected to lead to small polaron formation. Hole polarons named $V_k$ centers have been observed experimentally via electron paramegnetic resonance (EPR) spectroscopy~\cite{Castner_Kanzig_1957, Sidler_Aegerter_1973, Bailey_1964}; these polarons can be thought of as X$_2^-$ dimers aligned along the body diagonal of the rocksalt structure, cf. Fig.~\ref{Fig:alkali_halides}(a). In addition, femtosecond pump-probe spectroscopy in KI and RbI revealed a one-center small polaron, which can be thought of as a single $X^0$ atom with the hole wavefunction being the I-$5p$ orbital \cite{Iwai_Shluger_1996}, cf. Fig.~\ref{Fig:alkali_halides}(b). This $X^0$ one-center polaron is short-lived and converts into the $X_2^-$ two-center polaron over picosecond timescales.

\begin{figure*}
    \centering
    \includegraphics[width=0.99\linewidth]{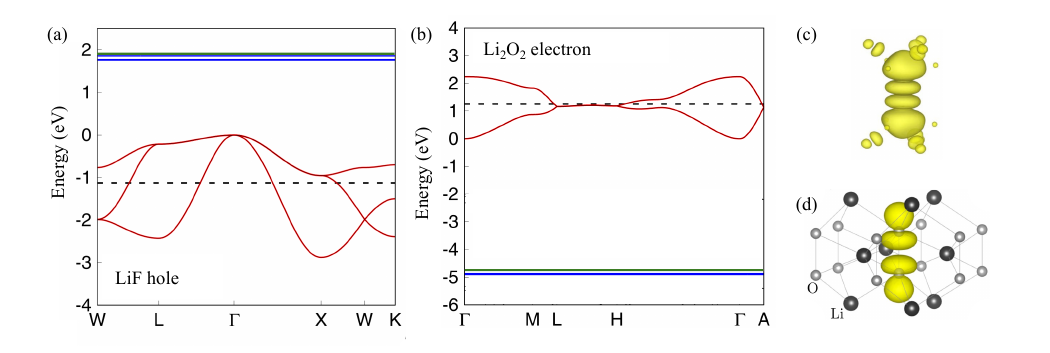}
    \caption{
    (a) Comparison of the polaron formation energies for $X_0$ hole polaron in \ch{LiF} calculated from the canonical transformation method (Sec.~\ref{Sec:bernardi_polaron}) and the \textit{ab initio} polaron equations (Sec.~\ref{Sec:PolaronEquations}).
    (b) Comparison of the polaron formation energies for small electron polaron in \ch{Li2O2} calculated from the two methods. 
    (c) and (d) displays the charge density isosurface of \ch{Li2O2} computed using \textit{ab initio} polaron equations~\cite{Sio_Giustino_2019a} and canonical transformation method~\cite{Luo_Bernardi_2022}, respectively.
    In (a) and (b), red lines represent the valence bands of \ch{LiF} and conduction bands of \ch{Li2O2}, respectively, while the other manifolds of bands are hidden for clarity. The black dashed lines represent the onsite energy of the chosen Wannier orbital in the canonical transformation method, blue lines represent the polaron formation energy determined from the canonical transformation method, and green lines represent the formation energy calculated from \textit{ab initio} polaron equations.
    Panels (a), (b), (d) adapted from \textcite{Lee_Bernardi_2021}; panel (c) adapted from \textcite{Sio_Giustino_2019b}.}
    \label{Fig:canonical_vs_plrneq}
\end{figure*}

Given their simplicity, alkali halides constitute a popular benchmark in polaron calculations. 
Hybrid functional calculations have frequently been employed to study hole polarons in alkali halides~\cite{Miceli_Pasquarello_2018, Tygesen_Garcia-Lastra_2023}; these calculations identified the two-center polaron. 
The specific type of the hybrid functional, e.g., PBE0, HSE, or HSEsol, does not alter the qualitative geometry of this polaron, but it is found that increasing the fraction of exact exchange leads to shorter $X_2^-$ bond lengths. \textcite{Tygesen_Garcia-Lastra_2023} also computed polaron hopping barriers across several alkali halides, and found that the fraction of exact exchange that yields the experimental band gaps also leads to good agreement for the hopping barriers~\cite{Pung_1979}. Although successful in the case of alkali halides, this strategy for choosing the fraction of exact exchange may not be transferable to other systems; for example it does not generalize to transition metal oxides, reviewed in Sec.~\ref{Sec:TMOs}. 

In addition to hybrid functionals, several pSIC schemes have been employed to study hole polarons in alkali halides~\cite{Sio_Giustino_2019b, Sadigh_Aberg_2015}. \textcite{Sadigh_Aberg_2015} used Eq.~\eqref{Eq:dft_eform_expr3} to compute the formation energy without self-interaction error, and leveraged the Janak's theorem to calculate Hellman-Feynman forces and perform structural optimization. Using this approach, they found the X$_2^-$ polaron for lithium and sodium halides. When using Eq.~\eqref{Eq:dft_eform_expr3} in combination with hybrid functionals, they found that the geometry and the formation energies of the polaron are insensitive to the fraction of the exact exchange. This finding confirms that the pSIC schemes of Eq.~\eqref{Eq:svpg-gw-ef} and \eqref{Eq:dft_eform_expr3} are indeed able to largely eliminate the SIE.

Small hole polarons in alkali halides were also investigated via the \textit{ab initio} polaron equations of Sec.~\ref{Sec:PolaronEquations}. In the case of LiF, this method yields the $X^0$ one-center polaron rather than the $X_2^-$ two-center polaron~\cite{Sio_Giustino_2019a}. The same result was obtained by \textcite{Lee_Bernardi_2021} via the canonical transformation method (cf.\ Sec.~\ref{Sec:bernardi_polaron}); furthermore, the formation energies obtained with these two methods are in quantitative agreement, cf.\ Fig.~\ref{Fig:canonical_vs_plrneq}(a). Similar results were obtained by \textcite{Britt_Siwick_2024}, who also calculated the experimental signatures of these polarons in ultrafast diffuse electron and X-ray scattering; and by \cite{Robinson_Reichman_2025} via the all-coupling variational canonical transformation method (cf.\ Sec.~\ref{Sec:Reichman}).

The reason why the methods of (\citeauthor{Sio_Giustino_2019a}, \citeyear{Sio_Giustino_2019a};  \citeauthor{Lee_Bernardi_2021}, \citeyear{Lee_Bernardi_2021}; \citeauthor{Robinson_Reichman_2025}, \citeyear{Robinson_Reichman_2025}) find the metastable one-center $X^0$ polaron as the ground state rather than the two-center $X_2^-$ state as in \cite{Tygesen_Garcia-Lastra_2023,Sadigh_Aberg_2015} is likely due to the neglect of anharmonic effects and nonlinear electron-phonon coupling in the former approaches, as discussed in Sec.~\ref{Sec:higher_order_effect}. It is interesting that, in structurally related systems such as MgO, the situation is inverted and the one-center polaron is the true ground state~\cite{Chae_Kioupakis_2025}, while the $X_2^-$ is unstable;
this delicate balance between one-center and two-center polarons in alkali halides offers a useful benchmark for future method development.

\begin{figure*}
    \centering
    \includegraphics[width=0.99\linewidth]{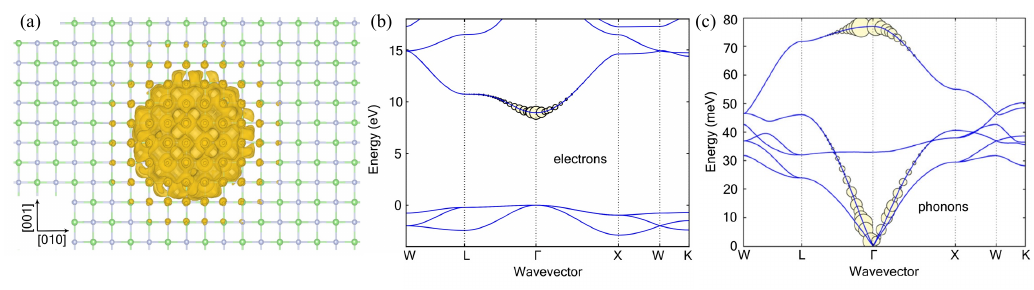}
    \caption{
    (a) Large electron polaron in LiF as obtained from the \textit{ab initio} polaron equations (Sec.~\ref{Sec:PolaronEquations}). (b) Spectral decomposition of the polaron in terms of KS states. The radius of the circles is proportional to $|A_{n\bk}|^2$. (c) Spectral decomposition of the polaron in terms of normal vibrational modes. The radius of the circles is proportional to $|B_{\bq\nu}|^2$. From \textcite{Sio_Giustino_2019b}.}
    \label{Fig:lif_polarons}
\end{figure*}

Calculations of electron polarons in alkali halides are more rare, since the relatively light band effective mass (0.3-0.9~$m_0$) leads to large polarons which require supercells with thousands of atoms. This challenge can be addressed using the \textit{ab initio} polaron equations in reciprocal space (Sec.~\ref{Sec:PolaronEquations}); the result for LiF is shown in Fig.~\ref{Fig:lif_polarons}  \cite{Sio_Giustino_2019a}. This figure shows the wavefunction of the electron polaron, and the decomposition of the wavefunction and the atomic displacements in terms of KS states and vibrational eigenmodes, respectively. From this decomposition, one can infer that the polaron draws weight primarily from the parabolic conduction band minimum and from long-wavelength longitudinal optical phonons, as well as longitudinal acoustic phonons. Except for the acoustic phonon contribution, this polaron constitutes a first-principles realization of the celebrated Fr\"ohlich polaron (Sec.~\ref{Sec:FrohlichLLP}). Similar results were obtained by \textcite{Robinson_Reichman_2025} using the all-coupling variational canonical transformation, and by \textcite{Vasilchenko_Gonze_2024} within a generalized \textit{ab initio} Fr\"ohlich model.

The large spatial extent of this electron polaron sets an upper bound to the carrier density at which polarons can in principle be observed. For example, \textcite{Sio_Giustino_2019b} estimate that beyond the Mott density of $\sim 10^{20}~{\rm cm^{-3}}$, individual polarons become unstable and delocalize into extended Bloch states. The metal-insulator transition (MIT) is discussed in more detail in Supplemental Note~2.

\subsection{Metal oxides} \label{Sec:TMOs}

Metal oxides (MOs), such as \ch{TiO2}, \ch{Fe2O3}, \ch{CuO}, \ch{NiO}, \ch{WO3}, and \ch{SrTiO3}, are known for their multifunctional properties and applications in catalysis, photosynthesis, charge transports, energy generation and storage, and more recently in neuromorphic computing~\cite{Yang_Qing_2008, Zhao_Zhao_2021, Wang_Strasser_2021, Oregan_Gratzel_1991, Strukov_Stanley_2008, Austin_Turner_1968, Schirmer_Salje_1980, Calvani_Peschiaroli_1993, vanMechelen_Mazin_2008, Liu_Dupuis_2022, Wiktor_Pasquarello_2018, Ambrosio_Wiktor_2025, Seo_Galli_2018, Seo_Choi_2025, Gerosa_Galli_2018,
Smart_Ping_2018}. Beyond these applications, MOs also serve as a playground to explore emergent physics, such as MITs, unconventional superconductivity, and altermagnetism~\cite{Kaur_Ping_2023, Bednorz_Muller_1986, Feng_Liu_2022}. As several of these phenomena and properties are related to polarons, numerous experimental and theoretical studies have been performed to elucidate polaron physics in these materials. In this section, we review representative efforts by discussing titanium dioxide, \ch{TiO2}~\cite{Lile_Zhang_2022}, hematite, \ch{Fe2O3}~\cite{Redondo_Setvin_2024}, and lithium peroxide, \ch{Li2O2}~\cite{Feng_Zaghib_2013, Garcia-Lastra_Vegge_2013, Ong_Ceder_2012, Radin_Siegel_2013}.

\ch{TiO2} crystallizes into several different phases: the most stable phase at ambient condition is the rutile phase~\cite{Hanaor_Charles_2012}, while the anatase phase is primarily found in nanocrystalline form and is the main polymorph of \ch{TiO2} for applications in artificial photosynthesis and solar photovoltaics~\cite{Yang_Qing_2008, Fujishima_Honda_1972}. Polarons play an important role in the functional properties of rutile and anatase \ch{TiO2}~\cite{Gong_Diebold_2006, Reticcioli_Franchini_2019}: for example, $n$-doped anatase \ch{TiO2} is a popular transparent conductor, while the corresponding rutile phase exhibits insulating characteristics~\cite{Zhang_Venkatesan_2007}; this difference can be explained in terms of the formation of small electron polarons in rutile~\cite{Janotti_Walle_2013}, and very large electron polarons in anatase~\cite{Dai_Giustino_2024c}. 

\begin{figure}
    \centering
    \includegraphics[width=0.99\linewidth]{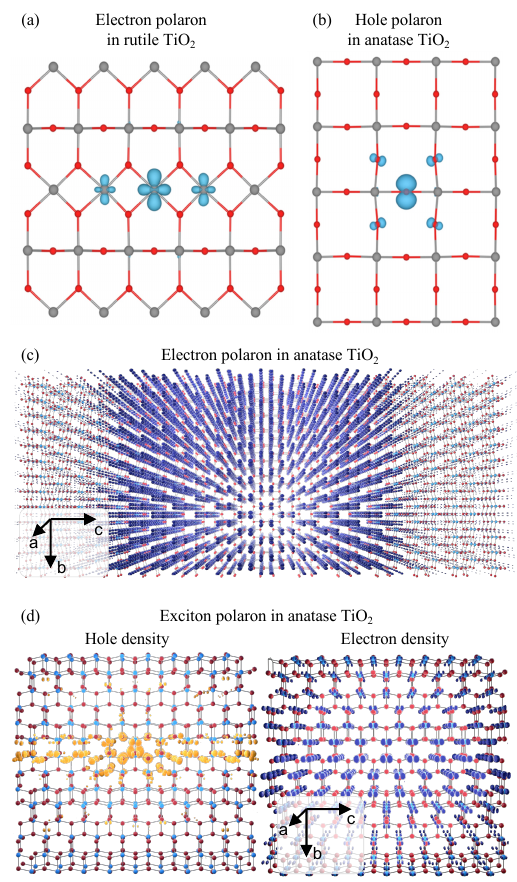}
    \caption{(a),(b) Small electron polaron in rutile \ch{TiO2} and small hole polaron in anatase \ch{TiO2}, respectively, calculated using hybrid functional DFT. From \textcite{Elmaslmane_McKenna_2018}. (c) Large electron polaron in anatase \ch{TiO2}, calculated using the \textit{ab initio} polaron equations. (d) Marginal hole density (upper panel) and marginal electron density (lower pannel) of the large exciton polaron in anatase \ch{TiO2}, as calculated using the exciton polaron equations. From \textcite{Dai_Giustino_2024c}.}
    \label{Fig:tio2_polarons}
\end{figure}

To calculate polarons in bulk \ch{TiO2}, hybrid functionals in combination with the Koopman's condition have been a popular choice (cf.\ Sec.~\ref{Sec:DFTPolaron_hubbard_hybrid}). In the bulk rutile phase, calculations using the HSE06~\cite{Kokott_Scheffler_2018} and PBE0~\cite{Elmaslmane_McKenna_2018} functionals found small electron polarons with formation energy in the range 0.1-0.4~eV, cf.\ Fig.~\ref{Fig:tio2_polarons}(a)\rev{. Hole polaron results vary: most studies report no evidence of their formation}~\cite{Deak_Frauenheim_2011, Elmaslmane_McKenna_2018, Kokott_Scheffler_2018}\rev{, whereas others suggest that small hole polarons can indeed form \cite{Varley_Walle_2012, Palermo_Pasquarello_2024}}. The localized nature of electron polarons in rutile is confirmed by transport experiments~\cite{Zhang_Venkatesan_2007} and optical experiments~\cite{Bogomolov_Mirlin_1968, Tian_Durrant_2025} In the anatase phase, the situation is reversed: hybrid functional calculations based on HSE06, PBE0, and B3LYP identified small hole polarons~\cite{Deak_Frauenheim_2011, Elmaslmane_McKenna_2018, DiValentin_Selloni_2006, Palermo_Pasquarello_2024}, cf.\ Fig.~\ref{Fig:tio2_polarons}(b), while electron polarons could not be found for meaningful choices of the fraction of exact exchange~\cite{Deak_Frauenheim_2011}. 

Calculations of polarons in \ch{TiO2} based on DFT+$U$ paint a more complicated picture. 
On one hand, these calculations support the existence of small electron polarons in rutile~\cite{Setvin_Diebold_2014, Morita_Walsh_2023}; on the other hand, \rev{several studies based on} DFT+$U$ \rev{calculations} also predict small hole polarons~\cite{Deskins_Dupuis_2009, Mcbride_Hautier_2025, Palermo_Pasquarello_2024}. 
In the case of anatase \ch{TiO2}, DFT+$U$ yields small hole polarons; some studies also report small electron polarons~\cite{Deskins_Dupuis_2007}, while other studies indicate no electron polarons~\cite{Setvin_Diebold_2014}. These contrasting observations likely originate from the sensitivity of Hubbard-corrected DFT to the choice of the Hubbard parameter $U$, the Hubbard projectors, and the atomic species included in these projectors (Sec.~\ref{Sec:DFTPolaron_hubbard_hybrid}); for example, by choosing a large $U$ value for the O atoms, one can promote the formation of small hole polarons in rutile \ch{TiO2}~\cite{Deskins_Dupuis_2009}. In addition to this sensitivity, different choices of $U$ may fulfill certain conditions at the expense of others: for instance, the value $U=7.8$~eV reproduces the experimental band gap of anatase TiO$_2$~\cite{Deskins_Dupuis_2007}, while the value $U=3.9$~eV fulfills Koopman's condition~\cite{Setvin_Diebold_2014}.

To overcome the sensitivity of polaron calculations on the fraction of exact exchange or the Hubbard parameter, one could employ the pSIC schemes reviewed in Sec.~\ref{Sec:DFTPolaron_SIC}. However, these schemes typically require dedicated software implementations, which somewhat hinders their widespread adoption. One exception is the study by \textcite{Kokott_Scheffler_2018}; these authors employed the pSIC scheme of \textcite{Sadigh_Aberg_2015}, which only requires post-processing of standard DFT calculations. They
investigated polarons in rutile \ch{TiO2} including finite-size corrections~(Sec.~\ref{Sec:DFT_difficulty}), and found small electron polarons, in line with prior work.  

A pSIC-based scheme was also used by \textcite{Dai_Giustino_2024c}, who studied polarons in anatase and rutile \ch{TiO2} by solving the \textit{ab initio} polaron equations (Sec.~\ref{Sec:PolaronEquations}). In this work, the authors could find the small electron polaron in rutile \ch{TiO2} and the small hole polaron in anatase \ch{TiO2}, and obtained formation energies in good agreement with previous hybrid functional calculations using the Koopman condition~\cite{Elmaslmane_McKenna_2018}. Since the \textit{ab initio} polaron equations do not require explicit supercell calculations, \textcite{Dai_Giustino_2024c} also computed large polarons in both polymorphs, successfully identified three new species, namely a large hole polaron in rutile, a large quasi-two-dimensional electron polaron in anatase, and a large exciton polaron
in anatase, cf.\ Fig.~\ref{Fig:tio2_polarons}(c) and (d). These findings provided evidence for the existence of large polarons in \ch{TiO2}.

Numerous studies investigated polarons at \ch{TiO2} surfaces. The earliest hybrid functional calculation for the rutile TiO$_2$(110) surface was reported by \citet{DiValentin_Selloni_2006}, who showed that oxygen vacancies lead to the localization of excess electrons on two nonequivalent Ti cations. These findings were later challenged by STM measurements by \textcite{Minato_Hou_2009}, which instead revealed charge delocalization over multiple cations surrounding the vacancy. Subsequent DFT+$U$ studies demonstrated that the activation barrier for polaron hopping between sites is small (0.2~eV), and \textit{ab initio} molecular dynamics simulations \cite{Kowalski_Marx_2010} suggested that thermal fluctuations enable excess charge from oxygen vacancies to dynamically sample different Ti sites. This picture was further refined by \citet{Yim_Thornton_2016}, who observed a symmetric charge distribution around the vacancy at high temperature, consistent with the findings by \citet{Minato_Hou_2009}, while at lower temperatures the charge localizes asymmetrically at a single site. Eventually, this scenario was confirmed by comparing experimental STM images with STM simulations based on DFT+$U$ calculations \cite{Reticcioli_Franchini_2018}.

\rev{
From calculations of polarons in \ch{TiO2}, it is evident that the results are rather sensitive to the specific method employed.
For example, different fractions of exact exchange in hybrid functionals or different values of Hubbard $U$ in DFT+U calculations could result in qualitatively different results. 
This variability suggests that it might be time for a community effort aimed at analyzing the qualitative and quantitative impact of different approaches, and at  establishing universally-accepted protocols for reliable and reproducible \textit{ab initio} calculations of polarons.
}

Among many other MOs for which polaron studies have been performed, we briefly mention hematite, $\alpha$-\ch{Fe2O3}, which is popular as the photoanode in solar water splitting. The phototacalytic properties of this material have been proposed to originate from the formation of self-trapped polarons upon photoexcitation~\cite{Redondo_Setvin_2024, Pastor_Bakulin_2019}. 
Hybrid functional DFT calculations~\cite{Redondo_Setvin_2024, Pastor_Bakulin_2019} as well as DFT+$U$ calculations~\cite{Adelstein_DeJonghe_2014,Smart_Ping_2017} indicate the existence of small electron and hole polarons; furthermore, Ti-doping was suggested to promote electron delocalization, which could be useful for controlling the reactivity. \textcite{Redondo_Setvin_2024} also employed non-contact atomic force microscopy (AFM) to individually create and manipulate electron and hole polarons at Fe$_2$O$_3$(1$\overline{1}$02) surface, and to measure their diffusion rate as a function of temperature; their measurements are in agreement with hybrid functional DFT calculations performed in the same study.

Beyond polarons, experimental studies suggest the possible existence of exciton polarons in $\alpha$-\ch{Fe2O3} upon photo-excitation~\cite{Fan_Yang_2021}. These measurements call for additional theoretical work to elucidate the competition between the formation of exciton polarons (Sec.~\ref{Sec:ExcitonPolaron}) and of charged polarons. 

To close this section, we mention calculations on lithium peroxide, \ch{Li2O2}, which is a close relative of alkali halide. This is a quasi-2D compound consisting of Li-intercalated \ch{LiO2} layers, and finds applications as  cathode in lithium-air batteries; it has been used as a computational benchmark to test various methods. Supercell-based, hybrid functional calculations indicate the existence of small electron polarons in \ch{Li2O2} \cite{Feng_Zaghib_2013}; in the same work, the authors investigated the dynamics of polaron formation within the framework of DFPT. The same system was considered by \textcite{Sio_Giustino_2019a} and \textcite{Lee_Bernardi_2021} to benchmark the \textit{ab initio} polaron equations and the canonical transformation method, respectively. Figure~\ref{Fig:canonical_vs_plrneq}(b) shows that both methods yield a deep polaron state, in quantitative agreement with each other and with the results of \cite{Feng_Zaghib_2013}.

\subsection{Halide perovskites}
\label{Sec:app_halide_perov}

\subsubsection{Single perovskites}
Halide perovskites are a popular family of materials in photovoltaic and light-emitting applications~\cite{Stranks_Snaith_2013, Burschka_Gratzel_2013, Kim_Lee_2021, Kim_Lee_2024}. These systems exhibit unique features, such as ultra-long carrier lifetimes which have been proposed to originate from the formation of polarons~\cite{Zhu_Podzorov_2015}. Experimental evidence for polarons comes from ultrafast X-ray diffraction experiments~\cite{Guzelturk_Lindenberg_2021} and time-resolved optical Kerr effect~\cite{Miyata_Zhu_2017}, among others.

The majority of first-principles calculations of polarons in halide perovskites focus on methylammonium lead tri-halides (\ch{MAPbX3})~\cite{Tantardini_Saidi_2022} and their inorganic counterparts (\ch{CsPbX3})~\cite{Meggiolaro_DeAngelis_2020}. The first studies of electron localization in halide perovskites were performed on charge-neutral system, rather then systems with an excess electron or hole. For example, \textcite{Ma_Wang_2015} performed \textit{ab initio} molecular dynamics simulations of \ch{MAPbI3} in the charge-neutral state. They found that, at room temperature, the wavefunctions of the valence and conduction band extrema exhibit localization, with localization lengths in the range 3-6~nm; in addition, they observed that the valence and conduction states tend to localize on different sublattices, which may help explain the long recombination lifetimes. \textcite{Quarti_DeAngelis_2015} reported similar observations. These early investigations hinted at the possible formation of polarons in halide perovskites. 

Subsequent work including excess carriers was performed by \textcite{Miyata_Zhu_2017} for \ch{MAPbI3} and \ch{CsPbBr3}. Using the hybrid functional PBE0 to mitigate polaron self-interaction, they showed that in a small 2$\times$2$\times$1 pseudocubic supercell of \ch{CsPbBr3}, the added charge induces lattice distortions but remain delocalized. To test whether a polaron forms, the supercell was enlarged to 2$\times$2$\times$8; in this setting, the hole charge density started to localize, with a localization length of 2.5~nm. A more recent study by \textcite{Qian_Duan_2023} also observed a moderate tendency for the hole to localize in \ch{CsPbBr3}.

In related work, \textcite{Osterbacka_Wiktor_2020} found small electron polarons in \ch{CsPbBr3} by carrying out hybrid functional calculations in combination with Koopman's condition (Sec.~\ref{Sec:DFTPolaron_hubbard_hybrid}). However, the authors noted the similar energy scale for the localized small polaron and delocalized solution in this material, and the small polaron becomes unstable upon including spin-orbit coupling (SOC). 

\textcite{Zheng_Wang_2019} used a DFT-based tight-binding model to study large polarons in \ch{MAPbI3}, incorporating polaronic stabilization via the effective potential energy of the Landau-Pekar model in Eq.~\eqref{Eq:elec_energy_to_psi}, and the wavefunctions obtained from the tight-binding eigenstates. The use of a sparse representation allowed them to perform calculations on large, 48$\times$48$\times$48 supercells, and demonstrate the stability of the large electron polaron. Taken together, all these studies support the notion that large polarons should exist in halide perovskites; however, a fully \textit{ab initio} treatment that includes all atomistic details is currently not feasible using supercells. The reciprocal-space formulation provided by the \textit{ab initio} polaron equations (Sec.~\ref{Sec:PolaronEquations}) could help shed some light on these questions.

\subsubsection{Double perovskites}

Beyond single halide perovskites of the type \ch{ABX3}, the related family of halide double perovskites A$_2$BB$^\prime {\rm X}_6$ attracted considerable interest recently, primarily in relation to their broadband light emission~\cite{Giustino_Snaith_2016, Luo_Tang_2018,Volonakis_Giustino_2016, Volonakis_Giustino_2017, Slavney_Karunadasa_2016, Mcclure_Woodward_2016}. 
For example, in vacancy-ordered double perovskites such as \ch{Cs2ZrBr6} and \ch{Cs2ZrCl6}, large Stokes shift have been observed. These shifts have been attributed to the formation of \rev{exciton polarons}~\cite{Abfalterer_Stranks_2020, Dai_Giustino_2024a}. 
Similarly, the double perovskite \ch{Cs2AgInCl6} is known to emit white light with near-unity quantum efficiency; \textit{ab initio} calculations indicated that this phenomenon may originate from the formation of \rev{exciton polarons}~\cite{Luo_Tang_2018, Jin_Galli_2024}. 

For a closely related double perovskite, \ch{Cs2AgBiBr6}, charged polarons were investigated in detail using both hybrid functionals~\cite{Baskurt_Wiktor_2023} and the \textit{ab initio} polaron equations~\cite{Lafuente_Giustino_2024}. These studies indicated the presence of both small and large polarons in \ch{Cs2AgBiBr6}, as well as periodic modulations of the electron density in the charged state, reminiscent of charged-density waves (CDWs). These computational studies support the ultrafast transient absorption measurements by \textcite{Wright_Herz_2021}, who reported the formation of large polarons in \ch{Cs2AgBiBr6} at short timescales (250~fs), and of small polarons at long timescales (1-5~ps).

\begin{figure}
    \centering
    \includegraphics[width=0.98\linewidth]{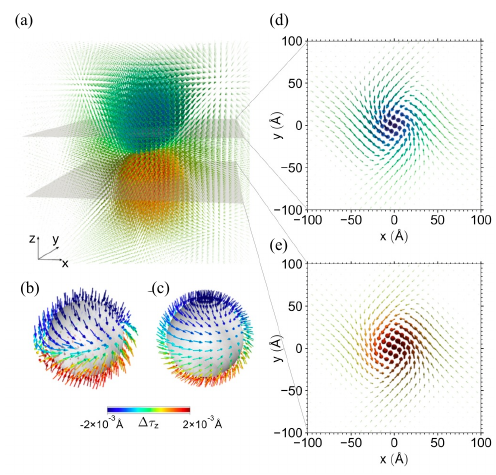}
    \caption{Topological polarons in \ch{Cs2AgBiBr6}. (a) 3D view of the atomic displacements associated with the large electron polaron in \ch{Cs2AgBiBr6}. For clarity, only \ch{Ag} displacements are shown. The color code represents the magnitude of the displacement in the $z$ direction. The polaron center coincides with the center of each panel. (b) Displacement patterns from (a), visualized on a sphere enclosing the polaron center. Displacements have been normalized for clarity. (c) Ideal Bloch point with the same topological charge, vorticity, polarity, and helicity, for comparison to the electron polaron in (b).  (d) and (e) Top view of the \ch{Ag} displacements from (a), along the cross-sections shown in (a). These vector fields have the same vortex patterns of magnetic merons~\cite{Fert_Cros_2017}. From \textcite{Lafuente_Giustino_2024}.}
    \label{Fig:topological_plrn}
\end{figure}

In calculations of the large electron and hole polarons in \ch{Cs2AgBiBr6}, \textcite{Lafuente_Giustino_2024} also noticed that their lattice distortions exhibit unusual patterns. For example, if one draws a sphere enclosing the center of the polaron, then the atomic displacements on this sphere take the form of a vortex, similar to magnetic skyrmions~\cite{Fert_Cros_2017}, cf.\ Fig.~\ref{Fig:topological_plrn}. A quantitative analysis of this texture can be performed by calculating several topological invariants such as the topological charge, the vorticity, and the helicity. \textcite{Lafuente_Giustino_2024} found that, using these invariants, both the electron and the hole polarons in \ch{Cs2AgBiBr6} can be classified as three-dimensional ``helical Bloch points'', which are essentially the 3D counterpart of 2D magnetic skyrmions. In particular, the electron polaron carries vortex character, while the hole polarons is an antivortex. These textures originate from the symmetry of the underlying crystal and of the electron and phonon band structures, and may be observable in ultrafast Huang diffuse scattering experiments~\cite{Dederichs_1973, Shimomura_Tokura_1999, Vasiliu-Doloc_Mitchell_1999, Filipetto_Nunes_2022, deCotret_Siwick_2022, Shi_Liang_2024}. 
\rev{Recent work by \textcite{Luo_Giustino_2026} established the symmetry principles governing the topology of polar textures in large polarons, and identified new types of topological polarons across a broad range of materials beyond halide perovskites.}
Future work will be needed to understand the potential implications of these topological textures 
\rev{in the optical and transport properties of polarons.}

\subsection{Two-dimensional materials} \label{Sec:2d_materials}

\begin{figure}
    \centering
    \includegraphics[width=0.99\linewidth]{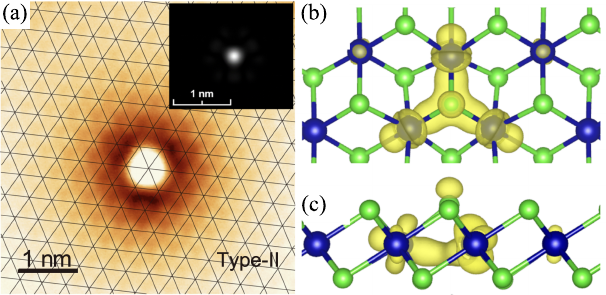}
    \caption{Polarons in 2D \ch{CoCl2}. 
    (a) STM image of \rev{Cl-centered} electron polarons in monolayer \ch{CoCl2}. (b)\rev{, (c)} Top and side views\rev{, respectively,} of crystal structure and isosurface of the calculated wavefunction \rev{of the Cl-centered electron polaron} in \ch{CoCl2}. From~\textcite{Cai_Fu_2023}.}
    \label{Fig:cocl2_polarons}
\end{figure}

Despite the tremendous progress that has been made in the science and technology of 2D materials, there are relatively few experimental and computational studies of polarons in these systems~\cite{Bai_Meng_2024, Liu_Wu_2023, Cai_Fu_2023, Cai_Fu_2025}. Among those, \textcite{Liu_Wu_2023} and \textcite{Cai_Fu_2023} used STM to inject individual electron polarons into a monolayer of \ch{CoCl2} on a highly-oriented pyrolytic graphite substrate, cf.\ Fig.~\ref{Fig:cocl2_polarons}(a). Both groups identified polaron-induced band-bending effects, and showed that these polarons could be created, erased, interconverted, and positioned individually using the STM tip. These studies were also supported by supercell-based, hybrid functional DFT calculations, which identified several possible polaron states, in qualitative agreement with the measurements, cf.\ Fig.~\ref{Fig:cocl2_polarons}(b).

Additional calculations on the same system were reported by \textcite{Li_Zhao_2024}; these authors found low transition barriers between different polaronic configurations, which potentially explains why these polarons can be interconverted or positioned using the STM tip. Furthermore, similar measurements were reported by \textcite{Hao_Gao_2024}, who proposed an alternative interpretation of the STM images based of the formation of localized charge-neutral dipoles; the microscopic nature of the proposed dipoles remains unclear at present.

Beyond CoCl$_2$, the generation and manipulation of individual polarons has recently been demonstrated also in the monolayer inorganic molecular crystal Sb$_2$O$_3$, and the experiments were supported by supercell DFT+$U$ calculations \cite{Zhang_Fu_2024}.

Except for the above reports, examples of polarons in 2D are far fewer than in 3D bulk materials. This asymmetry was analyzed by \textcite{Sio_Giustino_2023} using the \textit{ab initio} polaron equations (Sec.~\ref{Sec:PolaronEquations}). For the prototypical 2D polar material, monolayer hexagonal boron nitride (h-BN), they noted that the reduced dimensionality stabilizes the hole polaron as compared to bulk h-BN. However, this dimensionality effect is significantly less pronounced than for other excitations, such as for example excitons which exhibit very large binding energies in 2D~\cite{Qiu_Louie_2013}.

To rationalize the moderate increase in polaron formation energy from 3D to 2D, \textcite{Sio_Giustino_2023} developed an exactly soluble strong-coupling model in two dimensions. This model extends the Landau-Pekar framework of Sec.~\ref{Sec:LandauPekar}, \rev{and} employs the Fr\"ohlich electron–phonon coupling modified to account for the electrostatics of a finite-thickness 2D slab in vacuum. This approach goes slightly beyond earlier treatments that assumed vanishing slab thickness~\cite{Sarma_Mason_1985, Peeters_Devreese_1986, Hahn_Franchini_2018}, and is important for correctly capturing polaron energetics.
Their analysis indicates that, on the one hand, reduced dimensionality weakens electronic screening, which tends to increase binding; on the other hand, lattice screening is also weakened, thus reducing the binding. Overall, these two effects tend to offset one another, leading to only a modest enhancement of the polaron formation energy from 3D to 2D. Future work will be needed to establish whether this effect holds general validity across 2D materials, or it is limited to highly polar compounds.

The model of \textcite{Sio_Giustino_2023} also identifies a critical condition for the formation of polarons in 2D materials:
\begin{align}
    \label{Eq:2d_polaron_condition}
    \epsilon_\mathrm{ion} m^* d > 2 m_\mathrm{e} a_0,
\end{align}
where $\epsilon_\mathrm{ion}$ is the lattice contribution to the relative dielectric permittivity, $m_\mathrm{e}$ is the bare electron mass, $m^*$ is the electron effective mass, $d$ is the thickness of the 2D crystal, and $a_0$ is the Bohr radius.
The existence of such a critical condition is in stark contrast with the 3D Landau-Pekar model, which predicts polarons to always form. Equation~\eqref{Eq:2d_polaron_condition} might explain why polarons occur less frequently in 2D than in 3D materials, and why standard 2D systems like \ch{MoS2}, \ch{WS2} and similar do not seem to host stable polarons. Future work will be needed to generalize Eq.~\eqref{Eq:2d_polaron_condition} to additional electron-phonon coupling mechanisms and to more accurate polaron theories.

\subsection{\rev{Disordered and organic semiconductors}}
\label{Sec:Organics}

\rev{The examples discussed above mostly concern polarons in crystalline inorganic materials, where electronic bands, phonons, and electron-phonon matrix elements can be defined starting from a periodic reference structure. 
In disordered and organic semiconductors, this reference is often less straightforward: static disorder, thermal fluctuations, and variations in molecular geometry or intermolecular arrangement can all influence the formation of polarons.}

\rev{Conjugated polymers provide an interesting example \cite{Bredas_Street_1985}. 
Adding or removing charge can locally relax the polymer backbone, 
and form self-localized charged excitations with in-gap electronic states. 
In non-degenerate polymers these excitations are naturally described as local polarons, whereas in degenerate cases they can instead appear as solitons, i.e. domain walls between equivalent dimerization patterns \cite{Heeger_Su_1988}.}

\rev{Experimentally, polarons in polymers are usually identified through the electronic and vibrational fingerprints produced by this local relaxation. 
Optical and infrared spectroscopies reveal sub-gap absorption bands associated with polaronic states \cite{LeCroy_Salleo_2024},
while Raman spectroscopy detects changes in the vibrational modes of the polymer backbone \cite{Mansour_Koch_2022};
electron paramagnetic resonance provides complementary information on whether the charged excitation carries spin \cite{Enengl_Ehrenfreund_2016}.
A recent step forward is the real-space imaging of ordered polaronic textures in individual conjugated polymers by \textcite{Wu_Wang_2025}. 
They combined scanning probe microscopy, tip-enhanced Raman spectroscopy, and density functional theory calculations to identify polaron superlattices through electronic-density modulations, bond distortions, and vibrational frequency shifts in highly $n$-doped ethynylene-bonded polypentacenes on silver substrates.}

\rev{Organic molecular crystals provide a complementary example in which polaronic effects are governed by a soft molecular lattice \cite{Fratini_Sirringhaus_2020}.
Although these materials can be crystalline, weak intermolecular bonding leads to large thermal fluctuations of the intermolecular transfer integrals, which can transiently localize carriers \cite{Troisi_Orlandi_2006, Fratini_Ciuchi_2016}.
Consistent with this picture, 
\textcite{Giannini_Blumberger_2019} used fragment-orbital-based surface hopping to show that, 
at finite temperature, 
polarons in organic molecular crystals fluctuate with the molecular lattice and can delocalize over several molecules, 
up to about 10--20 in the most conductive systems.
Experimentally, this intermediate localization regime has been probed by electron spin resonance measurements in pentacene field-effect devices \cite{Marumoto_Iwasa_2006}.}

\rev{We note that complementary many-body approaches to polaronic effects in organic molecular crystals have already been discussed in Sec.~\ref{Sec:ManyBody},
including Lang-Firsov-like canonical transformations for polaronic bandwidth narrowing \cite{Hannewald_Hafner_2004, Ortmann_Hannewald_2009, Fetherolf_Berkelbach_2020},
and finite-temperature cumulant approaches for spectral functions and transport \cite{Chang_Bernardi_2022}.
\\}

\section{Transport properties of polarons} \label{Sec:transport}

Polaron transport has been studied in a wide variety of systems, from \rev{transition metal oxides} (TMOs) \cite{Zhang_Venkatesan_2007} and halide perovskites \cite{Buizza_Herz_2021} to manganites \cite{Salamon_Jaime_2001} and organic crystals \cite{Karl_2003}.
Two limiting regimes are commonly distinguished: band-like diffusive transport of large polarons, and hopping-like activated transport of small polarons~\cite{Deskins_Dupuis_2022}. We discuss these two limits separately.

\subsection{Large polarons: band-like diffusive transport} \label{Sec:bte}

The standard first-principles approach to charge transport in metals and semiconductors is based on the \textit{ab initio} Boltzmann transport equation (\textit{ai}BTE) (\citeauthor{Li_2015}, \citeyear{Li_2015}; 
\citeauthor{Zhou_Bernardi_2016},
\citeyear{Zhou_Bernardi_2016};
\citeauthor{Bernardi_2016}, \citeyear{Bernardi_2016}; 
\citeauthor{Ponce_Giustino_2018}, \citeyear{Ponce_Giustino_2018}, 
\citeauthor{Zhou_Bernardi_2018}, \citeyear{Zhou_Bernardi_2018}). The \textit{ai}BTE is a semiclassical method whereby one seeks to determine the response of the electron distribution function $f_{n\bk}$ to external electric and magnetic field. These fields are responsible for collisionless flow, and equilibrium is restored by considering carrier scattering by phonon absorption/emission and other mechanisms. At equilibrium in the absence of external fields, $f_{n\bk}$ reduces to the Fermi-Dirac distribution. In this approach, it is assumed that the electron and phonon band structures remain unchanged.

To the lowest order, including polaronic effects in the \textit{ai}BTE requires one to incorporate the phonon-induced renormalization of the electron bands (see Sec.~\ref{Sec:ManyBody}). \textcite{Kang_VandeWalle_2019} demonstrated this effect for anatase TiO$_2$, by computing the band structure renormalization from the Fan-Migdal self-energy, Eq.~\eqref{Eq:selfen_FM_kq}. Their calculations yielded a twofold reduction of the electron mobility upon inclusion of electron-phonon interactions, indicating that these effects are significant and must be taken into account in future calculations. In a similar spirit, \textcite{Ranalli_Franchini_2024} recently showed that phonon-induced effective-mass renormalization, combined with anharmonic effects and polymorphism, leads to significant mobility reduction in SrTiO$_3$ and KTaO$_3$.

In a more sophisticated approach, in addition to band renormalization one would take into account the redistribution of spectral weight from the quasiparticle peak to phonon satellites (cf.\ Sec.~\ref{Sec:Cumulant}). This approach was demonstrated by \citet{Zhou_Bernardi_2019} using \ch{SrTiO3} as a test case. These authors computed the spectral function $A_{n\mathbf{k}}(\omega)$, \rev{Eq.~\eqref{Eq:spectral_func}}, within the cumulant expansion method (cf.\ Sec.~\ref{Sec:Cumulant}), and used this function in the Kubo linear-response formula \cite{Mahan_2000}: 
\begin{align} \label{Eq:Kubo_bubble}
    \sigma(\omega) =&~  \frac{\pi \hbar e^2}{3\Omega_{\mathrm{uc}}}
    \int d\omega' \, \frac{f(\omega')-f(\omega'+\omega)}{\omega} \nonumber \\
    & \times \int \frac{d\mathbf{k}}{\Omega_{\mathrm{BZ}}} \, |\bv_{n\mathbf{k}}|^2 \, A_{n\mathbf{k}}(\omega') \, A_{n\mathbf{k}}(\omega'+\omega) ~,
\end{align}
where $\sigma$ is the optical conductivity, $f$ is the Fermi-Dirac function, $\bv_{n\mathbf{k}}$ is the band velocity, and we considered the isotropic average for simplicity. The carrier mobility is obtained from this expression by taking the low-frequency limit and dividing by the carrier density, $\mu = \sigma(\omega=0)/n e$. 
\textcite{Zhou_Bernardi_2019} showed that including spectral weight redistribution reduces the mobility by an order of magnitude as compared to \textit{ai}BTE calculations~\cite{Zhou_Bernardi_2018}, bringing theory into better agreement with experiment. 

In subsequent work, \citet{Chang_Bernardi_2022} applied the same methodology to a naphthalene crystal. For hole mobilities, Eq.~\eqref{Eq:Kubo_bubble} and the \textit{ai}BTE yielded comparable values, both in good agreement with experiment. In contrast, in the case of electrons, which exhibit heavier masses and stronger electron-phonon coupling, Eq.~\eqref{Eq:Kubo_bubble} is in better agreement with experiments. Furthermore, for carrier transport perpendicular to the molecular planes, neither approach could reproduce the measured ultra-low mobilities ($\mu < 1$~cm$^2$/Vs); this suggests that describing transport in this regime may require consideration of small polaron hopping, as discussed in Sec.~\ref{sec:smallpolhop}.

In recent work, \textcite{Lihm_Ponce_2025,Lihm_Ponce_2025b} extended Eq.~\eqref{Eq:Kubo_bubble} by incorporating vertex corrections to the current-current correlation function \rev{to make it fulfill the continuity equation and charge conservation}. Their results showed systematically improved agreement with numerically exact calculations for the Fr\"ohlich, Holstein, and Peierls/SSH models (cf.\ Sec.~\ref{Sec:History}) in the weak-to-intermediate coupling regimes. They also reported good agreement with experimental mobility data for metals, nonpolar semiconductors, and \rev{intermediate-coupling} polar materials. The strong-coupling regime, where activated transport is expected to dominate, was not addressed and remains a promising direction for future work.

One method that is able to capture \rev{both the weak and} the strong coupling limit is DMC (Sec.~\ref{Sec:DMC}). In this approach, transport properties can be investigated by computing the thermodynamic current-current correlation function:
\begin{equation} \label{Eq:current_current_iw}
    \chi_{\alpha\alpha'}(i\omega_{n})
    = 
    \frac{1}{\beta}\!
    \int_{0}^{\beta}\! d\tau\!\!
    \int_{0}^{\beta}\!d\tau'
    e^{i\omega_n (\tau-\tau')}
    \langle \hat{T}\hat{J}_{\alpha}(\tau)\hat{J}_{\alpha'}(\tau')\rangle ~,
\end{equation}
where $i\omega_n$ is the Matsubara frequency \cite{Mahan_2000}, $\beta$ is the inverse temperature as in Sec.~\ref{Sec:Feynman}, $\hat{T}$ is Wick's time-ordering operator as in Eq.~\eqref{Eq:green_el_N}, 
and $\hat{J}_\alpha= -e\sum_{n\mathbf{k}} v_{n\mathbf{k},\alpha} \hat{c}_{n\mathbf{k}}^{\dagger} \hat{c}_{n\mathbf{k}}$ is the current operator. The optical conductivity can be obtained from the analytic continuation of Eq.~\eqref{Eq:current_current_iw} to real frequencies \cite{Goulko_Svistunov_2017},
and also in this case the mobility is extracted from the zero-frequency limit. This approach has been implemented to study the temperature-dependent mobilities of polarons within the Holstein \cite{Mishchenko_Cataudella_2015} and Fr\"ohlich \cite{Mishchenko_Nagaosa_2019} models (cf.\ Sec.~\ref{Sec:History}). 
\textcite{Luo_Bernardi_2025} \rev{developed transport calculations using the \textit{ab initio} DMC method}, and investigated the electron mobility in anatase and rutile TiO$_2$, \rev{reproducing the three-regime temperature dependence expected from models} \cite{Mishchenko_Cataudella_2015} \rev{for small polaron transport in rutile TiO$_2$.}

\subsection{Small polarons: hopping-like activated transport}\label{sec:smallpolhop}

One of the key signatures of small polarons is their motion via thermally-activated hopping events between adjacent lattice sites. This mechanism is distinct from the band-like diffusive transport discussed in Sec.~\ref{Sec:bte}, and is often described within semiclassical theories inspired by chemical reaction kinetics. Activated transport is observed, for example, in TMOs \cite{Bosman_vanDaal_1970, Tuller_Nowick_1977} and in organic crystals \cite{Hulea_Morpurgo_2006}, and is associated with an increase of the carrier mobility with temperature.

The conceptual basis for small polaron transport goes back to the seminal work by \textcite{Holstein_1959b}, and the subsequently developments by \textcite{Emin_Holstein_1969}, \textcite{Holstein_1959b}, and \textcite{Austin_Mott_2001}, drawing from similar concepts used in the transition-state theory of electron transfer \cite{Marcus_1993}. The central quantity in this approach is the electron transfer rate $k_{\mathrm{ET}}$, which takes the Arrhenius-like form:
\begin{equation} \label{Eq:transfer_rate}
        k_{\mathrm{ET}} = \kappa \, \nu \exp(-E_{\rm a}/k_{\mathrm{B}}T) ~.
\end{equation}
In this expression, $\kappa$, $\nu$, and $E_{a}$ are called the transmission factor, the attempt frequency, and the activation barrier for hopping (cf.\ Supplemental Note 3). The dimensionless transmission factor $\kappa$ quantifies the level anti-crossing between the lowest-energy Born-Oppenheimer surfaces where the polaron reside; $\kappa \to 1$ indicates large separation and adiabatic transition, while $\kappa\to 0$ indicates small separation and non-adiabatic transition. The adiabatic vs.\ non-adiabatic nature of polaron transfer has been investigated, for example, by \citet{Park_Siegel_2018} in metal sulfides, and by \citet{Wu_Ping_2018, Morita_Walsh_2023, Wang_Bevan_2024, Lin_Ceder_2025} in TMOs; \rev{it has also been discussed in the context of real-time TDDFT simulations of polaron hoppings~\cite{Wang_Meng_2023}.} 
In the simplest approximation, the attempt frequency $\nu$ is taken to be the characteristic frequency of the normal modes that drive polaron formation \cite{Deskins_Dupuis_2007}; more sophisticated approaches involve calculations of vibrational frequencies at the initial configuration and the transition state configuration \cite{Adelstein_DeJonghe_2014, Wu_Ping_2018,Carey_McKenna_2021, Falletta_Pasquarello_2023, Palermo_Pasquarello_2024}, and are discussed in Supplemental Note~3. The activation energy $E_{\rm a}$ is the adiabatic energy barrier for a polaron to hop between adjacent sites, as shown in Fig.~\ref{fig:hopping_barrier}. 

Starting from the transfer rate in Eq.~\eqref{Eq:transfer_rate}, a simple expression for the diffusion coefficient is obtained by considering only hops to nearest neighbors:
  \begin{equation} \label{Eq:diff_approx}
  D = R^2\, n\, k_{\rm ET}~,
  \end{equation}
where $R$ is the distance to the nearest-neighbor site, and $n$ is the number of equivalent sites \cite{Deskins_Dupuis_2007}.
The corresponding mobility then follows from the Einstein-Smoluchowski relation,
  \begin{equation}\label{eq:transp-einstein}
  \mu = eD/k_{\rm B}T~.
  \end{equation}
This approximate model is very popular in calculations of small polaron mobilities in materials, including for example rutile \ch{TiO2} \cite{Deskins_Dupuis_2007,Morita_Walsh_2023,Dai_Giustino_2024c}, \ch{BiVO4} \cite{Kweon_Kim_2015,Liu_Dupuis_2020}, \ch{Fe2O3} \cite{Adelstein_DeJonghe_2014}, \ch{Ga2O3} \cite{Falletta_Pasquarello_2023}, and \ch{Cs2AgBiBr6} \cite{Lafuente_Giustino_2024}.

A more accurate description of activated transport can be achieved using the kinetic Monte Carlo method, which explicitly accounts for multiple hopping pathways~\cite{Wu_Ping_2018, Houmsi_Li_2025b}. For example, \textcite{Wu_Ping_2018} investigated the transport of electron polarons in \ch{BiVO4} by constructing a lattice model where the hopping events between adjacent sites are sampled stochastically, using the rates computed from Eq.~\eqref{Eq:transfer_rate}. From these simulations, they obtained the diffusion coefficient by monitoring the mean-squared displacement of the polaron:
  \begin{equation} \label{Eq:diff_kMC}
  D = \lim_{t\to\infty} \langle |\br(t)-\br(0)|^2 \rangle/2N_{\rm d}t~,
  \end{equation}
where $\br(t)$ is the polaron position at the time $t$, and $N_{\rm d}=3$ is the dimensionality of the lattice; the average $\< \cdots \>$ is performed over many trajectories. 
This approach revealed that polaron transport in \ch{BiVO4} is in an intermediate regime between the adiabatic and nonadiabatic limits; furthermore, it was found that, although the energy barriers for first- and second-nearest-neighbor hops are similar, only the former contributes significantly to polaron mobility, owing to the significantly lower transfer rate of the latter processes.

A similar approach was applied to hole polarons in \ch{TiO2} by \textcite{Palermo_Pasquarello_2024}; these authors found that, while hole transport in anatase is controlled by the pathway with lowest barrier, in the case of rutile it involves a delocalized intermediate state. This finding is consistent with the formation of large polarons reported by \cite{Dai_Giustino_2024c}, as well as with the very low formation energy of small hole polarons identified by \textcite{Mcbride_Hautier_2025} using Koopmans-compliant functionals, cf.\ Sec.~\ref{Sec:TMOs}. 

A systematic comparison of the mobilities obtained from Eq.~\eqref{Eq:diff_kMC} and the nearest-neighbor approximation in Eq.~\eqref{Eq:diff_approx} is still lacking. Similarly, a systematic benchmarking of procedures for calculating the effective attempt frequency is yet to be performed\rev{, and the effects of polaron-polaron interactions and finite polaron densities remain to be addressed}. More critically, the activation barrier $E_{\rm a}$ is very sensitive to the exchange-correlation functional \cite{Adelstein_DeJonghe_2014,Castleton_Kullgren_2019,Falletta_Pasquarello_2023,Tygesen_Garcia-Lastra_2023} (cf.\ Sec.~\ref{Sec:DFT_difficulty}). Since $k_{\rm ET}$ depends exponentially on this energy, small errors in the barrier can translate into large changes in the predicted mobility via Eq.~\eqref{eq:transp-einstein}. Addressing these issues will be important to achieve a quantitative \textit{ab initio} theory of polaron transport in the hopping regime. By way of comparison, in the related area of \textit{ai}BTE calculations for band-like transport, verification and validation efforts have matured to the point that independent implementations yield mutually consistent results and good agreement with experiment \cite{Ponce_Giustino_2020}. Comparable community standards will be needed to achieve similar predictive accuracy for small polaron transport.

At a more fundamental level, the boundary between band-like and hopping-like transport remains ill-defined. While the height of the activation barrier $E_{\rm a}$ provides a very natural descriptor, in most cases the choice between methods for band-like or hopping-like transport is largely empirical. A fully \textit{ab initio} criterion for selecting the appropriate framework, and even better a unified framework that could capture both limits at once, would be highly valuable developments.
\rev{Prior model-based studies on organic crystals appear promising in this regard \cite{Ortmann_Hannewald_2009}.}

As a word of caution, we also emphasize that the Arrhenius-like dependence of the mobility on temperature is not always an indication of small polaron hopping. In fact, the thermal ionization of impurities, or even simply the Fermi-Dirac function, also carry an exponential dependence on temperature \cite{Kang_Snyder_2018}. Developing reliable strategies to disentangle these effects will be important to achieve a predictive theory of polaron transport in materials.

\section{Translational invariance, broken symmetry, and decoherence}\label{Sec:BrokenSymmetry}

In this section, we address the interplay between electron self-localization and the translational invariance of polarons. This issue has generated some ambiguity in the literature, mostly  due to the apparent contradictions between variational results and the exact properties of many-body polaron eigenstates \cite{Gerlach_Lowen_1991}. A satisfactory answer to these questions requires a detour in the physics of decoherence and open quantum systems \cite{Leggett_Zwerger_1987}.

The question goes as follows: DFT-based calculations (Sec.~\ref{Sec:DFTPolaron}) provide a reliable description of the formation energy, electron wavefunction, and the structure of the polaron cloud; yet, the resulting electron wavefunctions and lattice distortions are not eigenstates of the full electron-phonon Hamiltonian. In fact, the exact many-body eigenstates of Eq.~\eqref{Eq:epi-hamilt} should be Bloch-like states since the Hamiltonian commutes with the lattice translation operator. The same difficulty arises when considering the Landau-Pekar model, cf.\ Sec.~\ref{Sec:LandauPekar}, the strong-coupling limit of the Holstein model, see Eq.~\eqref{Eq:Holstein_classical}, and even the field-theoretic approach of Sec.~\ref{Sec:FormalTheory}. What all these approaches have in common is the adiabatic decoupling between electronic and vibrational degrees of freedom, which removes electron-phonon entanglement from the picture.\footnote{In the Hedin-Baym approach, this decoupling takes place when one introduces the approximation by \textcite{Gillis_1970}, see Eq.~\eqref{Eq:tau_eqmotion_main}.} 

It would be tempting to conclude that electron self-localization is merely an artifact of the approximate treatment of the polaron Hamiltonian. However, experimental evidence suggests otherwise: for example, classic ESR experiments on color centers \cite{Castner_Kanzig_1957}
STM/STS studies \cite{Liu_Wu_2023,Yim_Thornton_2016}, and ultrafast XRD measurements \cite{Guzelturk_Lindenberg_2021} all point to the existence of localized, broken-symmetry states. 
The resolution of this apparent contradiction requires one to examine how classical-like broken-symmetry states emerge from symmetric quantum Hamiltonians. To this end, we consider two textbook examples, the hydrogen atom and the two-sites Holstein model.

The hydrogen atom provides the simplest proxy model to discuss localization in interacting electron-phonon systems \cite{Allcock_1956,Lafuente_Giustino_2022b}. In this analogy, the proton mimics the localized ionic distortion that accompanies the formation of a polaron. The eigenstates of the hydrogen Hamiltonian are:
  \begin{equation}\label{Eq:H.1}
    \Psi(\br_{\rm e},\br_{\rm p}) = \frac{1}{\sqrt{V}} \exp\left[i\bk\cdot \frac{m_{\rm e} \br_{\rm e} + m_{\rm p} \br_{\rm p}}{m_{\rm e}+m_{\rm p}}\right]\psi_{nlm}(\br_{\rm e} - \br_{\rm p}),
  \end{equation}
where $\br_{\rm e}$ and $\br_{\rm p}$ are the positions of the electron and proton, respectively, and $\psi_{nlm}$ is the textbook hydrogenic wavefunction. These states are also eigenstates of the translation operator that shifts both coordinates at once, $\hat T_\bR \Psi_\bk(\br_{\rm e},\br_{\rm p}) = \Psi_\bk(\br_{\rm e}-\bR,\br_{\rm p}-\bR)$. 
The square modulus of the wavefunction in Eq.~\eqref{Eq:H.1} is invariant under $\hat T_\bR$, therefore the atom is delocalized over the volume $V$ of the system. For example, the marginal probability density for the electron is uniform:
  \begin{equation}
    n(\br_{\rm e}) = \int d\br_{\rm p} |\Psi_\bk(\br_{\rm e},\br_{\rm p})|^2 = \frac{1}{V}.
  \end{equation}
At the same time, the conditional probability density for a given proton position is a localized function:
  \begin{equation}
    {\rm Prob}\,(\br_{\rm e}| \br_{\rm p} = \br_0) = \frac{1}{V} |\psi_{nlm}(\br_{\rm e} - \br_0)|^2.
  \end{equation}
In practice, the hydrogen atom is not typically observed in a delocalized state such as Eq.~\eqref{Eq:H.1}; instead, it is found as a localized particle. This apparent contradiction is resolved by considering the interaction of the atom with the environment: even weak coupling to the environment suppresses quantum coherence and selects a particular position of the proton. Only when quantum coherence is preserved can one observe delocalized wavelike behavior, as in neutral atom interferometers \cite{Cronin_Pritchard_2009}.

The above qualitative reasoning can be placed on formal ground by considering the simplest model that embodies the essential physics of electron self-trapping, i.e., the two-site Holstein model. This model is a simplified version of the Holstein Hamiltonian in Eq.~\eqref{Eq:Holstein2}, where one considers only two lattice sites:
  \begin{eqnarray}\label{Eq:holstein-dimer}
  \hspace{-4pt}\hH &=& -t (\hcd_1 \hc_2 \!+\! \hcd_2 \hc_1) + \hbar\w (\had_1\ha_1 \!+\!1/2)+ \hbar\w(\had_2\ha_2\!+\!1/2)
    \nonumber \\
      \hspace{-10pt}&+& g \hn_1 (\had_1 + \ha_1) + g\hn_2 (\had_2 + \ha_2) 
      + \epsilon (\hn_2 \!-\! \hn_1).\hspace{10pt} 
  \end{eqnarray}
The last term on the right is included to break the symmetry. In the absence of such a term, the Hamiltonian is symmetric with respect to the exchange of sites 1 and 2. This symmetry is the counterpart of the lattice translation symmetry in the extended Holstein Hamiltonian of Eq.~\eqref{Eq:Holstein2}. For $\epsilon=0$, the many-body eigenstates of this model must be even or odd with respect to a site swap. 

\begin{figure}
\centering
\includegraphics[width=0.8\linewidth]{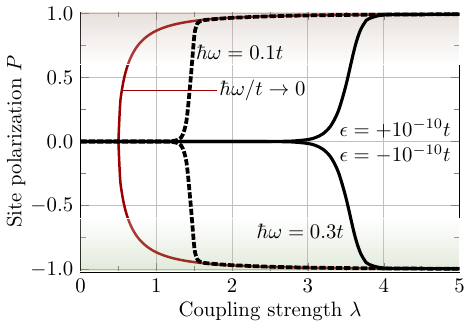}
  \caption{Site polarization $P=\<\hn_1\!-\!\hn_2\>/\<\hn_1\!+\!\hn_2\>$ in the two-site Holstein model, Eq.~\eqref{Eq:holstein-dimer-rel}. The black lines on the top and bottom half-planes have been calculated using $\epsilon = +10^{-10}t$ and $\epsilon = -10^{-10}t$, respectively, for two adiabaticity parameters, $\hbar\w/t = 0.1$ and 0.3. The black lines were obtained from exact diagonalization on the basis $|s,n\>$, where $s=1,2$ is the electron site index, and $n=0,1,\cdots,N_{\rm max}$ is the phonon number ($N_{\rm max}=1000$ in this plot).
  The red line is the site polarization in the adiabatic limit, and is given by 
  $P = [(2\lambda -\sqrt{4\lambda^2-1})^2-1]/[(2\lambda -\sqrt{4\lambda^2-1})^2+1\rev{]}$ for $\lambda \ge 1/2$, and $P=0$ for $\lambda<1/2$ \cite{Ceulemans_2022}. 
  }
  \label{Fig:dimer}
\end{figure}

Can the electron localize at either site? It can be shown that, for $\epsilon=0$, the electron must be delocalized over both sites; however, for any $\epsilon \ne 0$, the electron localizes at one site if the coupling strength exceeds a critical threshold. If the electron is initially localized at one site, and $\epsilon=0$, this state may be long-lived and effectively behave as a quasi-stationary state. To explore these effects, it is expedient to rewrite the Hamiltonian by moving to the phonon center-of-mass frame. This is achieved by introducing the sum and difference ladder operators $\hat{A} = (\ha_1+\ha_2)/\sqrt{2}$ and $\ha = (\ha_2-\ha_1)/\sqrt{2}$. With these definitions, Eq.~\eqref{Eq:holstein-dimer} decouples into $\hH = \hH_{\rm cm} + \hH_{\rm rel}$, where $\hH_{\rm cm}$ is a displaced harmonic oscillator that does not couple to the electron, and $\hH_{\rm rel}$ is given by:
  \begin{equation}
  \hH_{\rm rel} =  -t (\hcd_1 \hc_2 + \hcd_2 \hc_1) + \hbar\w \big(\had\ha + \frac{1}{2}\big)
    + \frac{g}{\sqrt{2}}(\had + \ha) (\hn_2- \hn_1) \vspace{-15pt}\nonumber
    \end{equation}
\begin{equation}\label{Eq:holstein-dimer-rel}
    \hspace{-4.4cm}+ \epsilon (\hn_2- \hn_1)~.
  \end{equation}
More details on this transformation are provided in Supplementary Note 4. Equation~\eqref{Eq:holstein-dimer-rel} describes one electron that can be at site 1 or 2, coupled to a single bosonic mode. Numerical exact diagonalization is a standard textbook exercise, and the phase diagram is shown in Fig.~\ref{Fig:dimer}. The figure reports the electron site polarization as a function of the coupling strength $\lambda$, $P = \< \hn_1 - \hn_2\>/\< \hn_1 + \hn_2\>$. This ``order parameter'' takes the values $\pm 1$ when the electron is fully localized at site 1 or 2, respectively. When there is no bias and the Hamiltonian is symmetric ($\epsilon = 0$), we have $P=0$ and the electron spends equal time on either site. However, upon introducing a vanishingly small bias, beyond a critical coupling $\lambda$ the electron preferentially occupies one site, which is the hallmark of polaron self-trapping in this toy model. The choice of site depends on the sign of the bias, as shown in Fig.~\ref{Fig:dimer}. The same figure also shows how, in the fully adiabatic limit ($\hbar\w \ll t$), this transition is sharp (red line). In this case, the critical parameter $\lambda = 1/2$ marks a bifurcation in the solutions and corresponds to the emergence of a double-well potential energy surface, similar to that found in the Landau theory of phase transition \cite{Landau_Lifshitz_1980}.

The localized solutions with the electron occupying preferentially either site are not eigenstates of the site-swap operator, and therefore they are not Hamiltonian eigenstates. However, the energy splitting $\Delta E \!=\! E_+\!-\!E_-$ between the symmetric and antisymmetric ground states of the system, $|\pm\> =  (|1\>\pm|2\>)/\sqrt{2}$, becomes vanishingly small at large $\lambda$, $\Delta E = 2t\, e^{-2\lambda t/\hbar\w}$ (cf.\ Supplemental Note 4). In this case, the localized states $|1\>$ and $|2\>$ can form by admixing $|+\>$ and $|-\>$ via arbitrarily small perturbations. Once these localized states are formed, they constitute quasi-stationary states. In fact, while not being eigenstates, the time required to tunnel between $|1\>$ and $|2\>$ grows exponentially with the coupling strenght, $2\pi \hbar /\Delta E \propto e^{2\lambda t/\hbar\w}$. At strong coupling, this time may be longer than the characteristic timescale of the given experiment.

The next question to be addressed is why the states $|1\>$ and $|2\>$ should even form in the first place. This is a fundamental question about symmetry breaking in nature, and one that is best answered in the words of \citeauthor{Weinberg_2013} (\citeyear{Weinberg_2013}, p.~206):

\begin{adjustwidth}{13pt}{10pt}
\textit{``Why should these broken-symmetry states be the ones realized in nature, rather than the true energy eigenstates, which are either even or odd under the symmetry? The answer has to do with the phenomenon of decoherence. The wave function will inevitably be subject to external perturbations, which for a thick barrier produce fluctuations in the phase of the wave function, with no correlation between the phase changes on the two sides of the barrier. These fluctuations cannot change a broken-symmetry wave function that is concentrated on one side of the barrier into a solution that is wholly or partly concentrated on the other side, but they rapidly change an even or odd wave function into one that is an incoherent mixture of even and odd wave functions. The states realized in the real world are the ones that are stable up to a phase under these fluctuations, and these are the broken-symmetry states.''}
\end{adjustwidth}

\noindent
A quantitative discussion of these issues would require an explicit analysis of the coupling between the system and its environment, including how environmental degrees of freedom effectively monitor the system and induce decoherence in the reduced density matrix. While such a discussion lies beyond the scope of this review, we refer the reader to the classic articles on these topics by \textcite{Zurek_2003} and \textcite{Leggett_Zwerger_1987}.

More practically, Fig.~\ref{Fig:dimer} shows that both quantum (black lines) and semiclassical (red lines) approaches lead to self-trapping, but the critical coupling strengths at which this occurs differ between these approaches. If we take the semiclassical result as the one that more closely reflects DFT-based calculations, we can speculate that DFT may to some extent overestimate the tendency to self-trapping. Future work will be needed to test this conjecture and its implications via fully-fledged first-principles calculations.

\section{Conclusions and Outlook}\label{Sec:Conclusions}

In this article, we have outlined the current state of \textit{ab initio} theories of polarons, tracing their development from seminal effective Hamiltonians to modern atomistic calculations for real materials. Stepping back, one takeaway from this review is that a wealth of theoretical frameworks now exist, each addressing the polaron problem from a distinct vantage point. This diversity reflects the intrinsic complexity of polaron physics, and suggests that there may not be a universal, one-size-fits-all approach.

The theoretical frameworks that we have reviewed encompasses effective Hamiltonian approaches, ranging from canonical transformations to diagrammatic Monte Carlo, and DFT calculations, DFT-based Fock-space Hamiltonians, and Green's function methods. It is both reassuring and satisfying that most of these approaches are tightly connected, and in several cases they can be derived from one another by improving or simplifying the description of select ingredients. In this spirit, we hope that Fig.~\ref{Fig:jacob_ladder} will serve as a useful map for readers navigating the maze of conceptual and computational methods for polarons. Likewise, we hope that the formulation of the polaron problem in terms of the Hedin-Baym equations in Sec.~\ref{Sec:FormalTheory} may provide a unifying perspective, linking fully \textit{ab initio} methods and effective Hamiltonians  from the variational polaron equations to diagrammatic Monte Carlo. These connections offer valuable sanity checks to ensure continued progress along distinct yet complementary directions.

Beyond summarizing the many achievements of this field, it is also important to highlight aspects that remain challenging or underexplored, and call for future work. Here, we limit ourselves to drawing the reader's attention to six open questions that in our view are the most urgent.

First, the first-principles theory of exciton polarons is still in its infancy. Recent developments in \textit{ab initio} methods have succeeded in addressing the weak- and strong-coupling limits of this problem, yet a unifying conceptual framework that seamlessly bridges these regimes is still lacking. One possible avenue may lie in the generalization of the Hedin-Baym equations to the two-particle electron-hole Green's function. Alternatively, one could establish connections between the exciton polaron equations of Sec.~\ref{Sec:ExcitonPolaron} and Fock-space exciton-phonon Hamiltonians, which could then be tackled via the canonical transformations of Sec.~\ref{sec:canonical} or even the diagrammatic Monte Carlo approaches of Sec.~\ref{Sec:DMC}, perhaps leveraging techniques originally developed for charged polarons.

Second, the role of temperature in polaron physics is only starting to be explored from first principles. Finite-temperature effects are essential to describe polaron formation and dynamics under realistic experimental conditions, yet most \textit{ab initio} calculations are still restricted to zero temperature. Promising developments include molecular dynamics simulations with machine-learned force fields, which enable access to long timescales, as well as non-adiabatic molecular dynamics simulations that can describe temperature-dependent transport processes. On the front of many-body methods, diagrammatic Monte Carlo methods naturally incorporate temperature through thermodynamic Green's functions, and similar strategies could be pursued within the Hedin-Baym framework. Some of these advances are reviewed in Supplemental Note~5. It should be mentioned, however, that state-of-the-art many-body approaches have been formulated within the approximation of linear electron-phonon couplings and harmonic lattice dynamics; generalizations to nonlinear couplings and anharmonic lattices will be required to directly compare with DFT-based molecular dynamics simulations.

Third, polaron-polaron interactions at finite carrier densities remain largely uncharted territory. While the behavior of a single polaron is increasingly well understood, collective effects such as polaron ordering and phase diagrams are far more difficult to study from first principles. The challenge here lies in the need for computational methods that capture both electron-phonon couplings and electron-electron correlations. Addressing this regime will be critical for connecting polaron physics with transport and optical phenomena in doped semiconductors and oxides, where finite densities are unavoidable (Supplemental Note~2).

Fourth, the formation of bound polaron pairs, i.e., bipolarons, raises a number fundamental questions and calls for systematic \textit{ab initio} studies. These quasiparticles play a central role in the long-standing proposal of bipolaronic superconductivity, yet their stability, binding energy, and effective mass have not been investigated from first principles. \textit{Ab initio} calculations of bipolarons in realistic materials would be very valuable to assess whether bipolaron condensates can compete with, coexist with, or enhance conventional phonon-mediated pairing. A more quantitative understanding of bipolarons may also shed light on their possible role in the rich phase diagrams of correlated oxides (Supplemental Note~6).

Fifth, the dynamics of polarons out-of-equilibrium is only starting to be investigated from first principles. Ultrafast diffraction and pump-probe experiments are making significant progress in the study of these phenomena, and new \textit{ab initio} frameworks to support these experiments are urgently needed. Non-adiabatic molecular dynamics, Ehrenfest dynamics, and real-time Green's function techniques are all promising directions, but applications are still scarce. These approaches will be necessary to capture how polarons form and evolve under photoexcitation or strong driving fields. We discuss some of the recent work in this area in Supplemental Note~7.

Sixth, the field of \textit{ab initio} polaron calculations is still relatively young compared to the broader landscape of electronic structure theory. In DFT, decades of development have established community standards, best practices, and benchmark datasets that have enabled rapid progress and universal reproducibility. A comparable effort will be essential for polarons. 
\rev{The sensitivity of the polaron calculations in \ch{TiO2} to the computational method employed (Sec.~IX.B) clearly demonstrates the urgency for this effort, as different polaron self-interacton-correction schemes or even different parameter choices within the same scheme could result in qualitatively different predictions. This sensitivity makes the interpretations of these predictions nontrivial.}
Establishing verification and validation protocols, curating reference datasets, and adopting conventions for reporting calculation metadata will be critical to ensure that results can be reproduced and compared across different codes and approaches. 

Looking ahead, the continued integration of advanced theory, computation, and experiment, building on foundational effective-Hamiltonian methods, modern \textit{ab initio} techniques and codes, and emerging AI/ML frameworks, promises not only to resolve long-standing questions about polarons but also to enable the discovery of new emergent phenomena in quantum matter.

\acknowledgments
We are very grateful to 
Timothy Berkelbach, 
Marco Bernardi,
Kirk Bevan,
Filippo De Angelis,
\rev{Adam Denchfield},
Michel Dupuis,
Claudia Draxl,
Cesare Franchini, 
Ying-Shuang Fu,
Giulia Galli,
Marco Grioni,
Sohrab Ismail-Beigi,
Sebastian Kokott,
\rev{Pavel Kornilovich},
Juan Maria Garc\'{i}a Lastra,
\rev{Jing Li},
\rev{Frank Marsiglio},
\rev{Andrey Mishchenko,}
Sheng Meng, 
Keith McKenna,
Frank Ortman,
Alfredo Pasquarello, 
Yuan Ping, 
Samuel Ponc\'e, 
David Reichman, 
Martin Setvin, 
\rev{Giancarlo Strinati},
Boris Svistunov,
Carla Verdi, 
Matthieu Verstraete,
\rev{Grigory Volovik},
Kehui Wu,
Wennie Wang, 
Julia Wiktor, 
and
Aron Walsh,
for providing thoughtful feedback to the original draft of this article.
We are also indebted to 
Phil Allen, 
Edoardo Baldini, 
Matteo Calandra, 
Fabio Caruso, 
Asier Eiguren, 
Peio Garcia-Goiricelaya,
Idoia G. Gurtubay, 
Donghwan Kim, 
Emmanouil Kioupakis, 
Zhenglu Li, 
Chao Lian,
Steven Louie, 
Kaifa Luo, 
Allan MacDonald, 
Roxana Margine, 
Andrea Marini, 
Nicola Marzari, 
Francesco Mauri, 
Enrico Perfetto,
Chih-Kang Shih,
Weng Hong Sio,
Young-Woo Son, 
Gianluca Stefanucci, 
Joe Subotnik, 
Chris van de Walle, 
and Marios Zacharias, 
for many stimulating discussions over the years, which directly or indirectly contributed to shaping this review article. 
We also thank 
Boris Svistunov,
Marco Bernardi,
Michel Dupuis, 
Cesare Franchini,
Ying-Shuang Fu,
Giulia Galli,
Marco Grioni,
Keith McKenna,
and
Aron Walsh,
for kindly allowing us to reproduce their figures. This work was supported by the Computational Materials Science program of U.S. Department of Energy, Office of Science, Basic Energy Sciences under Award DE-SC0020129; and the National Science Foundation, Office of Advanced Cyberinfrastructure under Grant No. 2513830.
J.L.-B. was supported by Grant No. IT-1527-22, funded by the Department of Education, Universities and Research of the Basque Government, and Grant no. PID2022-137685NB-I00, funded by MCIN/AEI/10.13039/501100011033/ and by “ERDF A way of making Europe”.


\bibliographystyle{apsrmp4-2_mod}
\bibliography{References}

\foreach \x in {1,...,8}
{
\clearpage
\includepdf[pages=\x]{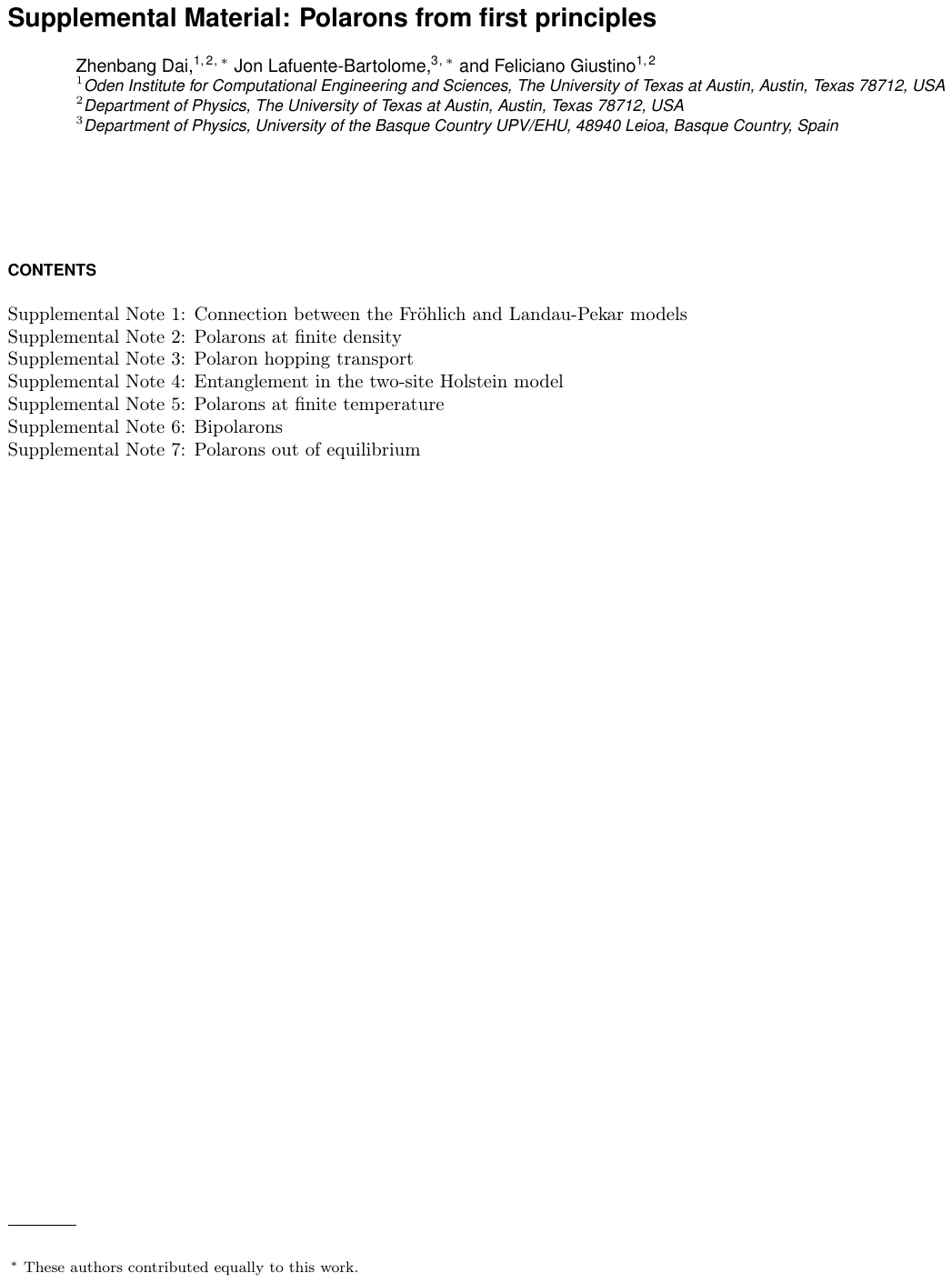}
}



\end{document}